# A Wideband Tunable, Nonreciprocal Bandpass Filter Using Magnetostatic Surface Waves with Zero Static Power Consumption

Xingyu Du[1], Yixiao Ding[1], Shun Yao[1], Yijie Ding[1], Dengyang Lu[1], Shuxian Wu[1], Chin-Yu Chang[1], Xuan Wang[1], Mark Allen[1], and Roy H. Olsson III[1*]

[1]Department of Electrical and Systems Engineering, University of Pennsylvania, Philadelphia, PA, USA

[*]Correspondence: Roy H. Olsson III (rolsson@seas.upenn.edu)

# Abstract

Modern wireless systems demand compact, power-efficient RF front-end components that support wideband tunability and nonreciprocity. We present a new class of miniature bandpass filter that achieves both continuously tunable frequency operation (4.0–17.7 GHz) and high nonreciprocity (>25 dB), all within a compact size of 1.07 $cm^3$. The filter employs a microfabricated 18 μm thick Yttrium Iron Garnet (YIG) waveguide with meander-line aluminum transducers, enabling low-loss unidirectional propagation via magnetostatic surface waves. Leveraging a benzocyclobutene (BCB) planarization fabrication process, this study enables a dispersion profile unique to thick YIG films, resulting in enhanced filter skirt performance with minimal spurious modes. Frequency tuning is enabled by a zero-static-power magnetic bias circuit using transient current pulses, eliminating continuous power consumption. The filter demonstrates low insertion loss (3–5 dB), high out-of-band rejection (>30 dB), narrow bandwidth (100–200 MHz), robust power handling (>10.4 dBm), and high linearity (IIP3 > 26 dBm).

Modern wireless communication systems face increasing demands, creating significant challenges for radio frequency (RF) front-end design, especially in achieving wideband tunable filtering [1, 2] and nonreciprocity [3, 4]. Figure 1(a) shows that integrating both capabilities permits simultaneous control of the signal's desired passband and directional isolation. Consequently, RF front-end module design can be substantially simplified and optimized.

Wideband tunable filtering enables dynamic adjustment of operating frequencies across a broad spectrum. This capability is essential for modern RF systems supporting diverse applications, from sub-6 GHz 5G New Radio for urban coverage [5], to satellite communication downlinks at 10.7-12.7 GHz [6]. As frequency allocations grow, managing interference becomes increasingly critical. For example, spectral overlap between sub-6 GHz 5G and C-band Very Small Aperture Terminals (VSATs) systems used in maritime and fixed satellite services poses a significant risk of unpredictable interference [7]. To address this, a combination of tuned and switched filter banks are often employed; however, this approach can incur additional loss from the additional signal paths and often hinders the creation of a multi-band filter response [8, 9]. Continuously tunable filters provide greater flexibility and are inherently upgradable to new bands and evolving interference environments. Some designs achieve tuning from 3.4 GHz to 11.1 GHz [10-12] —but they still fall short of covering the full bandwidths supported by today's wideband receivers [13-15].

An equally critical challenge lies is the implementation of nonreciprocal components, which are essential for protecting high-power RF sources from unwanted reflections and for isolating different sections of an RF transceiver [3, 4]. As shown in Fig. 1(a), a nonreciprocal bandpass filter in the transmitter path selectively allows only in-band frequency components to reach the power amplifier, while blocking both out-of-band signals in the forward direction and reflected in-band signals in the reverse direction from the amplifier or antenna. In the receiver path, the filter functions dually as a bandpass filter and an isolator, attenuating any reflected signals that could otherwise propagate back toward the antenna. Moreover, nonreciprocal components enable other applications, including full-duplex systems where the same frequency is being used for transmit and receive [4, 16], quantum information processing [4, 17], radars [18, 19], and biomedical sensing [4]. Traditional ferrite-based isolators and circulators are bulky, while alternative approaches using transistors, nonlinearity, or temporal modulation involve performance compromising trade-offs between power consumption, insertion loss, noise, and power handling [4, 19, 20].

In this study, we introduce a new class of RF components that simultaneously address the challenges of both wideband, continuously frequency tunable filtering and nonreciprocity. Unlike conventional solutions that require multiple discrete elements—such as RF switches, filters, and isolators—our approach combines these functions into a single device, significantly reducing overall system size and complexity. The filter operates with zero static power consumption and offers continuous tunability from 4.0 GHz to 17.7 GHz, with an insertion loss of 3–5 dB, out-of-band rejection exceeding 30 dB, and isolation ($S_{12}$–$S_{21}$) greater than 25 dB. An overall schematic is shown in Fig. 1(b), and an optical microscopy image of the fabricated device is shown in Fig. 1(c). The entire device occupies a compact footprint of just 20.0 mm × 16.7 mm × 3.2 mm and consists of two key components: a nonreciprocal Yttrium Iron Garnet (YIG) bandpass filter and a zero-static-power magnetic bias circuit.

The nonreciprocal YIG filter employs a thick YIG waveguide with meander-line aluminum transducers. It operates using magnetostatic surface waves (MSSW), where an in-plane magnetic bias is applied perpendicular to the direction of wave propagation in the YIG. The meander-line transducers efficiently excite and collect MSSW, while the thick YIG waveguide supports low-loss unidirectional propagation. We demonstrate a novel fabrication technique in Fig. 1(d) that enables scaling the YIG thickness from 3 μm, as reported in previous literature [10, 21, 22], to 18 μm. This substantial increase significantly improves power handling, sharpens the filter skirt, broadens the bandwidth, and enables unidirectional MSSW propagation. Additionally, the meander-line aluminum transducers further enhance nonreciprocity， out-of-band rejection, suppress spurious modes, and produce a flatter passband response.

The magnetic bias circuit leveraged our previous work [10, 23], but with improved materials and a redesigned structure. This updated design achieves a peak magnetic field of 5700 Gauss within a reduced volume of ~1.07 $cm^3$— a substantial improvement over the previously reported 3170 Gauss in 1.68 $cm^3$ [10]. The magnetic field, and therefore the filter center frequency, is tuned by applying current pulses to AlNiCo magnets, modifying their nonvolatile remanence. This method enables dynamic frequency tuning with only transient power consumption [23], eliminating the need for steady-state power typically required by the electromagnets in YIG filters to maintain the required magnetic field for filter operation [24, 25].

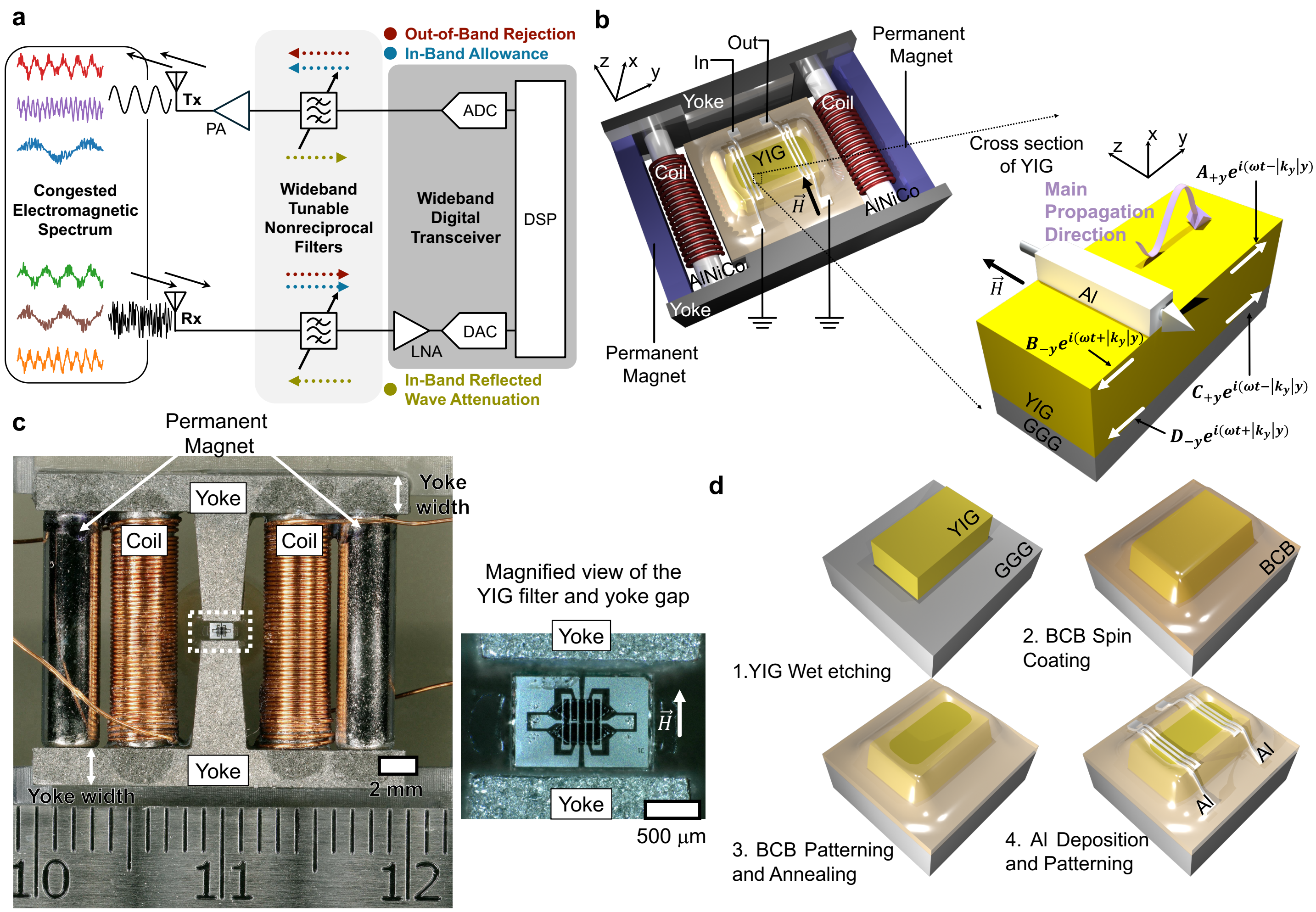

**Fig. 1 Overview of the wideband tunable nonreciprocal filter.** (a) Block diagram of a transceiver system incorporating wideband tunable, nonreciprocal filters. **(b)** 3D schematic of the filter, which integrates a YIG (yttrium iron garnet) filter with a zero-static-power magnetic bias circuit. An input aluminum meander-line transducer excites a unidirectional magnetostatic wave (MSW), which propagates through the YIG and is reconverted back to an electrical signal at the output transducer. The magnetic bias structure includes two permanent magnets, two shunt magnets wound with coils, and magnetically permeable yokes that focus the magnetic flux at the filter. **(c)** Optical microscope image of the fabricated device assembly with a magnified view of the white box region showing the YIG filter and yoke gap. **(d)** Fabrication process flow of the YIG filter, incorporating an innovative benzocyclobutene (BCB) planarization technique for thickness-scaled YIG filters.

# Thickness Scaled Yttrium Iron Garnet (YIG) Filter

Increasing YIG film thickness offers several advantages, including stronger MSSW coupling [26, 27], reduced theoretical propagation loss [27], improved power handling [28, 29], and enhanced unidirectionality [30]. Thicker films improve excitation and detection efficiency due to the larger volume of magnetic material interacting with the transducer antenna's magnetic field. In the setup shown in Fig. 1b, MSSWs propagate along the y-axis. At the top surface, waves are expressed as $A_{+y}e^{i(\omega t-|k_y|y)}$ for +y and $B_{-y}e^{i(\omega t+|k_y|y)}$ for -y directions. At the bottom surface, they are $C_{+y}e^{i(\omega t-|k_y|y)}$ and $D_{-y}e^{i(\omega t+|k_y|y)}$. Wong et al. showed that for a given spin wave wavelength, the amplitude ratio $A_{+y}/B_{-y}$ and $D_{-y}/C_{+y}$ are much higher in thick YIG films, indicating dominant MSSW propagation along $\vec{H} \times \vec{n}$ [30].

Fig. 2a shows the calculated dispersion relationship for 3 μm and 18 μm thick YIG films using the equations from [31], with demagnetization effect neglected. The calculation details can be found in Supplementary Note #1. For an infinitely wide YIG film with an infinitely far away ground plane, the MSSW frequency is bounded by $\omega_{min} = \gamma[H(H + 4\pi M_s)]^{1/2}$ and $\omega_{max} = \gamma(H + 2\pi M_s)$, where $\gamma =$ 2.8 MHz/Gauss, $H$ is the magnetic field applied in Gauss, and $M_s = 1780$ Gauss is the saturation magnetization of YIG. When the film has finite width, the wavevector in the width (z) direction becomes quantized, leading to discrete standing wave modes. Because the transverse quantization introduces a wavevector component parallel to the magnetization, the effective dipolar energy is reduced, causing these confined modes to appear at lower frequencies than the corresponding plane-wave MSSW mode. This finite-width confinement allows MSSWs to occur below $\omega_{min}$, entering the volume wave frequency range. As YIG thickness increases, these width effects become more pronounced, with higher-order modes showing larger frequency separation from lower-order modes and the wavenumber increases more sharply with frequency at small wavenumbers. Since the group velocity ($V_g$) is defined as $V_g = \frac{\partial \omega}{\partial k}$, the theoretical group delay can be calculated as $1/V_g$, as plotted in the Fig. 2b. Thicker YIG exhibits a steeper dispersion, resulting in higher group velocity and thus lower group delay. As prior studies indicate, propagation loss is nearly proportional to group delay [27, 32]. Notably, with a magnetic bias field of 2325 Gauss, the 18 μm YIG shows a flat and low group delay between 8.41 GHz and 8.95 GHz, followed by a sharp drop at ~8.96 GHz. In contrast, the 3 μm film shows a gradual increase in group delay from 8.65 GHz to 9.0 GHz. This sharp transition can be used to create steep filter skirts, achieving a "brick-wall" response at the upper band edge.

Despite the advantages of thicker YIG films, no prior studies have demonstrated microfabricated MSSW filters using etched YIG thicker than 3 μm, primarily due to challenges in microfabrication. Conventional MSSW filters are fabricated by patterning metal transducers, via lift-off or subtractive etching, on top of etched YIG waveguides [10, 21]. However, in 15-18 μm thick YIG, steep sidewalls hinder conformal metal deposition, leading to breaks (open) or incomplete removal (shorts) of the metal, as detailed in Supplementary Note #2. To overcome this, we implemented a planarization method using Benzocyclobutene (BCB), illustrated in Fig. 1d. After etching the YIG, a BCB layer was spin-coated and patterned via photolithography to remove BCB from the top surface of the YIG, minimizing the distance between the YIG and the aluminum transducers to ensure maximum energy coupling. This BCB process smooths the steep YIG sidewalls into gradual slopes. Supplementary Note #2 also confirms that the BCB layer does not introduce signal loss or alter MSSW propagation.

Fig. 2c and 2d show optical images of fabricated YIG filters with identical YIG waveguide shapes but YIG thicknesses of 18 μm and 3 μm, respectively. Both filters in (c) and (d) have a width (W) of 200 μm, length of 140 μm (L) and pitch (P) representing the distance between two aluminum transducers of 70 μm. Fig. 2e and 2f present the measured $S_{12}$ frequency response and group delay. The 18 μm-thick YIG filter exhibits significantly lower group delay and a sharp increase near 8.95 GHz. This results in a much steeper filter skirt at the upper passband edge compared to the 3 μm-thick YIG filter, indicating superior filtering performance. Supplementary Note #3 compares these two filters under various frequencies.

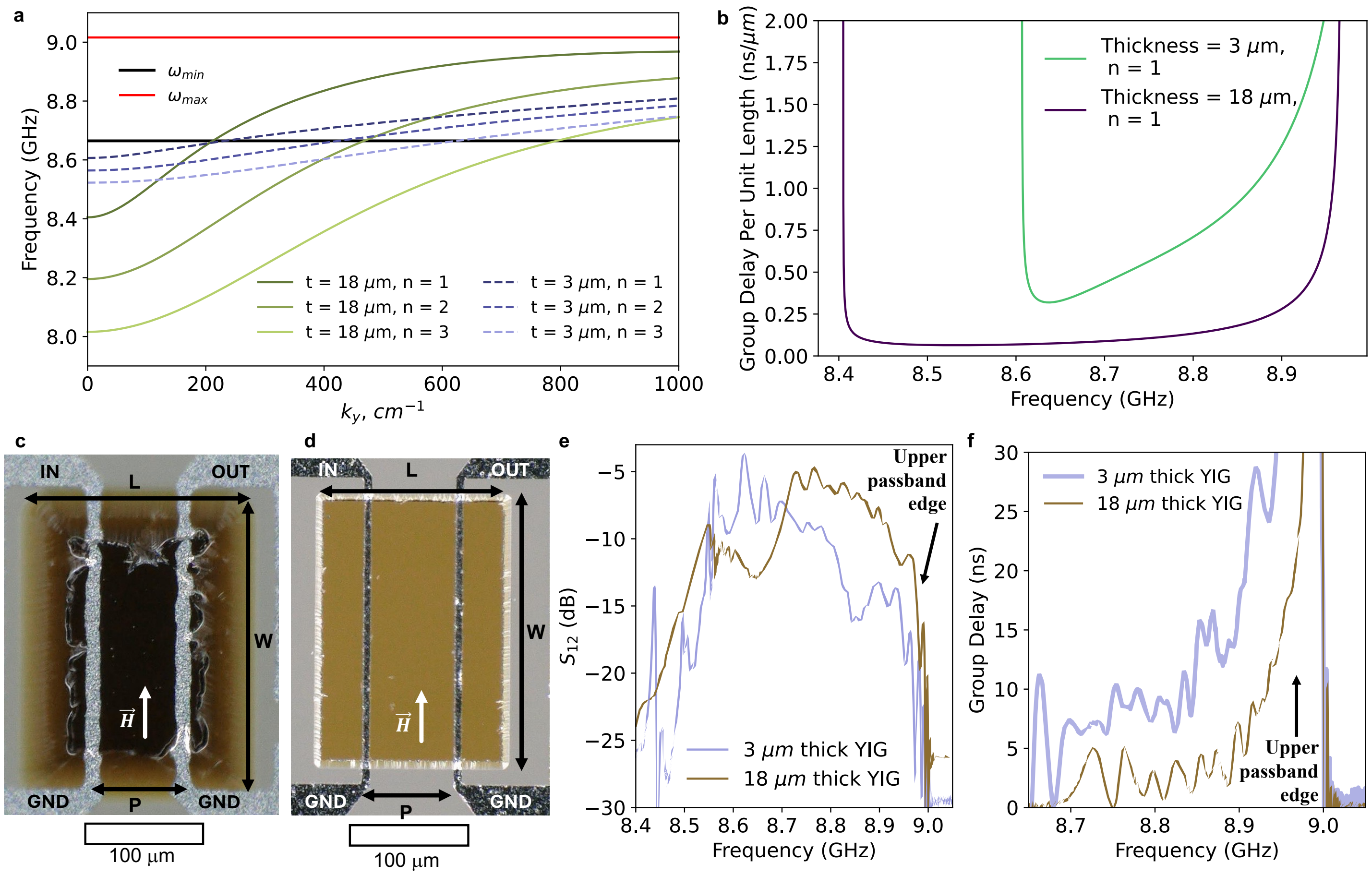

**Fig. 2 Thickness effect on YIG filters. (a)** Comparison of the dispersion relationships for 3 μm and 18 μm thick YIG film with a bias field of 2325 Gauss. **(b)** Calculated group delay per micrometer for different YIG film thicknesses with a bias field of 2325 Gauss. **(c)** Optical image of the fabricated YIG filter with an 18 μm thick film using BCB planarization. BCB is transparent and is not visible; aluminum transducers are 10 μm wide. **(d)** Optical image of the YIG filter with a 3 μm thick film. No BCB is used due to the thin YIG layer. Aluminum transducers are 5 μm wide. (e) Measured $S_{12}$ frequency response with applied magnetic flux density of around 2500 Gauss. (f) Measured $S_{12}$ group delay with applied magnetic flux density of around 2500 Gauss.

# YIG Filter with Meander Line Transducers

Although a thicker YIG offers improved filter skirt performance at the upper frequency band, further optimization is needed to achieve a flat filter passband. Meander-line transducers have been proposed in previous studies, showing improved radiation impedance compared to single straight-line transducers through theoretical calculations [33, 34], and demonstrating low insertion loss in filters utilizing 3 μm thick YIG [12]. However, a comprehensive study of YIG filters employing meander-line transducers is still needed, and their application in 18 μm thick YIG remains unexplored. A previous study determined the radiation resistance for a meander-line transducer as follows [33]:

$$T_m = \frac{R_m}{R_1 \times l} = \left[\frac{sin\,(\frac{W_T k_y}{2})}{\frac{W_T k_y}{2}}\right]^2 \left[\frac{sin\,(\frac{pNk_y}{2})}{cos\,(\frac{pk_y}{2})}\right]^2 \quad (1)$$

Here, $T_m$ is the transducer specific radiation impedance per unit length, $R_m$ is the radiation impedance, and $R_1$ is a function of YIG parameters and ground-plane spacing and is independent of transducer parameters, $l$ is the length of the meander-line transducer's arm, which also serves as the width of the YIG waveguide, $W_T$ is the width of the transducer, $p$ is the pitch, and $N$ is the number of arms in the meander-line.

Applying the previously calculated dispersion relation into equation (1), the $T_m$ for 3 μm and 18 μm thick YIG with meander-line transducers can be determined, as shown in Fig. 3a. In this case, $N = 4$, $W_T = 10\ \mu m$, and $p = 70\ \mu m$. Within the first width mode of 3 μm thick YIG, multiple peaks corresponding to the different order length mode harmonics are evident. These higher-order modes appear at higher frequencies and shorter wavelengths with reduced amplitude. However, the unique dispersion characteristics of the 18 μm thick YIG pushes the frequency of all the larger wavenumber modes close to $\omega_{max}$, beyond which MSSW propagation is forbidden. This significantly reduces the presence of higher-order length modes. An 18 μm thick YIG filter, detailed in Fig. 3b with a pitch of 70 μm, width of 200 μm, YIG waveguide total length of 420 μm, and meander-line with three arms and two bridges, demonstrates this effect. The arm is the section where the transducer is positioned on top of the YIG, and the bridge section connects the two arms. The measured $S_{12}$ frequency responses between 3 μm and 18 μm thick YIG waveguides is compared in Fig. 3c. The same YIG waveguide shape and transducers are utilized for both designs, except the 3 μm thick YIG filter has $W_T = 5\ \mu m$, while the 18 μm thick YIG filter has $W_T = 10\ \mu m$. The measurements confirm the substantial suppression of higher-order length mode spurious responses in the 18 μm thick YIG filter. Furthermore, the 18 μm thick YIG increases the 3dB bandwidth from 83 MHz to 142 MHz. This broader bandwidth and flatter passband are attributed to the flat region in the dispersion relationship between ~200-300 $cm^{-1}$, while the steep dispersion at larger wavenumbers contributes to the reduced length modes. Fig. 3d compares 18 μm thick YIG waveguides with meander-line and straight-line transducers, with an expanded comparison shown in Supplementary Note #4. This demonstrates that the meander-line YIG filter achieves an improved filter shape by selectively exciting specific wavelengths while attenuating others, whereas the straight-line YIG filter excites MSSWs over a broader range of wavelengths. Moreover, Supplementary Note #4 also compares the nonreciprocity of 3 μm and 18 μm thick YIG meander-line filters. It highlights that, with increasing film thickness and the resulting increased sidewall roughness, the YIG structure no longer supports resonant circulation. Instead, the spin waves excited at one port predominantly propagate in a single direction, leading to enhanced nonreciprocal behavior.

Previous studies [10, 35, 36] and equation (1) indicate a linear increase in radiation impedance with the width of the YIG waveguide (or overlap length between the YIG and Al transducer). Consequently, the meander-line YIG filter, featuring a 3x longer total arm length compared to a straight-line counterpart, exhibits approximately 3x higher impedance. This results in superior performance at 6.3 GHz under ~1500 Gauss, due to the better impedance matching to the 50 Ω port impedances. The meander-line with 18 μm thick YIG shows a low insertion loss of 4.4 dB compared to 7.4 dB for the straight-line YIG filter. The maximum filter input impedance, $Z_{11}$, values are 40.9 dB and 33.2 dB for the meander-line and straight-line YIG filters, respectively. The minimum return losses, determined by the full impedance matrix, are 21.5 dB and 10.4 dB for the meander-line and straight-line YIG filters, respectively. However, since the radiation impedance also increases with frequency [10, 35, 36], at 11.8 GHz under an applied magnetic flux density of ~3500 Gauss, the higher peak $Z_{11}$ observed in the meander-line filter (47.4 dB vs. 42.1 dB) results in a substantial impedance mismatch and increased insertion loss of 8.3 dB, compared to just 4.1 dB for the straight-line filter.

To reduce insertion loss and enable high-frequency filtering, a two-parallel meander-line transducer design was implemented, as shown in the Fig. 3e. The input current is evenly split between two identical meander-line transducers. Fig. 3f compares the performance of the one and two-parallel meander-line filters, while Supplementary Note #5 presents measurement results across different frequencies. At 6.3 GHz, introducing the parallel branch reduces the maximum $Z_{11}$ from 40.9 dB to 33.5 dB. This results in improved impedance matching and a reduction in insertion loss from 4.4 dB to 2.9 dB. This improvement is even more pronounced, at 11.8 GHz, where the single meander-line filter exhibits greater impedance mismatch and the insertion loss is reduced from 8.3 dB to 3.8 dB. In addition to lower insertion loss, the two-parallel transducer design also enhances out-of-band rejection by ~8 dB. This improvement arises because, at any moment, the currents in the upper and lower transducers flow in the opposite direction. The stray magnetic fields produced by these opposing currents cancel each other out far from the transducers, thereby increasing overall electromagnetic isolation. Furthermore, the use of a larger YIG waveguide in the two-parallel design contributes to 1-3 dB lower insertion loss by reducing sidewall propagation loss. Supplementary Note #6 compares the performance of two-parallel meander-line filters with single and dual YIG waveguides, showing that wider YIG waveguides exhibit lower insertion loss and reduced demagnetization effects. However, one drawback is the appearance of a more pronounced spurious mode between 5.7–5.9 GHz, as seen in Fig. 3f. This is likely due to the center bridge in the meander-line transducers, which can also excite magnetostatic backward volume waves (MSBVW), thereby introducing this spurious response. The pitch of the meander-line transducers can significantly influence the shape of the YIG filter. As discussed in Supplementary Note #7, a pitch of 70 μm is optimal for minimizing insertion loss and suppressing spurious modes.

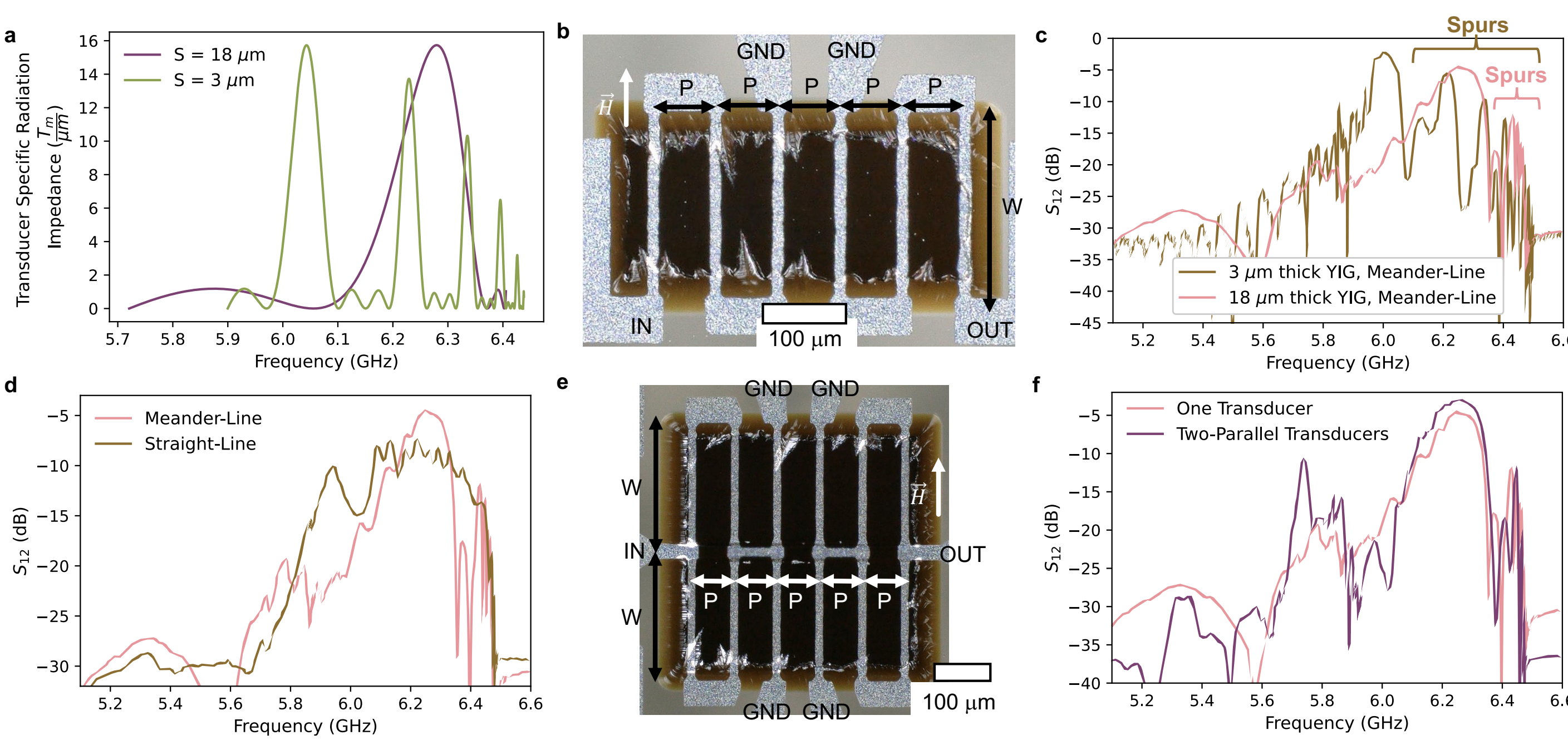

**Fig. 3 YIG filters with meander-line transducers. (a)** Comparison of the calculated transducer specific radiation impedance for meander-line transducers on 3 μm and 18 μm thick YIG films (first width mode, n = 1) at ~1500 Gauss. **(b)** Optical microscope image of a YIG filter with a meander-line transducer with a pitch of 70 μm and width of 200 μm. **(c)** Comparison of measured $S_{12}$ frequency responses between the 3 μm and 18 μm meander-line transducer filters under applied magnetic flux densities of approximately 1500 Gauss. **(d)** Comparison of measured $S_{12}$ frequency responses between the meander-line and straight-line transducer 18 μm thick YIG filters under applied magnetic flux densities of approximately 1500 Gauss. **(e)** Optical image of a YIG filter with two parallel meander-line transducers

with a pitch of 70 µm and width of 200 µm. **(f)** Comparison of measured $S_{12}$ frequency responses between filters with two parallel meander-line transducers and a single meander-line transducer at around 1500 Gauss.

As discussed earlier, meander-line transducers exhibit wavelength selectivity, providing enhanced excitation efficiency when the transducer period corresponds to an odd integral multiple of the MSSW wavelength, resulting in a higher radiation impedance [33, 34]. This property enables more effective control over the excited MSSW modes compared with straight-line transducers. They also contribute to improved out-of-band rejection. This enhancement arises because the currents in adjacent aluminum transducer lines flow in opposite directions, which helps confine the electromagnetic fields within the transducer region and reduces leakage into surrounding areas. Fig. 4a shows a meander-line YIG filter with increased spacing between the input and output transducers, introducing an additional separation of up to 210 µm. The measured $S_{12}$ frequency responses with different extra spacings are shown in the Fig. 4b. The out-of-band rejection at 9.5 GHz improves with increased spacing, measured at 32 dB, 42 dB, 48 dB, and 53 dB for additional spacings of 0 µm, 70 µm, 140 µm, and 210 µm, respectively. Meanwhile, the insertion loss remains relatively constant at 3.1 dB, 3.2 dB, 3.0 dB, and 3.4 dB for these spacings, respectively. This constant insertion loss is attributed to the high group velocity of MSSWs in thick YIG films, resulting in negligible propagation loss. Fig. 4c shows the group delay in the passband increases almost linearly with the extra spacings. A detailed comparison of different spacings at different frequencies are discussed in Supplementary Note #8.

Fig. 4d depicts a dual-hexagon YIG design that enhances nonreciprocity and suppresses the spurious mode around 8.4 GHz, while maintaining a nearly constant insertion loss. The measured $S_{12}$ and $S_{21}$ are showed in Fig. 4e and Fig. 4f, respectively. As discussed in the previous section, MSSWs propagate unidirectionally along the top and bottom surfaces of the YIG. By modifying the YIG edge from a straight edge to a triangular shape, the MSSWs on the top surface are effectively blocked from reflecting to the bottom surface, reducing internal MSSW circulation within the YIG waveguide. The oblique edges of the triangular YIG structure suppress standing-wave formation by scattering the propagating MSSW into directions that follow the MSBVW condition. Because these scattered waves lie within the stopband of the MSBVW, they rapidly decay [37-39]. Additionally, in the triangular regions, a large portion of the YIG lies near the etched sidewalls at both ends. The increased surface roughness of these sidewalls introduces extra damping and attenuation of MSSWs, preventing reflection toward the bottom interface and further reducing internal wave circulation [40-43]. This added loss mechanism enhances unidirectional propagation, leading to a significant improvement in nonreciprocity ($|S_{12} - S_{21}|$). Moreover, in a rectangle shape YIG filter, a strong spurious mode peak is observed near 8.4 GHz. This is attributed to the excitation of a higher-order width mode and MSBVW generated by the center bridge of the meander-line transducer, as this frequency lies below $\omega_{min}$ for MSSW waves in Fig. 1a. The dual-hexagon design mitigates this issue by introducing width variation along the YIG waveguide, which disrupts the formation of these spurious modes [44]. Supplementary Note #9 provides a comparison of these improvements across a range of operating frequencies.

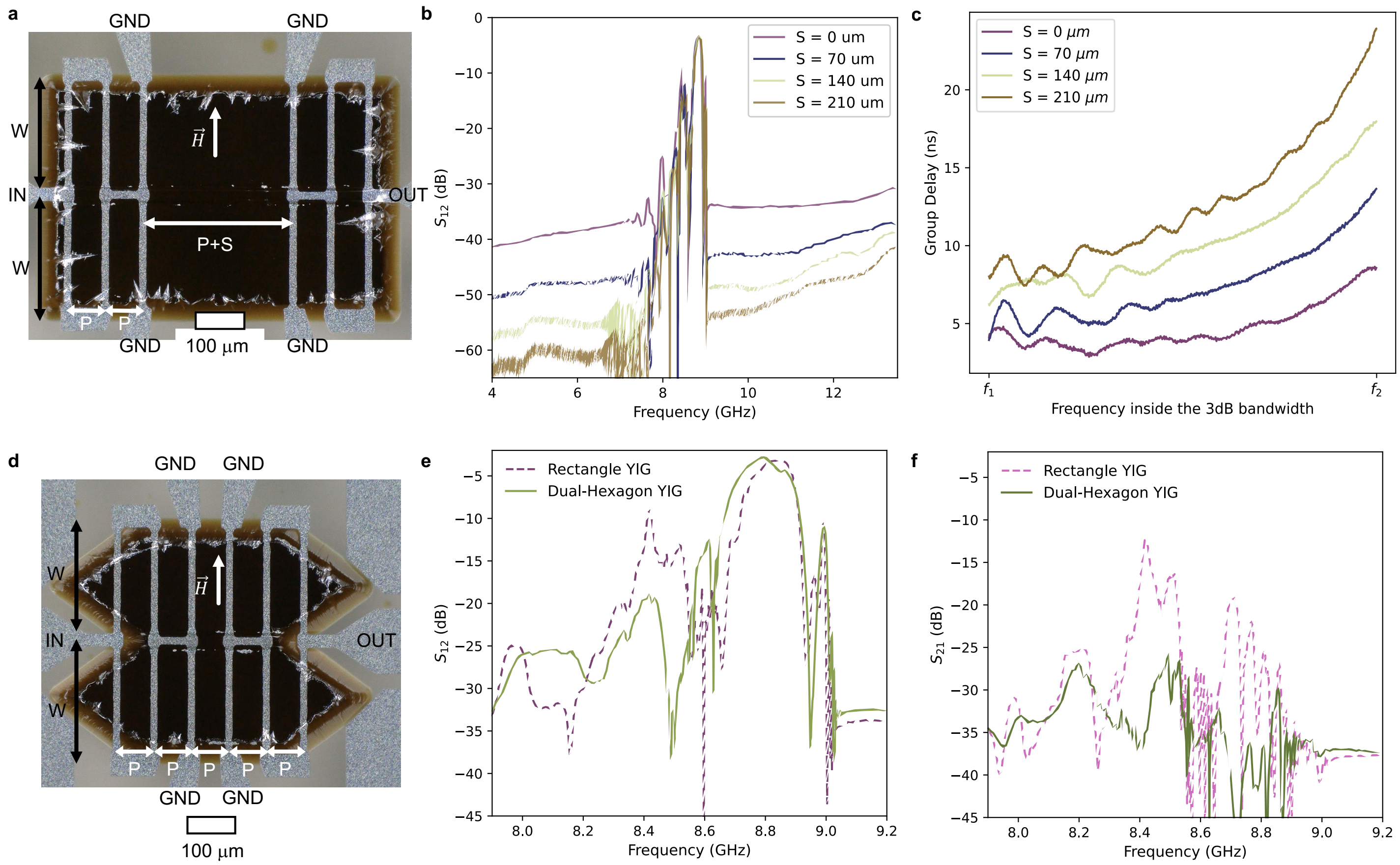

**Fig. 4 Meander-line transducer YIG filters with improved out-of-band rejection and non-reciprocity. (a)** Optical microscope image of a two parallel meander-line transducer YIG filter with pitch (P) of 70 µm, width (W) of 200 µm, and extra spacing (S) of 210 µm. **(b)** Measured $S_{12}$ frequency responses for different meander-line transducer spacings with magnetic flux density of approximately 2500 Gauss. **(c)** Measured passband group delay for different meander-line transducer spacings with magnetic flux density of approximately 2500 Gauss. **(d)** Optical microscope image of a dual-hexagon shaped YIG with two parallel meander-line transducers YIG filter with a pitch of 70 µm, width of 200 µm, and no extra spacings. **(e-f)** Comparison of measured **(e)** $S_{12}$ and **(f)** $S_{21}$ frequency responses between two parallel meander-line transducer YIG filters with dual-hexagon shaped YIG and rectangular shaped YIG at around 2500 Gauss.

# Integrated Devices

The tunable magnetic bias circuit comprises NdFeB permanent magnets, soft magnetic yokes, AlNiCo programmable magnets and solenoid coils wound around the AlNiCo magnets. The NdFeB magnets provide a constant magnetic flux, while the AlNiCo magnets, with lower coercivity, serve as a tunable source that can be magnetized or demagnetized by current pulses through the coils, retaining remanent flux after each pulse. A 3D magnetic field simulation of our previously reported bias circuit shows that only approximately 10% of the magnetic flux generated by the NdFeB and AlNiCo magnets passes through the yoke poles and is used for biasing the YIG device, while the remainder leaks into the air surrounding the yokes [23]. To improve the tuning range of the magnetic bias circuits, a possible approach is to reduce leakage flux. In this study, the surface area of the yoke pieces has been reduced by narrowing the width of the yokes, corresponding to the vertical dimension of the rectangular yoke base in Fig. 1c. At the same time, to prevent magnetic saturation in the smaller yokes, an iron-cobalt magnetic alloy with a high saturation flux density of 2.4 T is employed. The designed magnetic bias circuit is shown in Fig. 1c. Fig. 5a shows the measured tuning range of −25 Gauss to 5700 Gauss achieved, which is approximately twice that of a previous report [10], while the total volume of the bias circuit is reduced to 1.07 $cm^3$.

The YIG filter shown in Fig. 4d has been integrated into the magnetic bias circuit, with the complete integrated device shown in Fig. 1c. Fig. 5b shows the measured $S_{12}$ frequency response spanning from 4 GHz to 17.7 GHz. Fig. 5c focuses on the passband responses at four different frequency points, demonstrating a flat passband with over 25 dB of isolation. Supplementary Note #10 discusses the power handling capabilities of the integrated filter. The filter achieves a 1 dB compression point (P1dB) more than 10.4 dBm, significantly improved from the P1dB of -17 dBm for a YIG filter with a 3 μm thick film [10]. This enhancement results from the increased YIG thickness and a larger waveguide area, as similar meander-line YIG filters realized in 3 μm thick films exhibit P1dB of around 0 dBm. Supplementary Note #11 also shows the in-band linearity of the filter, which exhibits an input-referred third order intermodulation intercept point (IIP3) of more than +26 dBm. Supplementary Note #12 shows the details of the measured frequency response, including $S_{11}$, $S_{22}$, $S_{21}$, and group delay at various frequencies.

Supplementary Note #13 details the time and energy requirements for frequency tuning of the integrated device. The maximum switching duration is approximately 100 µs, requiring approximately 116 mJ to tune across the full 4–18 GHz range (0–5600 Gauss). Supplementary Note #14 examines the repeatability of the tunable filter. Although the AlNiCo magnet exhibits hysteresis, the filter demonstrates excellent frequency tuning repeatability when a reset (demagnetization) process is performed prior to setting the desired frequency.

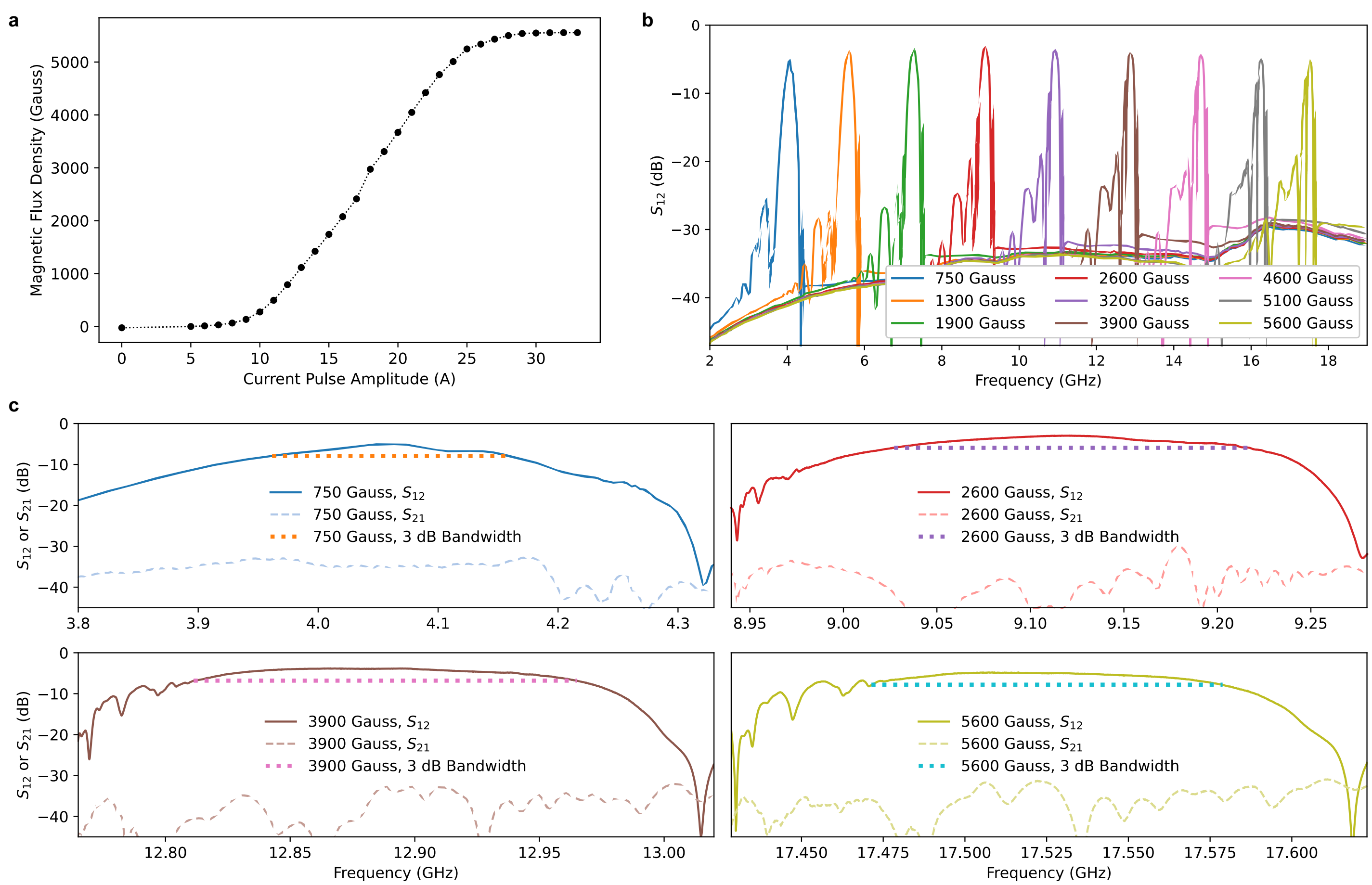

**Fig. 5 Integrated device. (a)** Measured maximum magnetic flux density at the center of the magnetic bias circuit gap under different applied current pulses. **(b)** Measured $S_{12}$ frequency responses with the magnetic field supplied by the zero static power consumption magnetic bias circuit. The YIG filter used in this study features a dual-hexagon design with an 18 µm thick YIG film and two-parallel meander-line transducers. The filter has a pitch of 70 µm, a waveguide width of 200 µm, and transducer widths of 10 µm, with no additional spacing between the input and output transducers. **(c)** Magnified measured $S_{12}$ and $S_{21}$ frequency responses showing the flat passband and large non-reciprocity of the tunable filter.

# Discussion

In this work, we have demonstrated a new class of miniature, narrowband, and tunable bandpass filters that offer zero static power consumption, low insertion loss (<5 dB), high out-of-band rejection (>30 dB), and substantial nonreciprocity (>25 dB). Figure 6 benchmarks the performance of this study with prior work on nonreciprocal filters [25, 38, 45-55], with a detailed comparison table available in Supplementary Note #15. To the best of our knowledge, all previous nonreciprocal filters have been limited to frequencies below 9 GHz. This study marks the first demonstration of nonreciprocal filters operating beyond 9 GHz —extending up to 18 GHz, while also introducing wideband frequency tunability. This advancement paves the way for potential deployment in the emerging FR3 Band (7 GHz to 24 GHz), which has already been used for satellite communication and is expected to be the band for future 6G wireless communications networks [5, 6].

This nonreciprocal bandpass filter plays an important role in both the transmit and receive paths, as illustrated in Fig. 1(a). In the receive path, wideband LNAs and wideband antennas often exhibit significant impedance mismatches across the 4–18 GHz range, resulting in partial reflections that can form standing waves. The reflected power can re-enter the LNA input, causing ripple and oscillations, ultimately degrading receiver sensitivity. By incorporating the nonreciprocal filter, reflected in-band signals from the amplifier or antenna are absorbed within the device rather than reflected, ensuring that unwanted energy is dissipated internally and does not re-enter the signal chain. This is confirmed by the high return loss ($S_{11}$) observed at the filter's output port. In the transmit path, the nonreciprocal bandpass filter effectively isolates the DAC from the PA by absorbing backward-propagating energy and suppressing out-of-band components. This absorptive isolation stabilizes the impedance between stages, minimizing distortion and feedback. The filter thus provides a compact and efficient solution that simultaneously achieves frequency selective transmission and absorptive isolation. Although placing the nonreciprocal filter between the PA and antenna could further enhance system stability, this configuration requires higher power handling capability. Future work will therefore focus on increasing the device's power tolerance from the current ~11 dBm to above 20 dBm to enable direct integration at the PA output.

While some prior works using active devices have reported lower insertion loss or even power gain (i.e., negative insertion loss) [53, 54], those solutions are not frequency tunable and suffer from the noise of the active circuits. To achieve wide frequency coverage, they must rely on multiple RF switches—an approach that significantly increases system complexity and contributes to additional insertion loss [8, 9]. Furthermore, our study offers the highest selectivity within the 2–18 GHz range, with a narrow bandwidth of 100–200 MHz. This level of selectivity is well-suited for modern communication systems. For instance, 5G systems specify a maximum bandwidth of 100 MHz in the sub-6 GHz range and up to 400 MHz in the millimeter-wave bands [56]. Moreover, although transistor based nonreciprocal filters can offer a smaller footprint, they typically require hundreds of milliwatts of DC power for operation [52, 54]. In contrast, the magnetic bias circuit developed in this study achieves a compact form factor, occupying 4 times less area and 82 times less volume than the electromagnets reported in [25], while eliminating the need for power-hungry electromagnets that consume 57.6 W of static power.

Even when disregarding the benefits of nonreciprocity, this study outperforms previous YIG-based reciprocal filters, as shown in Supplementary Note #16. It demonstrates lower insertion loss, superior out-of-band rejection, a sharper filter skirt, larger power handling and substantially higher linearity. Notably, this study achieved almost 30 dB of rejection when the frequency is 1.6 times the bandwidth above the center frequency and higher frequency spurious modes are eliminated. The enhanced filter skirt is attributed to the unique dispersion characteristics of the 18 μm thick YIG and the spurious mode attenuation provided by the dual-hexagon shaped YIG.

Given that the filter demonstrated in this work simultaneously achieves a narrow frequency band, excellent out-of-band rejection, a flat passband, and minimal group delay variation, the most comparable existing technology is the surface acoustic wave (SAW) delay line. Similar to how SAWs propagate from input-to-output interdigitated transducers (IDTs) on piezoelectric substrates, this study utilizes MSSW excited by aluminum meander-line transducers on a YIG film to achieve signal delay. However, this work offers several advantages beyond what SAW delay lines can provide: (1) Frequency turnability. SAW delay lines have fixed operational frequencies determined by lithographically defined IDTs and lack electrical tunability. (2) Nonreciprocity. While SAW delay lines can exhibit nonreciprocal behavior via the acoustoelectric effect [57] and helicity mismatch [58], these require additional fabrication steps and typically involve DC power consumption. (3) Low insertion loss. Even though SAW filters have been demonstrated with low insertion loss [59], state-of-the-art SAW delay lines with low loss are generally limited to below 6 GHz and exhibit insertion losses around 7 dB [60]. In contrast, this study demonstrates operation up to 18 GHz while maintaining an insertion loss below 5 dB.

Although only one thin YIG film (3 μm) and one thick YIG film (15 - 18 μm) were examined in this study due to limits on commercially available YIG thicknesses, the BCB planarization fabrication method demonstrated here can be readily adapted to YIG films of other thicknesses. Furthermore, this work provides the first comprehensive experimental and theoretical analysis of the thickness effect in microfabricated YIG films. The methodologies used here, including the dispersion relation, power handling, linearity, nonreciprocity, and propagation loss, can be directly applied to future investigations of YIG devices with intermediate or alternative thicknesses.

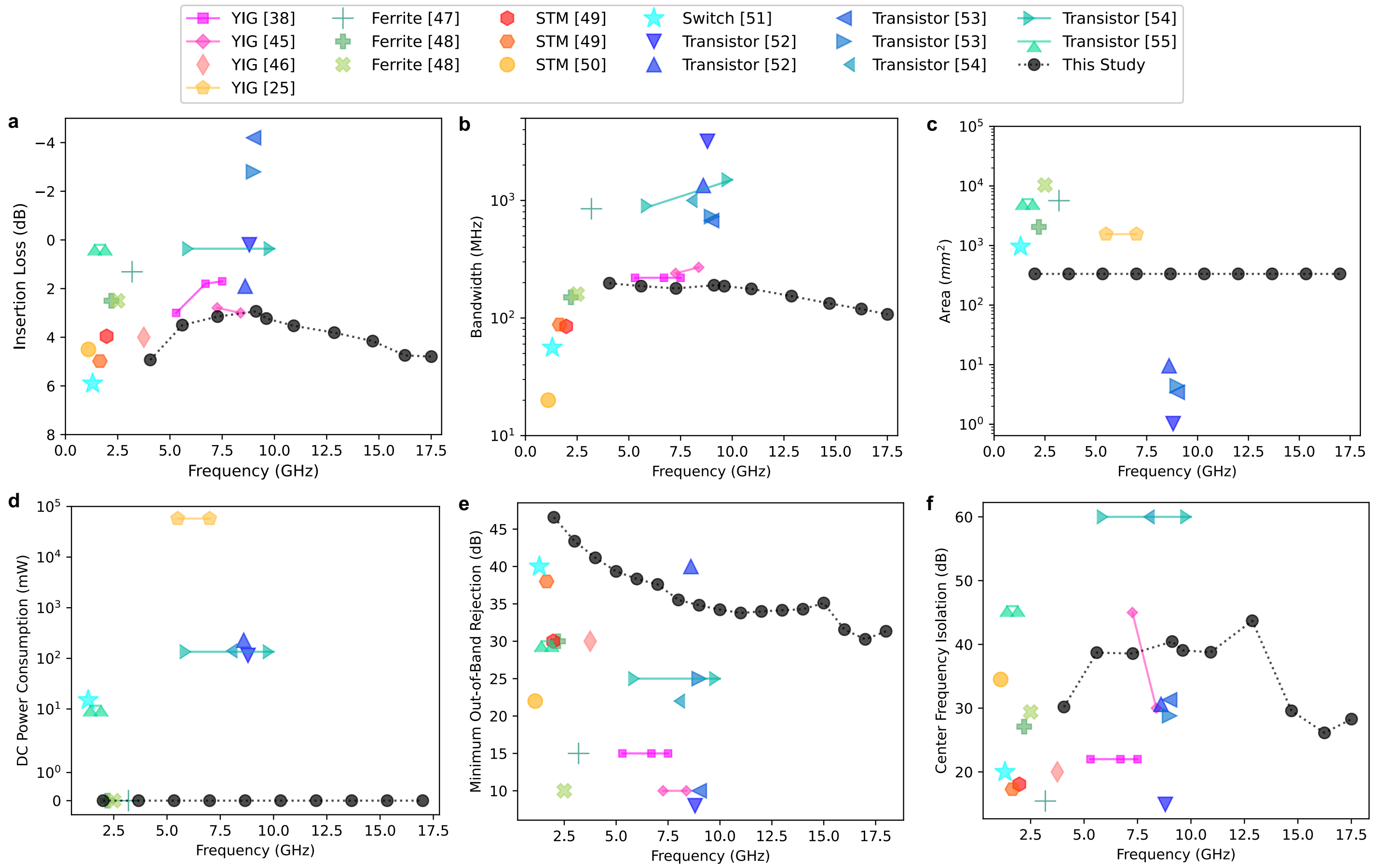

**Fig. 6 Comparison of this study with nonreciprocal filters based on YIG, Ferrites, Spatiotemporal Modulation (STM), PIN diode switches (Switch), and Transistors. (a)** Insertion loss **(b)** Bandwidth **(c)** Area **(d)** DC power consumption **(e)** Minimum out-of-band rejection — for each filter in references, this is plotted versus its center frequency. For this study, the out-of-band rejection is shown across the 2-18 GHz frequency range. **(f)** Center frequency isolation ($|S_{12} - S_{21}|$).

# Methods

**MSSW filter fabrication:** Initially, a <111> oriented 15-18 μm thick YIG was grown on a GGG substrate using liquid epitaxy, prepared by MTI Corporation, with a ferromagnetic resonance linewidth of 0.5–2.0 Oe. The difference between 15 μm and 18 μm YIG films is minimal and does not significantly influence the device performance. The YIG film used in Supplementary Note #2 (Thick YIG fabrication process) was approximately 15 μm thick, while all other experiments in the main text utilized the 18 μm film. In the first step, a 500 nm thick $SiO_2$ layer was deposited as a hard mask via atomic layer deposition using Bis(diethylamino)silane (BDEAS) and $O_3$ (Cambridge Nanotech S200). The sample was then annealed at 600°C for 30 minutes in a nitrogen environment. The hard mask was patterned using standard photolithography and $CHF_3$ dry etching (Oxford 80 Plus RIE). After removing the photoresist, the mask pattern was transferred to the YIG through wet etching, utilizing phosphoric acid at 140 °C. The remaining $SiO_2$ hard mask was stripped using Buffered Oxide Etchant (BOE) 6:1, a mixture of a buffering agent and hydrofluoric acid. To achieve surface planarization of the YIG substrate, photosensitive benzocyclobutene (BCB) (CYCLOTENE 4026-46) was spin-coated onto the YIG surface. Following spin coating, the BCB layer was exposed to UV light and patterned using AP3000 developer. The regions of BCB not covering the YIG were crosslinked upon UV light exposure and remained, while the unexposed BCB directly on top of the YIG was dissolved in the developer (see Fig. 1d). The patterned BCB was then thermally annealed at 350 °C for 2 hours to complete cross-linking and fully cure the film, forming a stable dielectric layer. To pattern RF electrodes, a 2 μm thick Al layer was deposited by sputtering at 1000 W from a 100 mm diameter Al target at a base pressure of $1e^{-7}$ mbar (Evatec Clusterline 200 II) at 150°C. This was followed by depositing a 300 nm thick $SiO_2$ layer using plasma-enhanced chemical vapor deposition (Oxford PlasmaLab 100). Standard photolithography was again employed to pattern the photoresist, followed by etching the $SiO_2$ using $CHF_3$ RIE dry etching. The photoresist layer was removed with 1165 solvent, and the Al layer was patterned using an ICP Etcher (Oxford Cobra ICP Etcher) with $Cl_2/BCl_3$ gases. Finally, the $SiO_2$ hard mask was removed using $CF_4$ RIE with the same RIE tool.

**Magnetic circuit fabrication:** The yoke pieces, the NdFeB magnets and the AlNiCo magnets were prepared separately and then assembled. The yokes were cut from a 3.2 mm thick Hiperco 50A cobalt-iron sheet using a wire EDM (electric discharge machining) process. Each yoke has a 20.0 mm × 2.0 mm rectangular base, with a 5.85 mm tall, tapered pole piece extending from the center. The pole piece is 3.0 mm wide at the base and narrows to 2.0 mm at the top. The NdFeB permanent magnets have a cylindrical shape with a diameter of 3.18 mm and a length of 12.7 mm, and were purchased from K&J Magnetics, Inc. The left and right NdFeB magnets are grades N42 and N52, respectively. The AlNiCo 5 magnets, identical in size to the NdFeB magnets, were purchased from DigiKey. A 49-turn coil of 32-gauge copper wire, coated with a polyimide insulation layer, was manually wound around each AlNiCo magnet. After preparing all the parts, the yokes, NdFeB magnets, and AlNiCo magnets were assembled on a 3D-printed substrate and fixed using cyanoacrylate adhesive.

**Measurement setup:** The YIG sample was first characterized using a magnetic probe station (MicroXact's MPS-1D-5kOe). The magnetic field was generated by electromagnets inside the magnetic probe station. A Gaussmeter (Model GM2, AlphaLab Inc) was used to calibrate the magnetic probe station. Due to potential variations in sample placement across measurements, the magnetic field experienced by the device may not have been entirely consistent in this study.

For the integrated device, the magnetic field was produced by the magnetic bias circuit, eliminating the need for the electromagnets. To tune the fabricated magnetic circuit, current pulses with different amplitudes and a consistent duration of 0.5 ms were applied to the coils wound around the AlNiCo magnets using a DC electronic load (EL34243A, Keysight) and a pre-charged large 65F supercapacitor (XVM-16R2656-R, Eaton). The two coils are connected in series, allowing a single current pulse to simultaneously magnetize both AlNiCo magnets. After each current pulse, the magnetic field within the air gap between the yoke tips was measured using a gaussmeter.

Filter performance was measured using a Keysight vector network analyzer (VNA) P5026B with a power level of -20 dBm with 50 Ω port impedances, unless otherwise noted. Prior to measurement, a two-port calibration to the probe tips was performed within the desired frequency range using the Short-Open-Load-Through (SOLT) method. The ground-signal-ground (GSG) probe used had a pitch of 150 μm from GGB industries. There is no on wafer de-embedding utilized in this study. Filter parameters were manually extracted from the measured S-parameter data.

# Data availability

The data that support the findings of this study are available from the corresponding author upon reasonable request.

# Acknowledgement

The authors would like to thank Dr. Todd Bauer, Dr. David Abe and Dr. Tim Hancock of the Defense Advanced Research Projects Agency (DARPA) and Dr. Michael Page of the Air Force Research Laboratory for their guidance and support of this work under the DARPA Wideband Adaptive RF Protection (WARP) program, contract FA8650-21-1-7010. The fabrication of devices was performed at the Singh Center for Nanotechnology, supported by the NSF National Nanotechnology Coordinated Infrastructure Program (No. NNCI-1542153).

# Author contributions

X.D. and R.O. developed the device concepts and experimental implementations. X.D., S.Y., Yijie D., S.W., and C.C. fabricated the YIG filters under the supervision of R.O. Yixiao D. designed and fabricated the magnetic bias circuit under the supervision of M.A. Yixiao D., D.L., and X.W. performed the magnetic bias circuit measurements and M.A supervised them. X.D., S.Y., D.L., and S.W. performed the filter measurements and R.O. supervised the measurements. X.D., Yixiao D. and R.O. analyzed all data and wrote the manuscript. All authors have given approval to the final version of the manuscript.

# References

[1] P. Wong and I. Hunter, "Electronically Tunable Filters," *IEEE Microwave Magazine,* vol. 10, no. 6, pp. 46-54, 2009, doi: 10.1109/mmm.2009.933593.

[2] J. Ge, T. Wang, Y. Peng, and G. Wang, "Electrically tunable microwave technologies with ferromagnetic thin film: recent advances in design techniques and applications," *IEEE Microwave Magazine,* vol. 23, no. 11, pp. 48-63, 2022.

[3] P. Dutta, G. A. Kumar, G. Ram, and D. S. Varma, "Spatiotemporal Nonreciprocal Filters: Theoretical Concepts and Literature Review," *IEEE Microwave Magazine,* vol. 23, no. 6, pp. 85-101, 2022, doi: 10.1109/MMM.2022.3157970.

[4] A. Kord, D. L. Sounas, and A. Alu, "Microwave nonreciprocity," *Proceedings of the IEEE,* vol. 108, no. 10, pp. 1728-1758, 2020.

[5] R. R. Paulo, E. B. Teixeira, and F. J. Velez, "Service quality of the urban microcellular scenario in the sub-6 GHz frequency bands," *IEEE Access,* 2024.

[6] T. E. Humphreys, P. A. Iannucci, Z. M. Komodromos, and A. M. Graff, "Signal structure of the Starlink Ku-band downlink," *IEEE Transactions on Aerospace and Electronic Systems,* vol. 59, no. 5, pp. 6016-6030, 2023.

[7] F. De Paolis, "Satellite filters for 5G/6G and beyond," in *2021 IEEE MTT-S International Microwave Filter Workshop (IMFW)*, 2021: IEEE, pp. 148-150.

[8] D. Lu, J. Liu, and M. Yu, "Highly selective bandpass switch block with applications of MMIC SPDT switch and switched filter bank," *IEEE Solid-State Circuits Letters,* vol. 5, pp. 190-193, 2022.

[9] B. Zhu, X. Zhu, X. Li, P.-L. Chi, and T. Yang, "A 1.9–18-GHz filter bank with improved passband flatness based on asymmetrical low-loss SP7T switch," *IEEE Transactions on Microwave Theory and Techniques,* 2024.

[10] X. Du *et al.*, "Frequency tunable magnetostatic wave filters with zero static power magnetic biasing circuitry," *Nature Communications,* vol. 15, no. 1, p. 3582, 2024.

[11] X. Du *et al.*, "Magnetostatic wave notch filters frequency tuned via a zero DC power magnetic bias circuit," in *2024 IEEE International Microwave Filter Workshop (IMFW)*, 2024: IEEE, pp. 176-179.

[12] X. Du *et al.*, "Meander Line Transducer Empowered Low-Loss Tunable Magnetostatic Wave Filters with Zero Static Power Consumption," in *2024 IEEE/MTT-S International Microwave Symposium-IMS 2024*, 2024: IEEE, pp. 42-45.

[13] A. Ahmed and G. M. Rebeiz, "A 8–30 GHz passive harmonic rejection mixer with 8 GHz instantaneous IF bandwidth in 45RFSOI," in *2022 IEEE Radio Frequency Integrated Circuits Symposium (RFIC)*, 2022: IEEE, pp. 19-22.

[14] H. Razavi and B. Razavi, "A 0.4–6 GHz receiver for cellular and WiFi applications," *IEEE Journal of Solid-State Circuits,* vol. 57, no. 9, pp. 2640-2657, 2022.

[15] C. Han and X. Luo, "A-10.1 dBm IIP3, 0.3-40GHz receiver using hybrid-path band-selection with reduced LO coverage bandwidth supporting 480Mb/s 4096-QAM and 7.2 Gb/s 64-QAM modulation," in *2024 IEEE Custom Integrated Circuits Conference (CICC)*, 2024: IEEE, pp. 1-2.

[16] A. Sabharwal, P. Schniter, D. Guo, D. W. Bliss, S. Rangarajan, and R. Wichman, "In-Band Full-Duplex Wireless: Challenges and Opportunities," *IEEE Journal on Selected Areas in Communications,* vol. 32, no. 9, pp. 1637-1652, 2014, doi: 10.1109/jsac.2014.2330193.
[17] L. Ranzani and J. Aumentado, "Circulators at the Quantum Limit: Recent Realizations of Quantum-Limited Superconducting Circulators and Related Approaches," *IEEE Microwave Magazine,* vol. 20, no. 4, pp. 112-122, 2019, doi: 10.1109/mmm.2019.2891381.
[18] H.-C. Kuo *et al.*, "A Fully Integrated 60-GHz CMOS Direct-Conversion Doppler Radar RF Sensor With Clutter Canceller for Single-Antenna Noncontact Human Vital-Signs Detection," *IEEE Transactions on Microwave Theory and Techniques,* vol. 64, no. 4, pp. 1018-1028, 2016, doi: 10.1109/tmtt.2016.2536600.
[19] A. Nagulu and H. Krishnaswamy, "Non-magnetic non-reciprocal microwave components—State of the art and future directions," *IEEE Journal of Microwaves,* vol. 1, no. 1, pp. 447-456, 2021.
[20] A. Nagulu, N. Reiskarimian, and H. Krishnaswamy, "Non-reciprocal electronics based on temporal modulation," *Nature Electronics,* vol. 3, no. 5, pp. 241-250, 2020/05/01 2020, doi: 10.1038/s41928-020-0400-5.
[21] C. Devitt, R. Wang, S. Tiwari, and S. A. Bhave, "An edge-coupled magnetostatic bandpass filter," *Nat Commun,* vol. 15, no. 1, p. 7764, Sep 5 2024, doi: 10.1038/s41467-024-51735-6.
[22] S. Tiwari, A. Ashok, C. Devitt, S. A. Bhave, and R. Wang, "High-performance magnetostatic wave resonators based on deep anisotropic etching of gadolinium gallium garnet substrates," *Nature Electronics,* vol. 8, no. 3, pp. 267-275, 2025/03/01 2025, doi: 10.1038/s41928-025-01345-x.
[23] Y. Ding, X. Wang, and M. G. Allen, "A Tunable Magnetic Bias Circuit With Zero Static Power Consumption," *IEEE Magnetics Letters,* vol. 16, pp. 1-5, 2025, doi: 10.1109/lmag.2025.3541915.
[24] Micro Lambda Wireless Inc, "MLFP 4 stage filter data sheet," *MLFP-42018,* Datasheet.
[25] C. S. Tsai and Q. Gang, "Wideband Microwave Filters Using Ferromagnetic Resonance Tuning in Flip-Chip YIG-GaAs Layer Structures," *IEEE Transactions on Magnetics,* vol. 45, no. 2, pp. 656-660, 2009, doi: 10.1109/tmag.2008.2010466.
[26] V. Castel, N. Vlietstra, B. van Wees, and J. Ben Youssef, "Yttrium iron garnet thickness and frequency dependence of the spin-charge current conversion in YIG/Pt systems," *Physical Review B,* vol. 90, no. 21, p. 214434, 2014.
[27] K. O. Levchenko, K. Davídková, J. Mikkelsen, and A. V. Chumak, "Review on spin-wave RF applications," *arXiv preprint arXiv:2411.19212,* 2024.
[28] J. D. Adam and F. Winter, "Magnetostatic Wave Frequency Selective Limiters," *IEEE Transactions on Magnetics,* vol. 49, no. 3, pp. 956-962, 2013, doi: 10.1109/tmag.2012.2227994.
[29] W. S. Ishak and K.-W. Chang, "Tunable microwave resonators using magnetostatic wave in YIG films," *IEEE transactions on microwave theory and techniques,* vol. 34, no. 12, pp. 1383-1393, 1986.
[30] K. L. Wong *et al.*, "Unidirectional propagation of magnetostatic surface spin waves at a magnetic film surface," *Applied Physics Letters,* vol. 105, no. 23, p. 232403, 2014/12/08 2014, doi: 10.1063/1.4903742.
[31] V. E. Demidov and S. O. Demokritov, "Magnonic waveguides studied by microfocus Brillouin light scattering," *IEEE Transactions on Magnetics,* vol. 51, no. 4, pp. 1-15, 2015.
[32] D. D. Stancil, "Phenomenological propagation loss theory for magnetostatic waves in thin ferrite films," *Journal of applied physics,* vol. 59, no. 1, pp. 218-224, 1986.
[33] J. Sethares, "Magnetostatic surface-wave transducers," *IEEE Transactions on Microwave Theory and Techniques,* vol. 27, no. 11, pp. 902-909, 1979.

[34] J. P. Parekh and H. S. Tuan, "Meander line excitation of magnetostatic surface waves," *Proceedings of the IEEE,* vol. 67, no. 1, pp. 182-183, 1979, doi: 10.1109/PROC.1979.11221.
[35] F. Vanderveken, V. Tyberkevych, G. Talmelli, B. Soree, F. Ciubotaru, and C. Adelmann, "Lumped circuit model for inductive antenna spin-wave transducers," *Sci Rep,* vol. 12, no. 1, p. 3796, Mar 8 2022, doi: 10.1038/s41598-022-07625-2.
[36] A. K. Ganguly and D. C. Webb, "Microstrip Excitation of Magnetostatic Surface Waves: Theory and Experiment," *IEEE Transactions on Microwave Theory and Techniques,* vol. 23, no. 12, pp. 998-1006, 1975, doi: 10.1109/tmtt.1975.1128733.
[37] H. Yu *et al.*, "Magnetic thin-film insulator with ultra-low spin wave damping for coherent nanomagnonics," *Scientific reports,* vol. 4, no. 1, p. 6848, 2014.
[38] J. Wu, X. Yang, S. Beguhn, J. Lou, and N. X. Sun, "Nonreciprocal tunable low-loss bandpass filters with ultra-wideband isolation based on magnetostatic surface wave," *IEEE transactions on microwave theory and techniques,* vol. 60, no. 12, pp. 3959-3968, 2012.
[39] W. Ishak, "Magnetostatic surface wave devices for UHF and L band applications," *IEEE Transactions on Magnetics,* vol. 19, no. 5, pp. 1880-1882, 1983, doi: 10.1109/TMAG.1983.1062691.
[40] N. Kanazawa *et al.*, "Spin wave absorber generated by artificial surface anisotropy for spin wave device network," *AIP Advances,* vol. 6, no. 9, 2016.
[41] J. H. Collins, D. M. Hastie, J. M. Owens, and C. V. Smith, Jr., "Magnetostatic wave terminations," *Journal of Applied Physics,* vol. 49, no. 3, pp. 1800-1802, 1978, doi: 10.1063/1.324870.
[42] V. L. Taylor, J. C. Sethares, and C. V. Smith, "MSW Terminations," in *1980 Ultrasonics Symposium*, 5-7 Nov. 1980 1980, pp. 562-570, doi: 10.1109/ULTSYM.1980.197461.
[43] H. Glass, L. Adkins, and F. Stearns, "Attenuation layer for magnetostatic wave (MSW) absorbers," *IEEE Transactions on Magnetics,* vol. 20, no. 5, pp. 1235-1237, 2003.
[44] W. S. Ishak, "Magnetostatic wave technology: a review," *Proceedings of the IEEE,* vol. 76, no. 2, pp. 171-187, 1988, doi: 10.1109/5.4393.
[45] Y. Zhang *et al.*, "Nonreciprocal isolating bandpass filter with enhanced isolation using metallized ferrite," *IEEE Transactions on Microwave Theory and Techniques,* vol. 68, no. 12, pp. 5307-5316, 2020.
[46] S. Odintsov, S. Sheshukova, S. Nikitov, E. H. Lock, E. Beginin, and A. Sadovnikov, "Nonreciprocal spin wave propagation in bilayer magnonic waveguide," *Journal of Magnetism and Magnetic Materials,* vol. 546, p. 168736, 2022.
[47] Y. Yang, Y. Wu, and W. Wang, "Design of nonreciprocal multifunctional reflectionless bandpass filters by using circulators," *IEEE Transactions on Circuits and Systems II: Express Briefs,* vol. 70, no. 1, pp. 106-110, 2022.
[48] A. Ashley and D. Psychogiou, "Ferrite-based multiport circulators with RF co-designed bandpass filtering capabilities," *IEEE Transactions on Microwave Theory and Techniques,* vol. 71, no. 6, pp. 2594-2605, 2023.
[49] G. Chaudhary and Y. Jeong, "Frequency tunable impedance matching nonreciprocal bandpass filter using time-modulated quarter-wave resonators," *IEEE Transactions on Industrial Electronics,* vol. 69, no. 8, pp. 8356-8365, 2021.
[50] C. Cassella *et al.*, "Radio Frequency Angular Momentum Biased Quasi-LTI Nonreciprocal Acoustic Filters," *IEEE Trans Ultrason Ferroelectr Freq Control,* vol. 66, no. 11, pp. 1814-1825, Nov 2019, doi: 10.1109/TUFFC.2019.2931121.
[51] M. A. Khater, A. Fisher, and D. Peroulis, "Switch-based non-reciprocal filter for high-power applications," in *2024 IEEE International Microwave Filter Workshop (IMFW)*, 2024: IEEE, pp. 180-182.

[52] A. Ashley and D. Psychogiou, "MMIC GaAs isolators and quasi-circulators with co-designed RF filtering functionality," *IEEE Journal of Microwaves,* vol. 3, no. 1, pp. 102-114, 2022.
[53] A. Ashley and D. Psychogiou, "X-band quasi-elliptic non-reciprocal bandpass filters (NBPFs)," *IEEE Transactions on Microwave Theory and Techniques,* vol. 69, no. 7, pp. 3255-3263, 2021.
[54] D. Simpson and D. Psychogiou, "GaAs MMIC nonreciprocal single-band, multi-band, and tunable bandpass filters," *IEEE Transactions on Microwave Theory and Techniques,* vol. 71, no. 6, pp. 2439-2449, 2023.
[55] K. Li and D. Psychogiou, "Multifunctional and tunable bandpass filters with RF codesigned isolator and impedance matching capabilities," *International Journal of Microwave and Wireless Technologies,* pp. 1-15, 2024.
[56] "5G FR(Frequency Range) / Operating Bandwidth in Detail." ShareTechnote. (accessed 2025).
[57] L. Hackett *et al.*, "S-band acoustoelectric amplifier in an InGaAs-AlScN-SiC architecture," *Applied Physics Letters,* vol. 124, no. 11, p. 113503, 2024, doi: 10.1063/5.0178912.
[58] A. R. Will-Cole *et al.*, "Chiral microwave nonreciprocity demonstrated via Rayleigh and Sezawa modes supported in an Al 0.58 Sc 0.42 N/4 H-Si C platform," *Physical Review Applied,* vol. 23, no. 3, p. 034058, 2025.
[59] K. Yang *et al.*, "SV-SAW RF filters based on low-cost 128° Y LiNbO3/SiO2/poly-Si/Si substrate for 6G cmWave wireless communications," *Microsystems & Nanoengineering,* vol. 11, no. 1, p. 79, 2025.
[60] Z.-Q. Lee *et al.*, "6 GHz lithium niobate on insulator low-loss SAW delay line adapting non-leaky composite waveguide mode," in *2025 IEEE 38th International Conference on Micro Electro Mechanical Systems (MEMS)*, 2025: IEEE, pp. 1149-1152.

# Supplementary Information

# A Wideband Tunable, Nonreciprocal Bandpass Filter Using Magnetostatic Surface Waves with Zero Static Power Consumption

Xingyu Du[1], Yixiao Ding[1], Shun Yao[1], Yijie Ding[1], Dengyang Lu[1], Shuxian Wu[1], Chin-Yu Chang[1], Xuan Wang[1], Mark Allen[1], and Roy H. Olsson III[1*]

[1]Department of Electrical and Systems Engineering, University of Pennsylvania, Philadelphia, PA, USA

[*]Correspondence: Roy H. Olsson III (rolsson@seas.upenn.edu)

# Table of Contents

## Supplementary Note #1: Magnetostatic Surface Wave Dispersion Relationship

Kalinikos et al. calculated the dispersion relationship of spin waves taking into account both the dipole and exchange interactions and surface pinning conditions. This relationship can be written as [1, 2]:

$$f^2 = \gamma^2[H + M_s(1 - P + \alpha k^2)]\left[H + M_s\left(P\frac{k_y^2}{k^2} + \alpha k^2\right)\right] \quad \text{(S1)}$$

where $P = 1 - (1 - \exp(-kS))/(kS)$, $k^2 = k_z^2 + k_y^2$, $\gamma =$ 2.8 MHz/Gauss which is the gyromagnetic ratio, $H$ is the magnetic field applied in Gauss, and $M_s = 1780$ Gauss is the saturation magnetization of YIG, $\alpha = 5.18 \times 10^{-18} m^2$which is the exchange stiffness, and $S$ is the film thickness. The coordinate system is shown in Fig. 1 (c), where $k_z$ is in the width (z) direction and $k_y$ is in the length (y) or magnetostatic surface wave (MSSW) propagation direction.

Ignoring the effects of internal static magnetic field non-uniformity caused by edge demagnetization in the stripe waveguide, spin-wave confinement can be understood as the quantization of plane spin waves. For MSSW propagating in infinitely wide YIG films, $k_z$ is zero and $k_y$ can be vary continuously. For MSSW propagating inside a stripe waveguide, the $k_z$ component must be discrete. This results in the formation of resonant standing waves across the waveguide width. These discrete values $k_x^n$ are governed by the boundary conditions of the dynamic magnetization at the waveguide edges, which depends on the thickness-to-width ratio $\frac{S}{W}$. When $\frac{S}{W} \ll 1$, the magnetic wall boundary conditions can be given by [2]:

$$k_x^n = \frac{n\pi}{w}, n = 1,2,3, \dots \quad \text{(S2)}$$

Another method to calculate the MSSW dispersion is reported by O'Keeffe et al [3], which can be expressed by:

$$exp(2MS) = \frac{\Omega_m M + \Omega_k + (\Omega_H^2 - \Omega^2)(M - N)}{\Omega_H M - \Omega_k + (\Omega_H^2 - \Omega^2)(M + N)} \times \frac{\Omega_m M - \Omega_k + (\Omega_H^2 - \Omega^2)(M - Ntanh(Nt)}{\Omega_m M + \Omega_k + (\Omega_H^2 - \Omega^2)(M + Ntanh(Nt)} \quad \text{(S3)}$$

Inside the film, we have

$$M^2 = k_Z^2 = \frac{(\frac{n\pi}{w})^2}{\mu_1} + k_y^2 \quad \text{(S4)}$$

and outside

$$N^2 = k_Z^2 = (\frac{n\pi}{w})^2 + k_y^2 \quad \text{(S5)}$$

where $\mu_1 = 1 - \Omega_H/(\Omega^2 - \Omega_H^2)$ , $\Omega = \omega/\gamma 4\pi M_s$, $\Omega_H = H/4\pi M_s$, and $t$ is the distance between the YIG film and a ground plane, which is infinite here.

**Supplementary Figures 1 and 2** compare the two methods applied to both 3 µm and 18 µm thick YIG films. For O'Keeffe's method applied to the 18 µm thick YIG film, the $M^2$ term in the dispersion relation for higher-order width modes changes rapidly from positive to negative near the $\omega_{min}$ frequency, which complicates the calculation. To address this, a straight-line approximation was used to estimate the corresponding frequency near the $\omega_{min}$ frequency. Despite this complexity, both Kalinikos' and O'Keeffe's methods exhibit very similar trends overall, although Kalinikos' approach incorporates more physical effects.

One effect included in Kalinikos' method is the exchange interaction. This interaction can be neglected under two conditions: (1) when the YIG film thickness (S) and waveguide width (W) are of the same order, and (2) when both S and W are significantly larger than the YIG exchange length, $\lambda_{ex} \approx 13$ nm. The critical waveguide width, above which exchange dynamics become negligible, can be estimated by $w_{cr} = 2.2S + 6.7\lambda_{ex} \approx 40$ µm [4]. In this case, the waveguide width is 200 µm—well above $w_{cr}$—so the exchange interaction can be ignored.

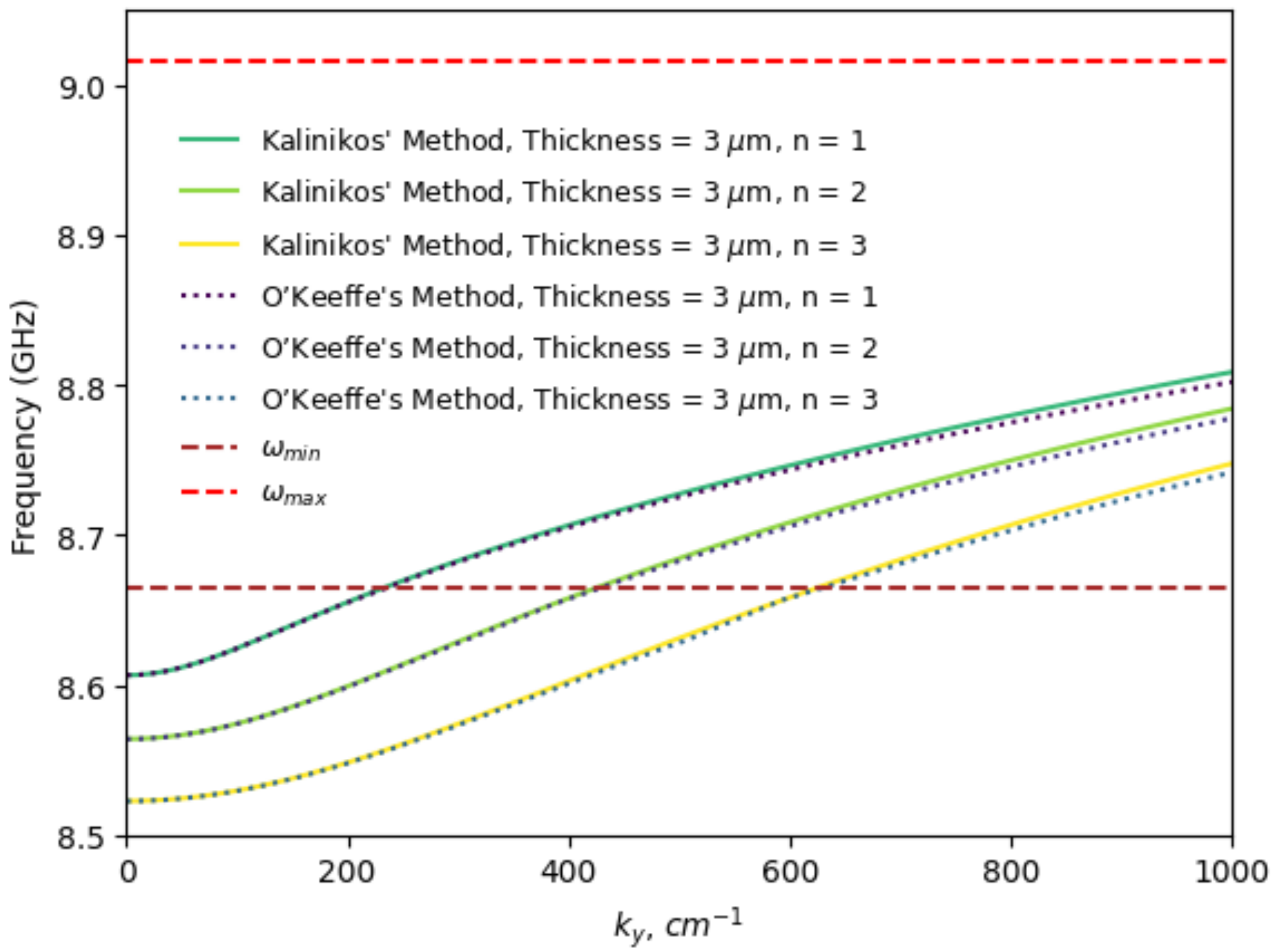

**Supplementary Figure 1**. Comparison of the 3 μm thick YIG dispersion relationship between Kalinikos' method and O'Keefee's method. The YIG waveguide width is 200 μm. The applied magnetic field is 2325 Gauss.

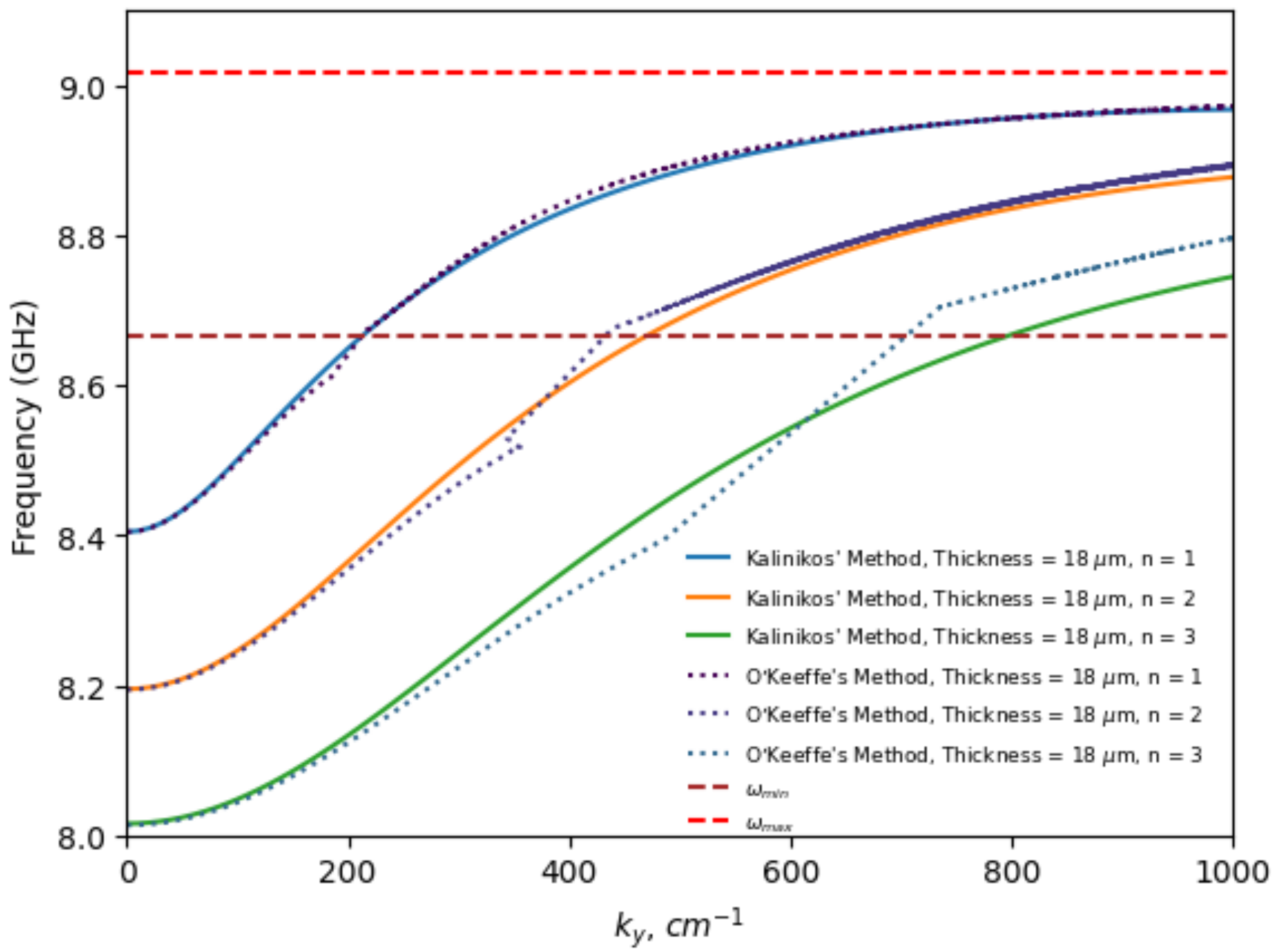

**Supplementary Figure 2**. Comparison of the 18 µm thick YIG dispersion relationship between Kalinikos' method and O'Keefee's method. The YIG waveguide width is 200 µm. The applied magnetic field is 2325 Gauss.

# Supplementary Note #2: Thick YIG Fabrication Process

Previous studies have primarily utilized conventional MSSW filters fabricated by patterning metal transducers, either through lift-off or subtractive etching, on top of etched YIG cavities [5, 6]. Typically, the YIG film is first patterned using wet etching or ion milling. If etching is used, a metal layer is deposited on the YIG surface, followed by photolithography to define the metal pattern, and then metal etching. Alternatively, in a lift-off process, a photoresist is first spin-coated, the metal is deposited, and then the excess is lifted off. These methods work well for thin YIG films, typically around 3 μm thick. However, when scaling up to thicker YIG films—such as 18 μm—the process encounters several challenges.

**Supplementary Figure 3** presents an example of a device where the conventional process was successful, though the overall fabrication yield was extremely low. The steep sidewalls of the thick YIG waveguide are clearly visible in the images. To aid aluminum (visible as black regions) in conformally coating the YIG sidewalls, diamond-shaped features were added near the waveguide edges. These structures help guide the metal over the steep sidewalls and prevent breakage during the etching process, which improves the yield slightly. In this successful device, the YIG sidewalls appear clean, indicating complete removal of aluminum from undesired regions. Additionally, the width of the aluminum transducers are increased to 20 μm, which contributed to a slight improvement in fabrication yield.

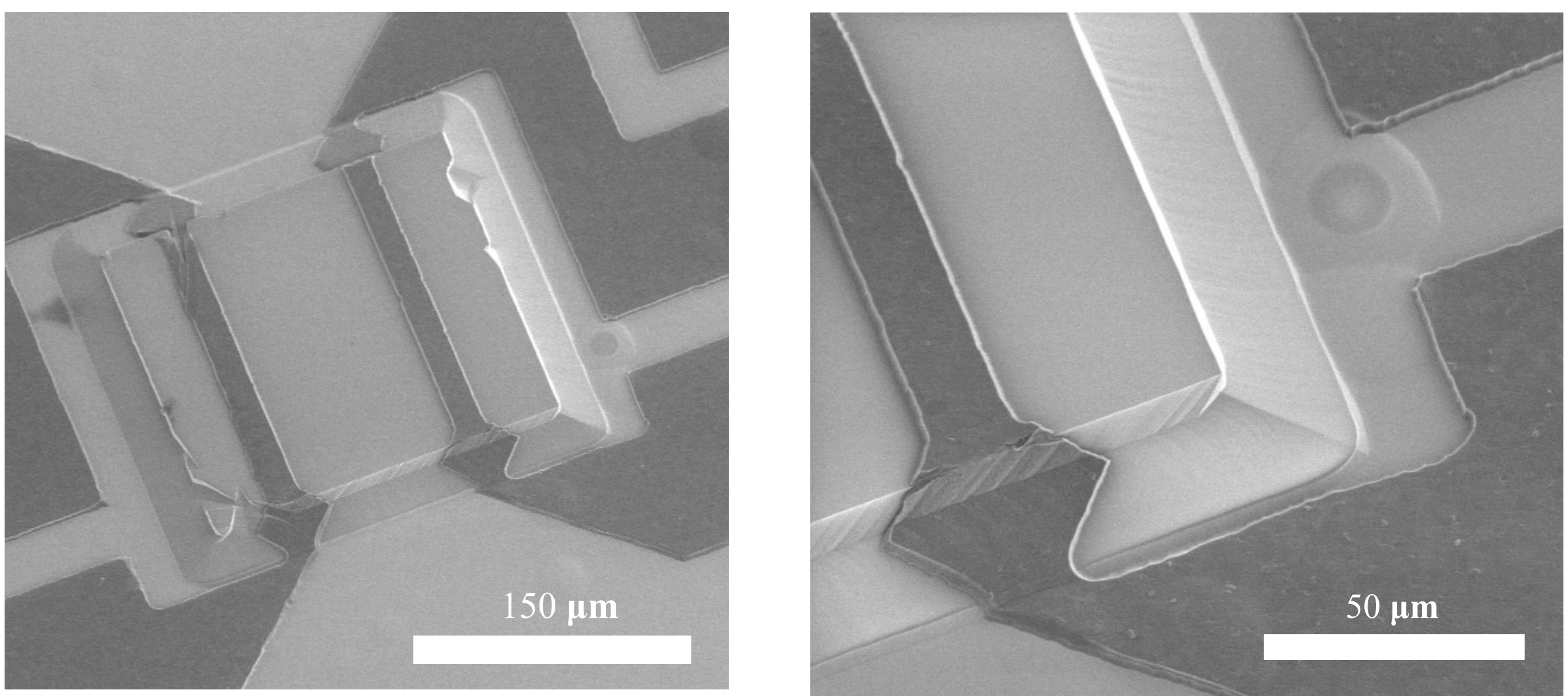

(a) (b)

**Supplementary Figure 3. (a-b)** Scanning electron microscope (SEM) image of a successful thick YIG filter using a previous fabrication process without BCB. The thickness of the YIG is 15 μm. The aluminum transducer width is 20 μm. The width of the YIG is 200 μm and the length is 220 μm. The pitch of the transducer is 110 μm.

**Supplementary Figure 4** shows another example device where the conventional fabrication method failed—a common outcome for most devices using this approach. In this case, the aluminum layer (black regions) barely connects to the aluminum deposited on top of the YIG waveguide. This poor connection introduces a large series resistance in the aluminum transducers, significantly hindering efficient excitation of the MSSW. Additionally, a substantial amount of residual aluminum remains on the YIG sidewalls. This not only causes electrical shorting between the input and output transducers—thereby degrading out-of-band rejection—but also obstructs current flow into the transducer located on top of the YIG.

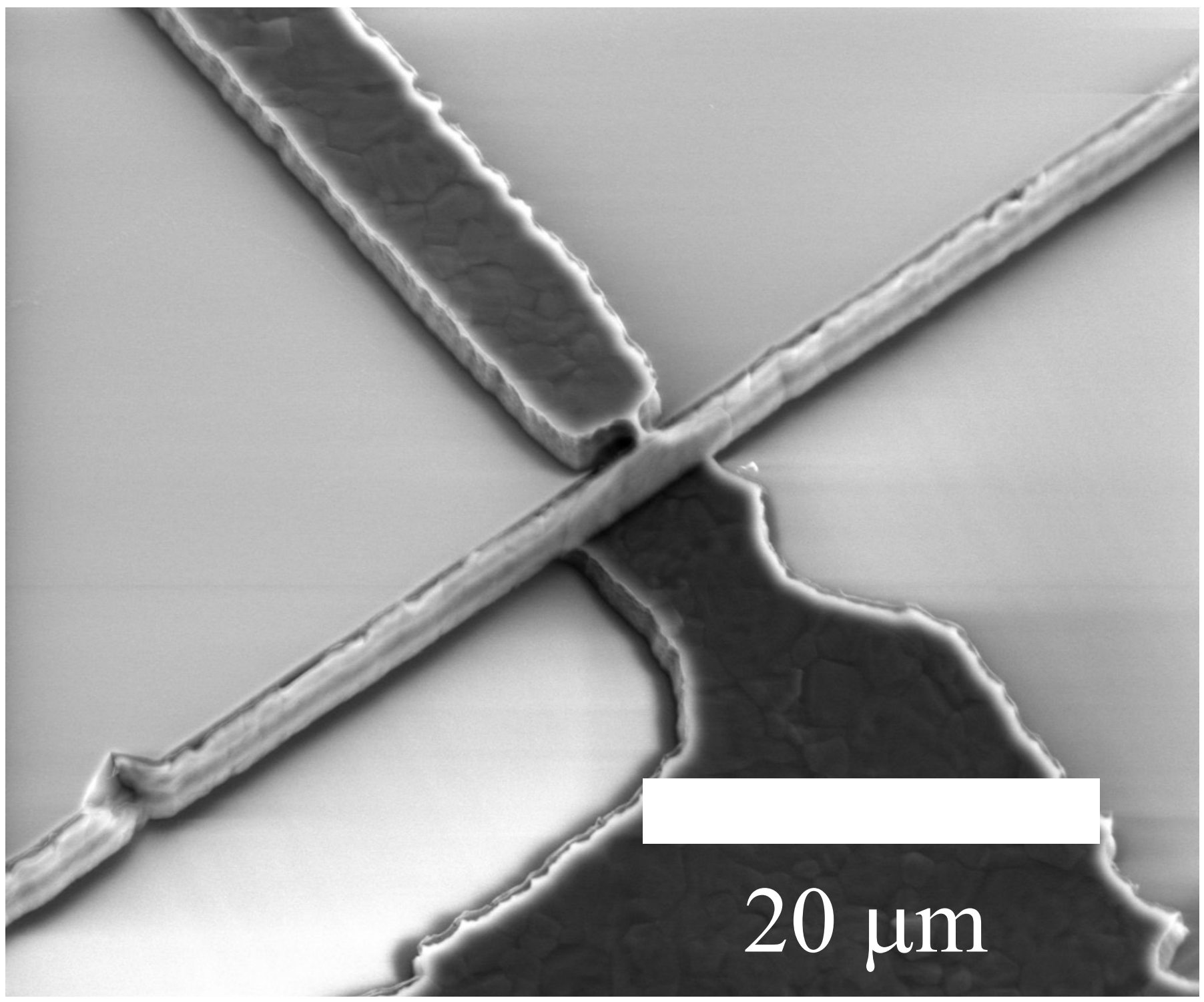

**Supplementary Figure 4.** SEM image of a failed YIG filter. The aluminum layer (dark regions) makes poor contact with the aluminum deposited on top of the YIG waveguide, leading to a weak electrical connection. Significant residual aluminum is also visible on the YIG sidewalls, contributing to electrical shorts and degraded device performance.

The key to the previously described problem lies in the steep sidewall of the YIG waveguide. A KLA Tencor P7 2D profilometer was used to measure the YIG sidewall before and after Benzocyclobutene (BCB) patterning and annealing, as shown in the **Supplementary Figure 5.** The use of BCB changes the 15 μm steep sidewall into a much more gradual sidewall with a height difference around 7 μm.

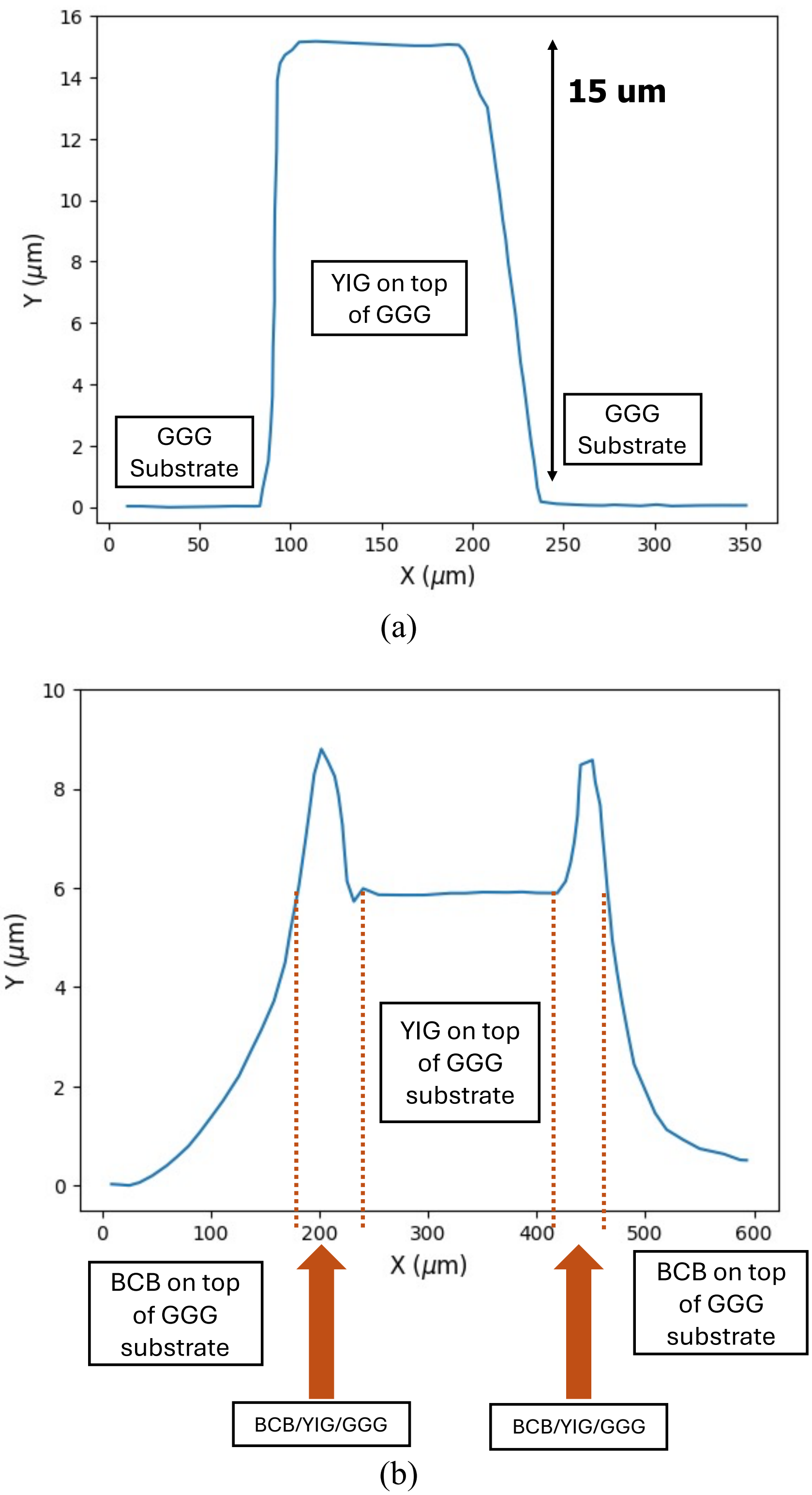

**Supplementary Figure 5**. Profilometer measurement of the YIG **(a)** before BCB spin coating and **(b)** after BCB patterning and annealing.

**Supplementary Figure 6** compares the frequency responses of two YIG filters: one fabricated with BCB planarization and the other using the previous method without any BCB layers. The fabrication process for YIG filters with BCB planarization shows significantly higher yield. Both devices share identical transducer layouts and YIG geometries, and they exhibit nearly identical frequency responses. This indicates that the introduction of the BCB layer does not contribute to additional RF loss or alter MSSW propagation. Previous studies have demonstrated that BCB is an excellent polymer for high-frequency RF applications, offering a minimal loss tangent [7].

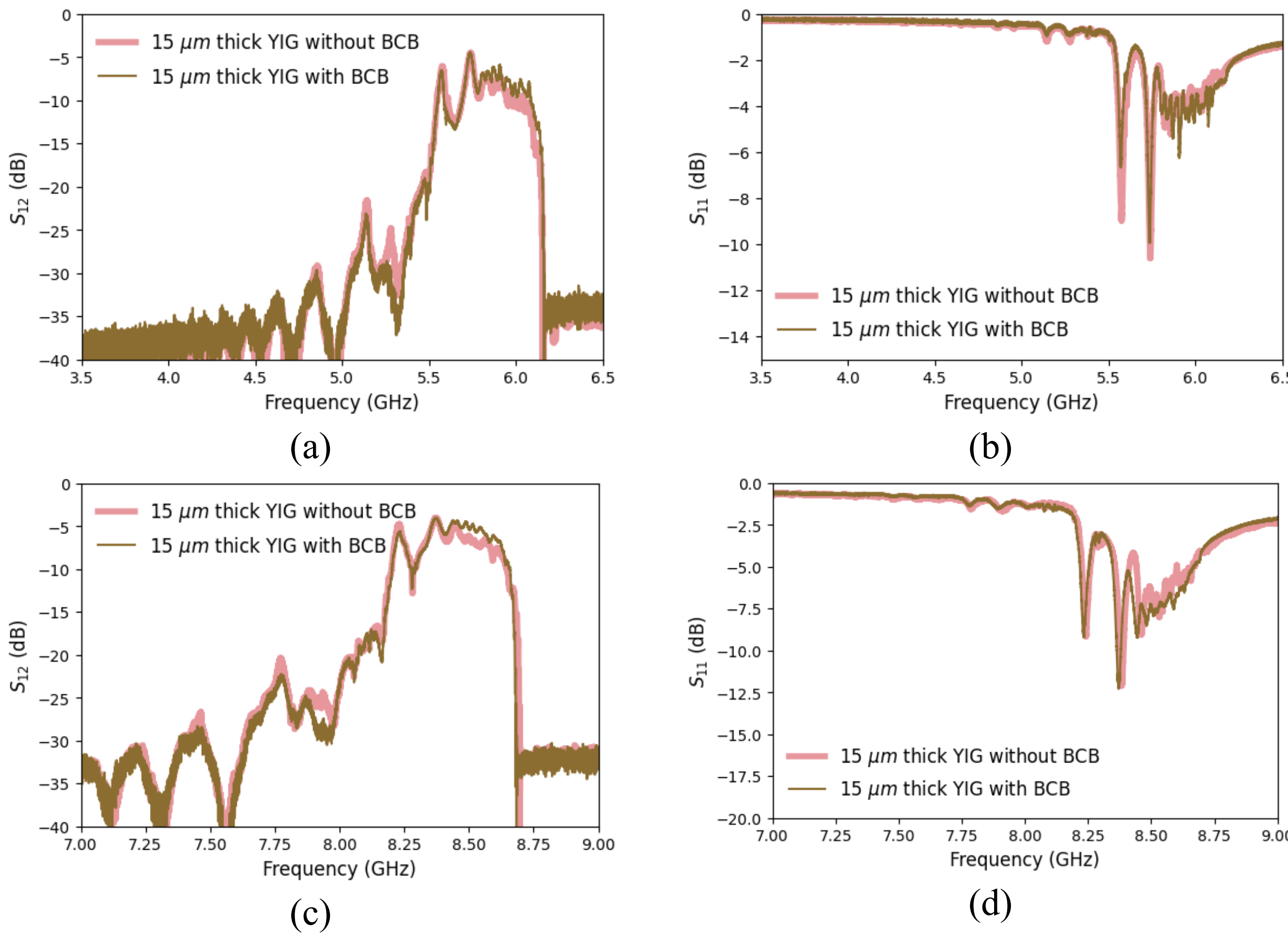

**Supplementary Figure 6**. Comparison of measured **(a and c)** $S_{12}$ and **(b and d)** $S_{11}$ frequency responses between YIG filters with and without the BCB planarization technique. The aluminum transducer width is 20 μm. The width of the YIG is 200 μm and the length is 220 μm. The pitch of the transducer is 110 μm. The thickness of the YIG is 15 μm. The applied magnetic field is around **(a and b)** 1500 Gauss, and **(c and d)** 2500 Gauss.

Overall, this fabrication process not only boosted the overall yield for the 18 μm thick film YIG filter but also facilitates more intricate transducer designs, such as meander-line transducers, which require the aluminum transducer traversing the YIG waveguide multiple times.

## Supplementary Note #3: Comparison of YIG Filters with 3 µm and 18 µm Thicknesses

**Supplementary Figure 7** compares the frequency responses of YIG filters with thicknesses of 3 µm and 18 µm tuned to three different frequencies. Both filters share the same waveguide width of 200 µm, length of 140 µm, and transducer pitch of 70 µm representing the center-to-center distance between the two aluminum transducers. A straight-line aluminum transducer design is used in both cases. For the 3 µm thick YIG filter, the aluminum transducer width is 5 µm, while a wider 10 µm transducer is used for the 18 µm thick YIG filter. The narrower transducer in the 3 µm filter helps to reduce insertion loss, particularly at 6 GHz. Additionally, increasing the YIG waveguide width or length may further reduce insertion loss, as seen in **Supplementary Figure 6**, where a shorter length of 110 µm contributed to improved performance. Overall, both filters demonstrate comparable insertion losses, especially at higher frequencies. However, the 18 µm thick YIG filter exhibits a much sharper filter skirt at the upper frequency edge, attributed to differences in their MSSW dispersion characteristics.

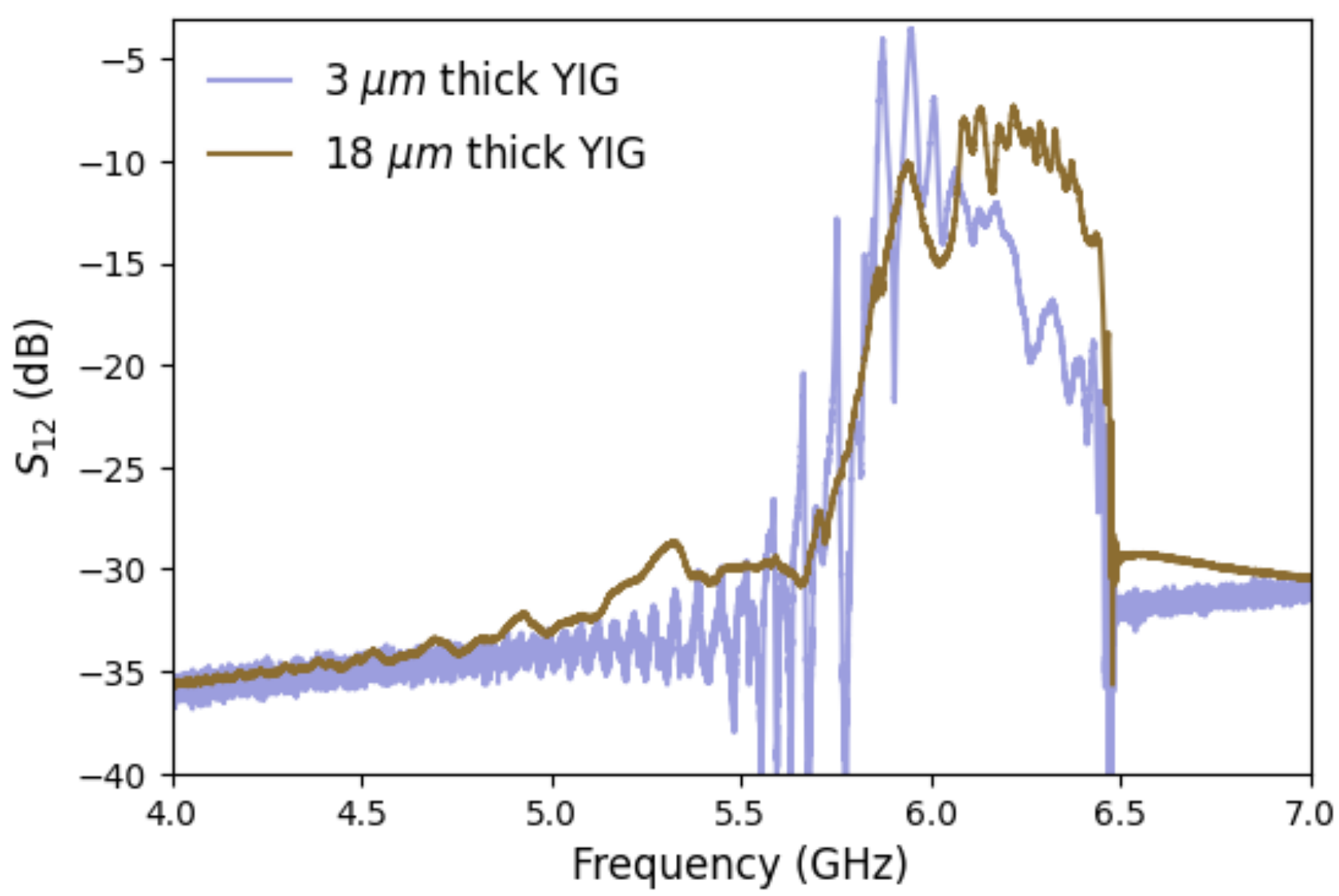

(a)

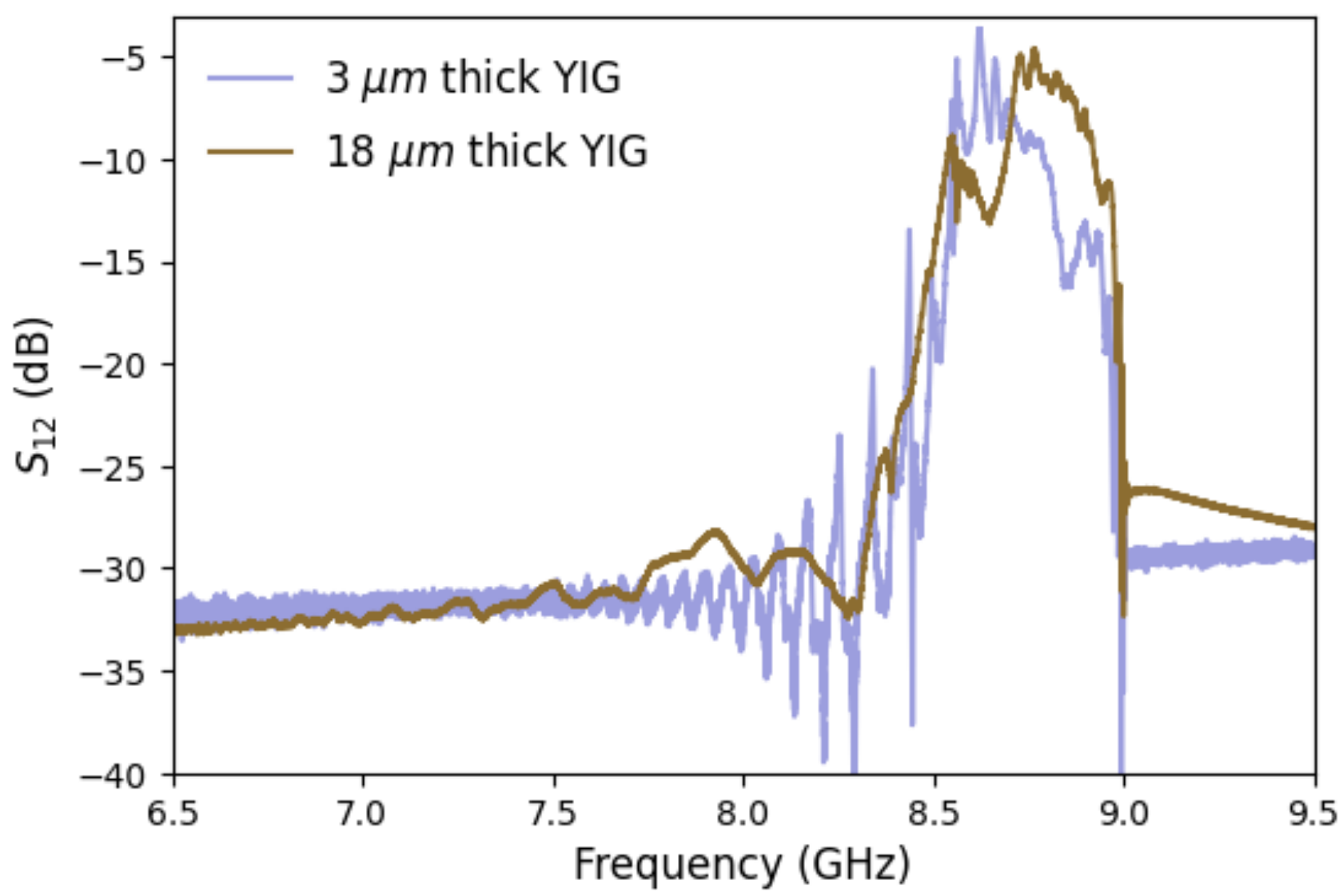

(b)

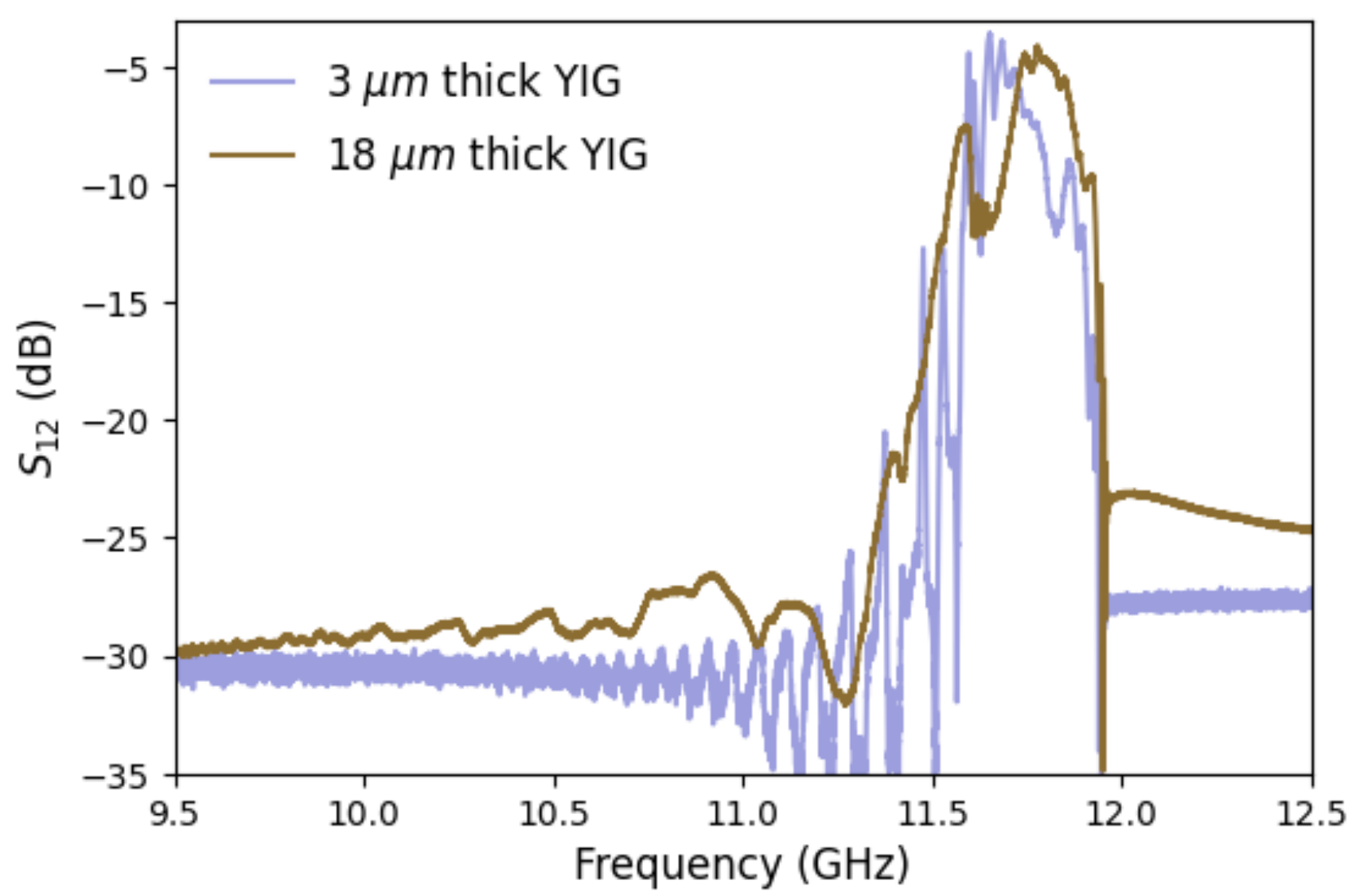

(c)

**Supplementary Figure 7**. Comparison of measured $S_{12}$ frequency responses between the 3 μm and 18 μm thick YIG filters under applied magnetic flux densities of approximately **(a)** 1500 Gauss, **(b)** 2500 Gauss, and **(c)** 3500 Gauss.

## Supplementary Note #4: Comparison of YIG Filters with Straight Line and Meander-Line Transducers

**Supplementary Figures 8, 9, and 10** present a comparative analysis of the frequency responses of 18 µm thick YIG filters featuring straight-line and meander-line transducers under three different applied magnetic flux densities of 1500, 2500, 3500 Gauss, respectively. Both filters share the same waveguide width of 200 µm, transducer pitch of 70 µm, and transducer width of 10 µm. The waveguide length of the meander-line YIG filter is 420 µm, as shown in the **Fig. 3b.** The waveguide length of the straight-line YIG filter is 140 µm, as shown in the **Fig. 2c.** The straight-line YIG filter is the same one detailed in the **Supplementary Note #3.**

At 6 GHz, the meander-line YIG filter exhibits superior performance, with an insertion loss of 4.4 dB compared to 7.4 dB for the straight-line configuration. The maximum $Z_{11}$ values are 40.9 dB and 33.2 dB for the meander-line and straight-line YIG filters, respectively. The corresponding minimum return losses are 21.5 dB and 10.4 dB for meander-line and straight-line YIG filters, respectively. Smith chart analysis (**Supplementary Figure 8c**) shows that the meander-line filter provides a higher radiation impedance and improved impedance matching to a 50 Ω port impedance, contributing to its reduced insertion loss.

However, as frequency increases, radiation impedance also increases, as observed in the Smith plots of **Supplementary Figure 8c, 9c, and 10c**. This leads to over-coupling in the meander-line filter and results in greater insertion loss. At 11.8 GHz under an applied field of ~3500 Gauss, the meander-line filter exhibits a peak $Z_{11}$ of 47.4 dB, compared to 42.1 dB for the straight-line filter, leading to significant impedance mismatch with the 50 Ω filter terminations and a corresponding increase in insertion loss to 8.3 dB, versus 4.1 dB for the straight-line configuration.

**Supplementary Figure 11** highlights the improved nonreciprocity observed in the 18 µm thick YIG filters with meander-line transducers compared to their 3 µm thick counterparts. As shown in Fig. 3c, both filters exhibit distinct passbands in the forward transmission ($S_{12}$). However, in **Supplementary Figure 11**, the backward transmission ($S_{21}$) differs markedly: the 3 µm filter shows less than 5 dB attenuation, whereas the 18 µm filter exhibits strong suppression, indicating pronounced nonreciprocal behavior.

The difference arises from the underlying propagation mechanisms. In the 3 µm-thick YIG filter, the device functions similarly to a resonator, where the two straight edges act as reflective boundaries[6, 8]. The magnetostatic surface waves (MSSWs) are reflected back and forth between the edges, forming circulating wave patterns that yield nearly reciprocal transmission.

In contrast, the 18 µm-thick YIG filter no longer supports such resonant circulation. The spin waves excited at one port propagate primarily in a single direction and experience minimal reflection. One key factor is the increased sidewall roughness of the thicker YIG film, as visible in the optical micrographs (Fig. 2c and Fig. 2d). The roughened edges, which result from the greater etching depth required for thick YIG, act as scattering centers that suppress spin-wave reflection. This phenomenon is consistent with previous studies that used rough and lossy edge terminations [9-12], nonuniform magnetic fields [13], or additional spin wave absorbers [11, 14, 15] to minimize reflection and enhance nonreciprocity. Furthermore, increasing the film thickness enhances surface confinement and unidirectionality of MSSWs on a single surface, further improving nonreciprocal transmission.

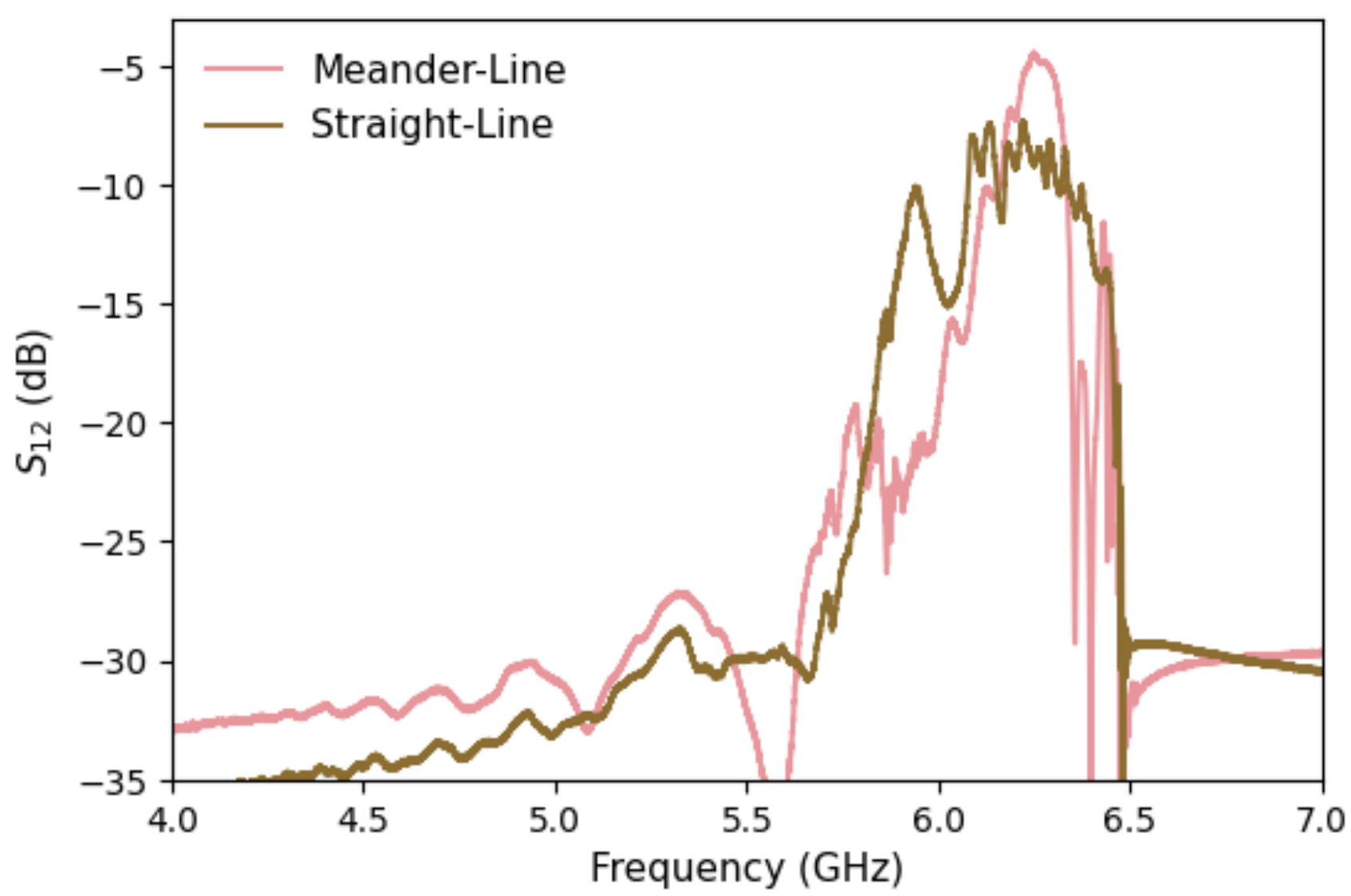

(a)

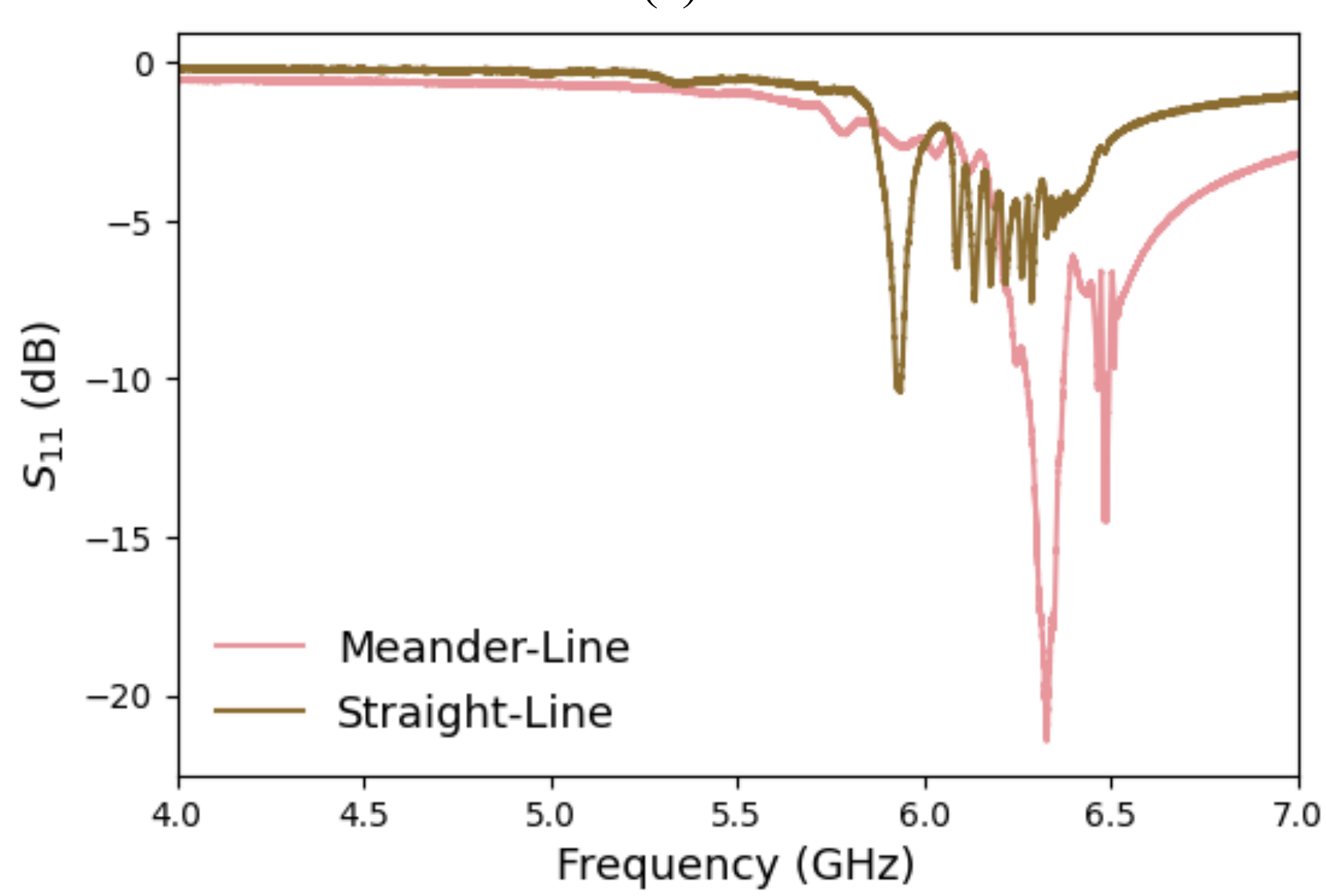

(b)

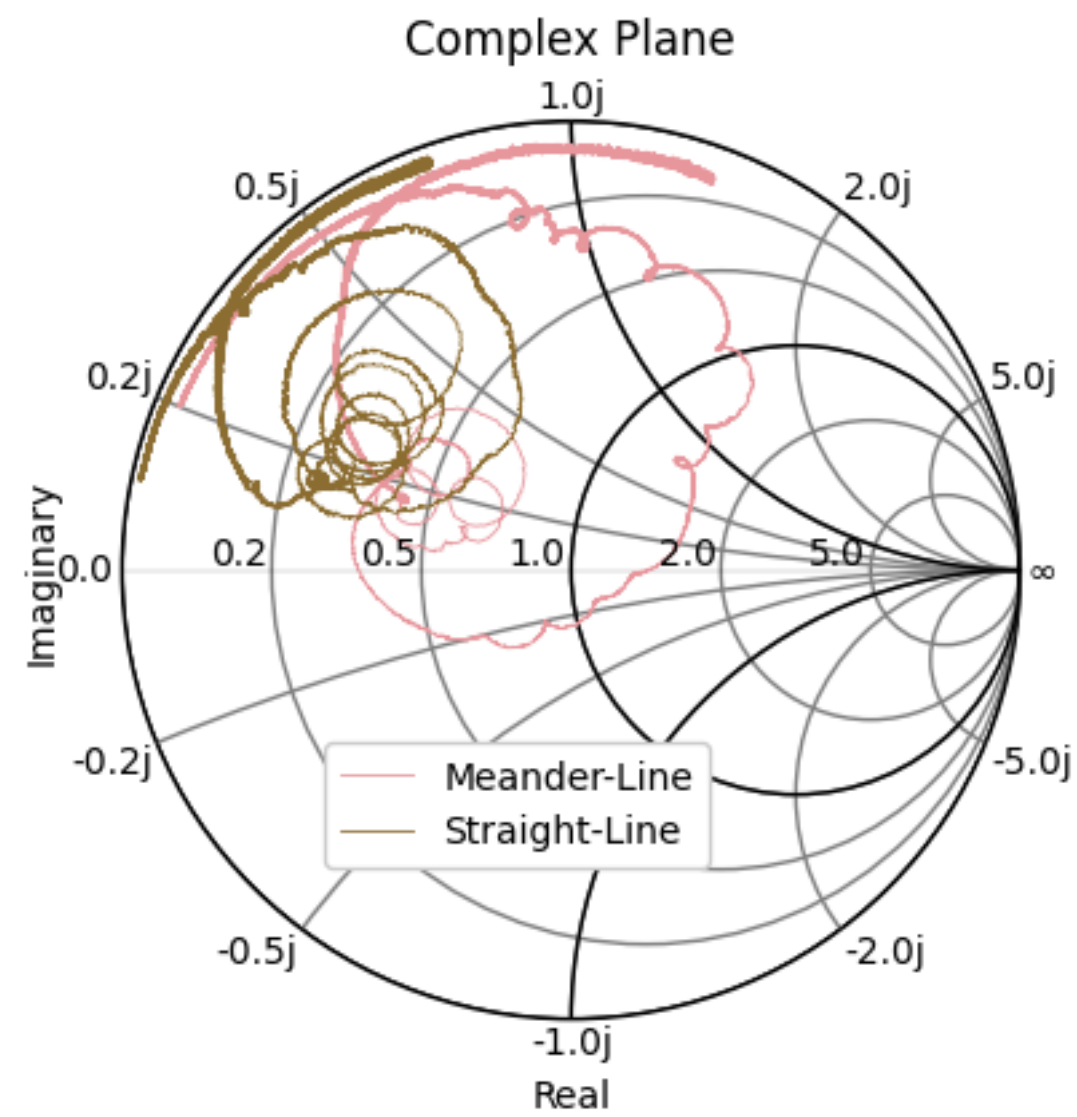

(c)

**Supplementary Figure 8**. Comparison of measured **(a)** $S_{12}$, **(b)** $S_{11}$, and **(c)** $S_{11}$ in Smith graph between meander-line and straight-line transducer YIG filters under applied magnetic flux densities of approximately 1500 Gauss.

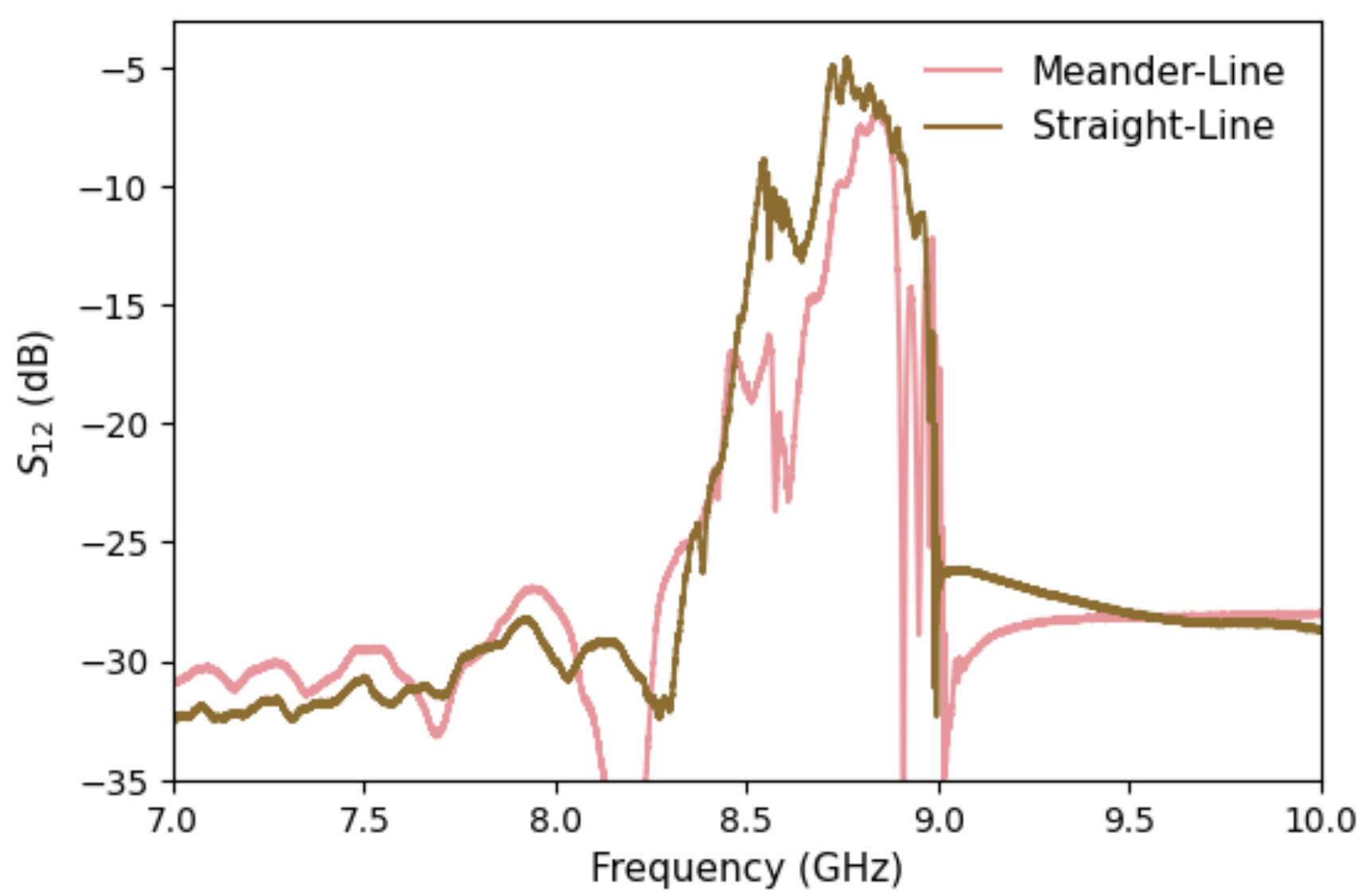

(a)

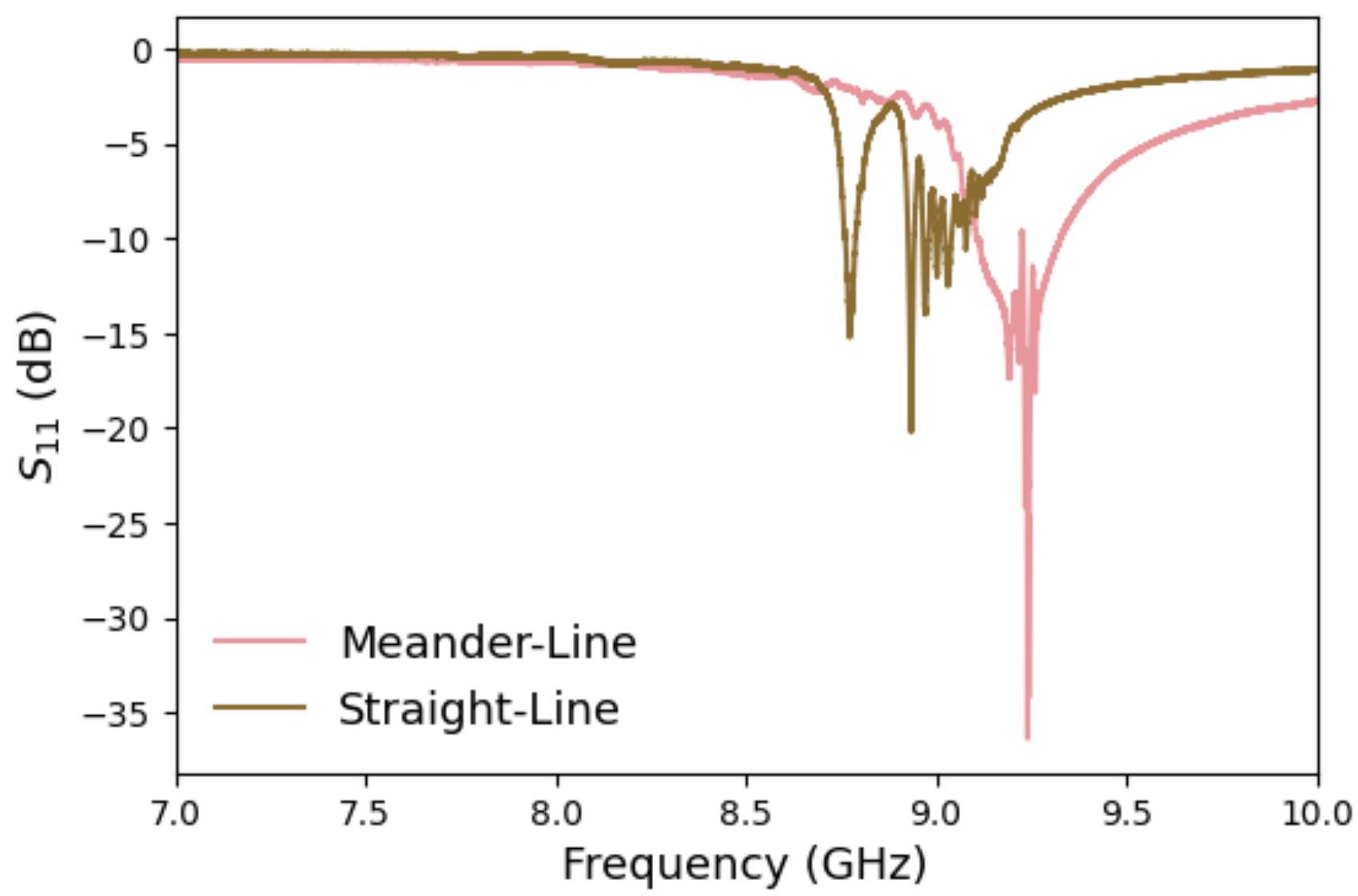

(b)

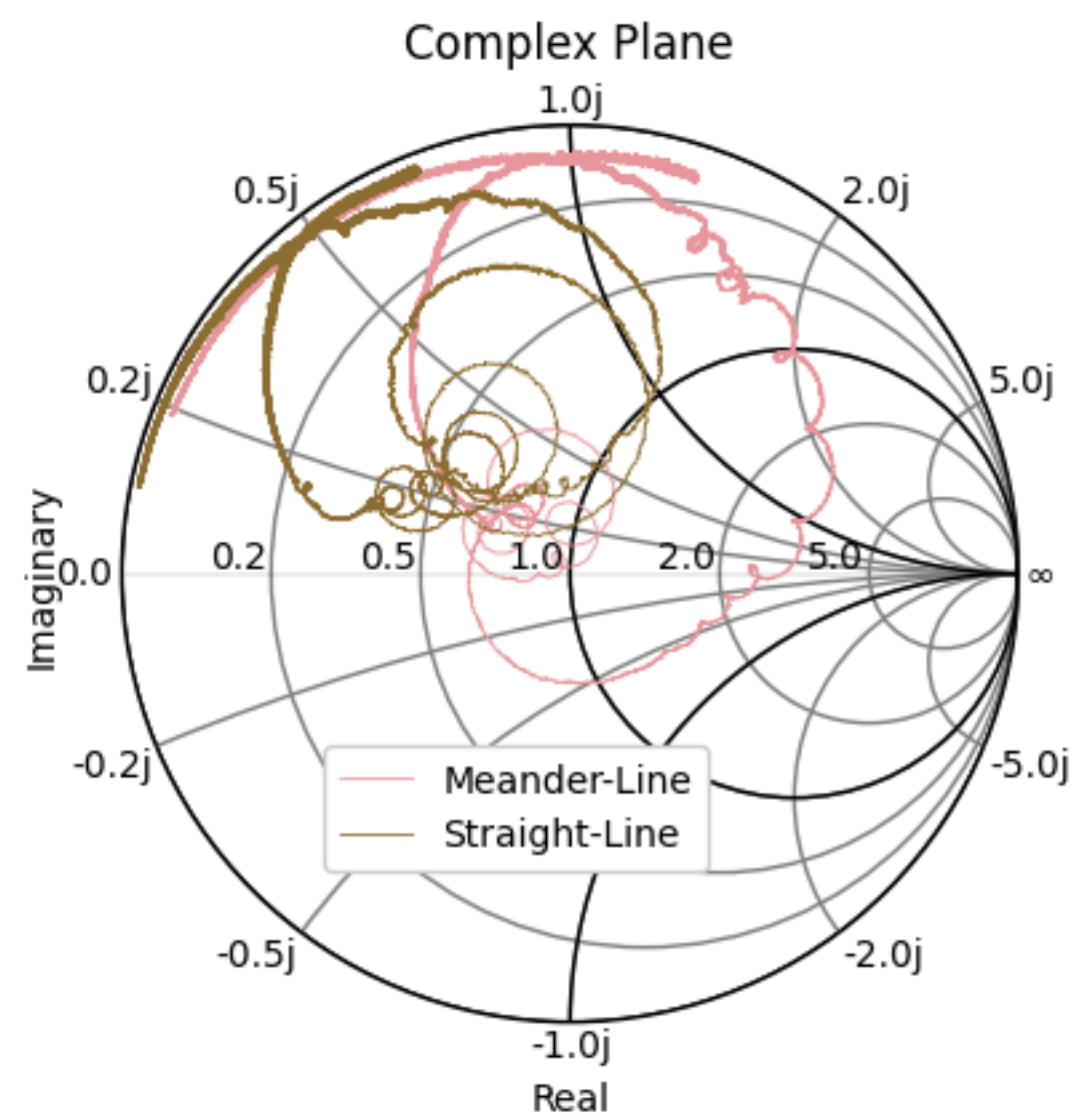

(c)

**Supplementary Figure 9.** Comparison of measured **(a)** $S_{12}$, **(b)** $S_{11}$, and **(c)** $S_{11}$ in Smith graph between meander-line and straight-line transducer YIG filters under applied magnetic flux densities of approximately 2500 Gauss.

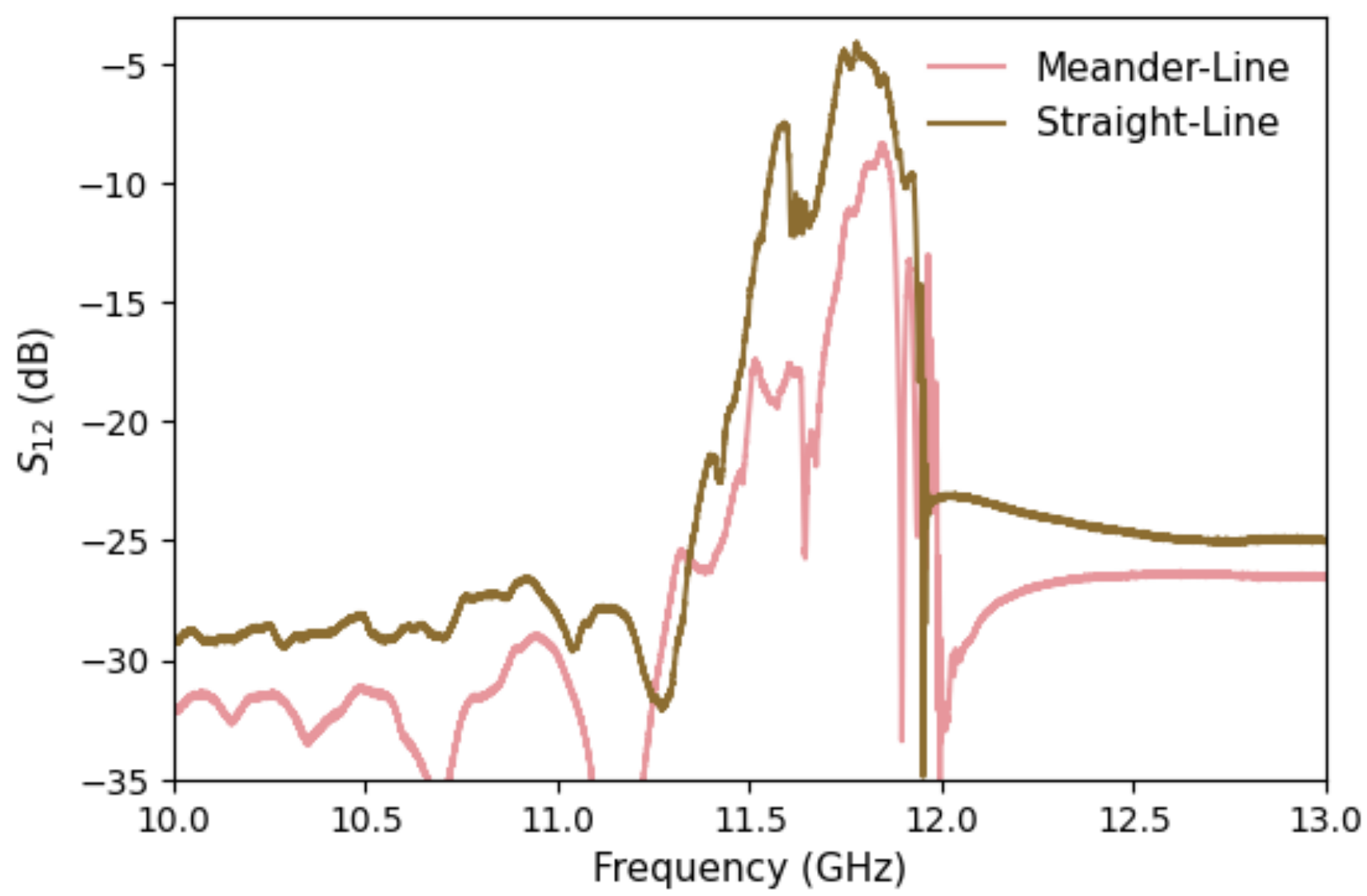
Meander-Line
Straight-Line
S12 (dB)
Frequency (GHz)
−5
−10
−15
−20
−25
−30
−35
10.0
10.5
11.0
11.5
12.0
12.5
13.0

(a)

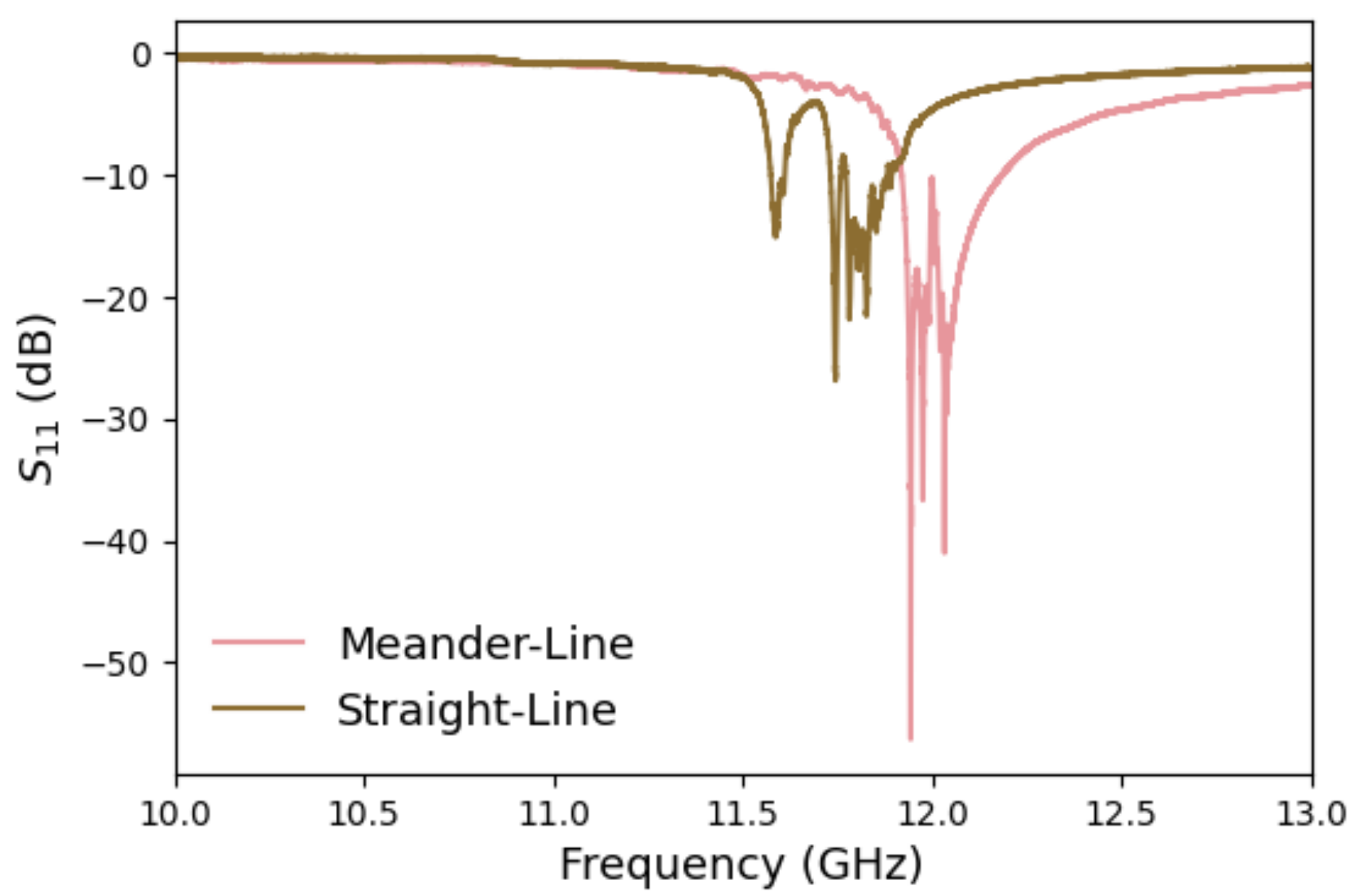
Meander-Line
Straight-Line
S11 (dB)
Frequency (GHz)
0
−10
−20
−30
−40
−50
10.0
10.5
11.0
11.5
12.0
12.5
13.0

(b)

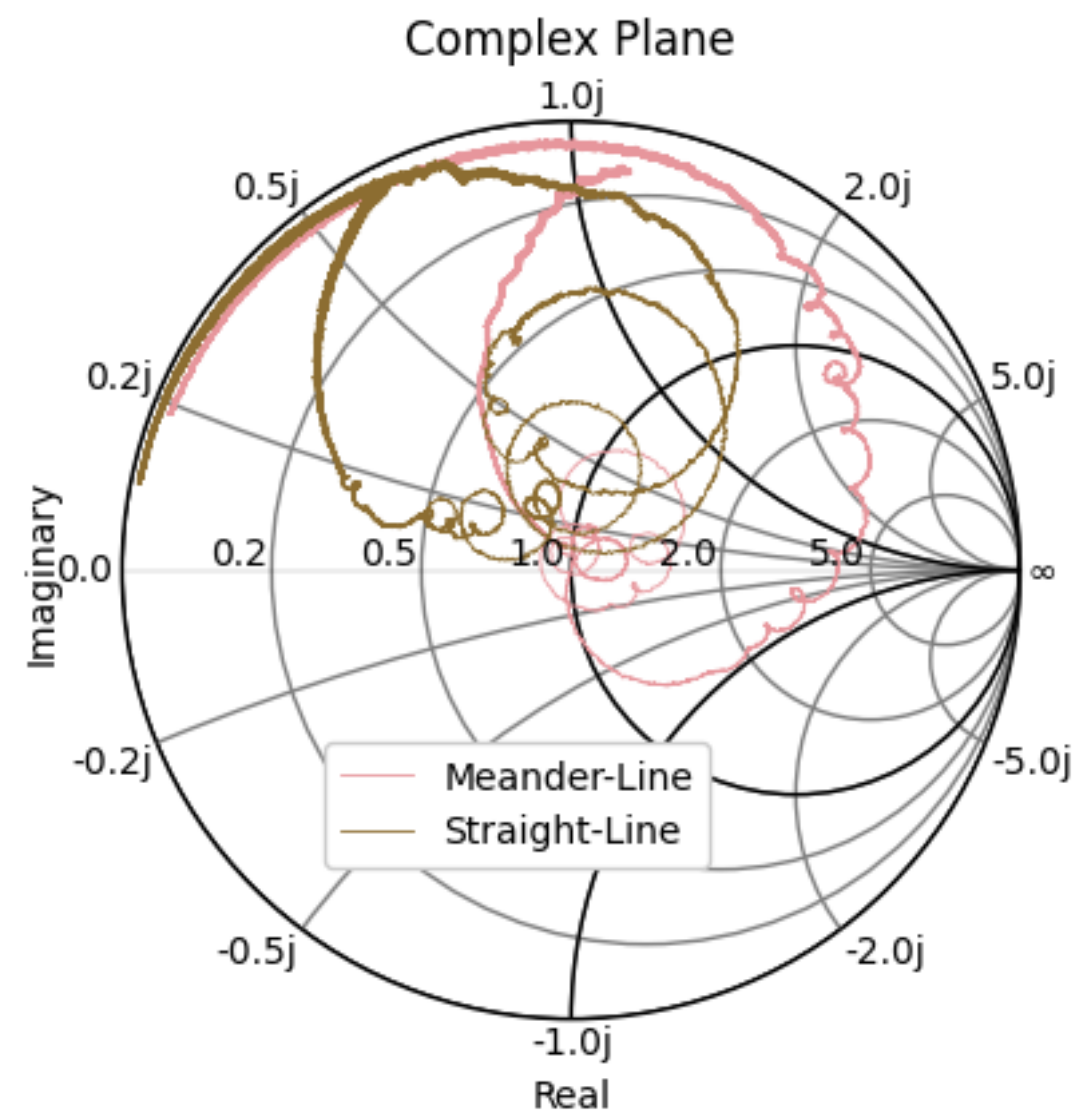

(c)

**Supplementary Figure 10.** Comparison of measured **(a)** $S_{12}$, **(b)** $S_{11}$, and **(c)** $S_{11}$ in Smith graph between meander-line and straight-line transducer YIG filters under applied magnetic flux density of approximately 3500 Gauss.

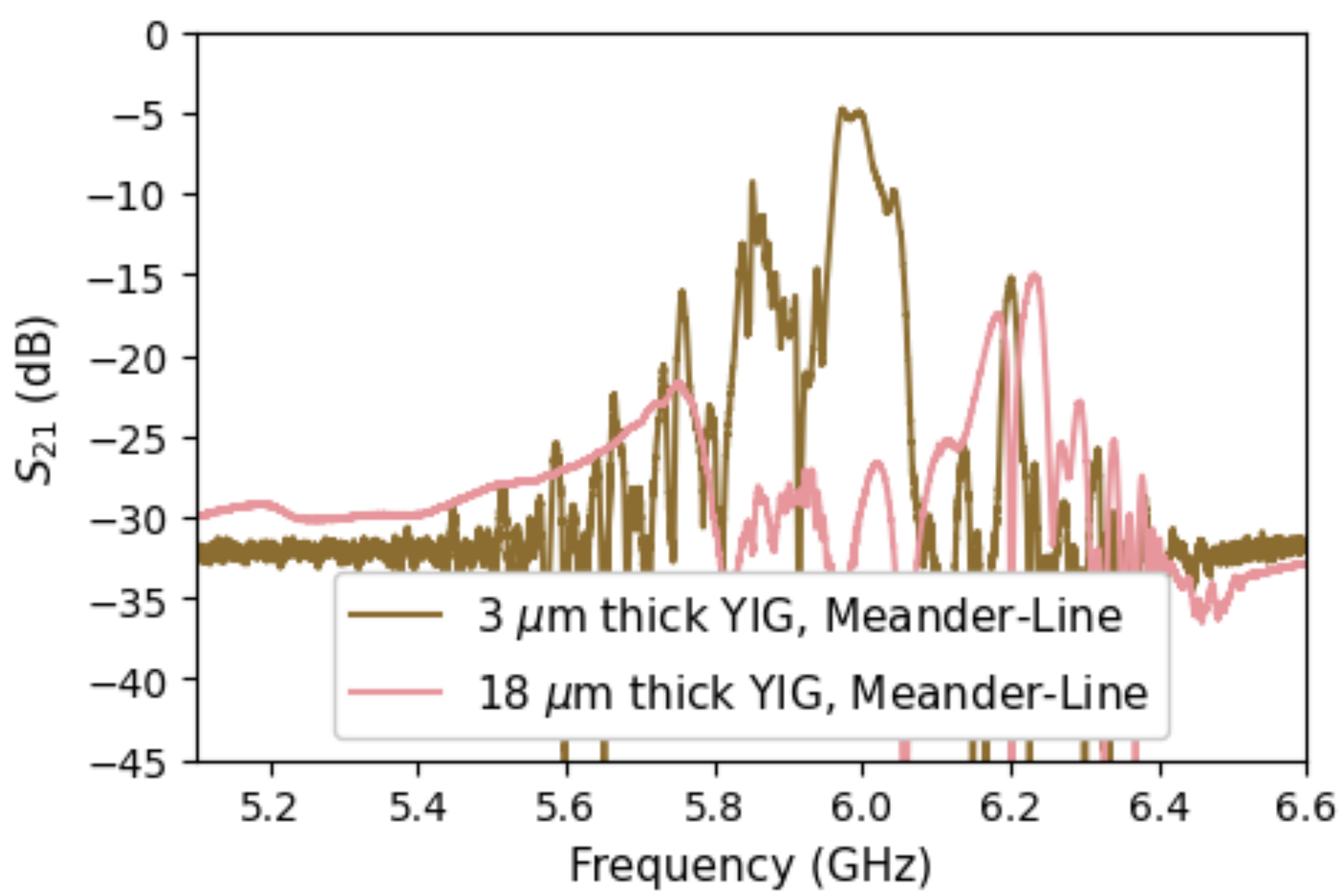

**Supplementary Figure 11.** Comparison of measured $S_{21}$ frequency responses between the 3 µm and 18 µm thick YIG meander-line transducer filters under applied magnetic flux densities of approximately 1500 Gauss. The $S_{12}$ frequency response comparison is shown in the Fig. 3 in the main text.

# Supplementary Note #5: YIG Filters with Two Parallel Transducers

**Supplementary Figure 12, 13, and 14** present a comparative analysis of the frequency responses of 18 µm thick YIG filters featuring one meander-line and two-parallel meander-line transducers under three different applied magnetic flux densities of 1500, 2500, 3500 Gauss, respectively. Both filters share an identical transducer pitch of 70 µm, transducer width of 200 µm, and YIG waveguide length of 420 µm. The two-parallel meander-line filter has a total width approximately double that of the single meander-line design (200 µm), plus the width of the central bridge section, as illustrated in **Fig. 3e.** The single meander-line transducer YIG filter is the same one detailed in **Supplementary Note #4.**

At 6.3 GHz, the inclusion of a parallel transducer branch reduces the peak $Z_{11}$ from 40.9 dB to 33.5 dB, improving impedance matching and decreasing the insertion loss from 4.4 dB to 2.9 dB. This better matching is confirmed by the Smith chart analysis in **Supplementary Fig. 12c.** At higher frequencies, this benefit becomes even more significant. For instance, at 11.8 GHz, the insertion loss drops from 8.3 dB (single meander-line) to 3.8 dB (two-parallel), due to better impedance matching, as seen in **Supplementary Figure 13c and 14c.**

However, a drawback of the parallel configuration is the emergence of pronounced spurious modes at slightly lower frequencies—approximately 5.7 GHz (1500 Gauss), 8.7 GHz (2500 Gauss), and 11.5 GHz (3500 Gauss). These undesired responses are likely introduced by the center bridge section, which can also excite magnetostatic backward volume waves (MSBVWs).

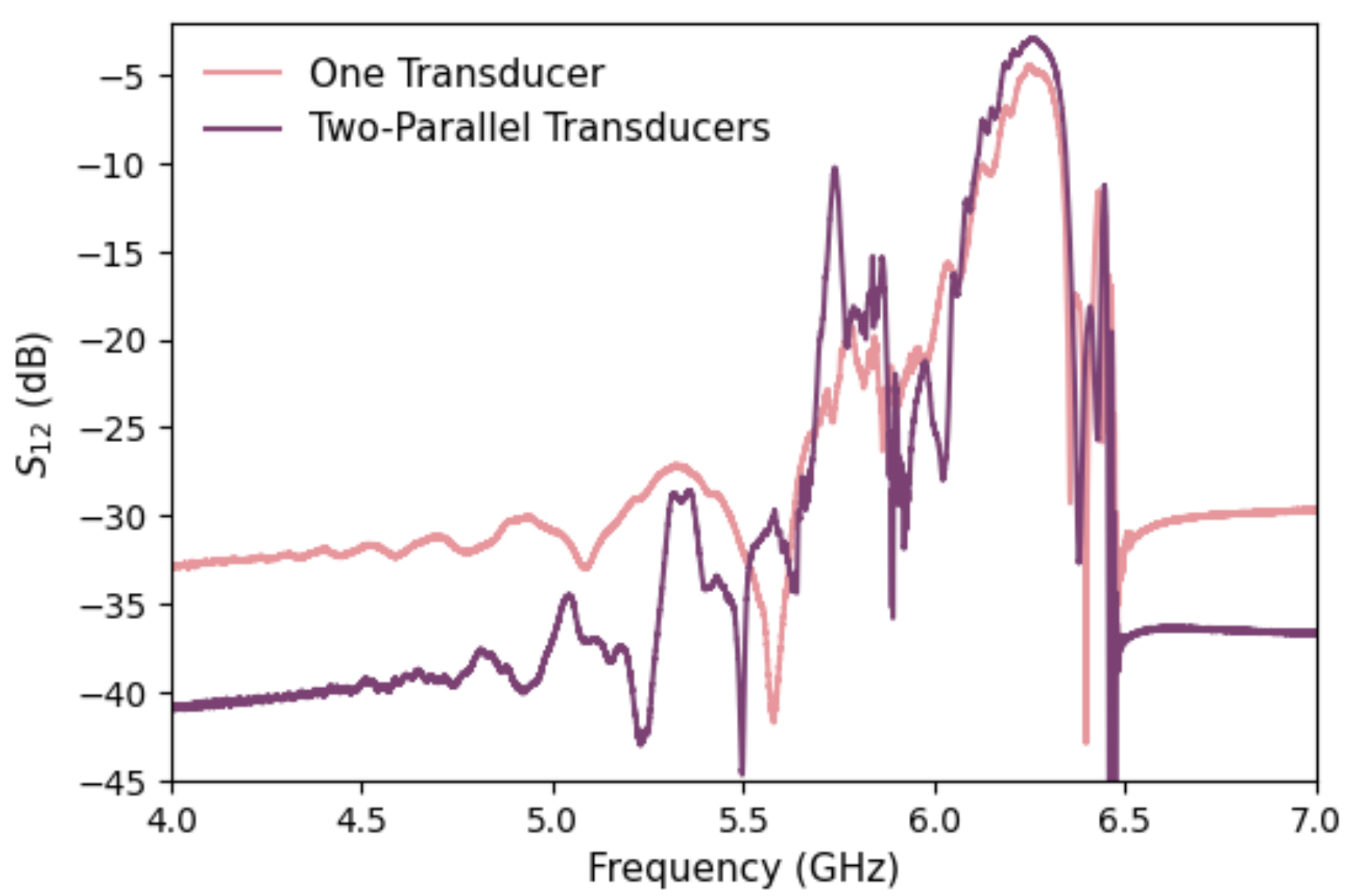

(a)

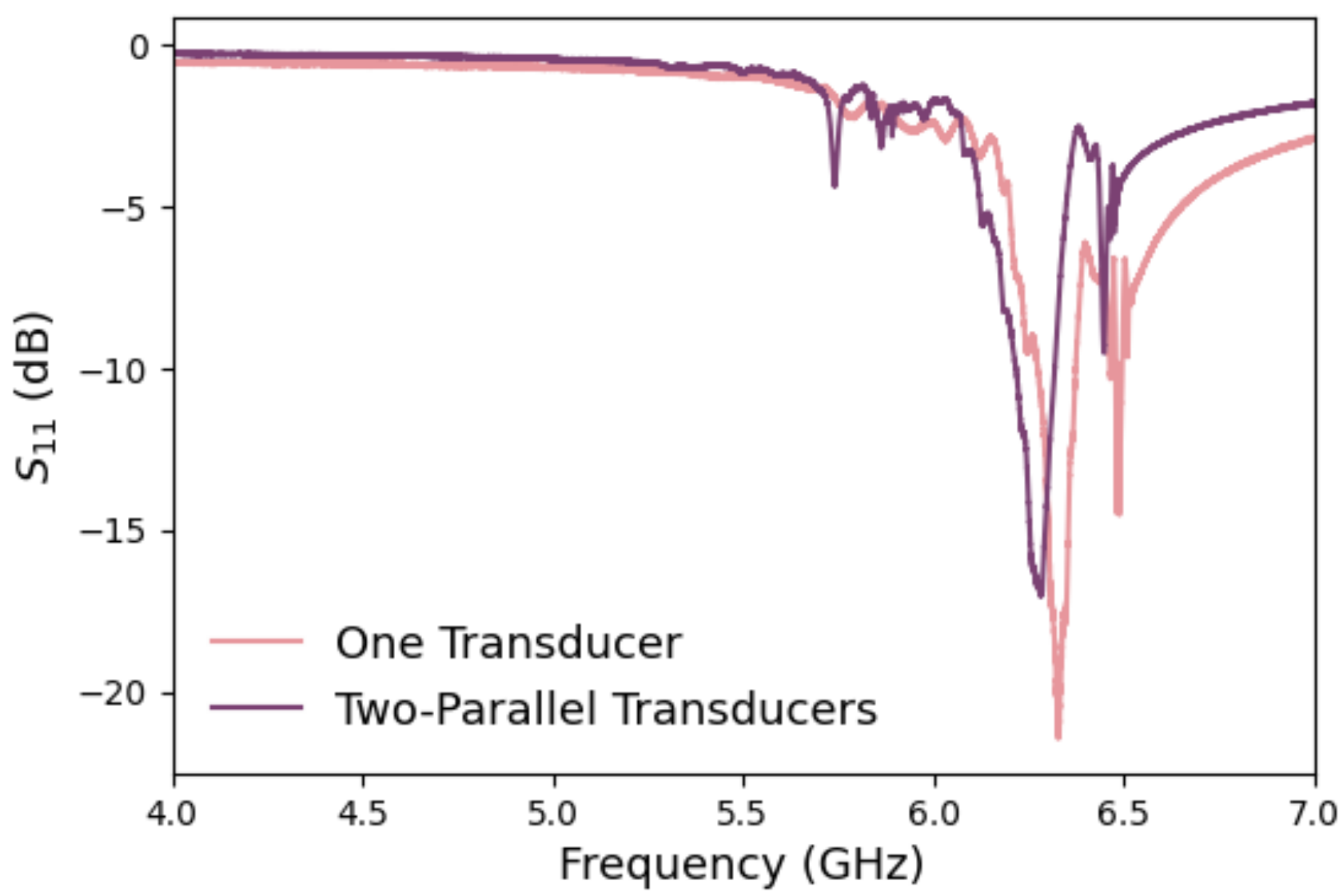

(b)

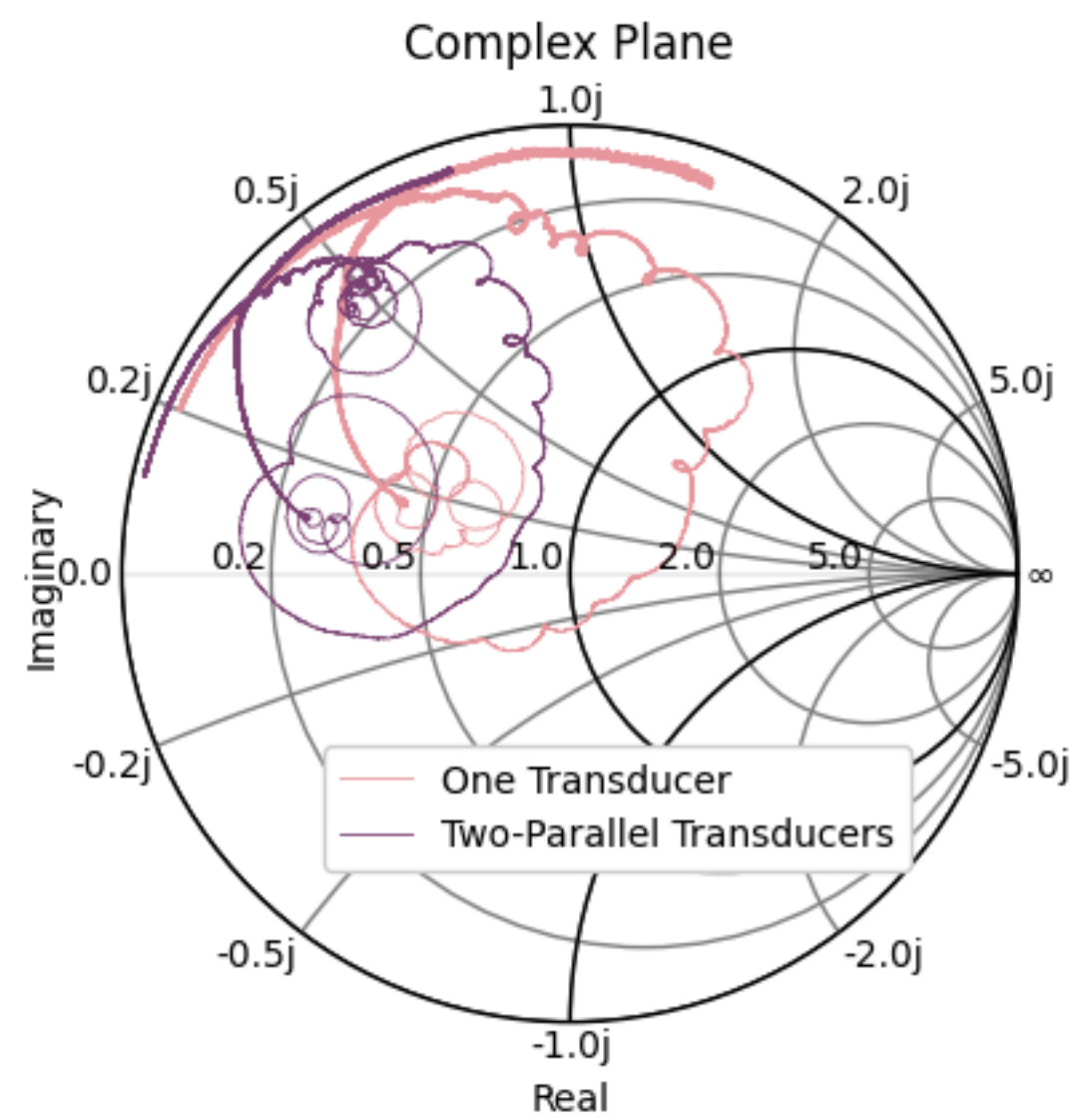

(c)

**Supplementary Figure 12**. Comparison of measured **(a)** $S_{12}$, **(b)** $S_{11}$, and **(c)** $S_{11}$ in Smith graph between two parallel meander-line transducer and a single meander-line transducer YIG filters under applied magnetic flux densities of approximately 1500 Gauss.

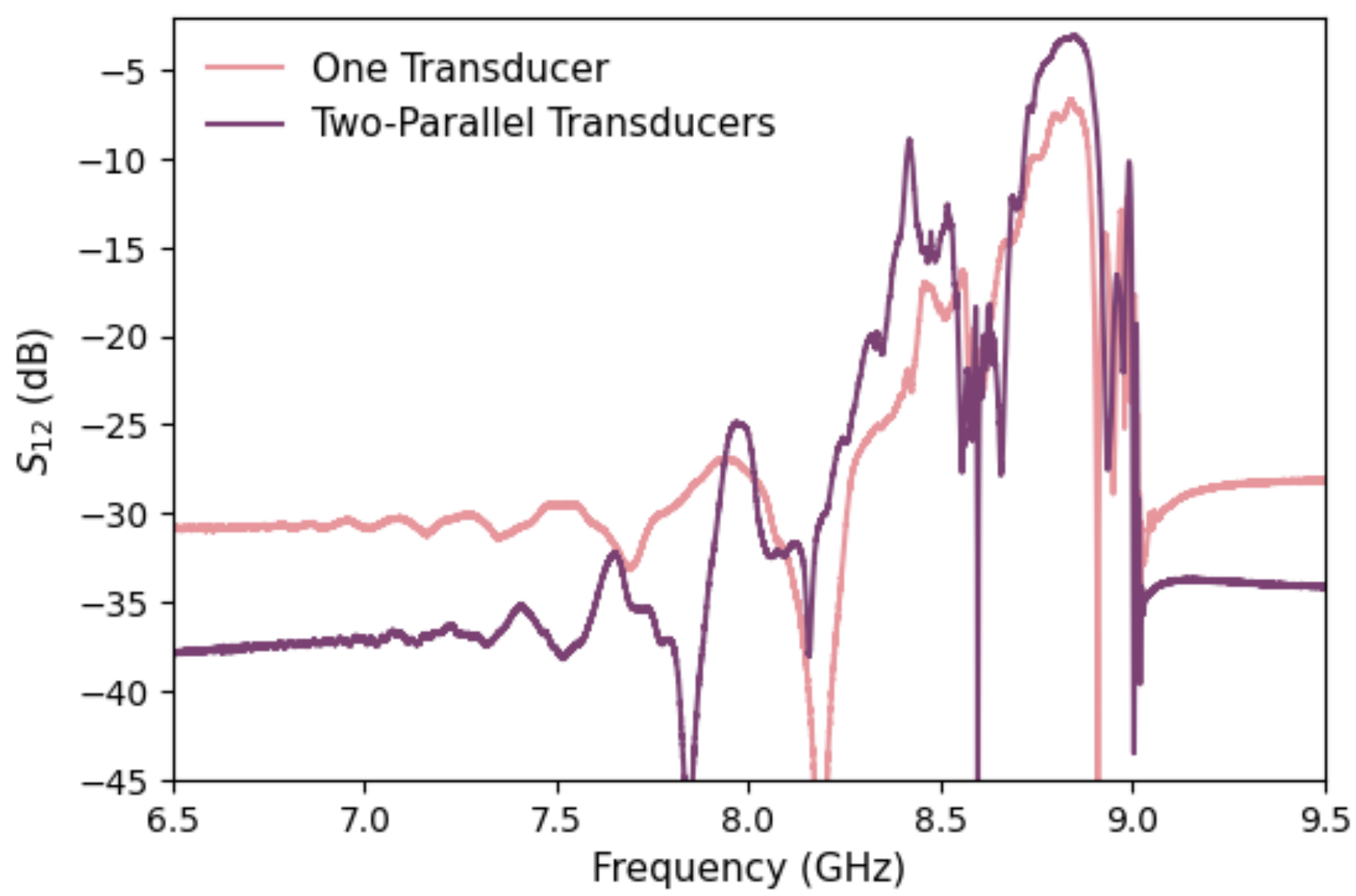
One Transducer
Two-Parallel Transducers
$S_{12}$ (dB)
−5
−10
−15
−20
−25
−30
−35
−40
−45
6.5
7.0
7.5
8.0
8.5
9.0
9.5
Frequency (GHz)

(a)

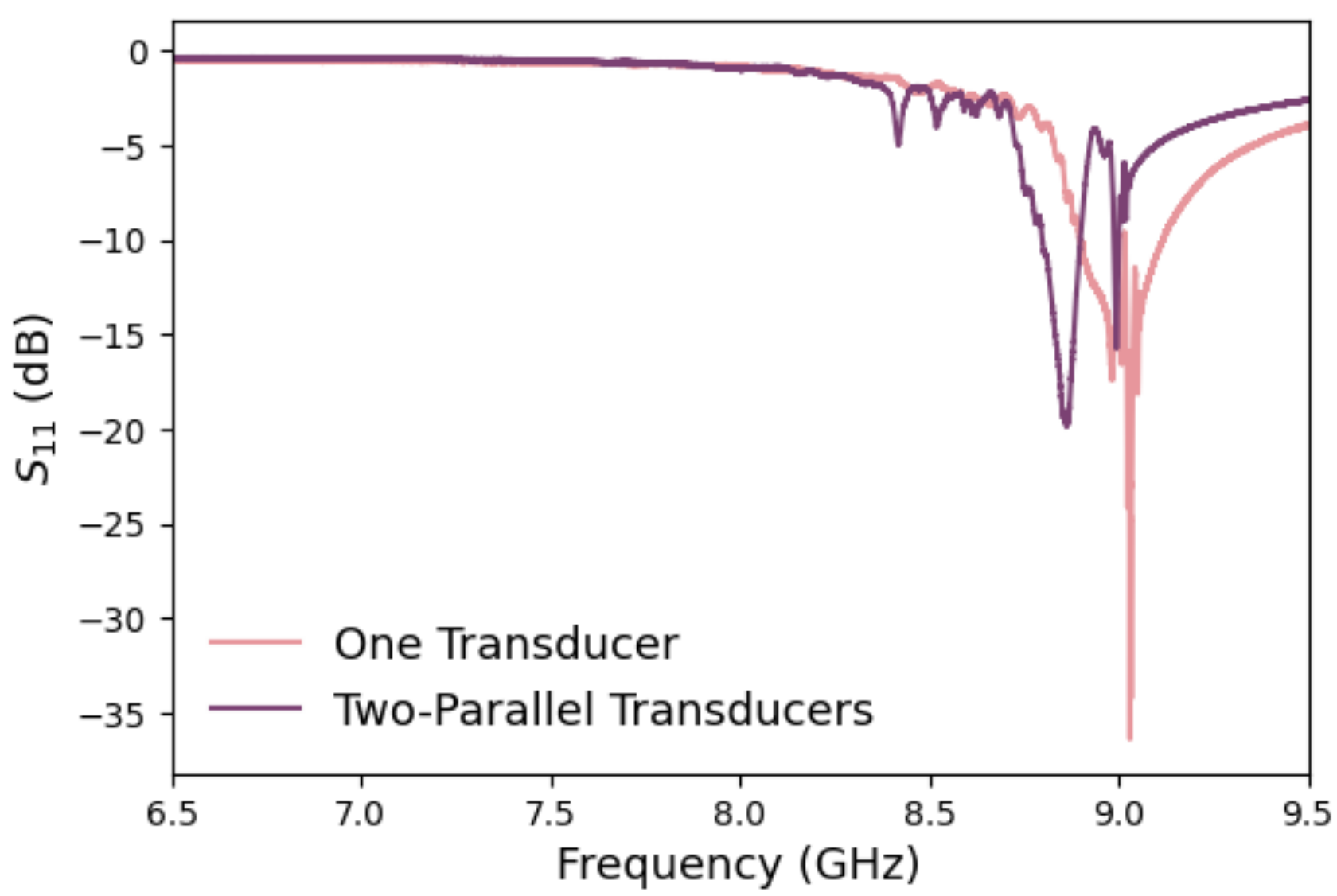
One Transducer
Two-Parallel Transducers
$S_{11}$ (dB)
0
−5
−10
−15
−20
−25
−30
−35
6.5
7.0
7.5
8.0
8.5
9.0
9.5
Frequency (GHz)

(b)

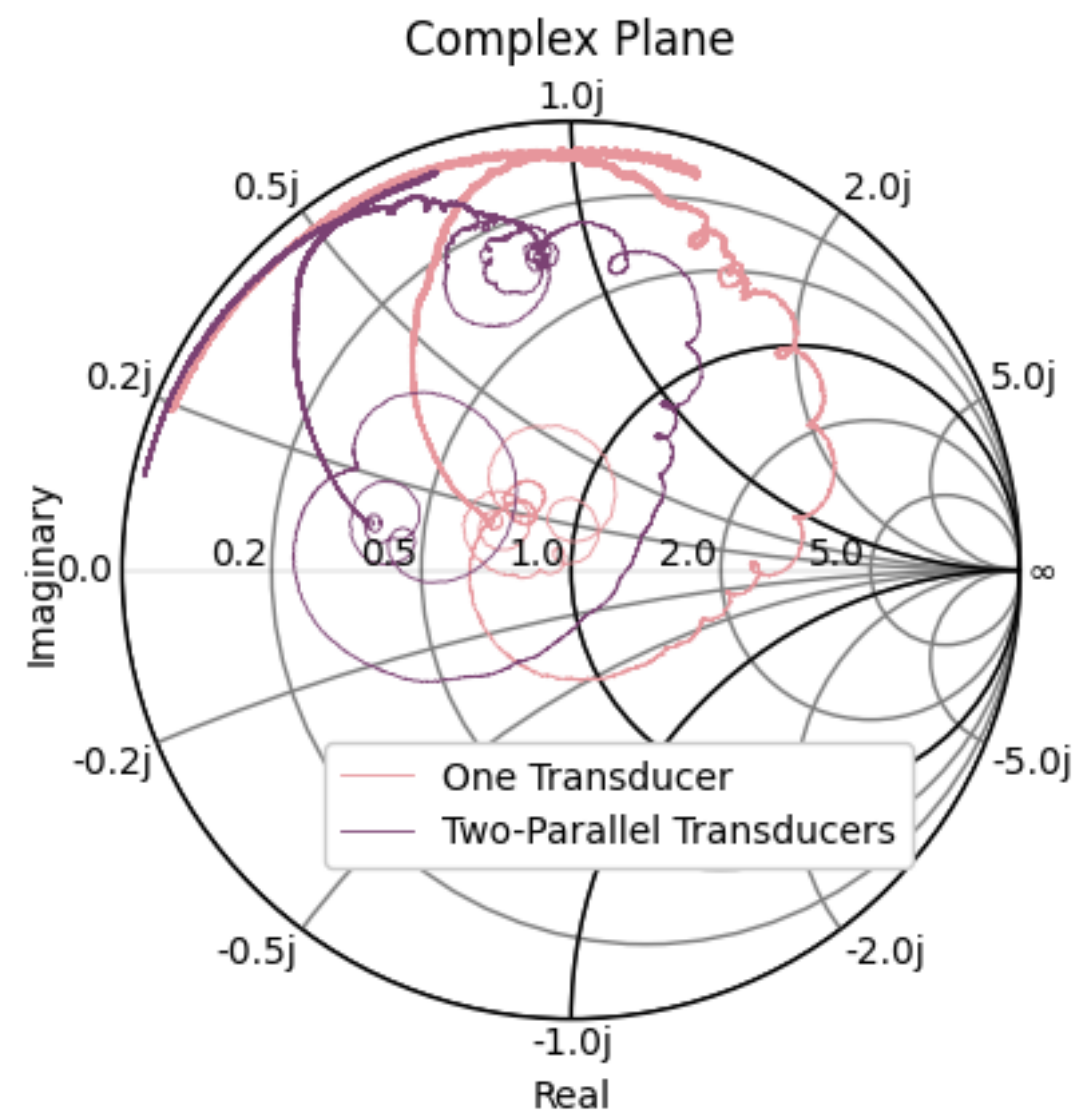

(c)

**Supplementary Figure 13.** Comparison of measured **(a)** $S_{12}$, **(b)** $S_{11}$, and **(c)** $S_{11}$ in Smith graph between two parallel meander-line transducer and a single meander-line transducer YIG filters under applied magnetic flux densities of approximately 2500 Gauss.

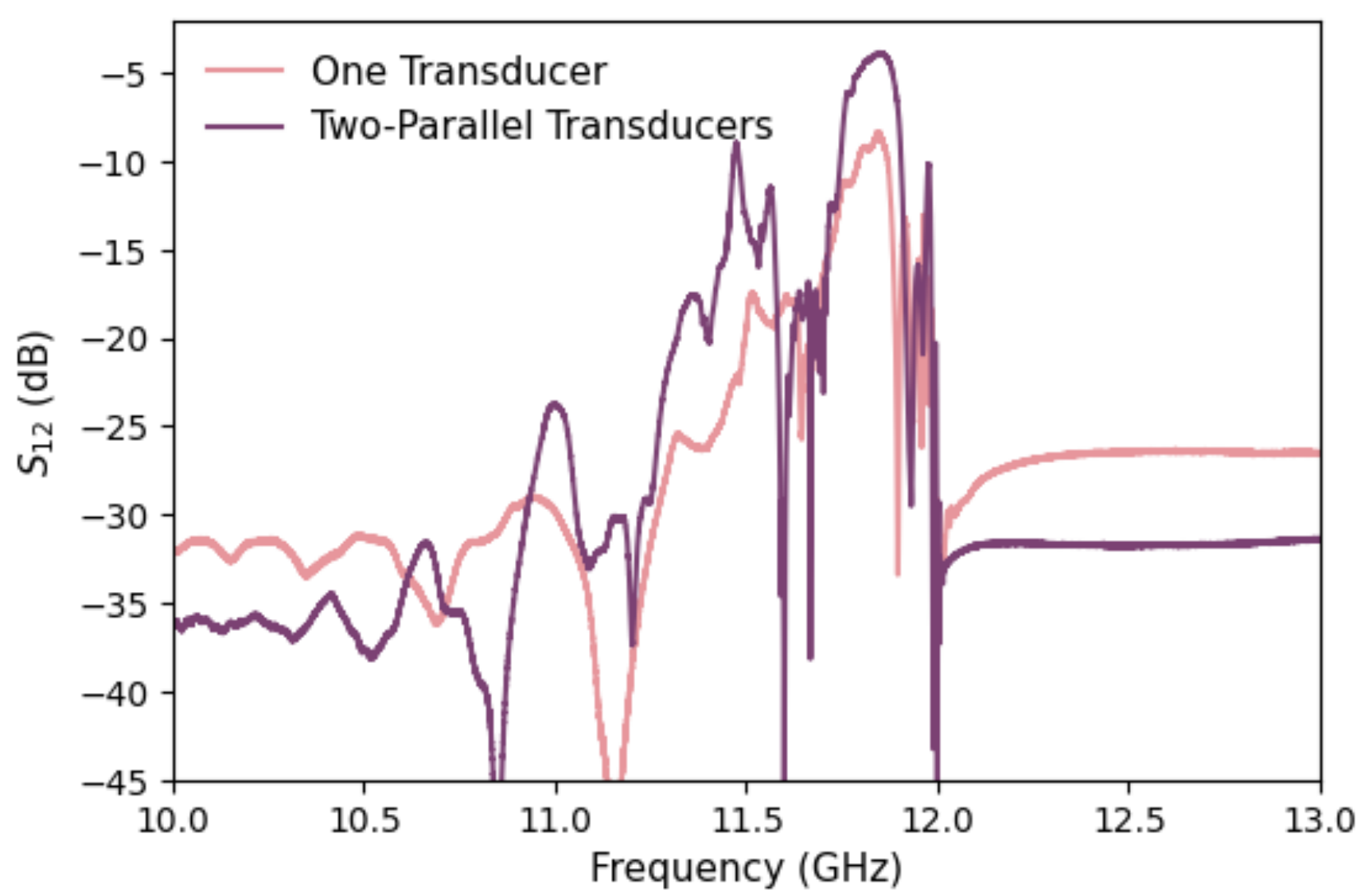

(a)

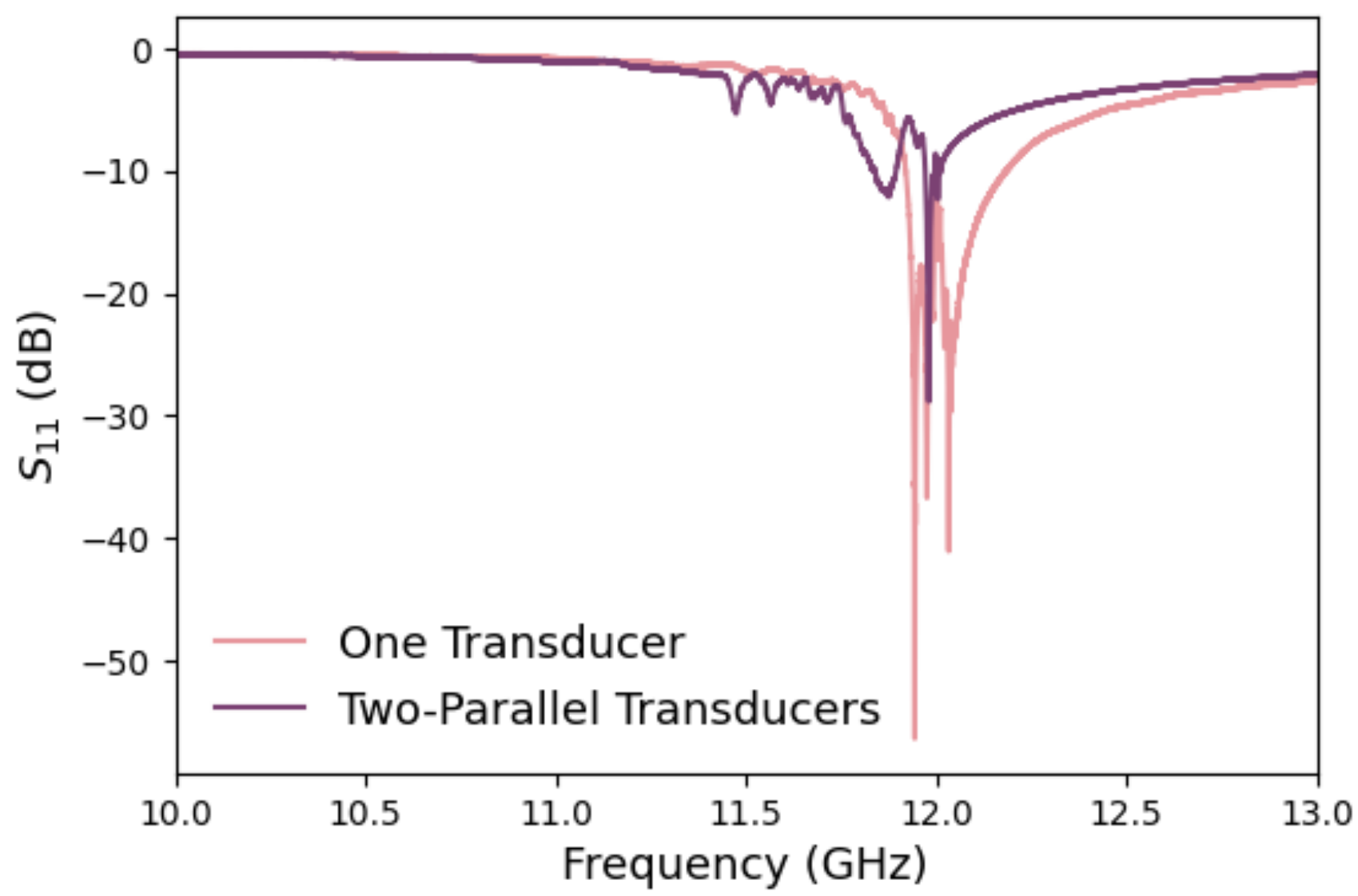

(b)

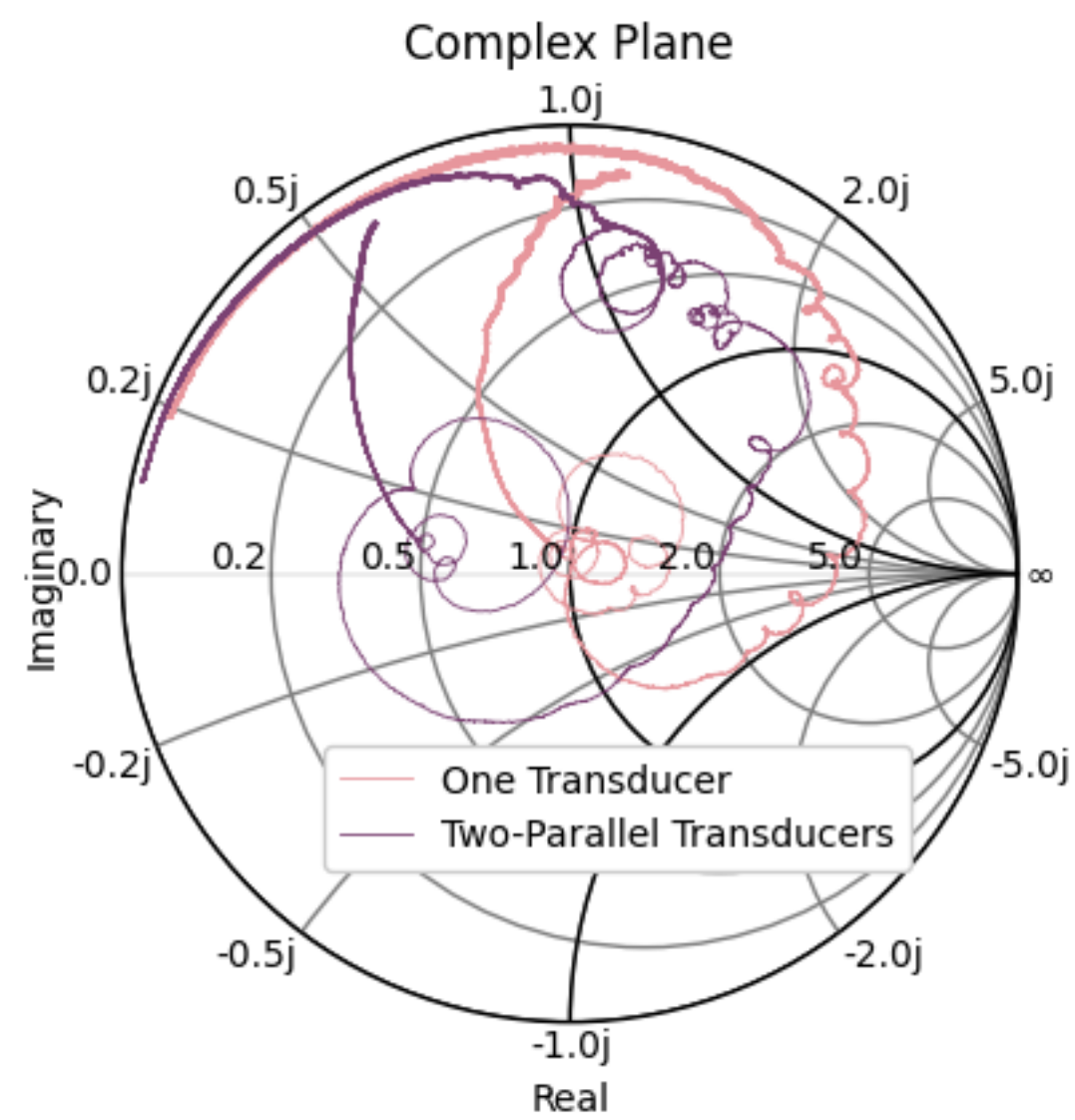

(c)

**Supplementary Figure 14.** Comparison of measured **(a)** $S_{12}$, **(b)** $S_{11}$, and **(c)** $S_{11}$ in Smith graph between two parallel meander-line transducer and a single meander line transducer YIG filters under applied magnetic flux densities of approximately 3500 Gauss.

## Supplementary Note #6: Comparison of YIG Filters with Dual Waveguide and Single Waveguide Designs

**Supplementary Figure 15** shows a two-parallel meander-line filter featuring dual YIG waveguides, where each YIG waveguide follows the same design as the single meander-line filter. Compared to the two-parallel meander-line filter with a single YIG waveguide (**Fig. 3e**), the dual YIG waveguide configuration positions the center bridge section of the meander-line above the BCB/GGG where there is no YIG. As a result, the dual YIG waveguide filter reduces the contact between the bridge section of the meander-line to the YIG waveguide and reduces the excitation of magnetostatic backward volume waves (MSBVWs).

**Supplementary Figure 16, 17, and 18** provide a comparative analysis of the frequency responses of two-parallel meander-line YIG filters with single and dual YIG waveguides under magnetic bias fields of 1500, 2500, and 3500 Gauss, respectively. The single-waveguide filter exhibits more pronounced spurious modes in both $S_{12}$ and $S_{11}$.

By moving the bridge section away from the YIG, the dual-waveguide configuration effectively suppresses the large spurious peaks that occur at frequencies slightly below the passband: 5.7 GHz (1500 Gauss), 8.7 GHz (2500 Gauss), and 11.5 GHz (3500 Gauss). However, the dual-waveguide design was not adopted in this study primarily due to its higher insertion loss. Specifically, the insertion loss increased from 2.9 dB to 4.2 dB at 6.3 GHz, from 3.1 dB to 4.9 dB at 8.8 GHz, and from 3.8 dB to 6.5 dB at 11.8 GHz. The increase in insertion loss can also be observed in the Smith chart analysis, where the dual-waveguide filter exhibits a smaller resonance circle, indicating weaker resonances. This increased loss is suspected to arise from the larger total edge region of the YIG waveguides in the dual-waveguide design. In contrast, the single-waveguide configuration minimizes the edge region, thereby reducing edge-related MSSW scattering and MSSW propagation loss.

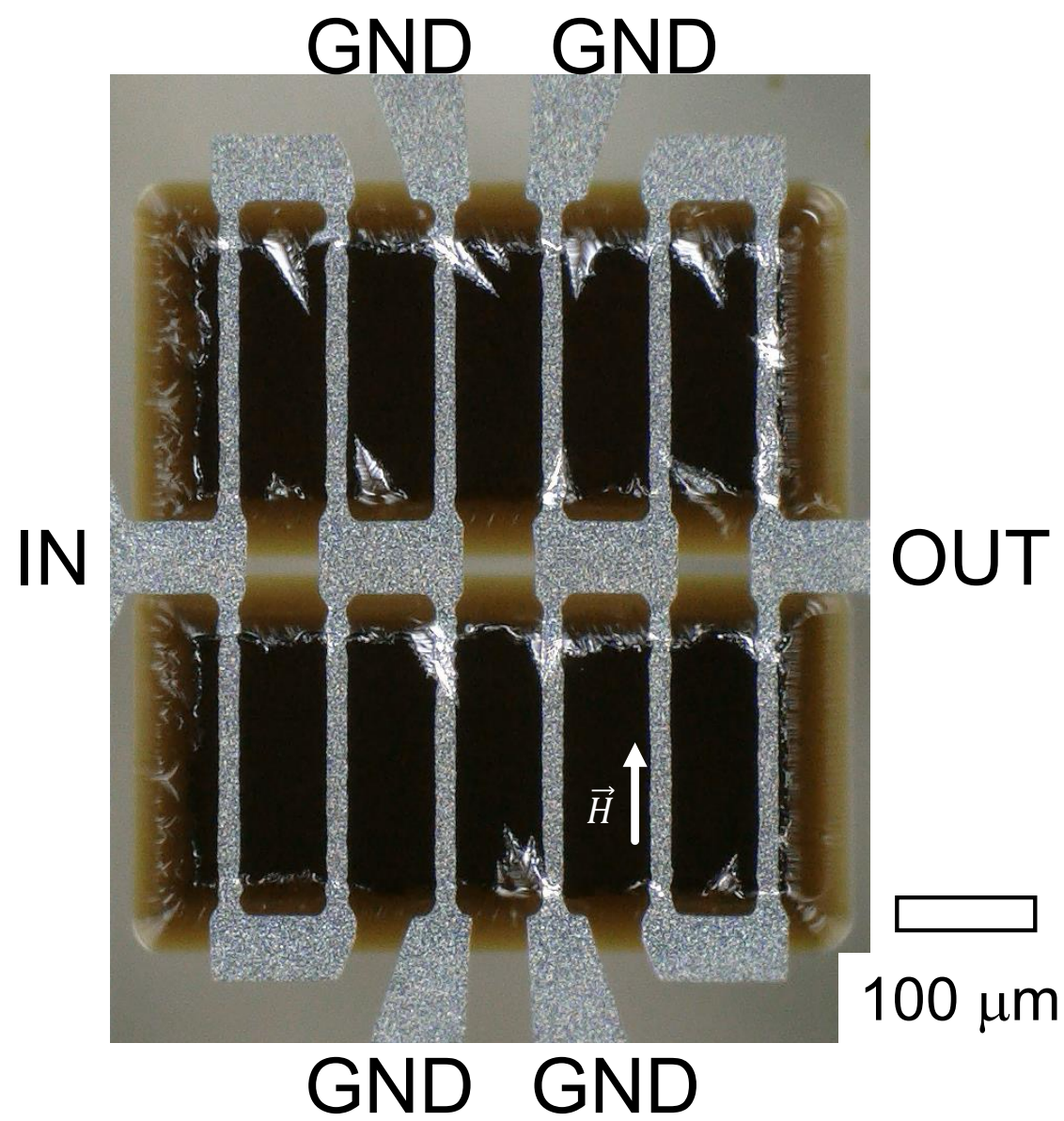

**Supplementary Figure 15.** Optical microscope image of a two parallel meander-line transducer dual-waveguide YIG filter with pitch of 70 μm and width of 200 μm.

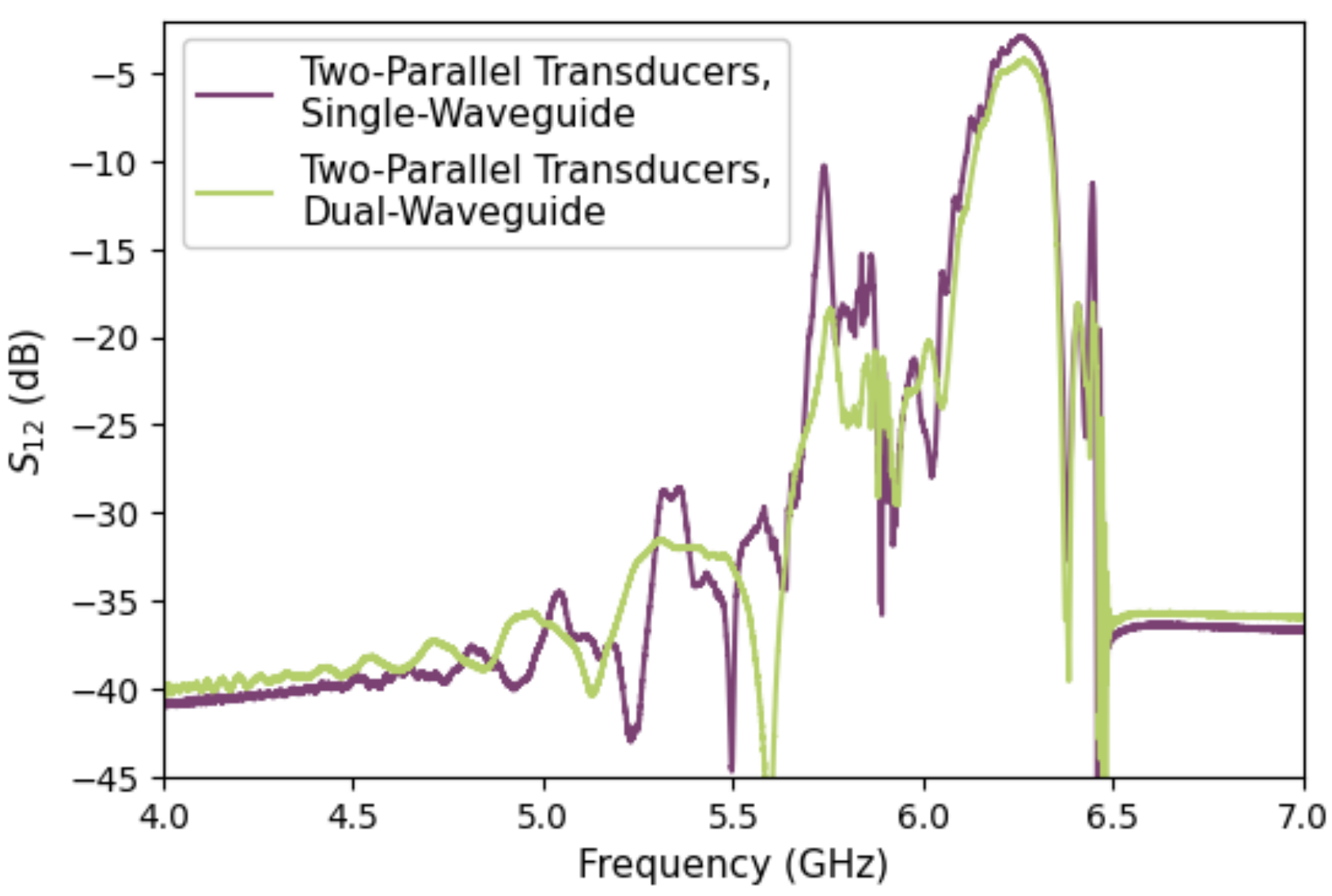

(a)

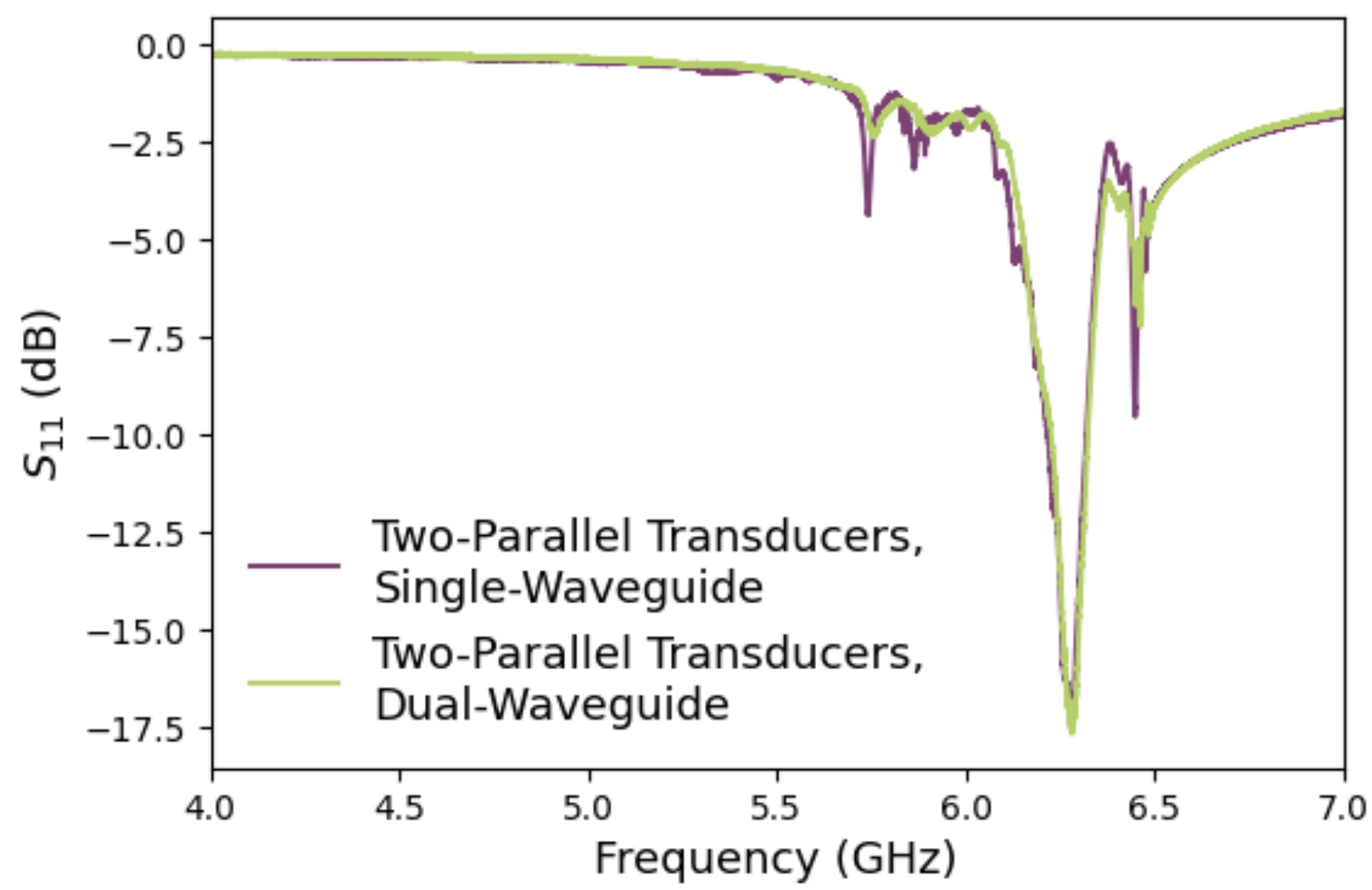

(b)

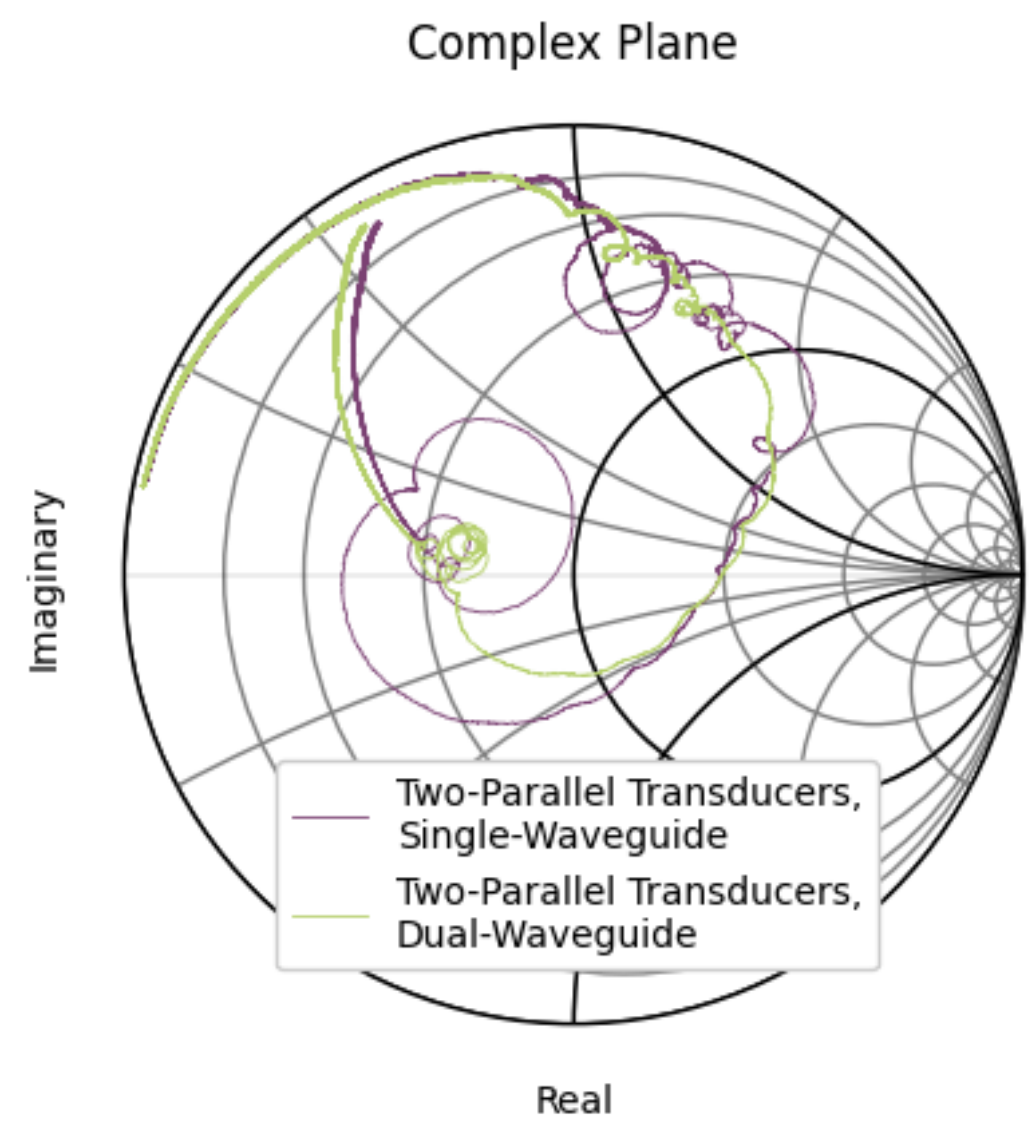

(c)

**Supplementary Figure 16**. Comparison of measured **(a)** $S_{12}$, **(b)** $S_{11}$, and **(c)** $S_{11}$ in Smith graph between two parallel meander-line transducer dual and single waveguide YIG filters under applied magnetic flux densities of approximately 1500 Gauss.

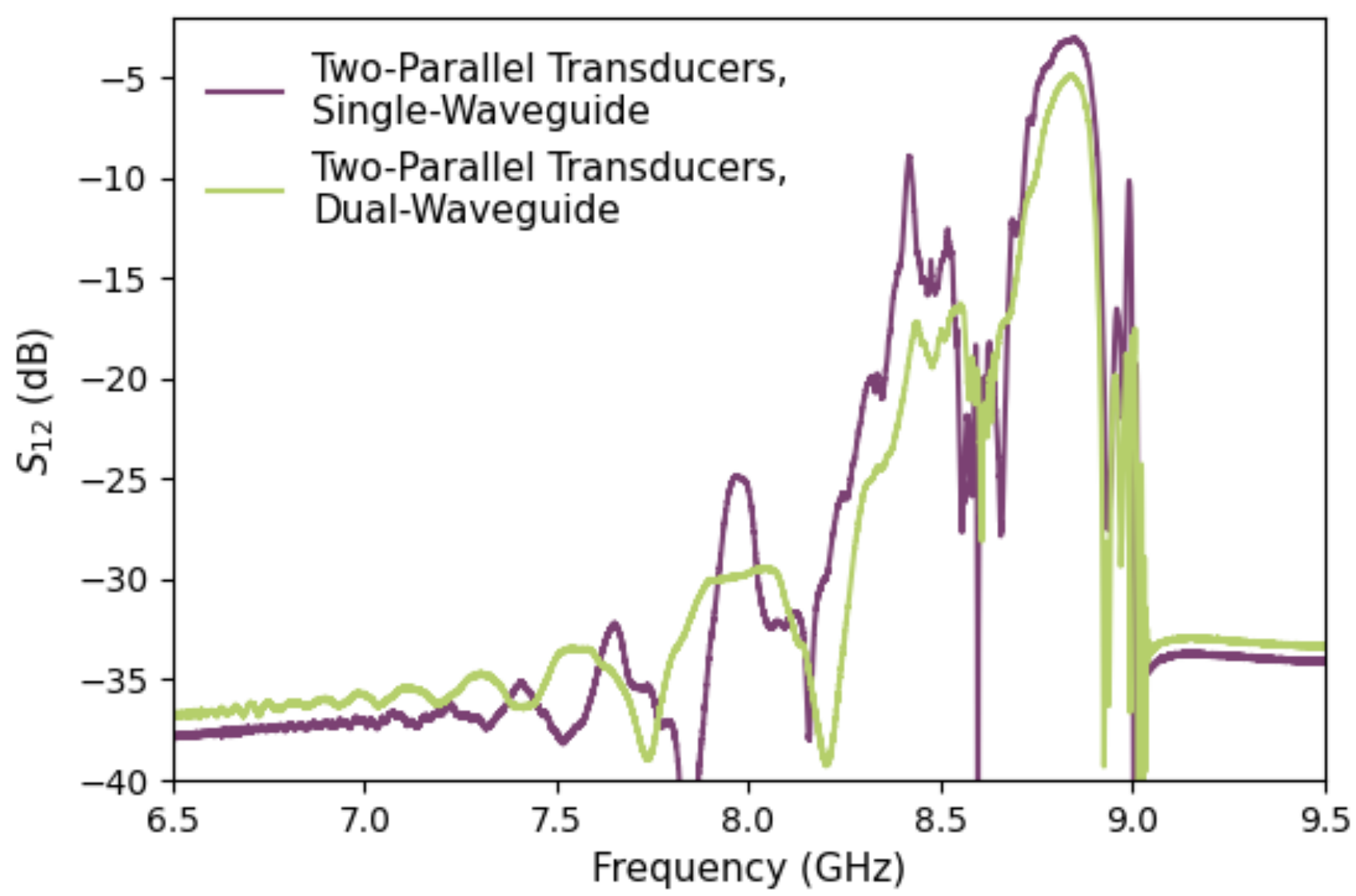

Two-Parallel Transducers, Single-Waveguide
Two-Parallel Transducers, Dual-Waveguide
$S_{12}$ (dB)
Frequency (GHz)

(a)

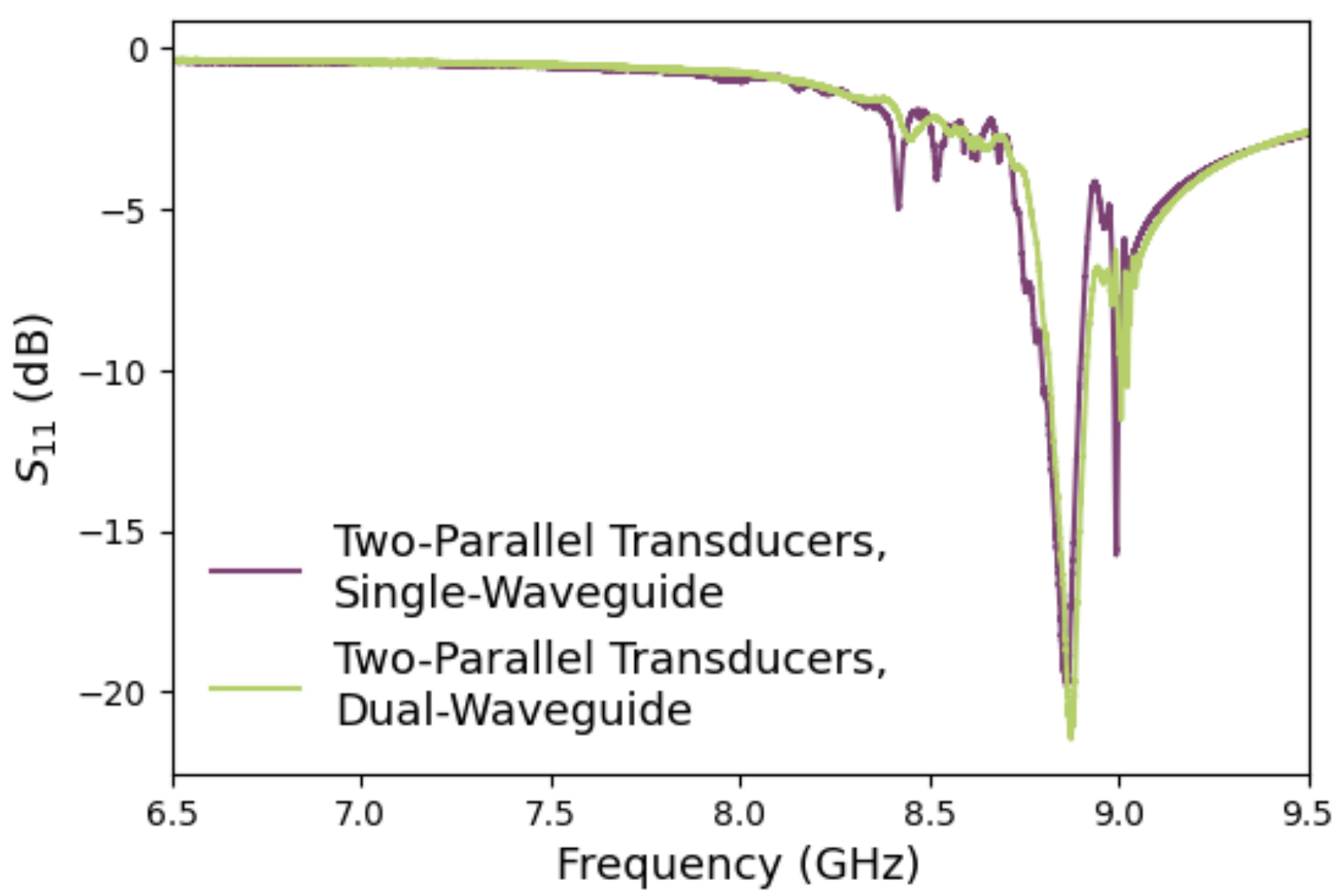

Two-Parallel Transducers, Single-Waveguide
Two-Parallel Transducers, Dual-Waveguide
$S_{11}$ (dB)
Frequency (GHz)

(b)

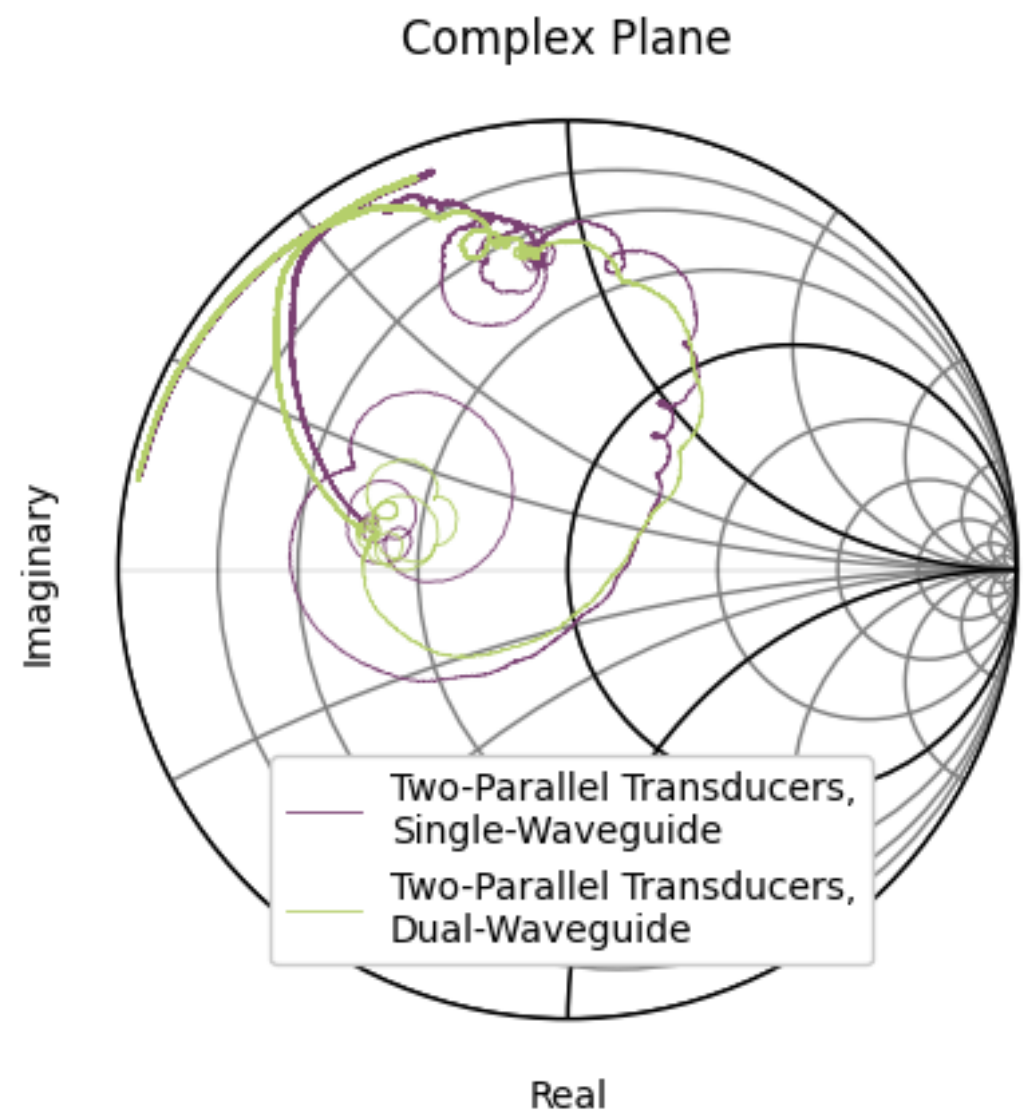

(c)

**Supplementary Figure 17.** Comparison of measured **(a)** $S_{12}$, **(b)** $S_{11}$, and **(c)** $S_{11}$ in Smith graph between two parallel meander-line transducer dual and single waveguide YIG filters under applied magnetic flux densities of approximately 2500 Gauss.

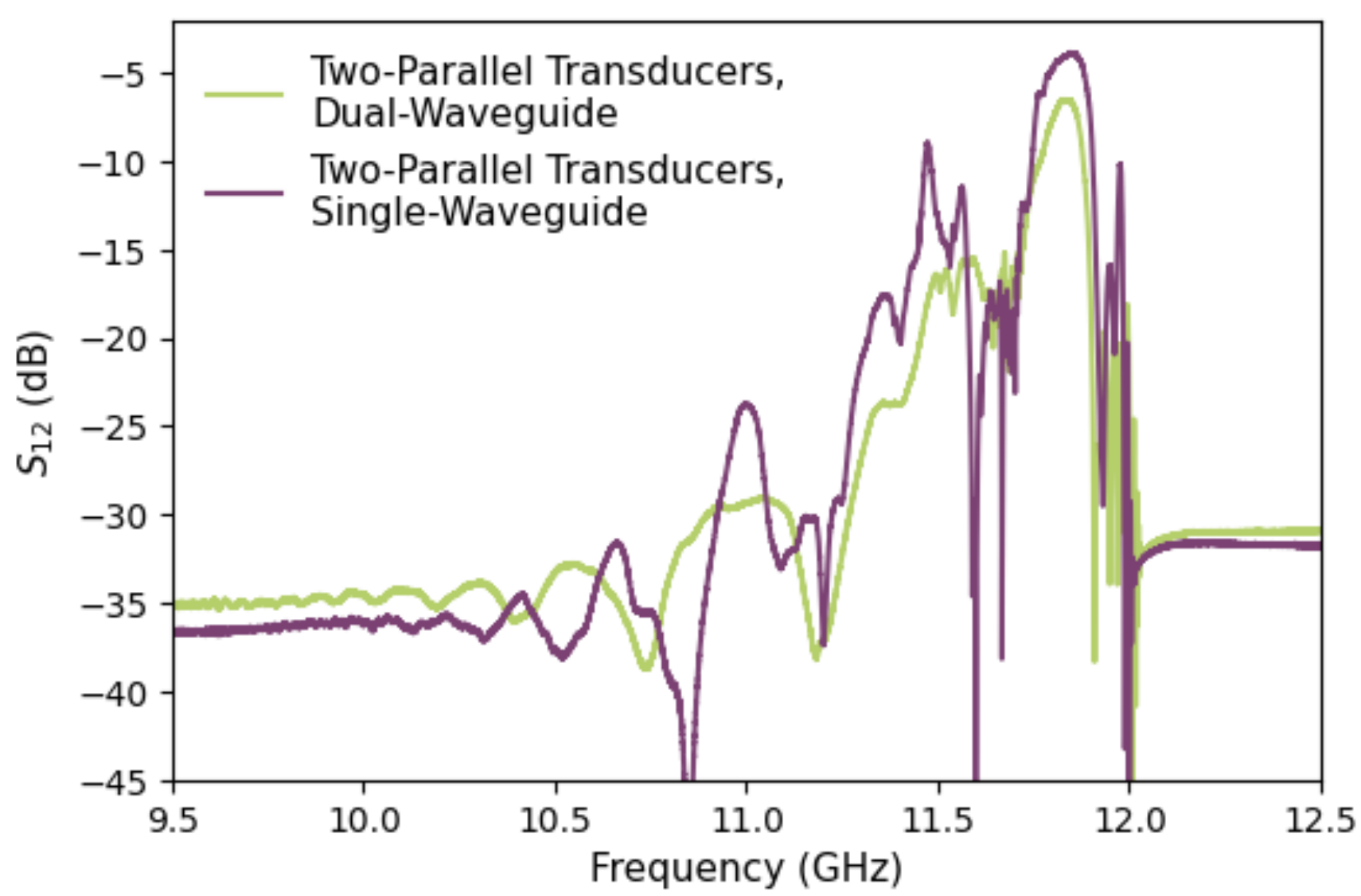

(a)

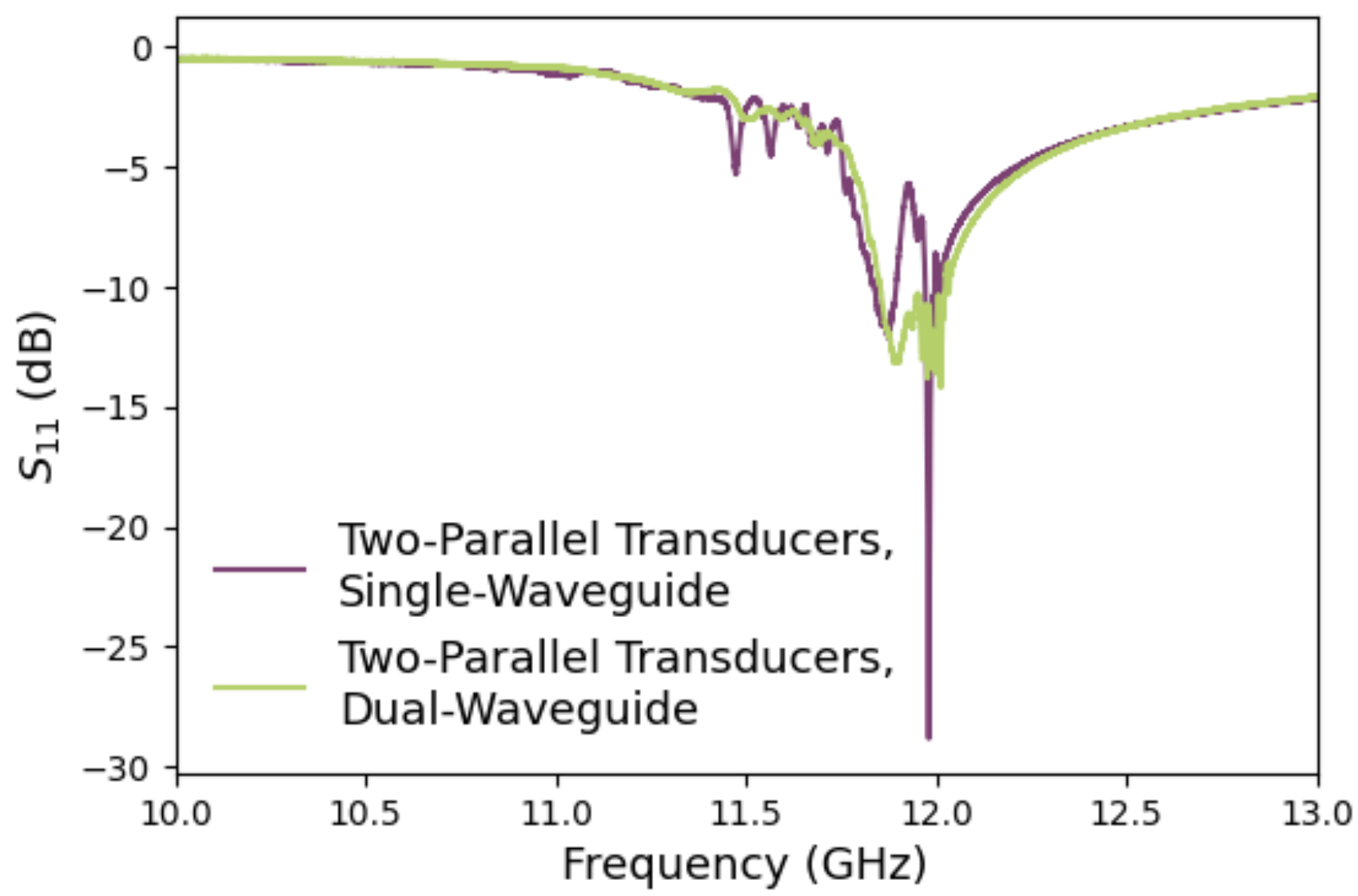

(b)

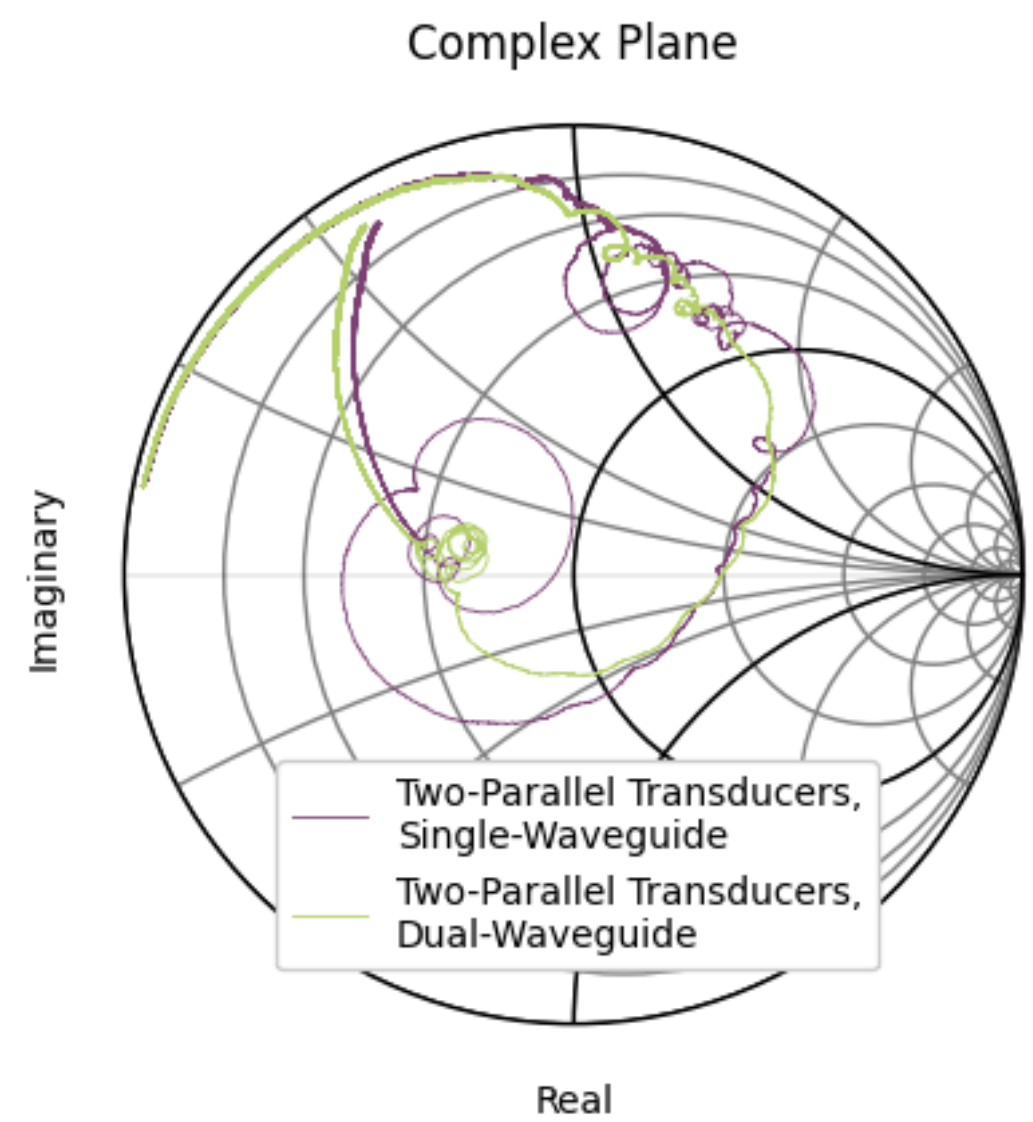

(c)

**Supplementary Figure 18.** Comparison of measured **(a)** $S_{12}$, **(b)** $S_{11}$, and **(c)** $S_{11}$ in Smith graph between two parallel meander-line transducer dual and single waveguide YIG filters under applied magnetic flux densities of approximately 3500 Gauss.

Although the dual-waveguide YIG filter suppresses several strong spurious responses near the resonance frequency, it exhibits more, albeit smaller, spurious peaks at frequencies significantly lower than the main passband. For instance, additional spurious resonances are observed in the ranges of 4–5 GHz (1500 Gauss), 6.5–7.5 GHz (2500 Gauss), and 9.5–10.5 GHz (3500 Gauss). This behavior is attributed to enhanced demagnetization effects in the dual-waveguide configuration. Previous studies have shown that the internal magnetic field can be calculated as[2]:

$$H(x) = H - \frac{M_s}{\pi}\left[arctan\left(\frac{S}{2x+w}\right) - arctan\left(\frac{S}{2x-w}\right)\right] \quad \text{(S6)}$$

Here, $H(x)$ is the internal magnetic field, $H$ is the applied magnetic field, $M_s$ is the saturation magnetization, $S$ is the thickness of the YIG, $w$ is the width of the YIG,

As illustrated in **Supplementary Figure 19**, the demagnetizing effect significantly reduces the internal magnetic field near the edges of the YIG waveguide. This could lead to the formation of field-induced channels where spin waves can become localized. These localized spin wave modes are typically concentrated near the edges and appear at much lower frequencies than those supported in the center region of the waveguide.

This demagnetization effect is more pronounced in the 18 µm thick YIG waveguide used in this study, compared to previously reported 3 µm thick waveguides. Both the edge-induced channel width and the reduction in the central internal field are more severe in thicker waveguides. Furthermore, **Supplementary Figure 19 (b)** suggests that increasing the width of the YIG waveguide can mitigate this effect—explaining why the single-waveguide design, which has a wider YIG region, shows fewer and weaker low-frequency spurious modes.

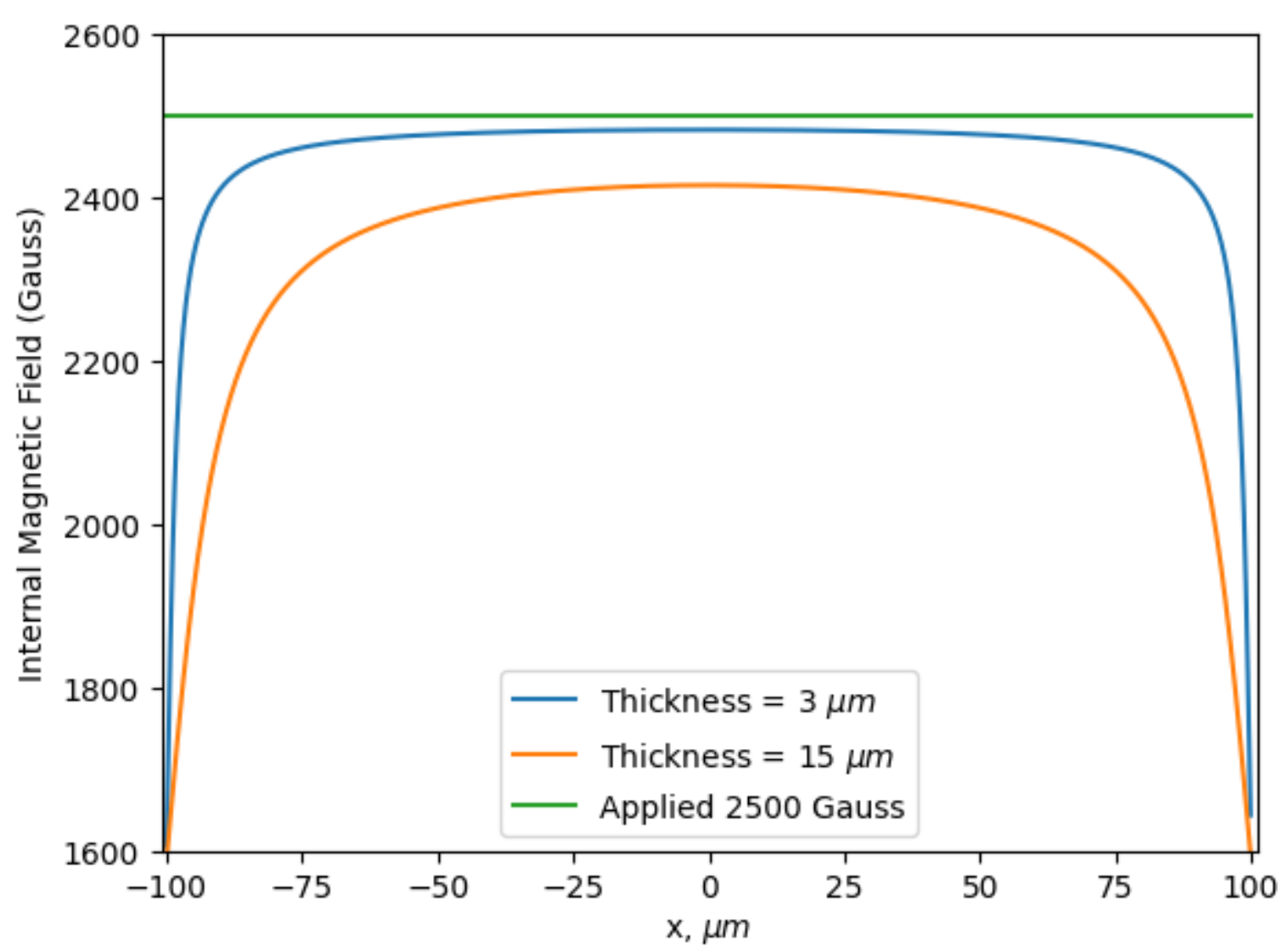

(a)

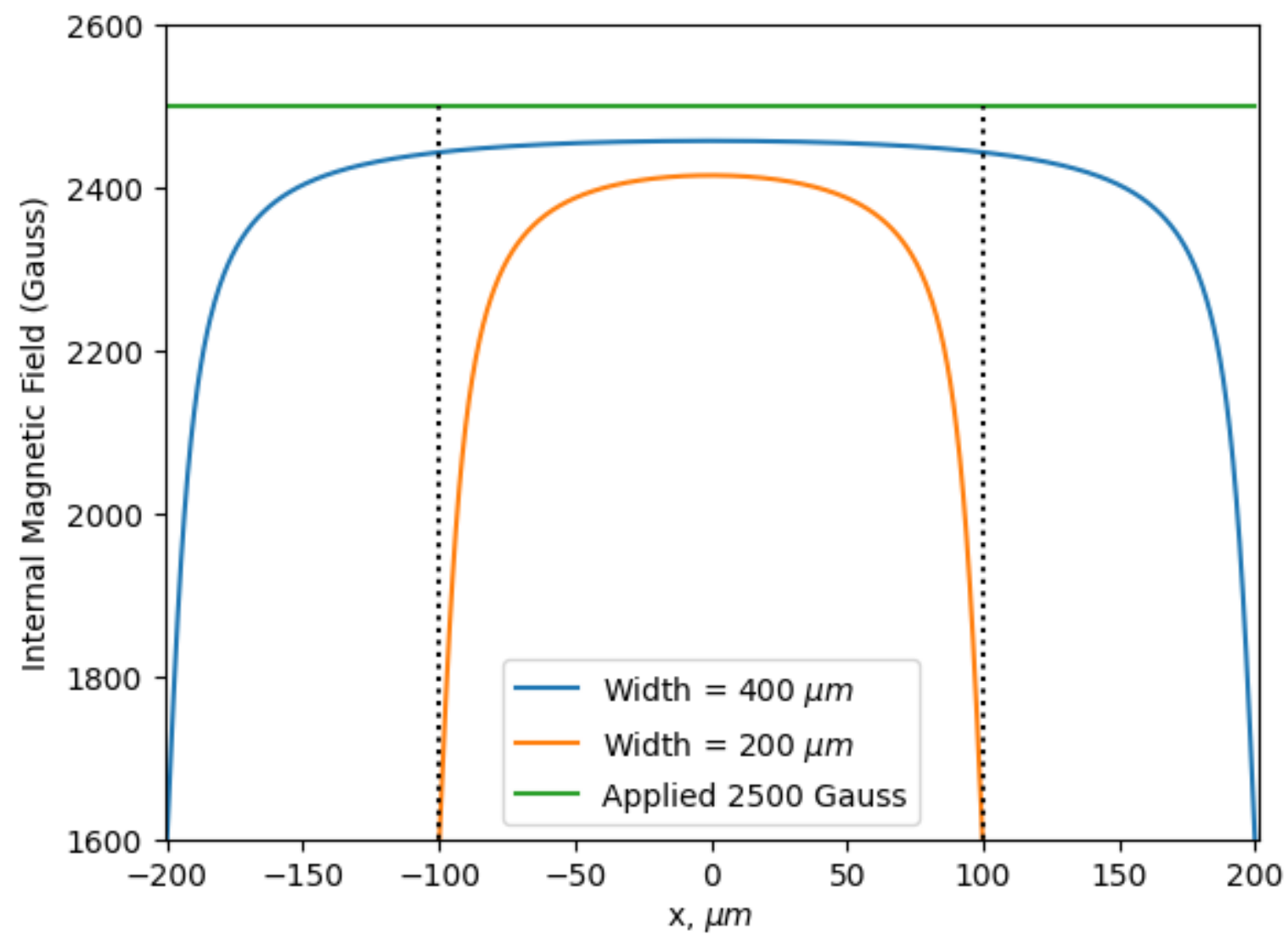

(b)

**Supplementary Figure 19.** Theoretical distribution of the internal static magnetic field across the width of a YIG waveguide. **(a)** Comparison between YIG waveguides with thickness of 3 µm and 15 µm, both with a fixed width of 200 µm. **(b)** Comparison between waveguide widths of 200 µm and 400 µm, with a fixed thickness of 15 µm. The green line indicates the value of the applied external magnetic field.

# Supplementary Note #7: Effect of Pitch on YIG Filters with Meander Line Transducers

**Supplementary Figure 20** presents a two-parallel meander-line filter incorporating a single 18 µm thick YIG waveguide with a transducer pitch of 130 µm. All other design parameters remain identical to the YIG filter previously discussed.

The impact of transducer pitch can be understood through the transducer-specific radiation impedance per unit length, as described by Equation (1). A theoretical comparison of different pitch values is provided in **Supplementary Figure 21**, considering only the first-order width mode (n = 1). Although higher-order width modes also exist at lower frequencies, they share similar profiles with the first-order mode and are thus not individually analyzed.

For a small pitch of 40 µm, the filter exhibits only a single fundamental mode near the $\omega_{max}$ frequency. As the pitch increases, the radiation impedance slightly rises and begins to saturate around 70 µm, beyond which the difference between 70 µm and 130 µm becomes negligible. However, the number and magnitude of the higher-order length modes increases significantly with increasing pitch.

**Supplementary Figure 22** shows the measured $S_{12}$ result, which validates the theoretical predictions. All filters in this comparison share identical transducer dimensions except for pitch. The filter with a 70 µm pitch also has been featured in **Fig. 3e** and discussed in **Supplementary Notes #4 and 5.** This comparison shows that the pitch of 70 µm is the best design as it has both low insertion loss and high spurious mode attenuation. Moreover, the out-of-band rejection also increases with respect to the pitch of the meander line transducer.

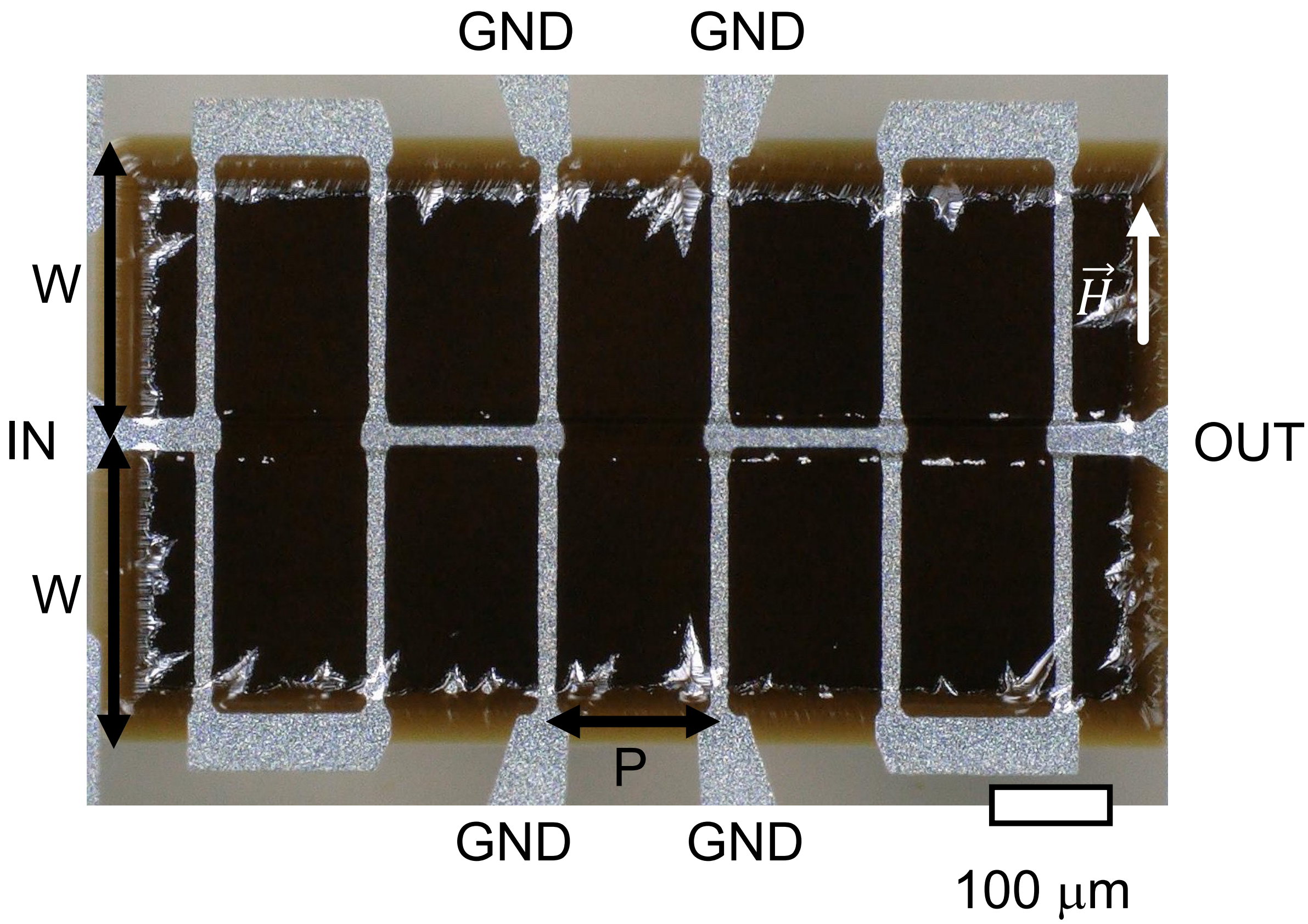

**Supplementary Figure 20.** Optical image of a YIG filter with two parallel meander-line transducers with a larger pitch of 130 μm and width of 200 μm

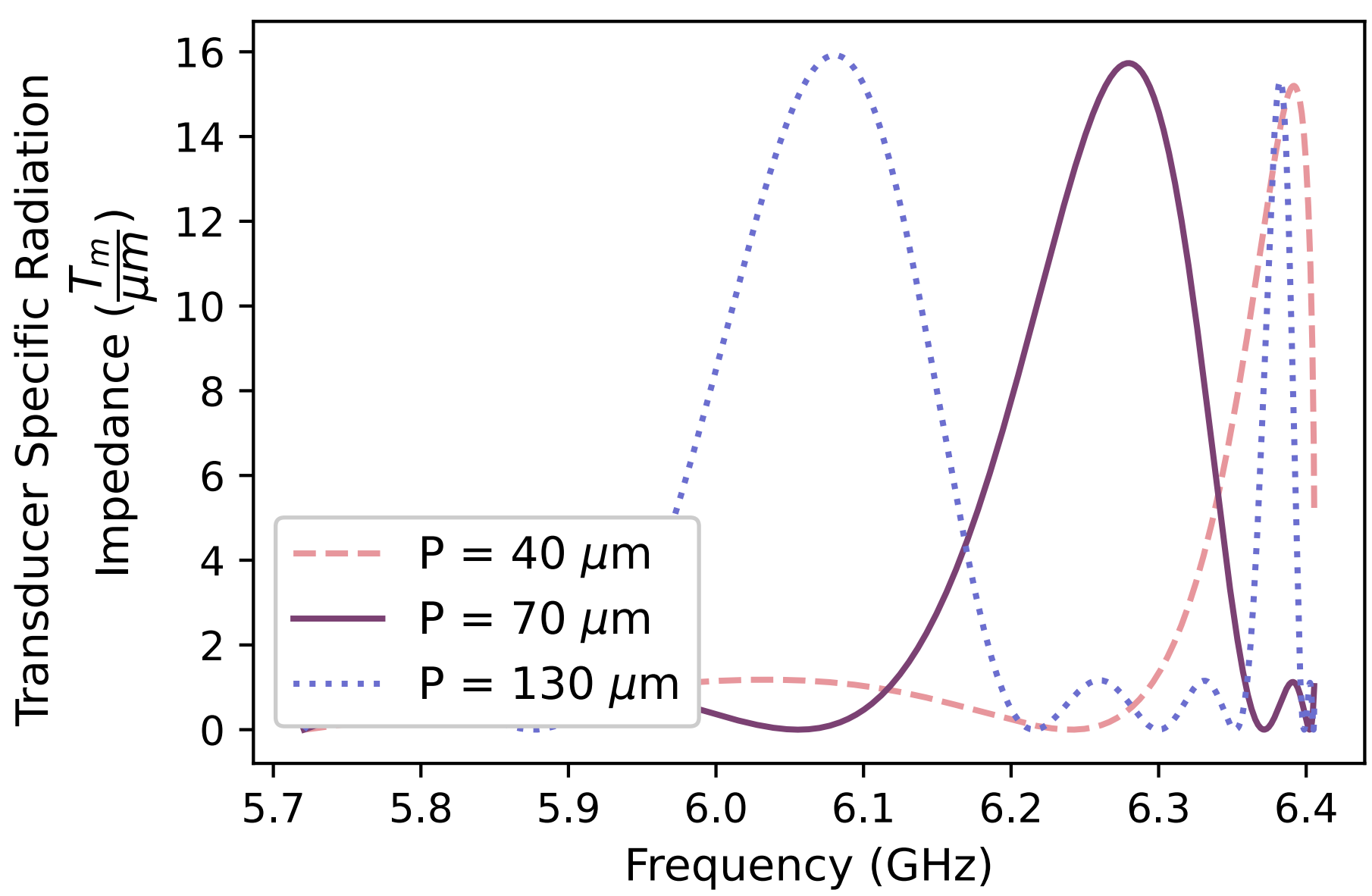

**Supplementary Figure 21.** Comparison of the calculated transducer specific radiation impedance for different meander-line transducer pitches.

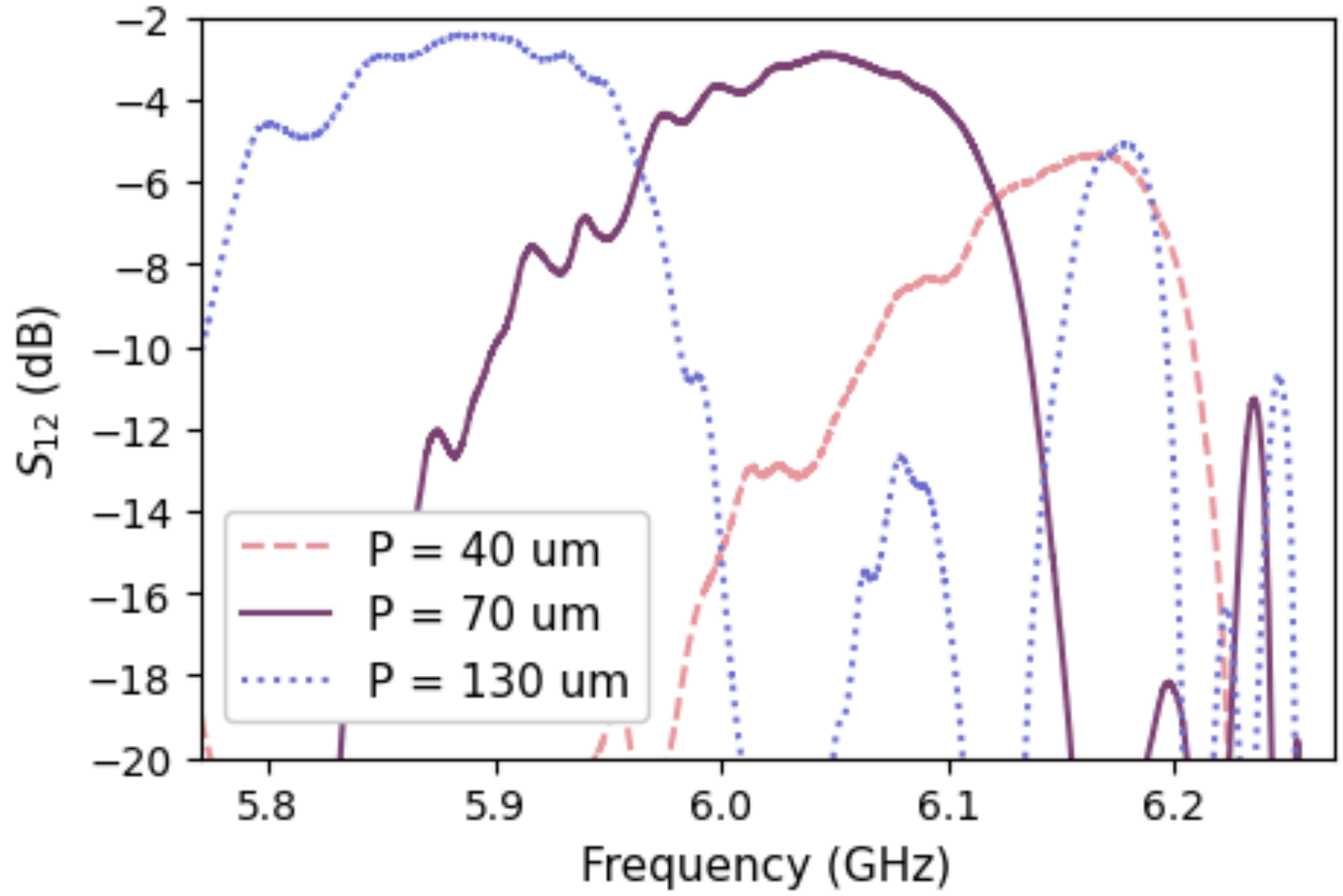

(a)

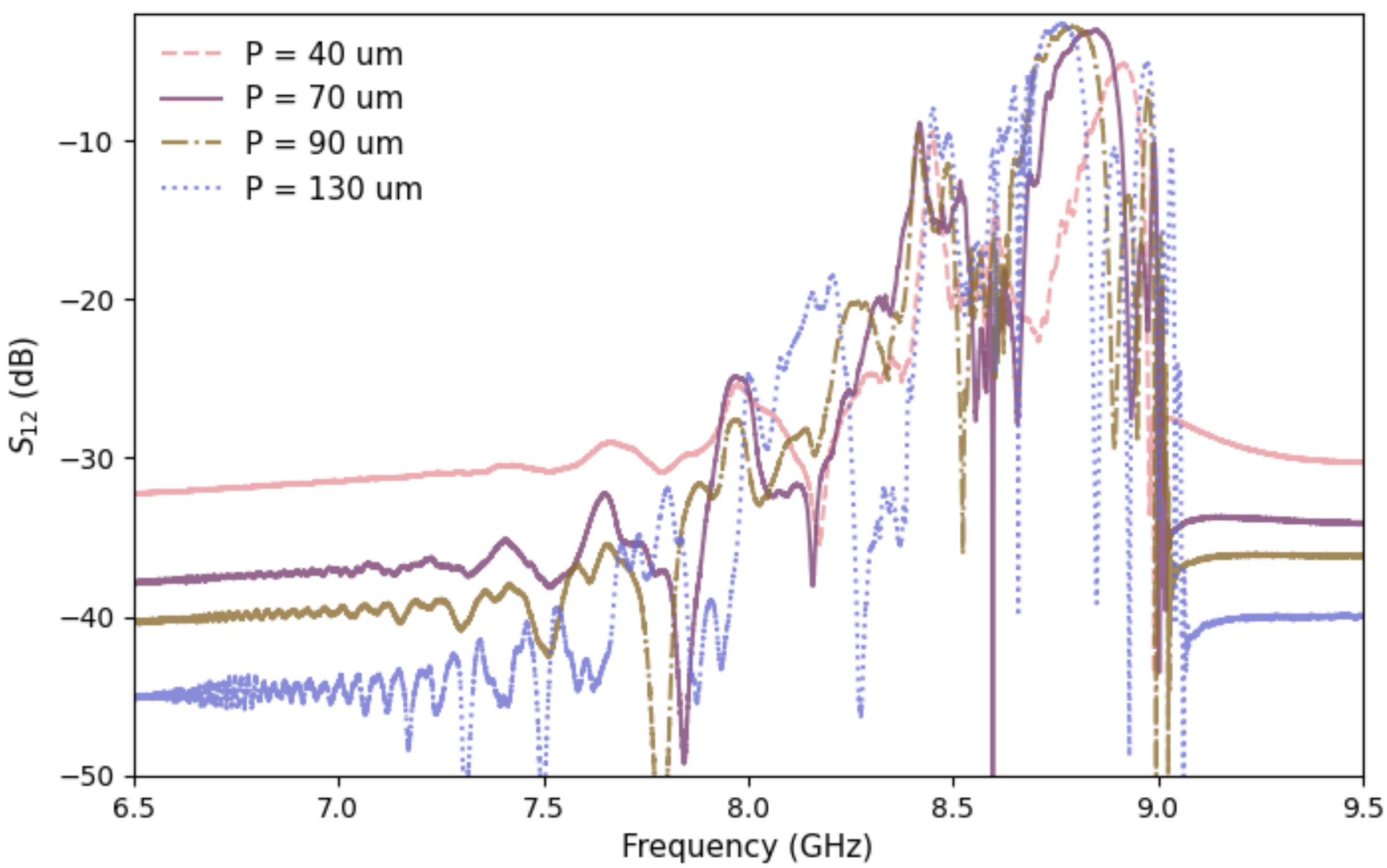

(b)

**Supplementary Figure 22.** Measured $S_{12}$ frequency responses for different meander-line transducer pitches with magnetic flux density of approximately 2500 Gauss. **(a)**

Zoomed-in view showing the passband and higher-order spurious length modes. **(b)** Full frequency response across the entire measurement range.

## Supplementary Note #8: Effect of Spacing on YIG Filters with Meander Line Transducers

**Supplementary Figure 23** present a comparative analysis of the frequency responses of 18 µm thick, two-parallel meander-line single-waveguide YIG filters with varying transducer spacings under applied magnetic flux densities of 1500, 2500, and 3500 Gauss, respectively. All filters share the same waveguide width (200 µm), transducer pitch (70 µm), and transducer width (10 µm). The filter without additional transducer spacing, also featured in **Fig. 3e** and discussed in **Supplementary Note #4**, serves as a baseline for comparison.

These results reveal that increasing the spacing between transducers significantly improves the out-of-band rejection, while maintaining a nearly constant insertion loss across the measured frequency range.

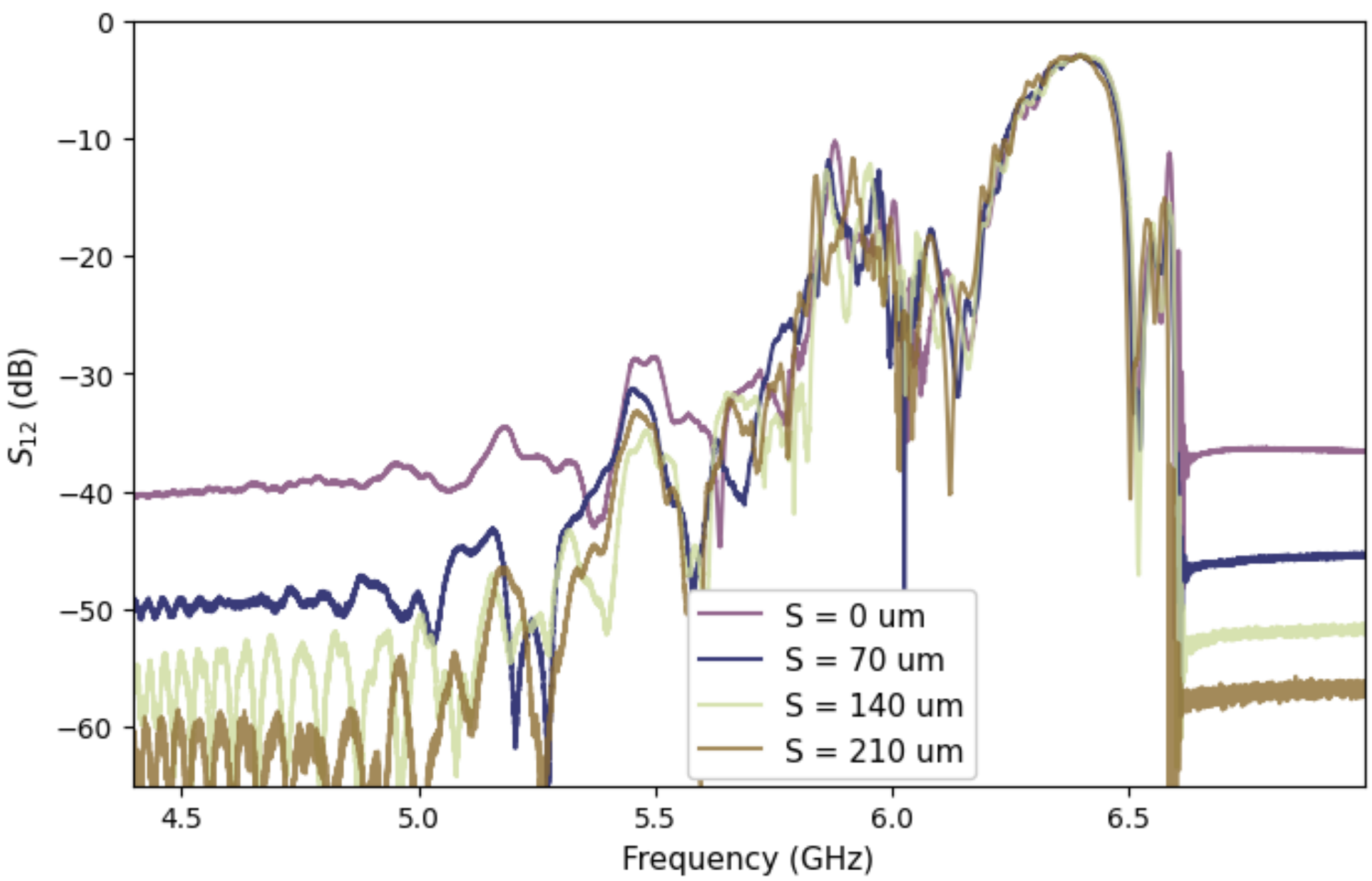

(a)

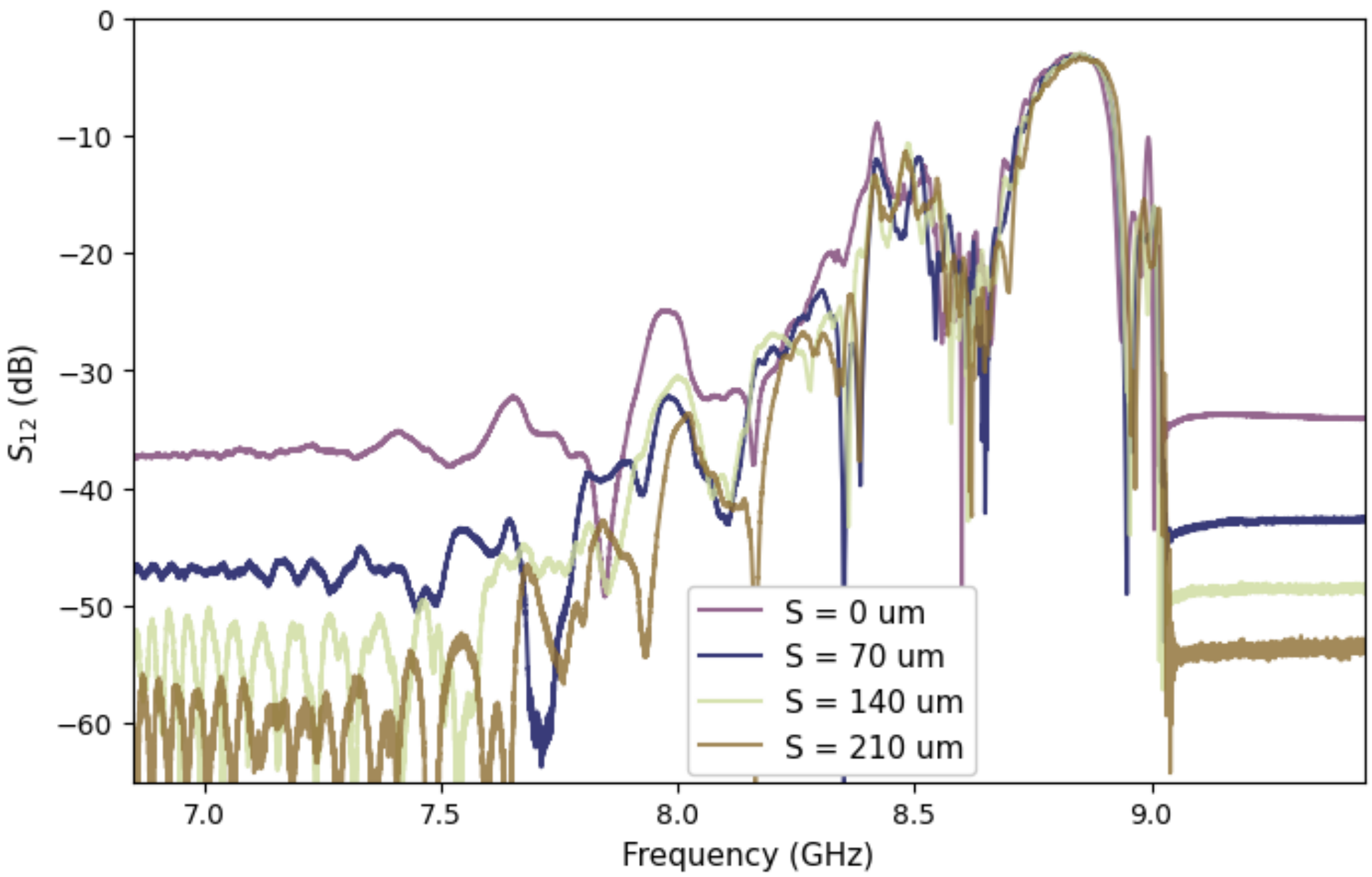

(b)

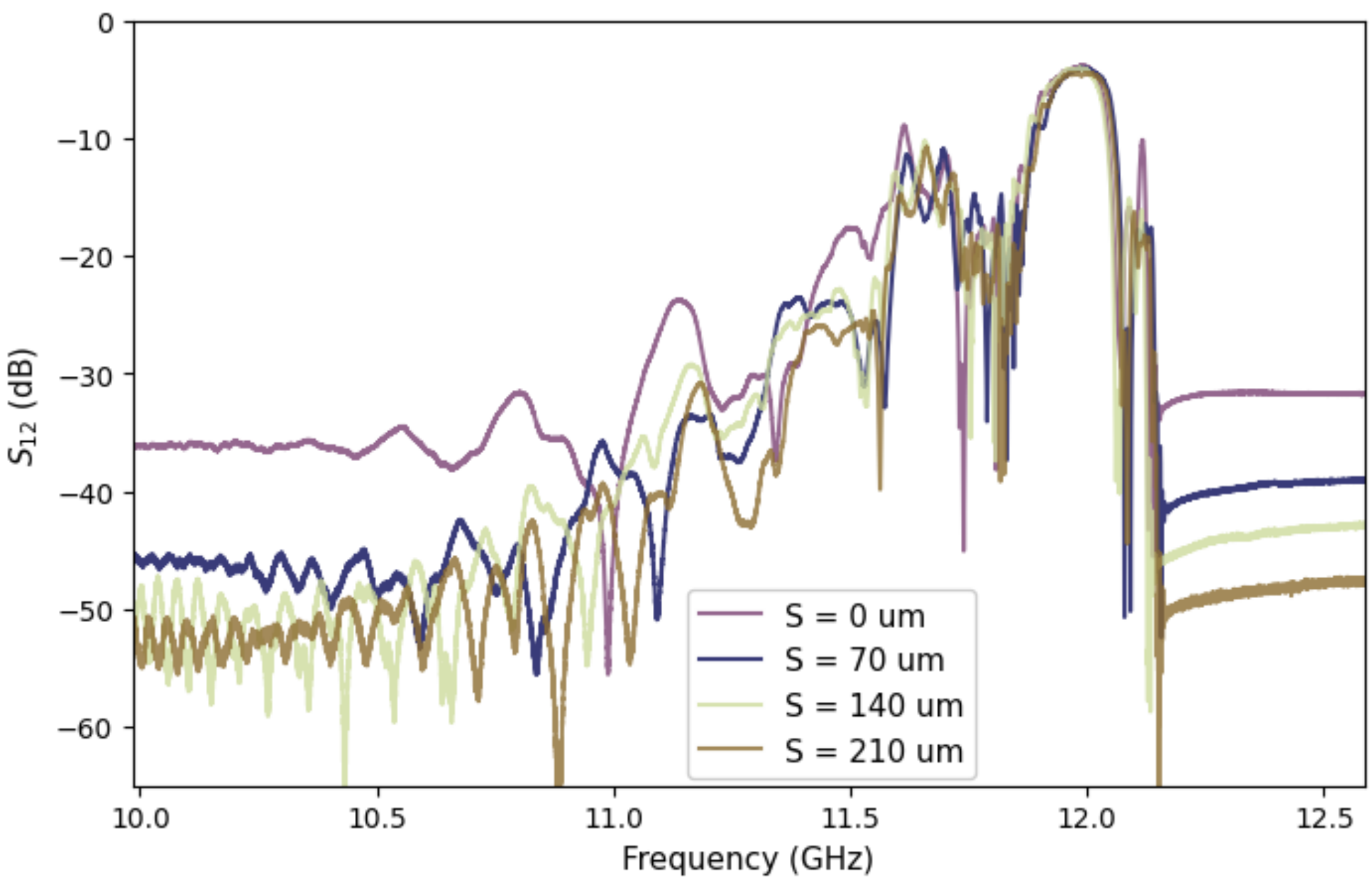

(c)

**Supplementary Figure 23.** Comparison of the measured $S_{12}$ frequency responses for YIG filters with different spacings under three different applied magnetic flux densities of **(a)** ~1500, **(b)** ~2500, and **(c)**~3500 Gauss, respectively.

# Supplementary Note #9: Dual-Hexagon-Shape YIG Filters

**Supplementary Figure 24** presents a comparative analysis of the frequency responses of 18 µm thick, two-parallel meander-line single-waveguide YIG filters with a dual-hexagon shaped YIG and a rectangular shaped YIG under applied magnetic flux densities of 1500, 2500, and 3500 Gauss, respectively. Both filters share the same waveguide width (200 µm), transducer pitch (70 µm), and transducer width (10 µm). The filter with a rectangular shape ,also featured in **Fig. 3e**, serves as a baseline for comparison. The dual-hexagon shaped YIG filter is highlighted in **Fig. 4e**.

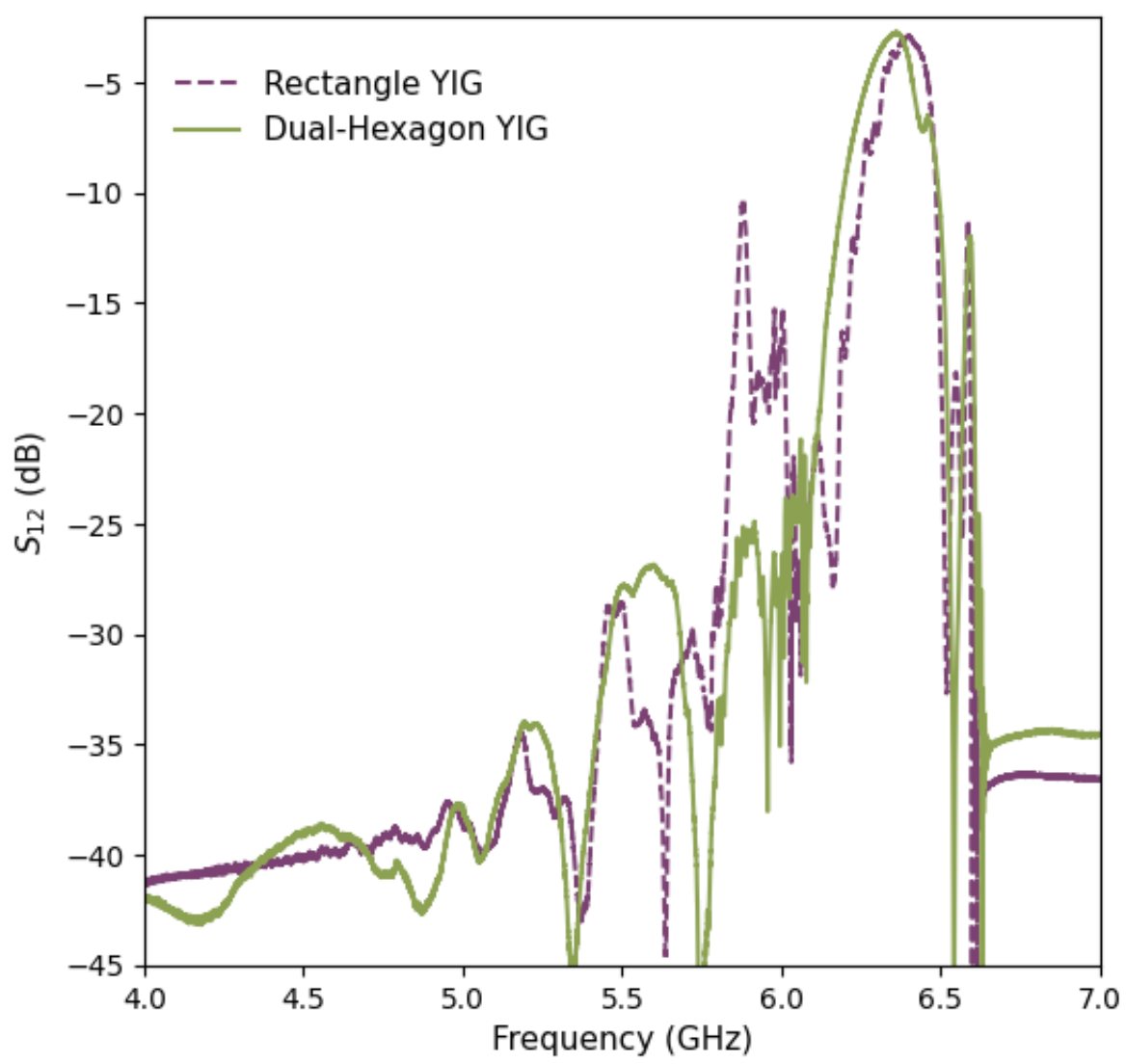

(a)

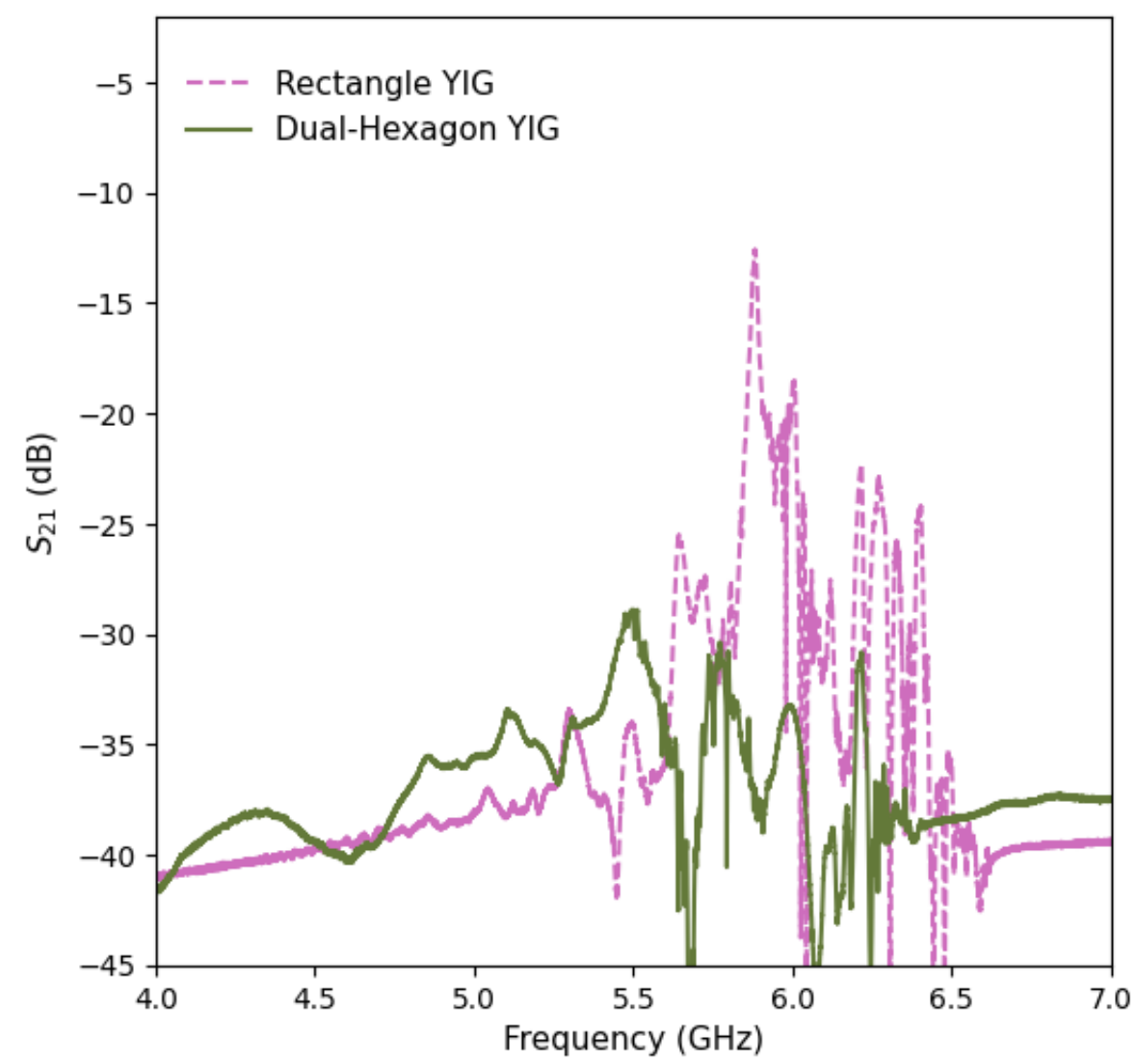

(b)

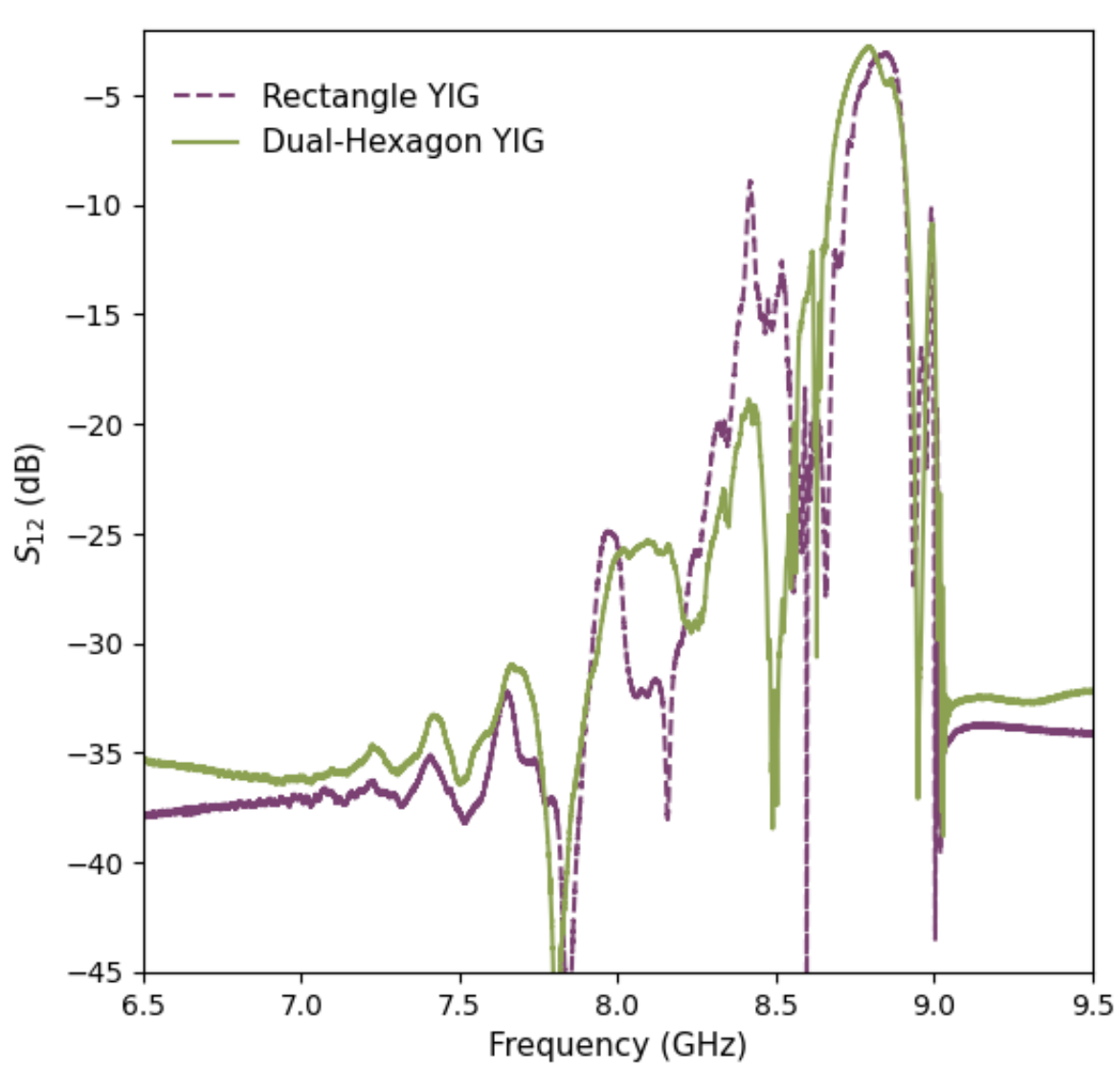

(c)

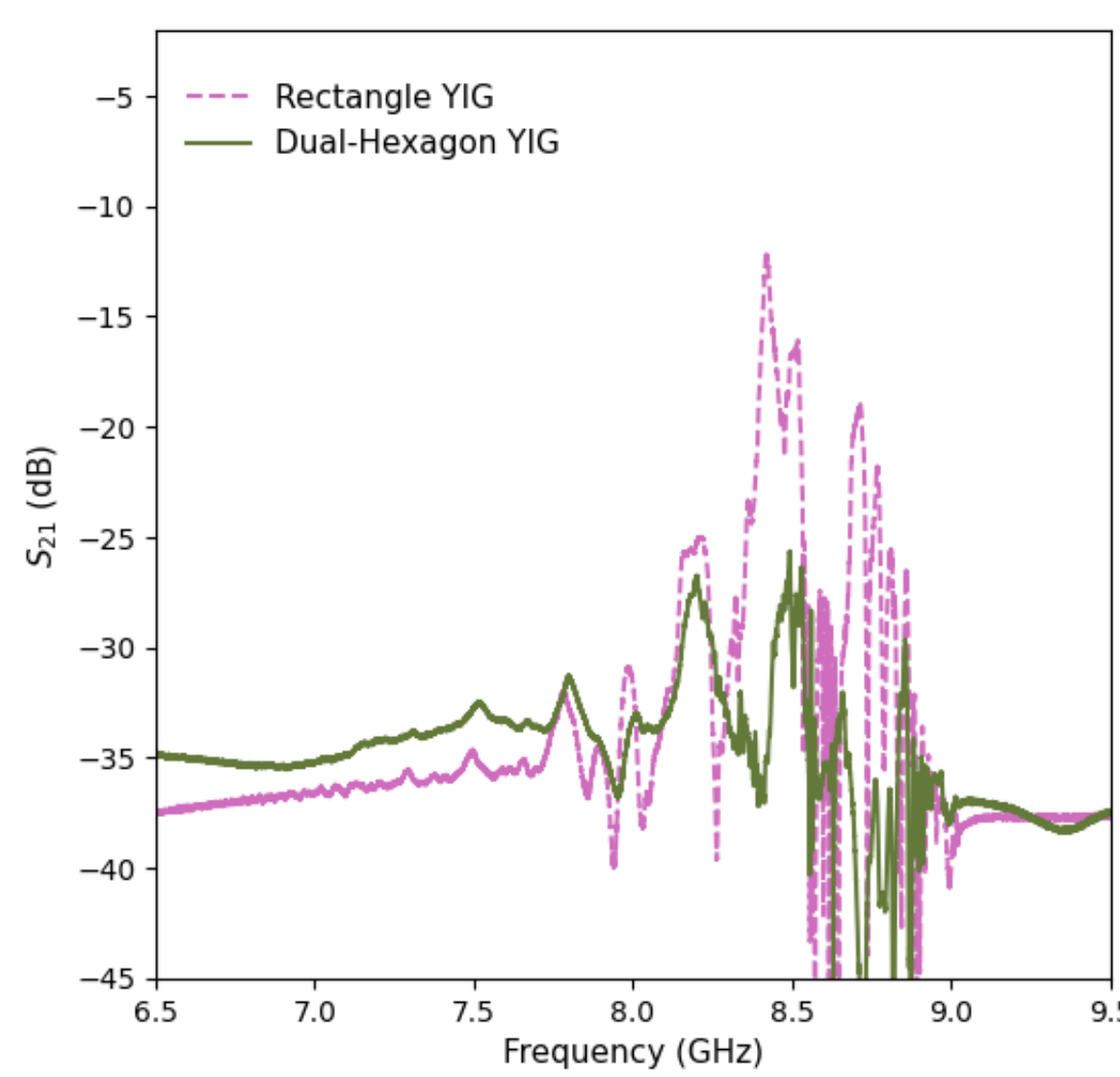

(d)

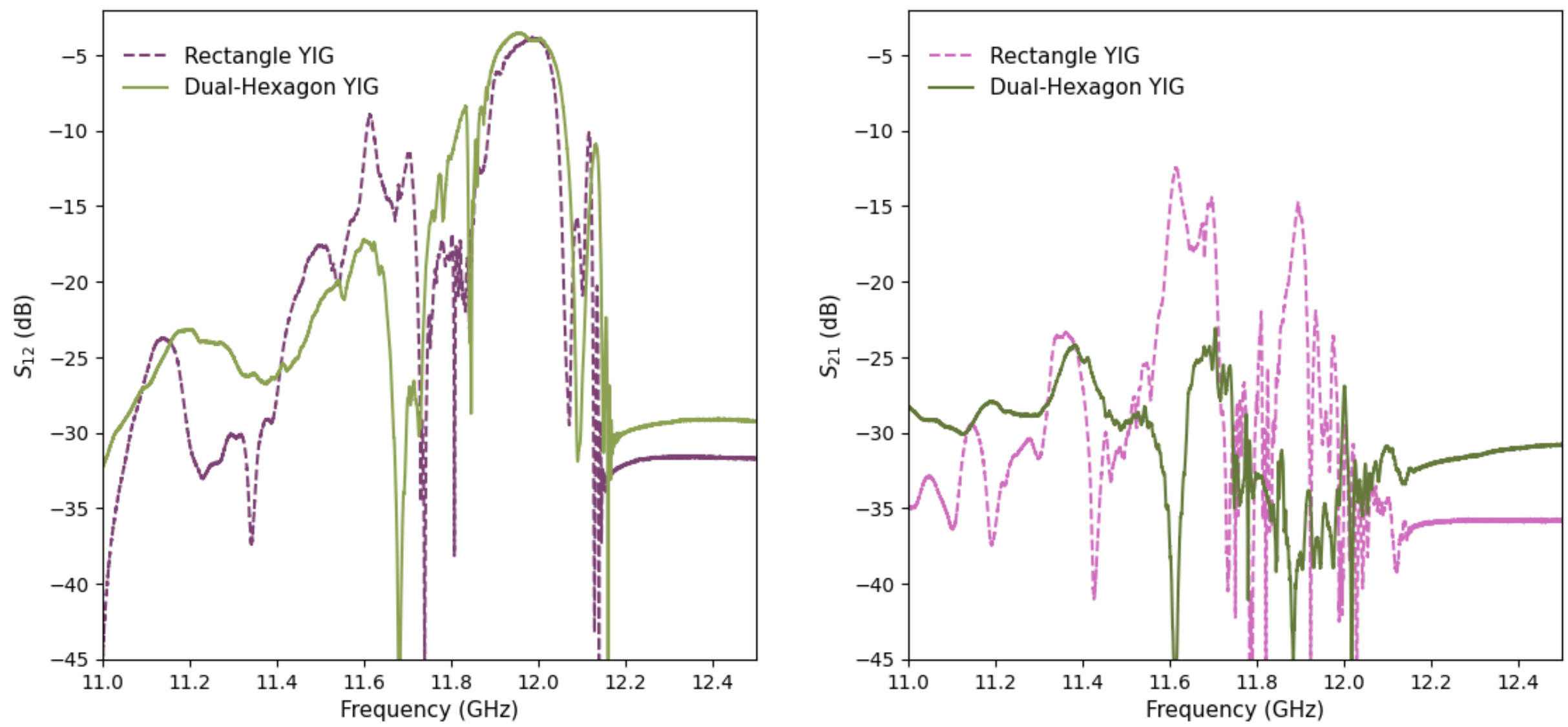

(e) (f)

**Supplementary Figure 24.** Comparison of the measured $S_{12}$ frequency responses for YIG filters with dual-hexagon shaped YIG and rectangle shaped YIG under three different applied magnetic flux densities of **(a)** ~1500, **(b)** ~2500, and **(c)** ~3500 Gauss, respectively.

# Supplementary Note #10: Power Handling Characteristics

**Supplementary Figures 25, 26, and 27** present a comparative analysis of the frequency responses of the filter with different input power levels, ranging from -30 dBm to 16 dBm. The filter under test is an 18 µm thick dual-hexagon-shaped YIG filter with two-parallel meander-line transducers. It features a waveguide width of 200 µm, a transducer pitch of 70 µm, and a transducer width of 10 µm.

The filter response remains unchanged up to an input power of approximately 8 dBm. The measured input 1 dB compression points (P1dB) are 10.4 dBm, 11.3 dBm, 11.5 dBm, 12.0 dBm, 12.0 dBm, 11.9 dBm, 14.3 dBm, and 13.5 dBm at 4.9 GHz, 6.2 GHz, 7.5 GHz, 9.1 GHz, 10.2 GHz, 11.9 GHz, 13.3 GHz, and 14.7 GHz, respectively. This large power handling capability has been significantly improved from a filter implemented using a straight-line transducer in a 3 µm thick YIG film, with P1dB of -17 dBm to -14 dBm reported in a previous study [6]. **Supplementary Figure 28** reports the power handling of a meander-line transducer filter realized in a 3 µm thick YIG film with a P1dB near 0 dBm. The design of this type of filter was reported in [16]. It can be concluded that the enhanced power handling is attributed primarily to the increased YIG thickness. Moreover, the filter reported in this work maintains consistent out-of-band rejection and nonreciprocity characteristics across the tested power range.

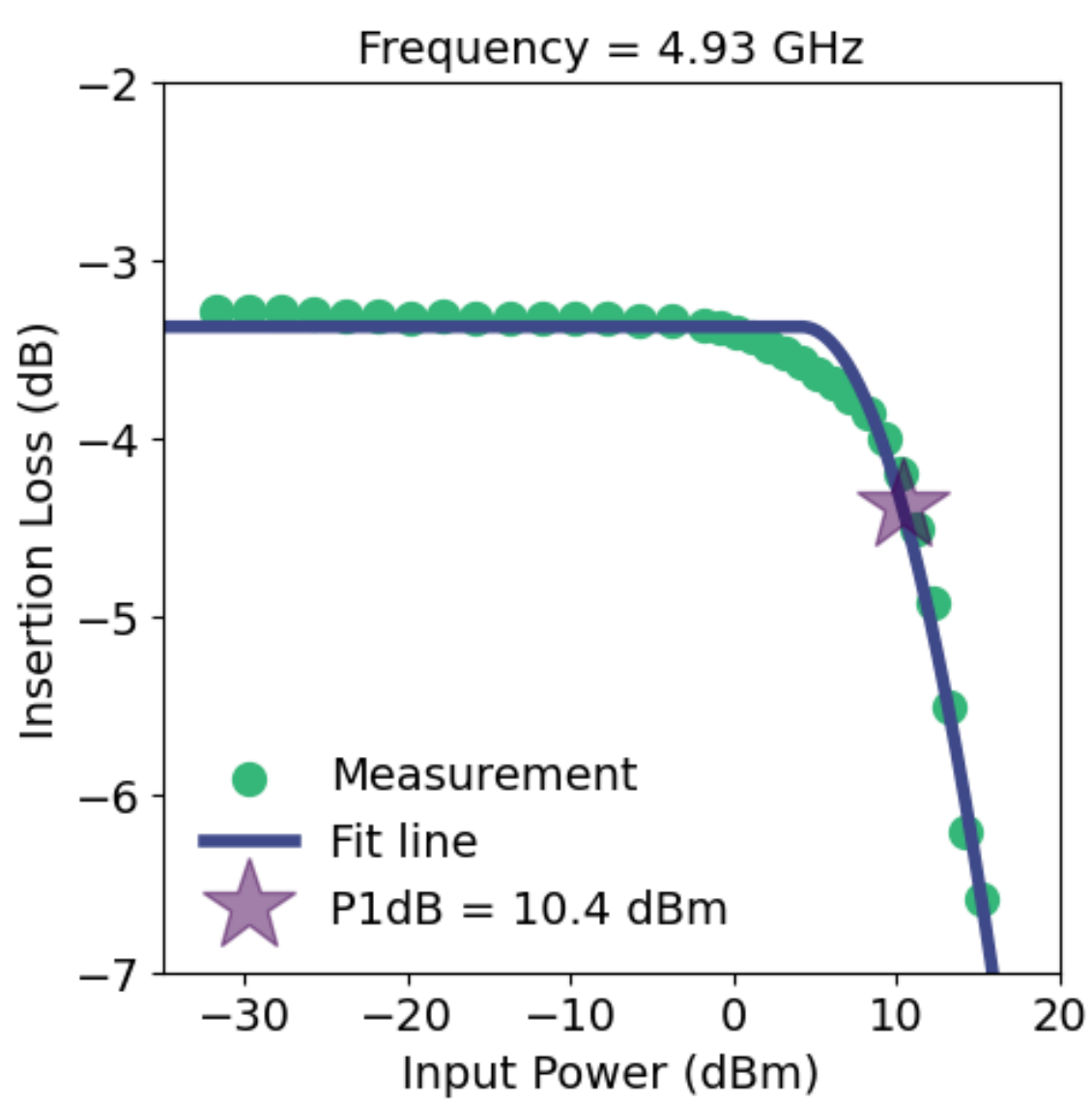

(a)

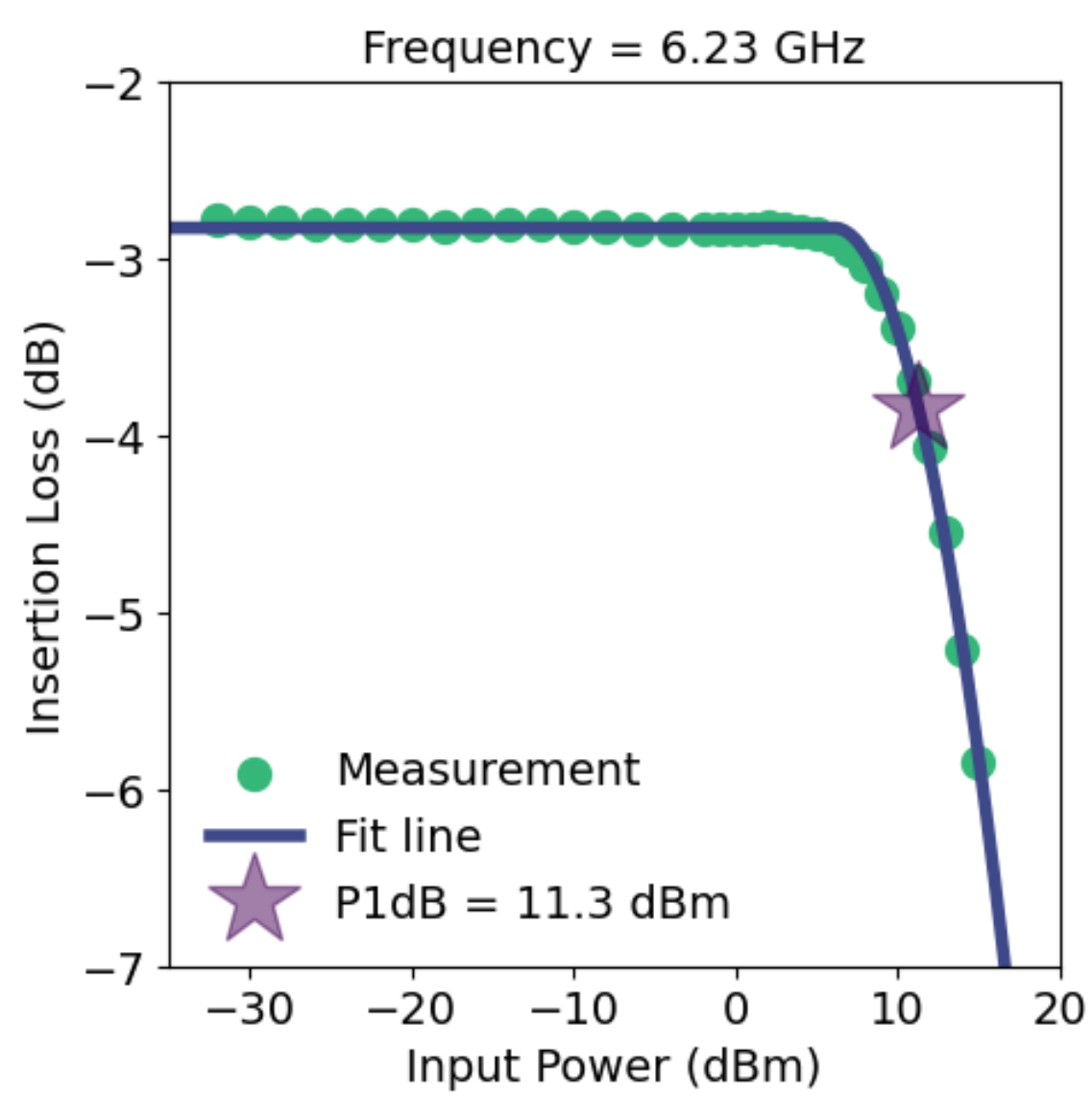

(b)

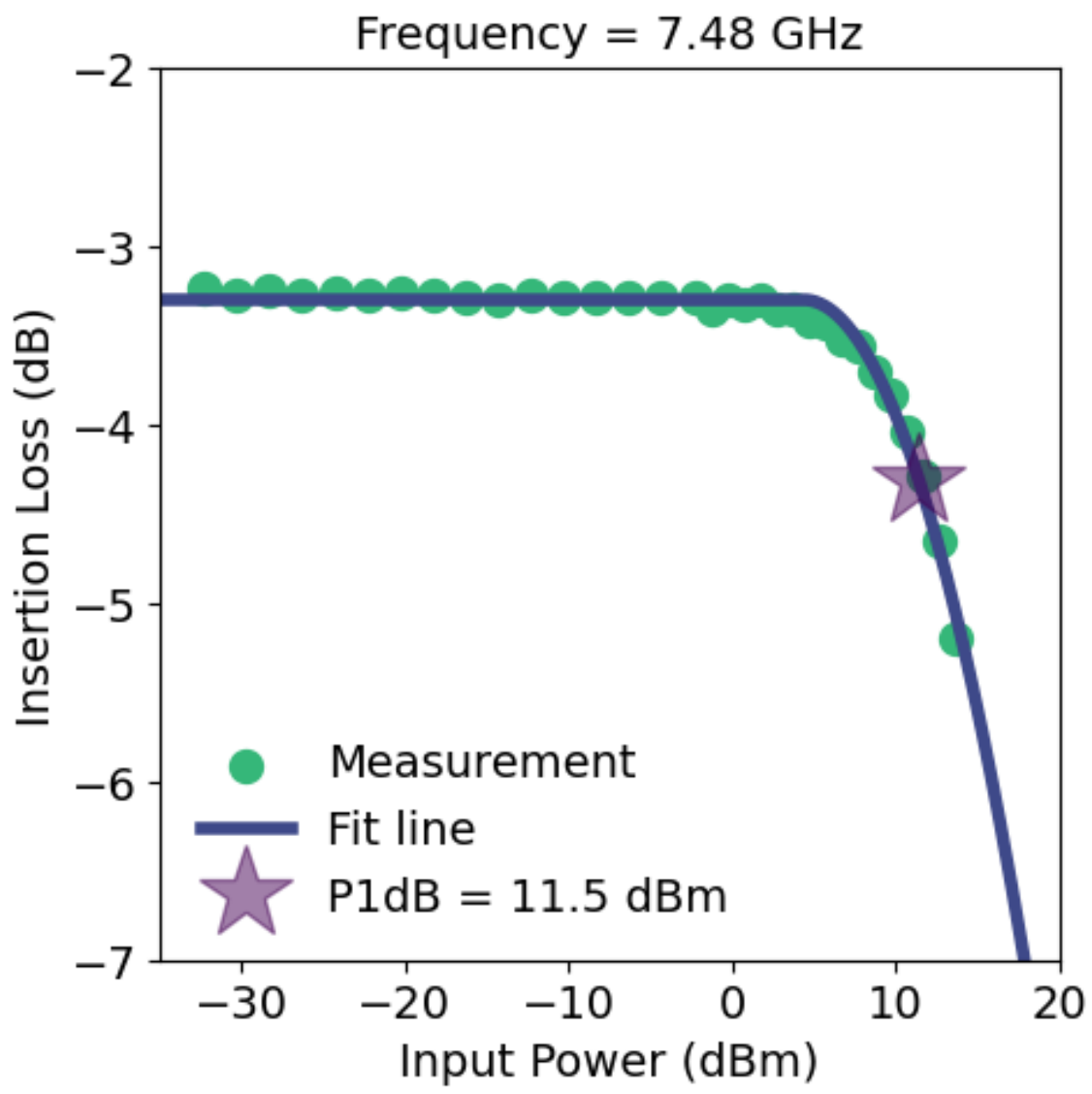
Frequency = 7.48 GHz
Insertion Loss (dB)
Input Power (dBm)
Measurement
Fit line
P1dB = 11.5 dBm
−2
−3
−4
−5
−6
−7
−30
−20
−10
0
10
20

(c)

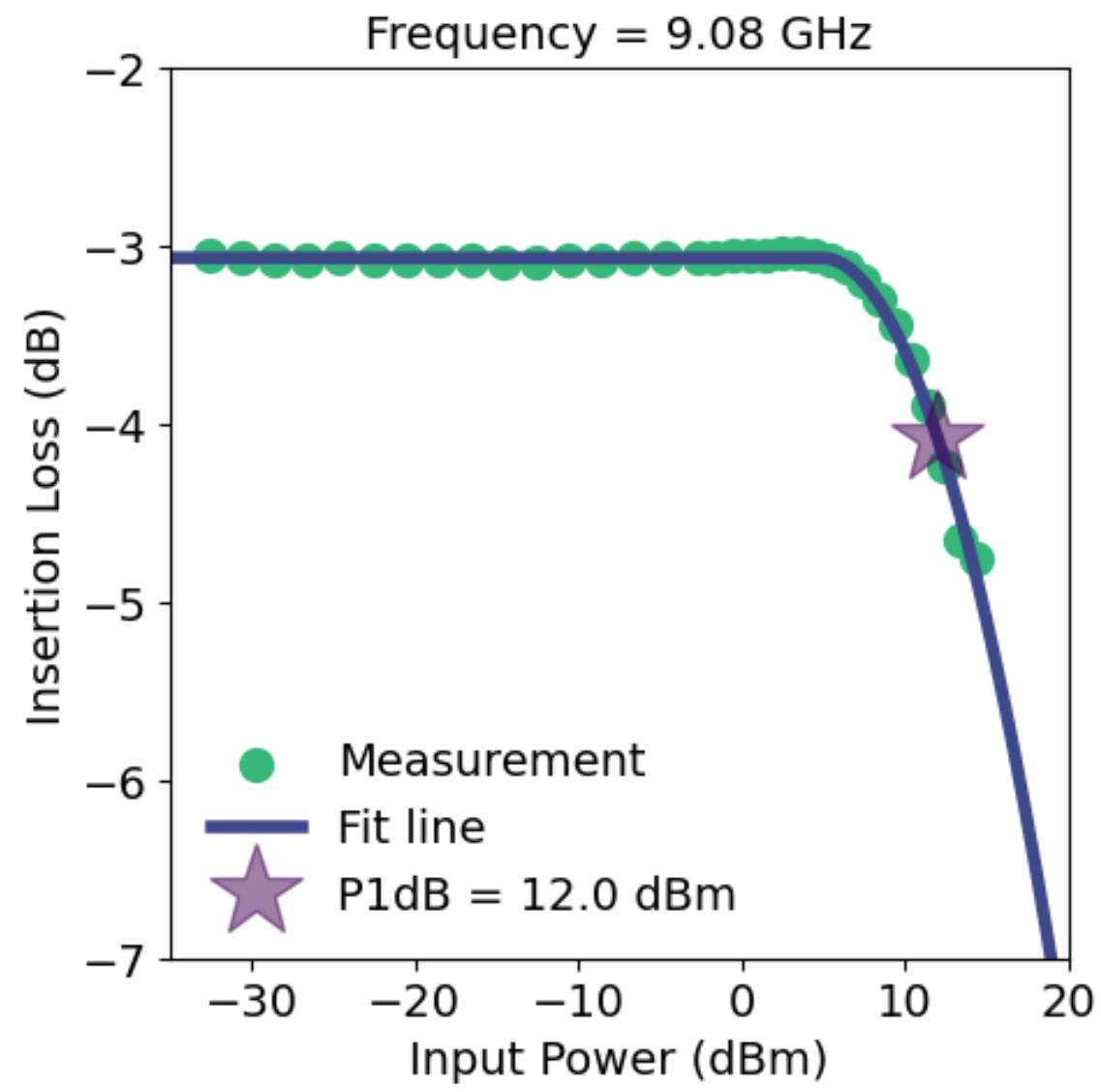
Frequency = 9.08 GHz
Insertion Loss (dB)
Input Power (dBm)
Measurement
Fit line
P1dB = 12.0 dBm
−2
−3
−4
−5
−6
−7
−30
−20
−10
0
10
20

(d)

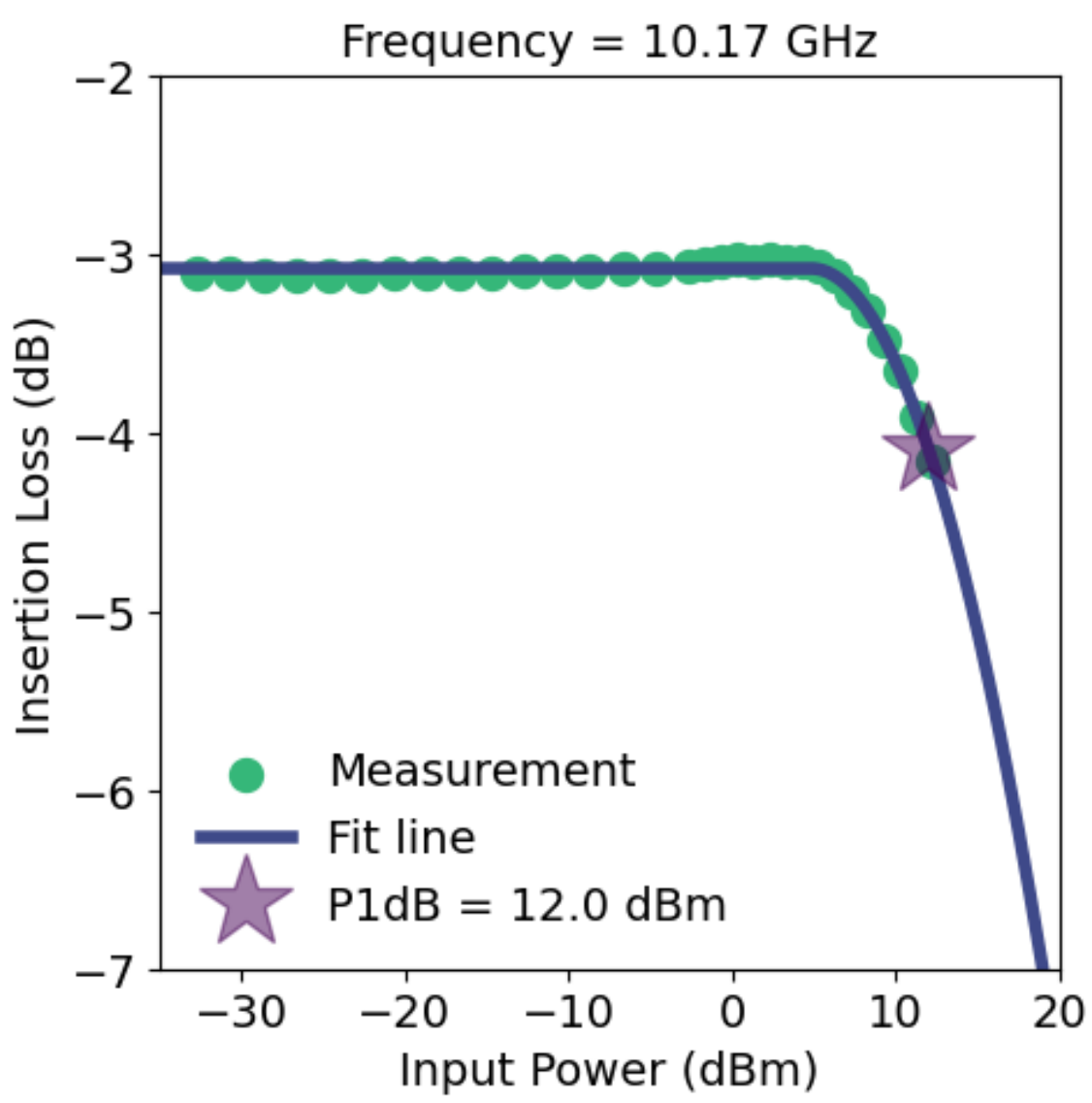
Frequency = 10.17 GHz
Insertion Loss (dB)
Input Power (dBm)
Measurement
Fit line
P1dB = 12.0 dBm
−2
−3
−4
−5
−6
−7
−30
−20
−10
0
10
20

(e)

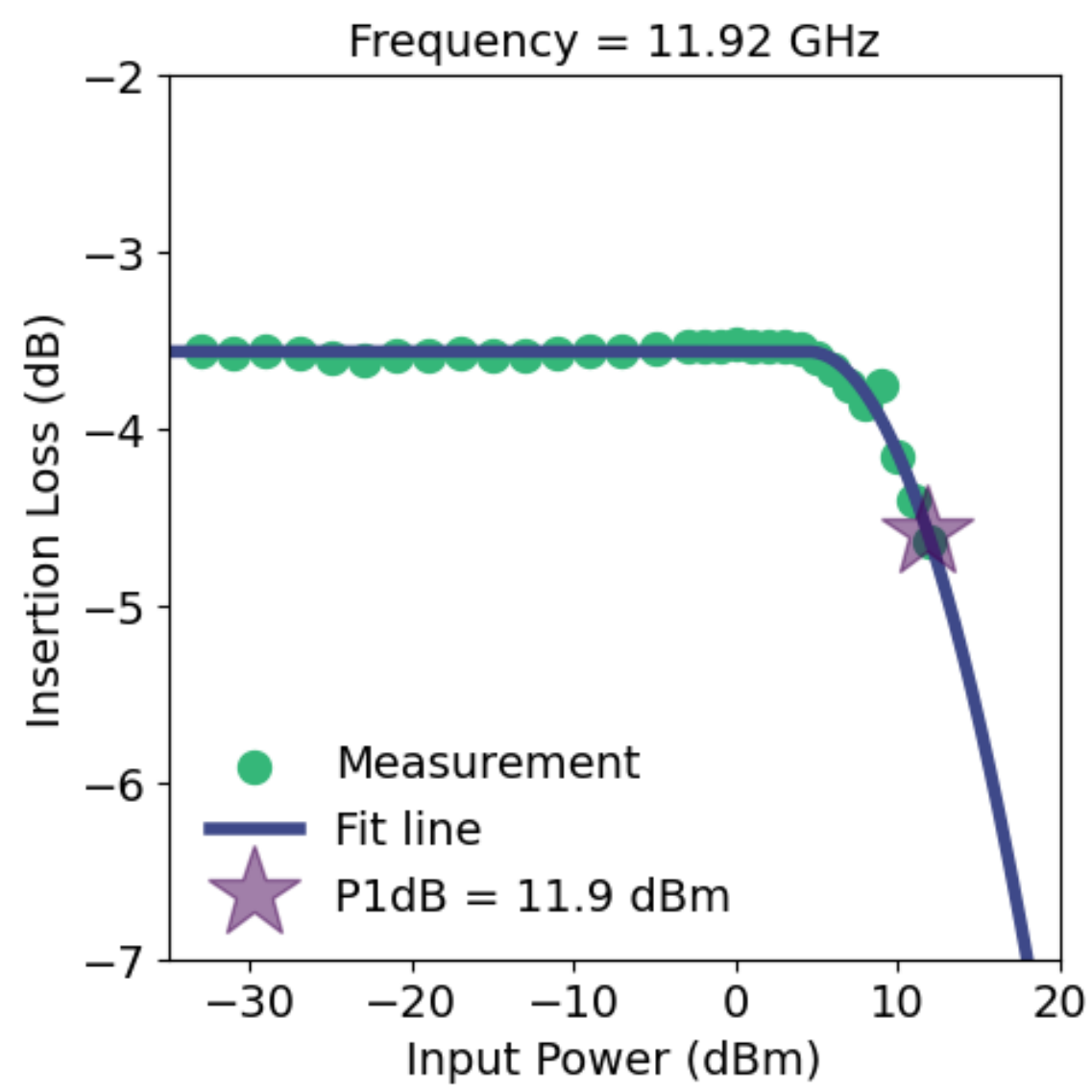
Frequency = 11.92 GHz
Insertion Loss (dB)
Input Power (dBm)
Measurement
Fit line
P1dB = 11.9 dBm
−2
−3
−4
−5
−6
−7
−30
−20
−10
0
10
20

(f)

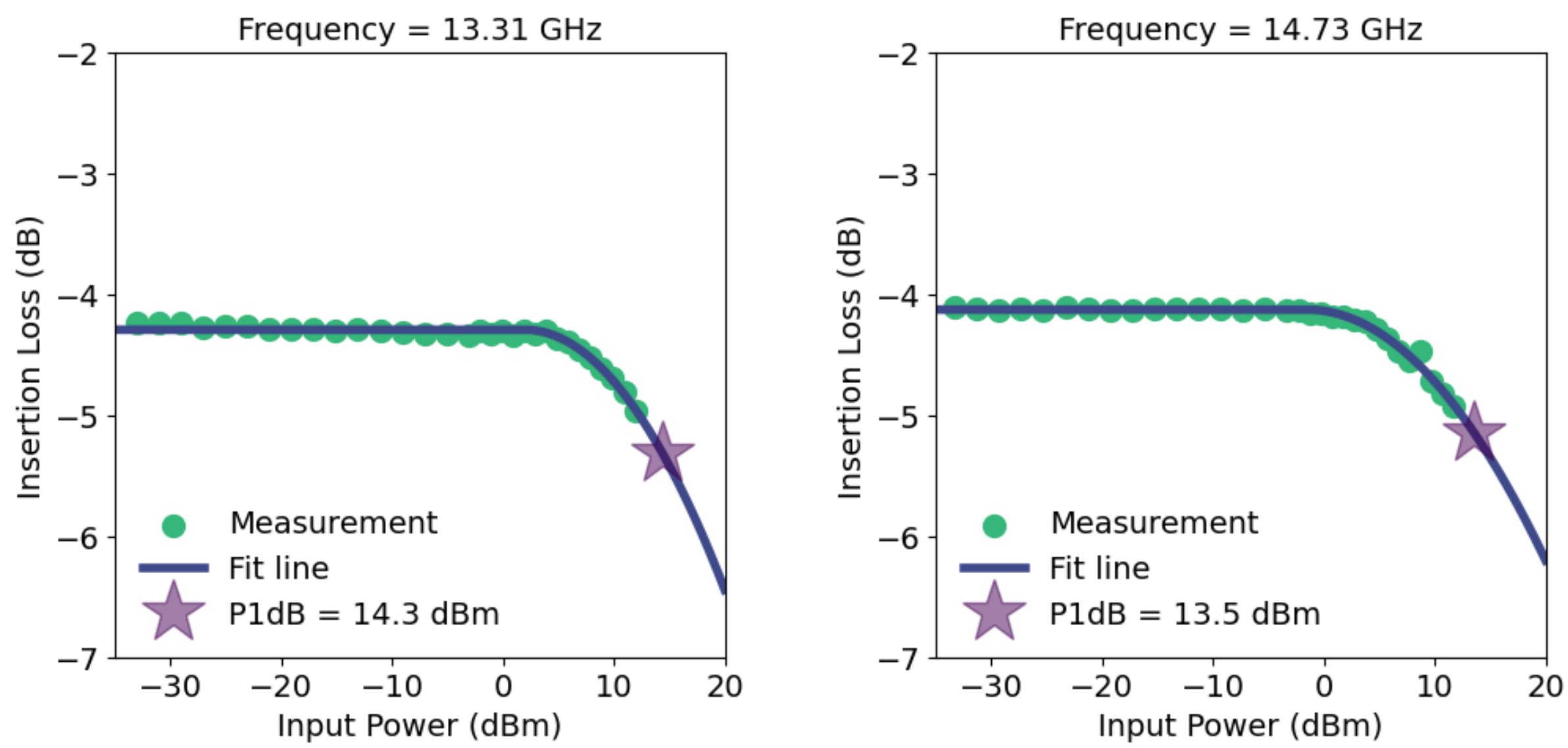

(g) (h)

**Supplementary Figure 25.** Comparison of the insertion loss with different input powers with bias field of **(a)** ~1000 Gauss, **(b)** ~1500 Gauss, **(c)** ~2000 Gauss, **(d)** ~2500 Gauss, **(e)** ~3000 Gauss, **(f)** ~3500 Gauss, **(g)** ~4000 Gauss, **(h)** ~4500 Gauss.

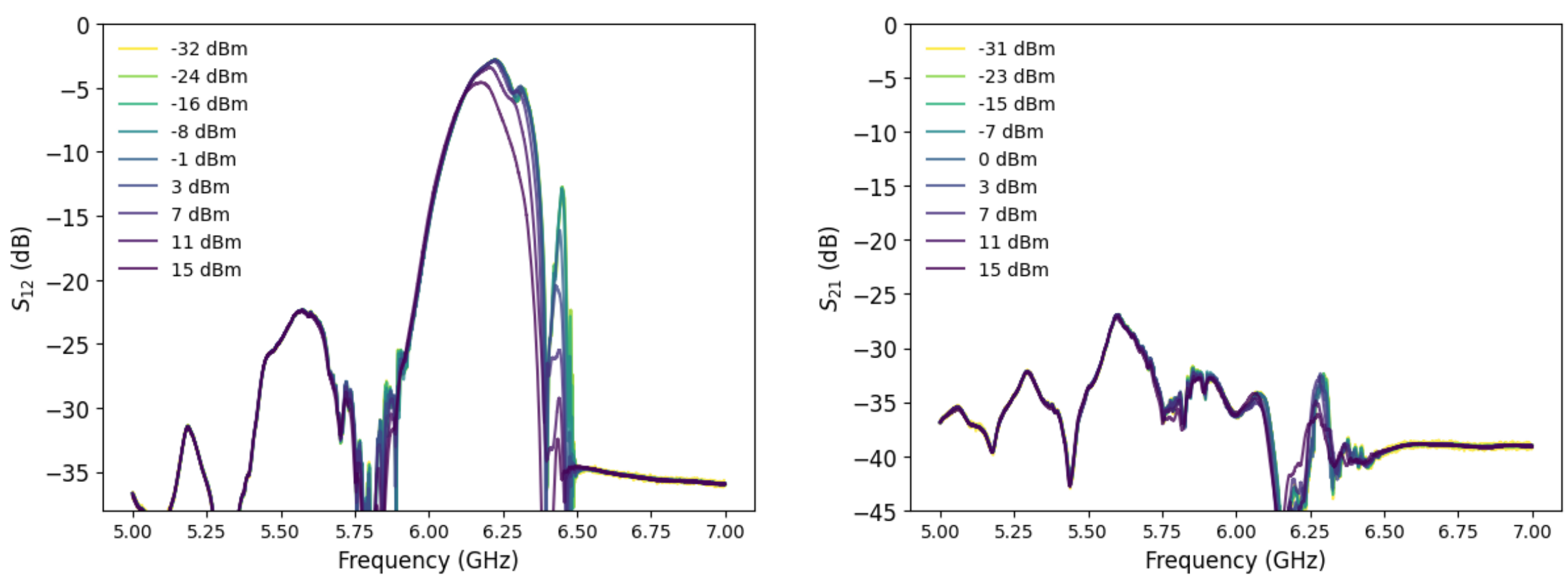

(a) (b)

**Supplementary Figure 26.** Comparison of the measured (a) $S_{12}$ and (b) $S_{21}$ frequency responses with different input power with a bias field of ~1500 Gauss.

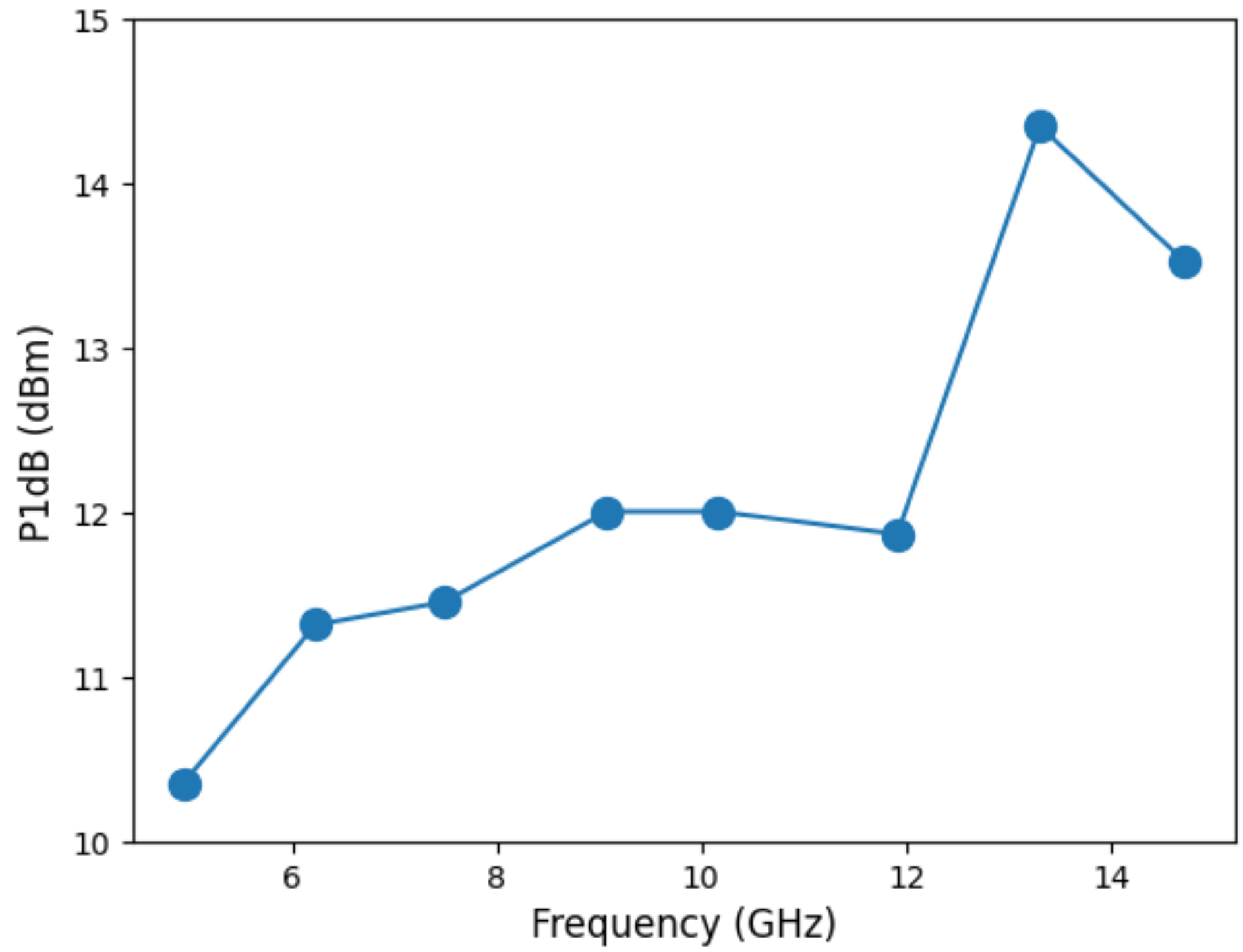

**Supplementary Figure 27.** Summary of the P1dB results as a function of frequency.

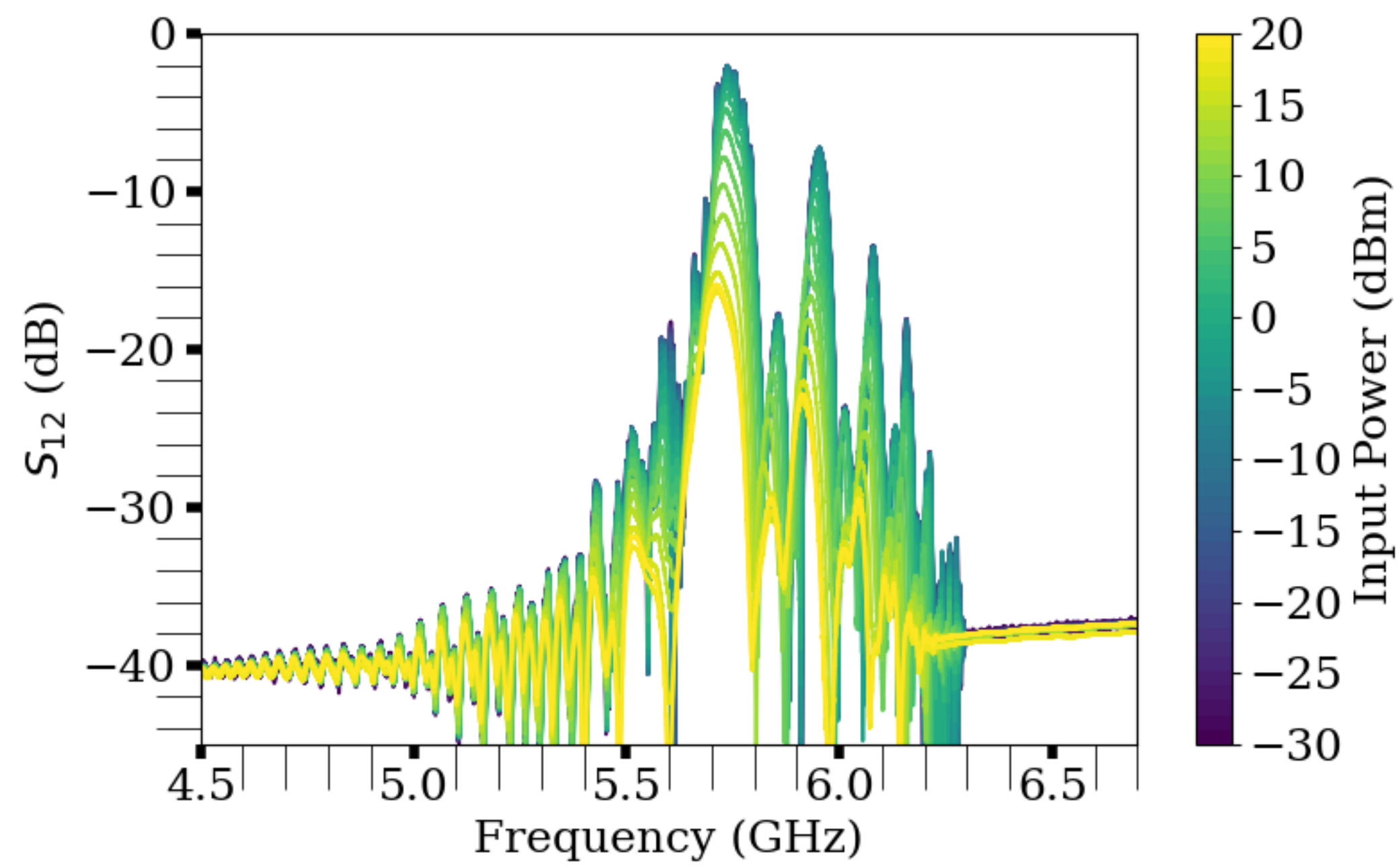

(a)

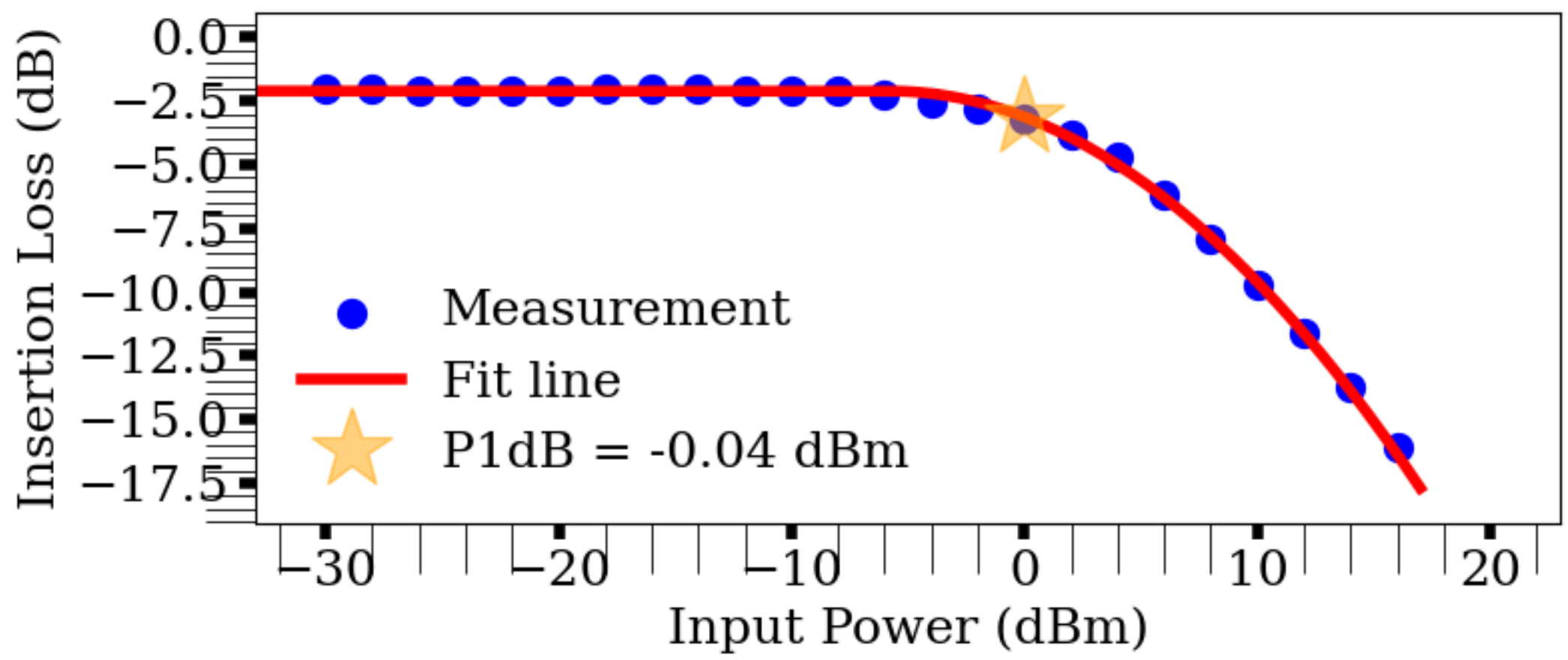

(b)

**Supplementary Figure 28.** Comparison of the measured **(a)** $S_{12}$ frequency responses and **(b)** insertion loss with different input power at 5.8 GHz. This 3 µm thick YIG meander-line filter was previously reported in [16].

## Supplementary Note #11: In-Band Linearity

The in-band third order intermodulation intercept point (IIP3) measurement setup followed the procedure described in a previous report [6], as shown in **Supplementary Figure 29**. Two-tone signals ($f_1$ and $f_2$) were generated using two signal generators (Keysight P5024B). The combined signal was amplified by a Minicircuits ZX60-24-S+ amplifier. To suppress unwanted harmonics and spurious tones, two bandpass filters (Minicircuits ZBSS-12G-S+) and two circulators (DiTom Microwave DMC6018) were employed to clean the amplified signals. The two tones were then combined using a power combiner (Krytar 6020265), passed through an additional circulator, and applied to the device under test (the YIG filter). The output of the YIG filter was routed through another circulator and a 10 dB attenuator before being connected to a spectrum analyzer (Keysight P5024B).

**Supplementary Figure 30** presents the resulting in-band IIP3 data measured at 7 GHz and 11 GHz for tone spacings of 1 MHz, 10 MHz, and 100 MHz. For 1 MHz and 10 MHz tone spacings the third order intermodulation (IM3) products fall fully within the filter passband. Over all conditions the IIP3 remains above +26 dBm, a significant improvement over prior reports [5, 6, 17] demonstrating the improve linearity associated with realizing filters in the thicker YIG films.

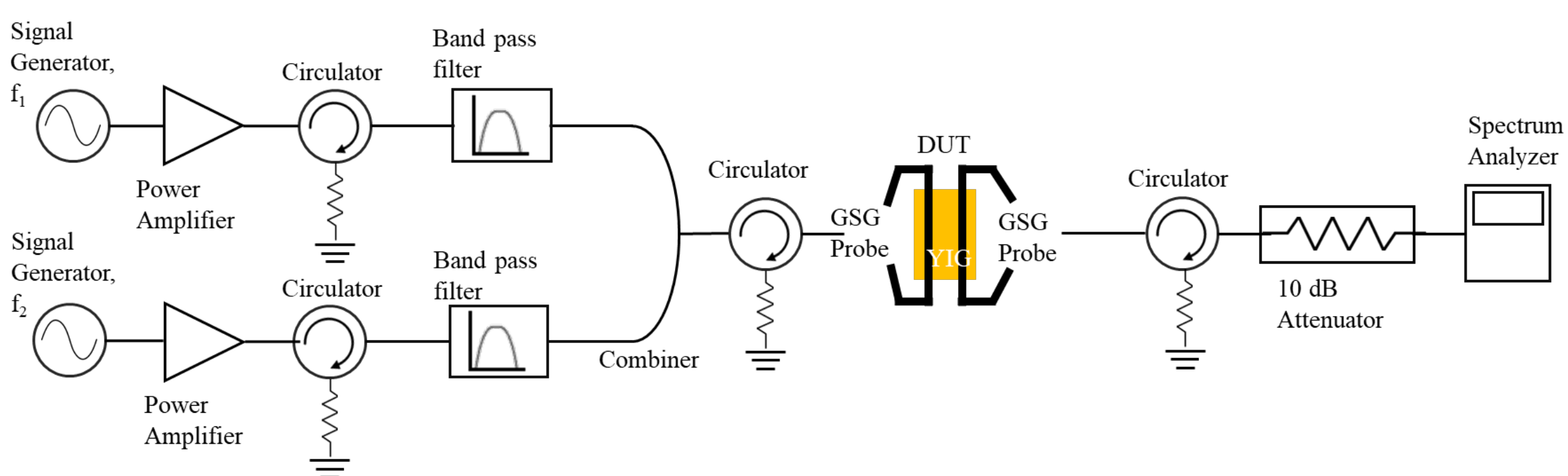

**Supplementary Figure 29.** In-band third order intermodulation intercept point (IIP3) measurement setup schematic.

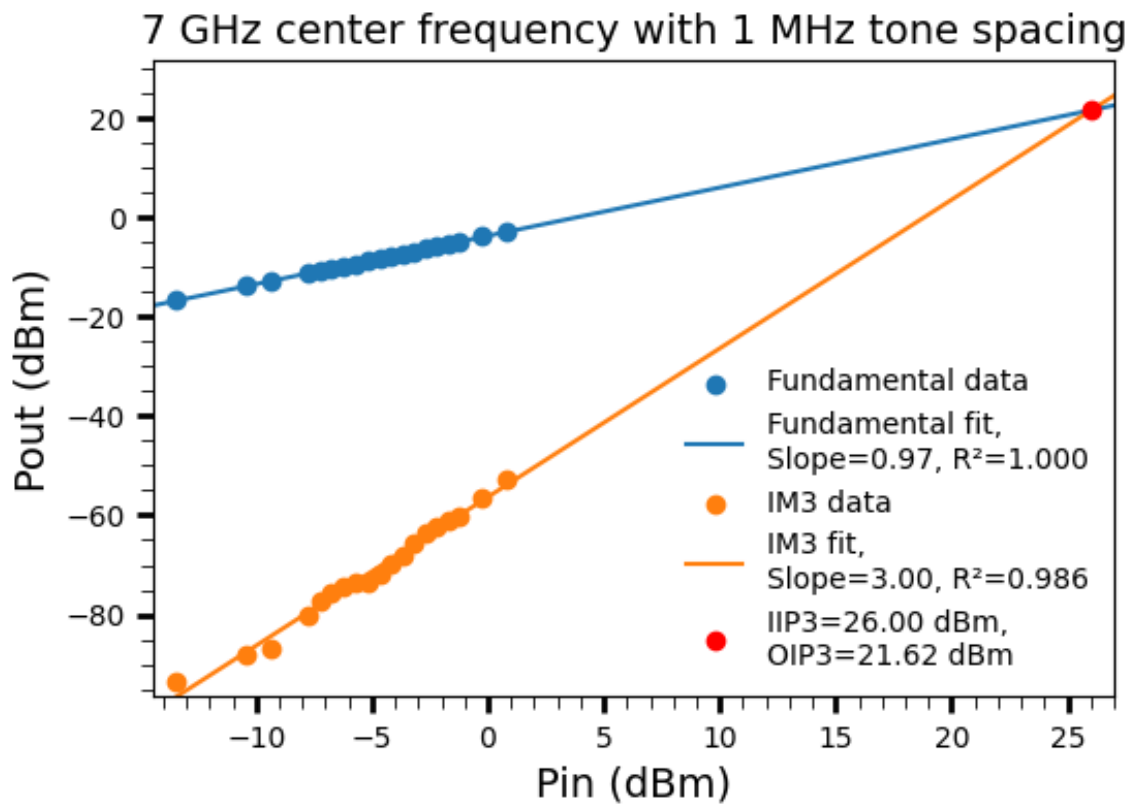
7 GHz center frequency with 1 MHz tone spacing
Pout (dBm)
Pin (dBm)
Fundamental data
Fundamental fit,
Slope=0.97, R²=1.000
IM3 data
IM3 fit,
Slope=3.00, R²=0.986
IIP3=26.00 dBm,
OIP3=21.62 dBm

(a)

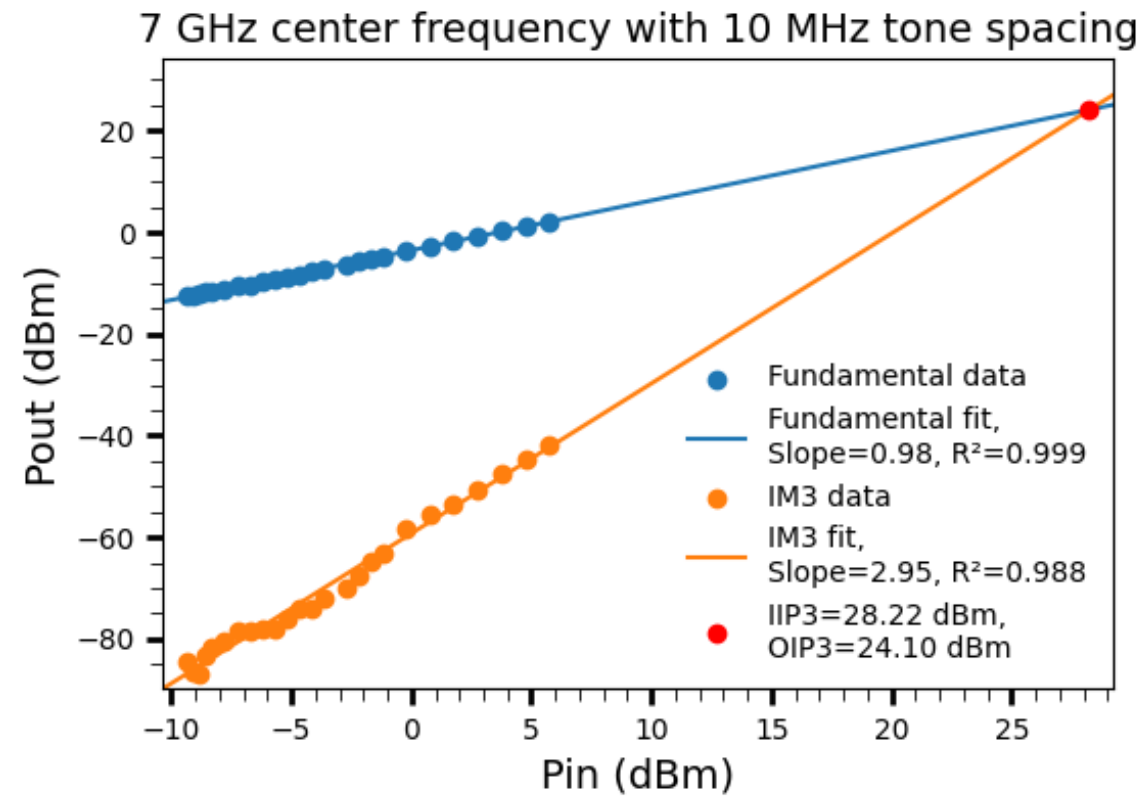
7 GHz center frequency with 10 MHz tone spacing
Pout (dBm)
Pin (dBm)
Fundamental data
Fundamental fit,
Slope=0.98, R²=0.999
IM3 data
IM3 fit,
Slope=2.95, R²=0.988
IIP3=28.22 dBm,
OIP3=24.10 dBm

(b)

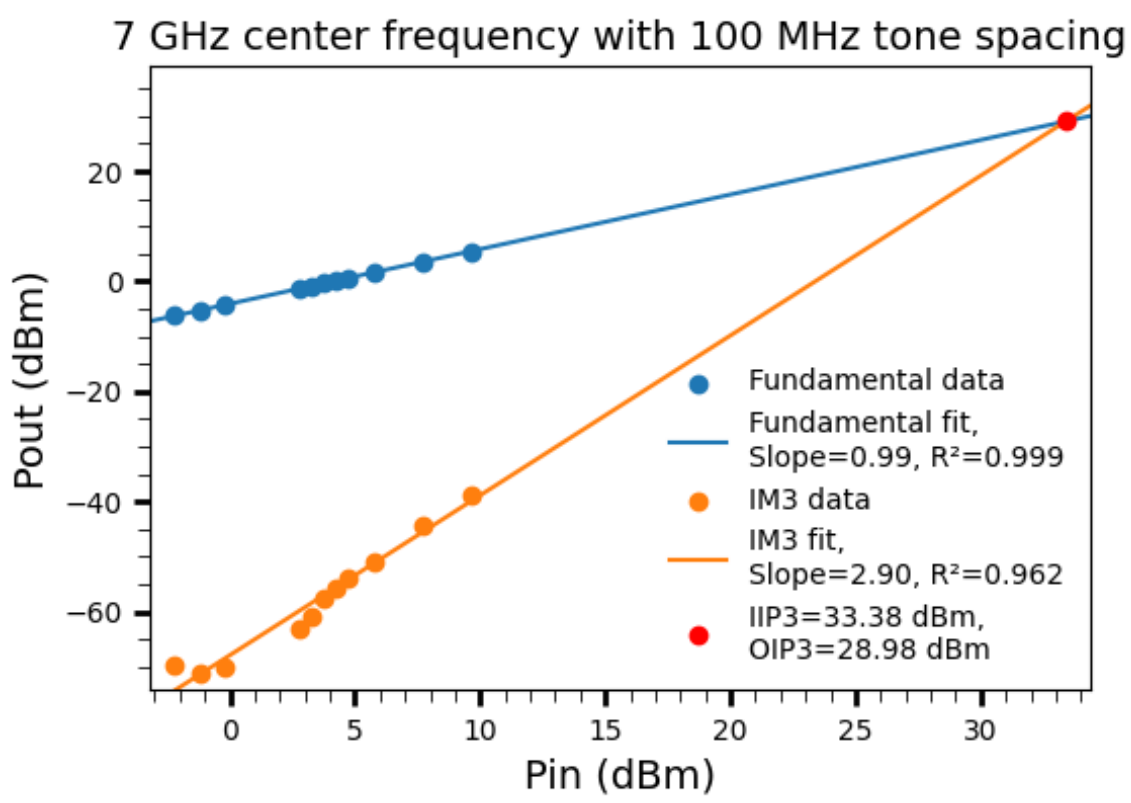
7 GHz center frequency with 100 MHz tone spacing
Pout (dBm)
Pin (dBm)
Fundamental data
Fundamental fit,
Slope=0.99, R²=0.999
IM3 data
IM3 fit,
Slope=2.90, R²=0.962
IIP3=33.38 dBm,
OIP3=28.98 dBm

(c)

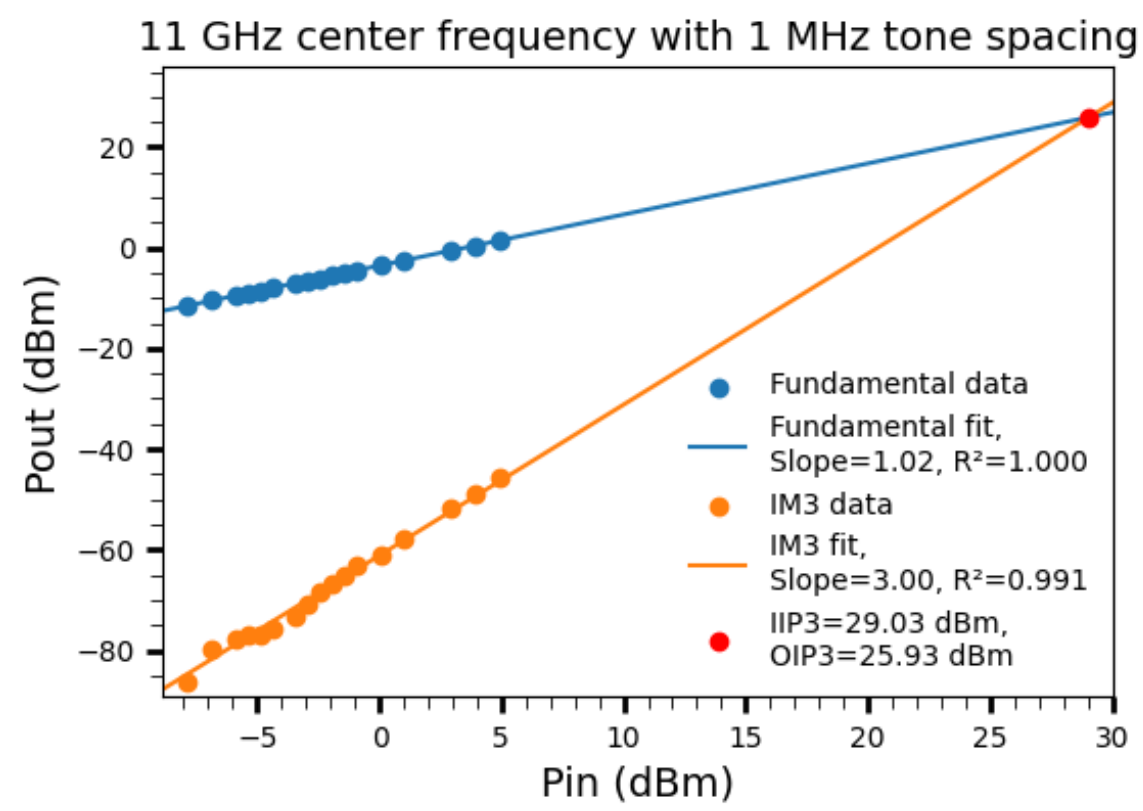
11 GHz center frequency with 1 MHz tone spacing
Pout (dBm)
Pin (dBm)
Fundamental data
Fundamental fit,
Slope=1.02, R²=1.000
IM3 data
IM3 fit,
Slope=3.00, R²=0.991
IIP3=29.03 dBm,
OIP3=25.93 dBm

(d)

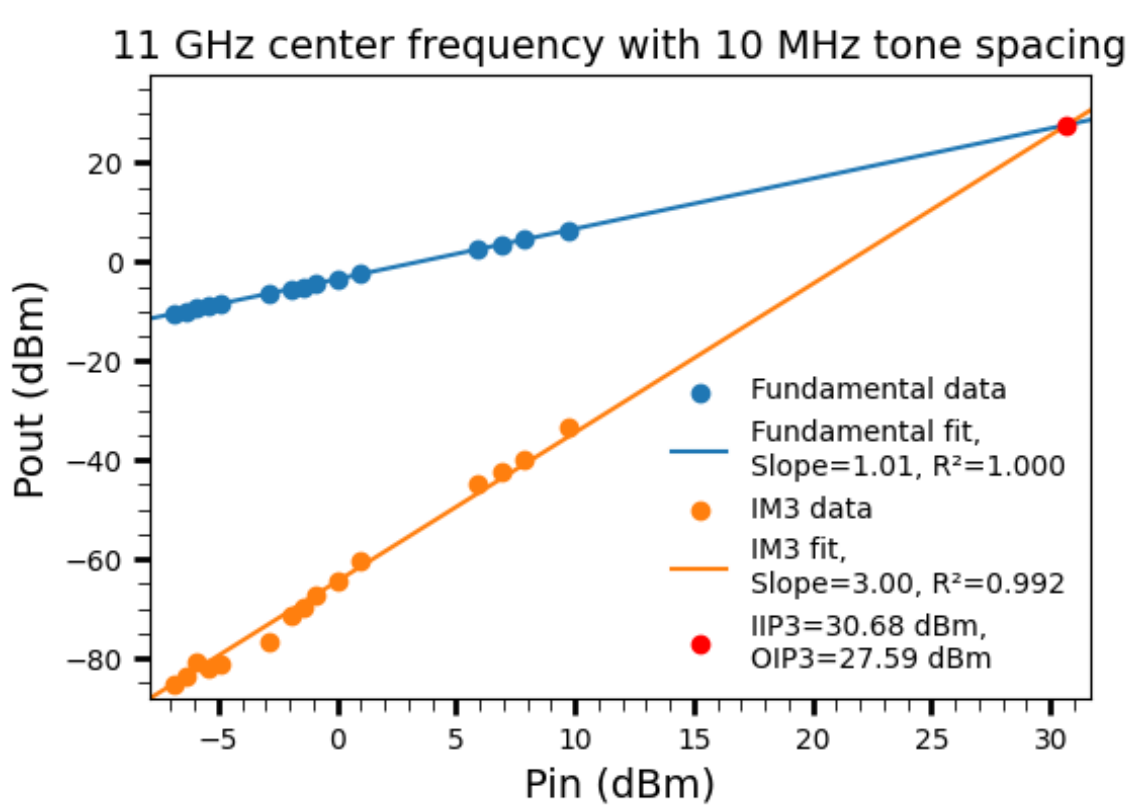
11 GHz center frequency with 10 MHz tone spacing
Pout (dBm)
Pin (dBm)
Fundamental data
Fundamental fit,
Slope=1.01, R²=1.000
IM3 data
IM3 fit,
Slope=3.00, R²=0.992
IIP3=30.68 dBm,
OIP3=27.59 dBm

(e)

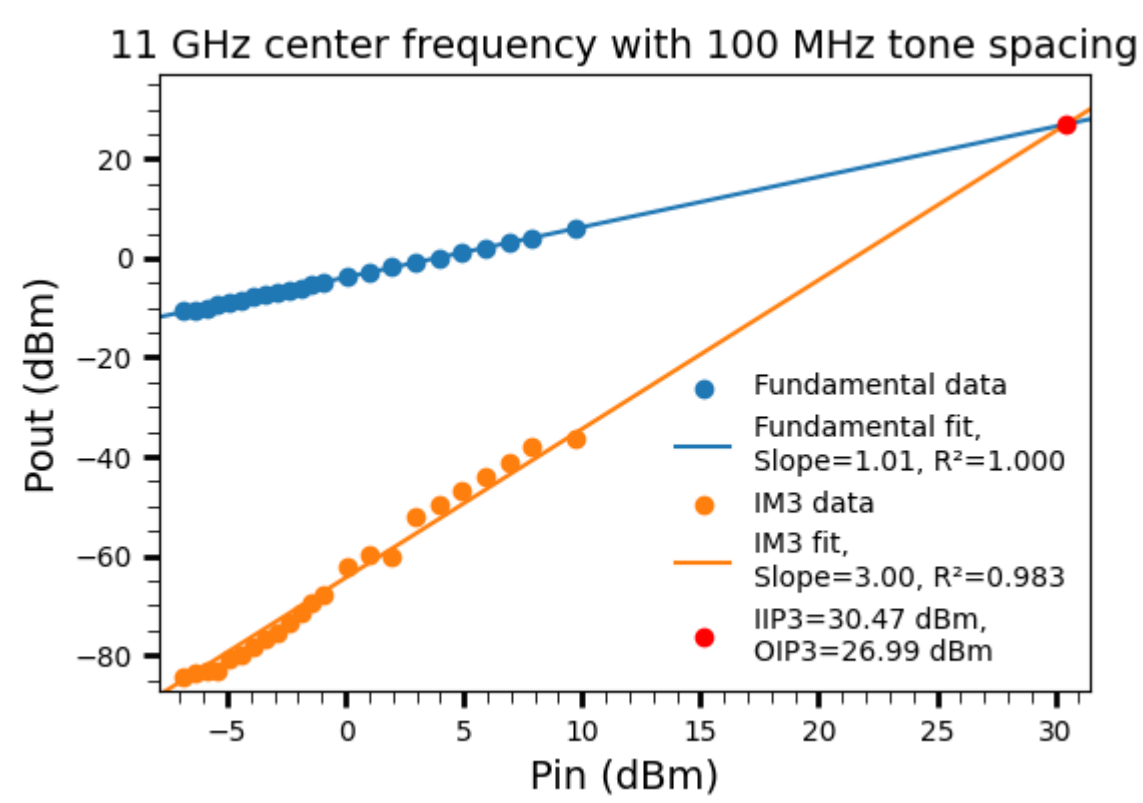
11 GHz center frequency with 100 MHz tone spacing
Pout (dBm)
Pin (dBm)
Fundamental data
Fundamental fit,
Slope=1.01, R²=1.000
IM3 data
IM3 fit,
Slope=3.00, R²=0.983
IIP3=30.47 dBm,
OIP3=26.99 dBm

(f)

**Supplementary Figure 30.** In-band input third order intercept point (IIP3). The device is measured at **(a, b, and c)** 7 GHz and **(d, e, and f)** 11 GHz with tone spacings of **(a and d)** 1 MHz, **(b and e)** 10 MHz, and **(c and f)** 100 MHz

# Supplementary Note #12: Integrated Filter

**Supplementary Figure 31** shows the measured $S_{21}$, $S_{22}$ and $S_{11}$ frequency response across 4-18 GHz.

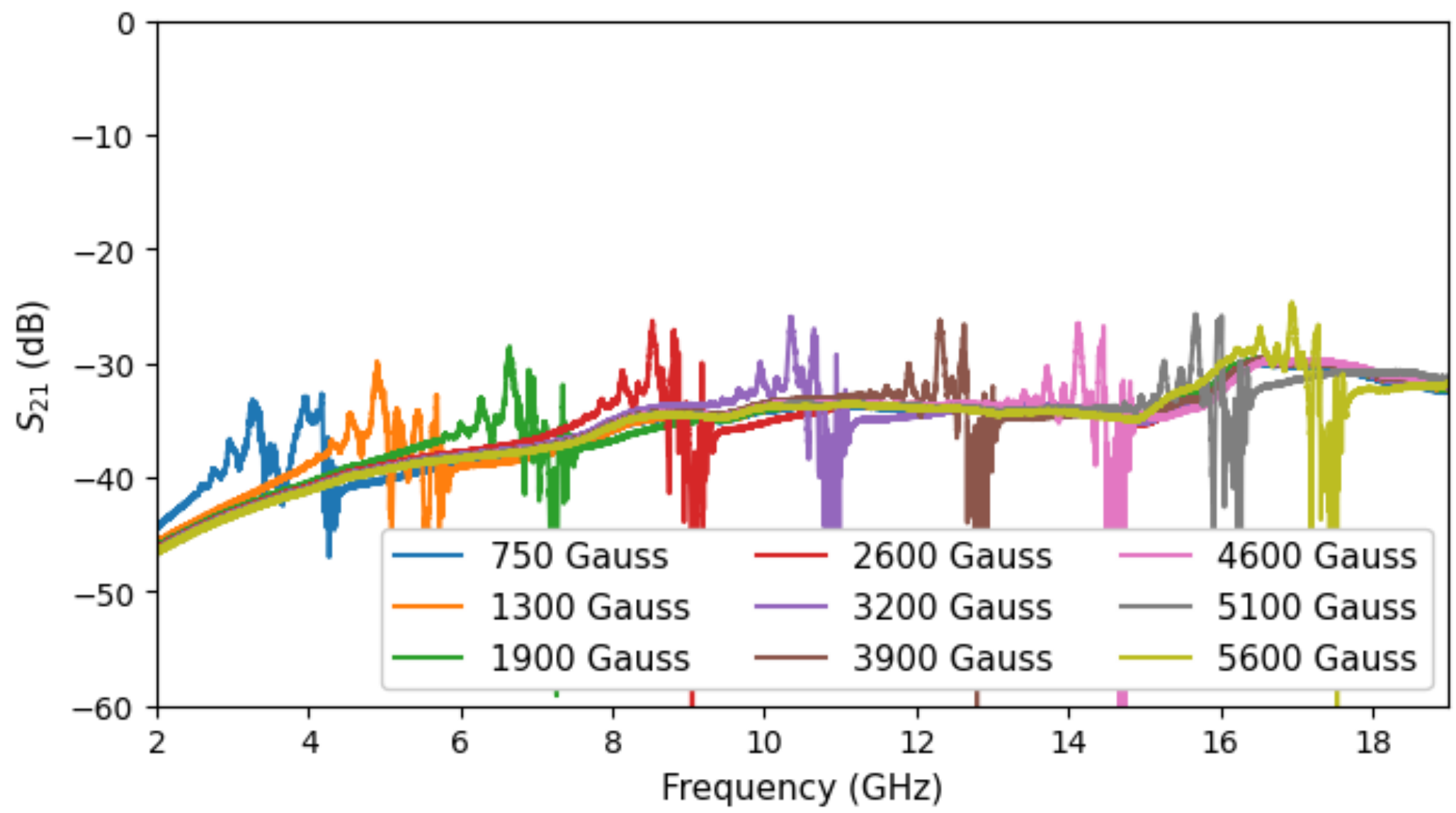

(a)

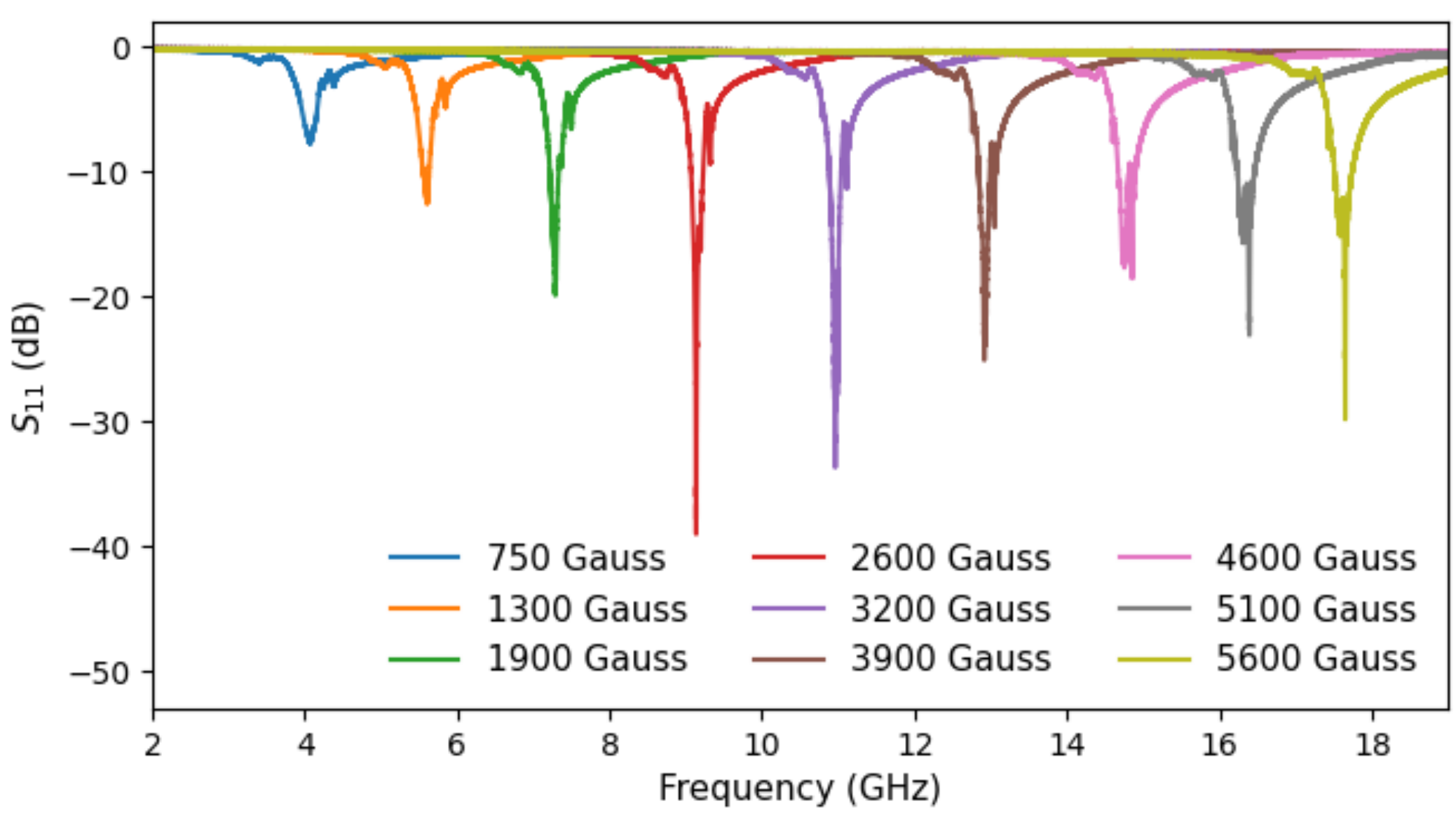

(b)

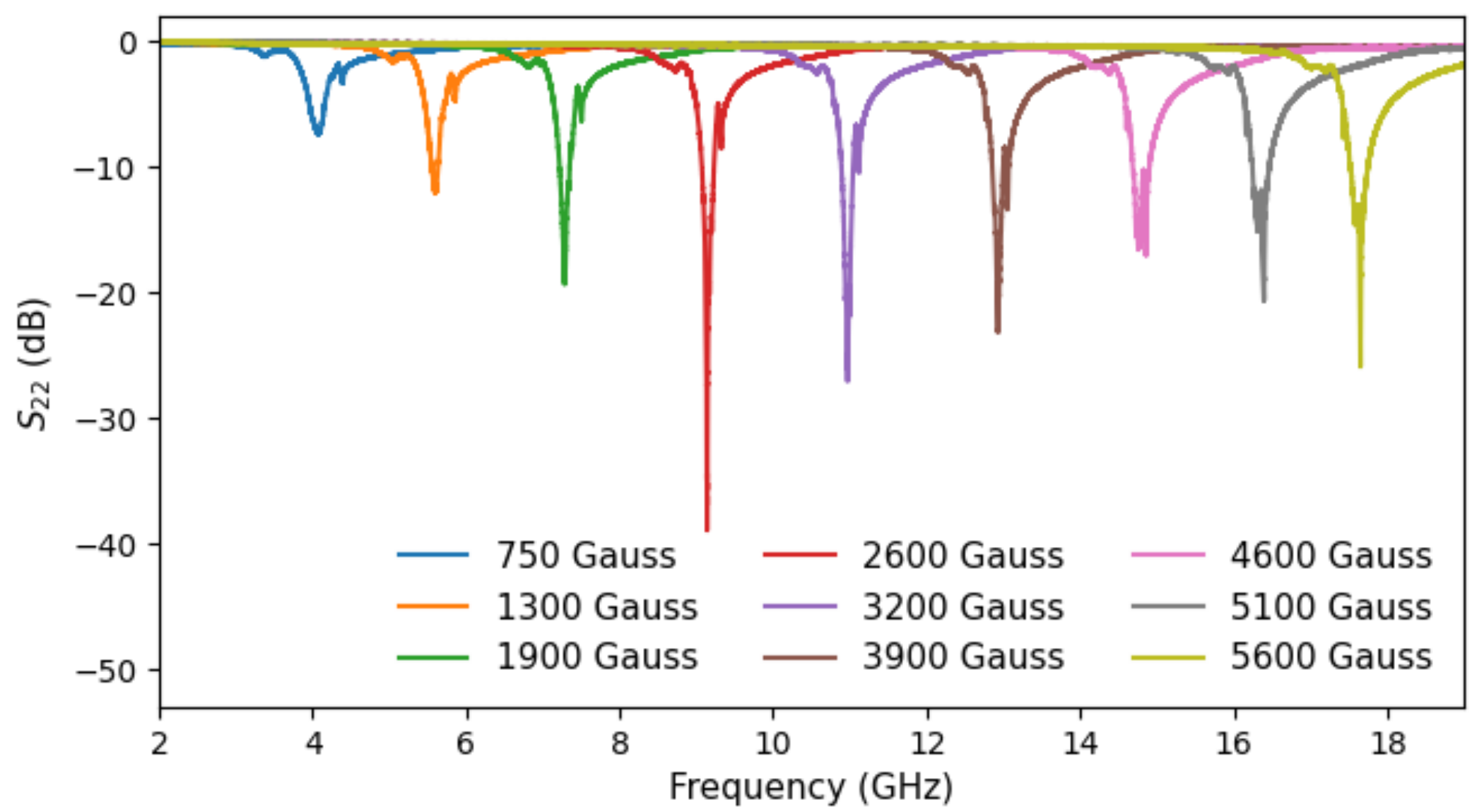

(c)

**Supplementary Figure 31.** Measured **(a)** $S_{21}$, **(b)** $S_{11}$, and **(c)** $S_{22}$ frequency responses with the magnetic field supplied by the zero static power consumption magnetic bias circuit.

**Supplementary Figure 32** highlights the small passband group delay variation of the filter. At 7.2 GHz, a minimum group delay variation of 1.0 ns is observed, with the group delay ranging from 2.7 ns to 3.7 ns. The largest group delay variation occurs at 16.3 GHz, where the delay ranges from 1.3 ns to 7.4 ns, resulting in a maximum variation of 6.0 ns. These results demonstrate the filter's ability to maintain relatively flat group delay performance across a wide tuning range.

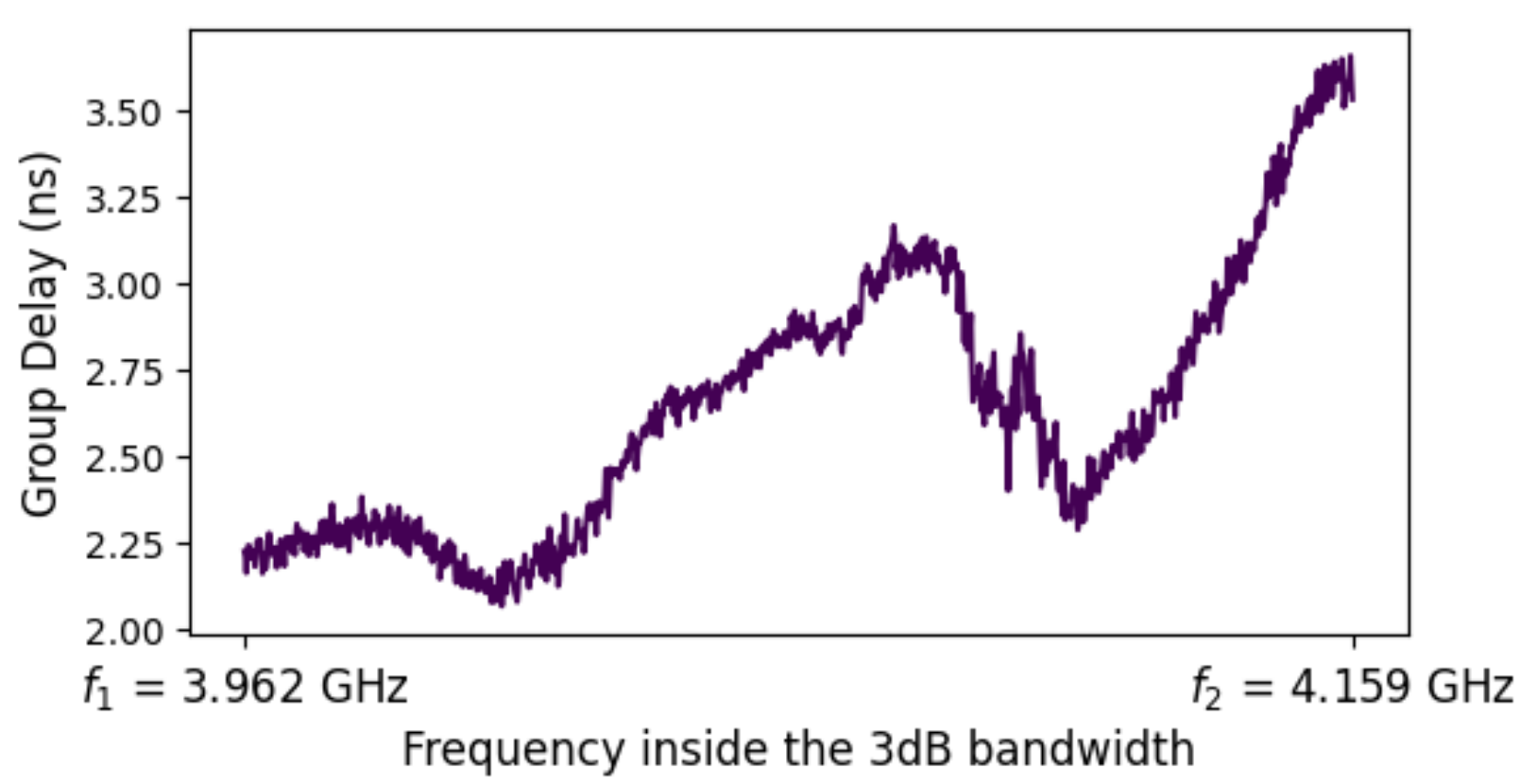

(a)

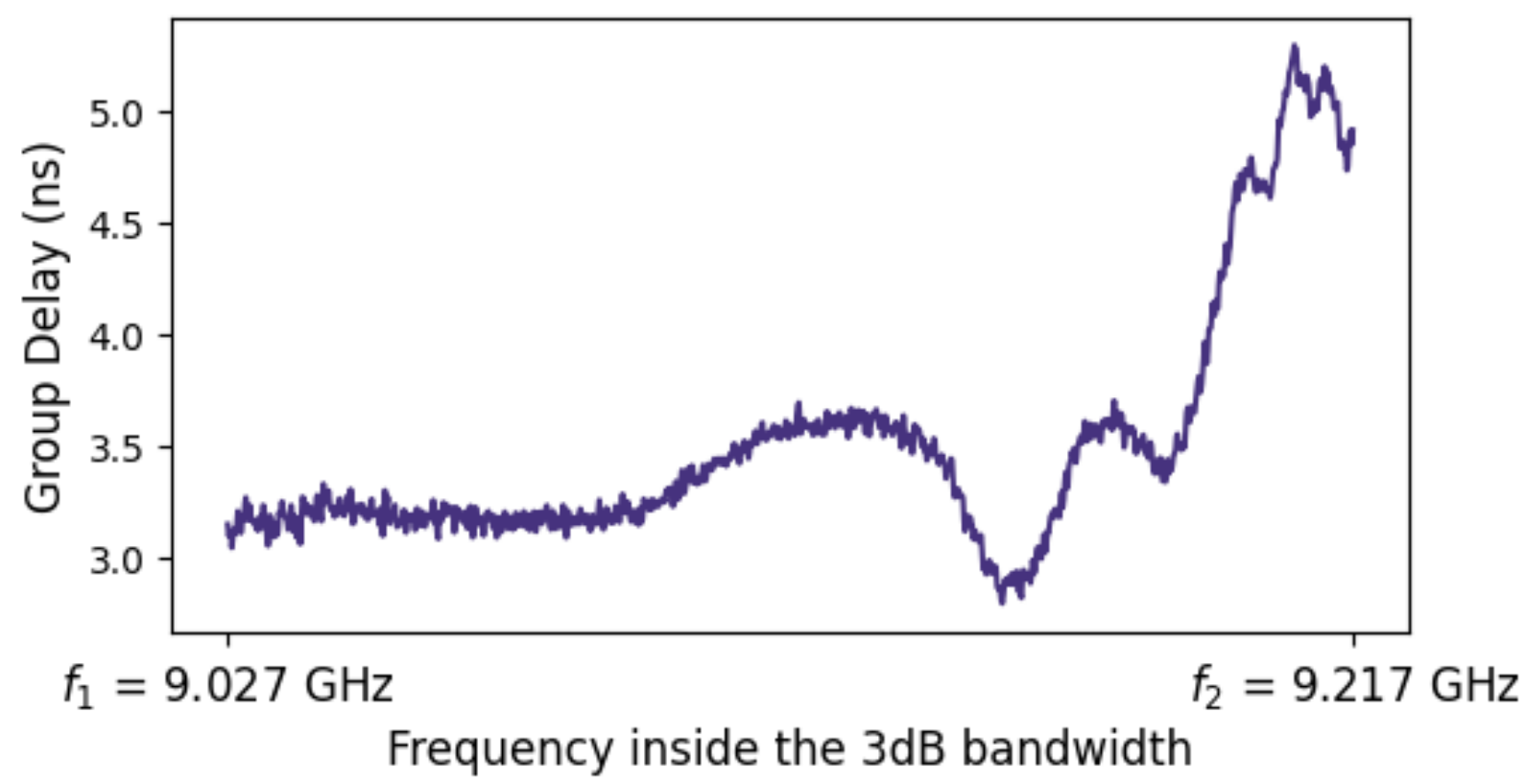

(b)

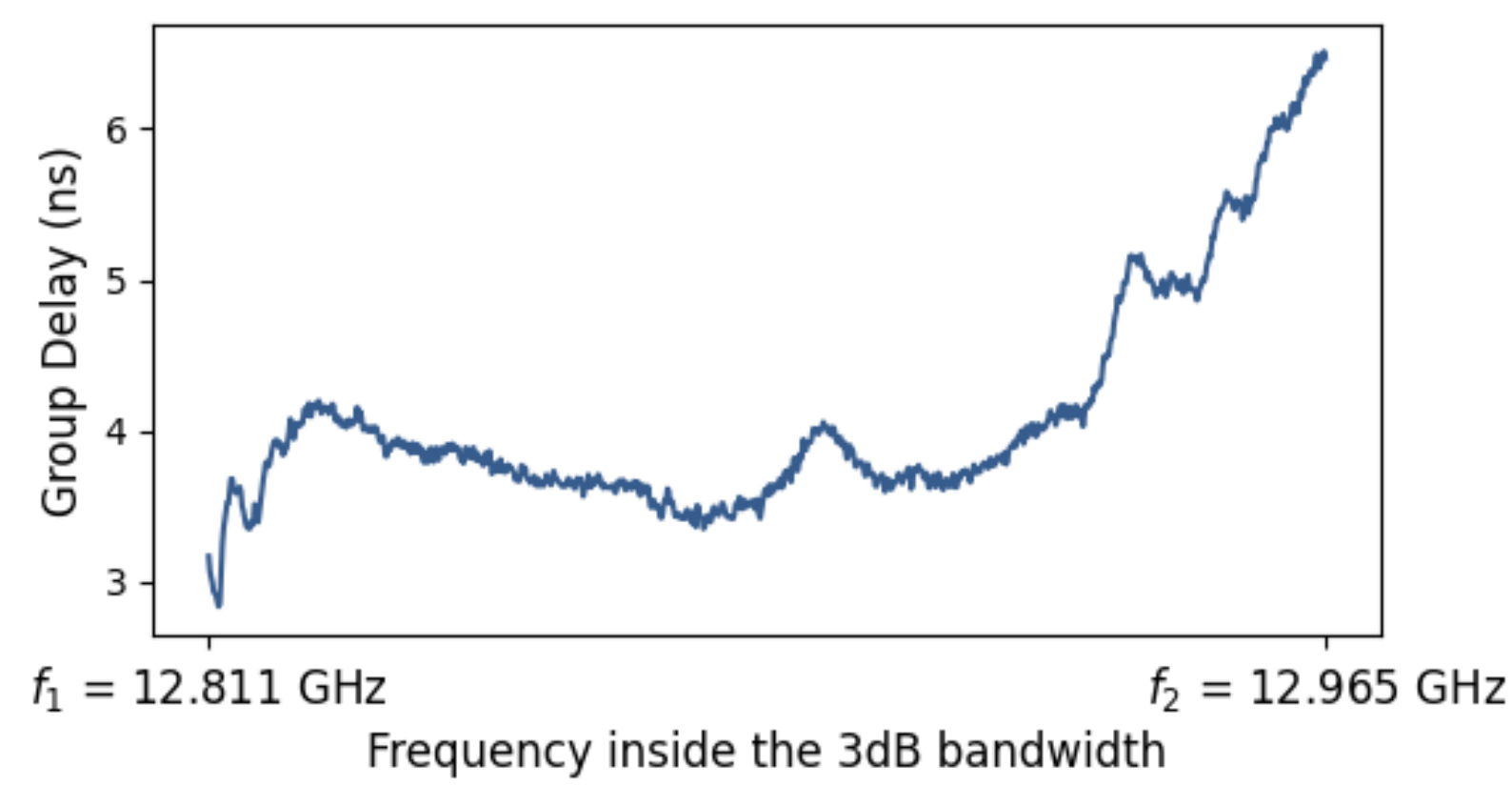

(c)

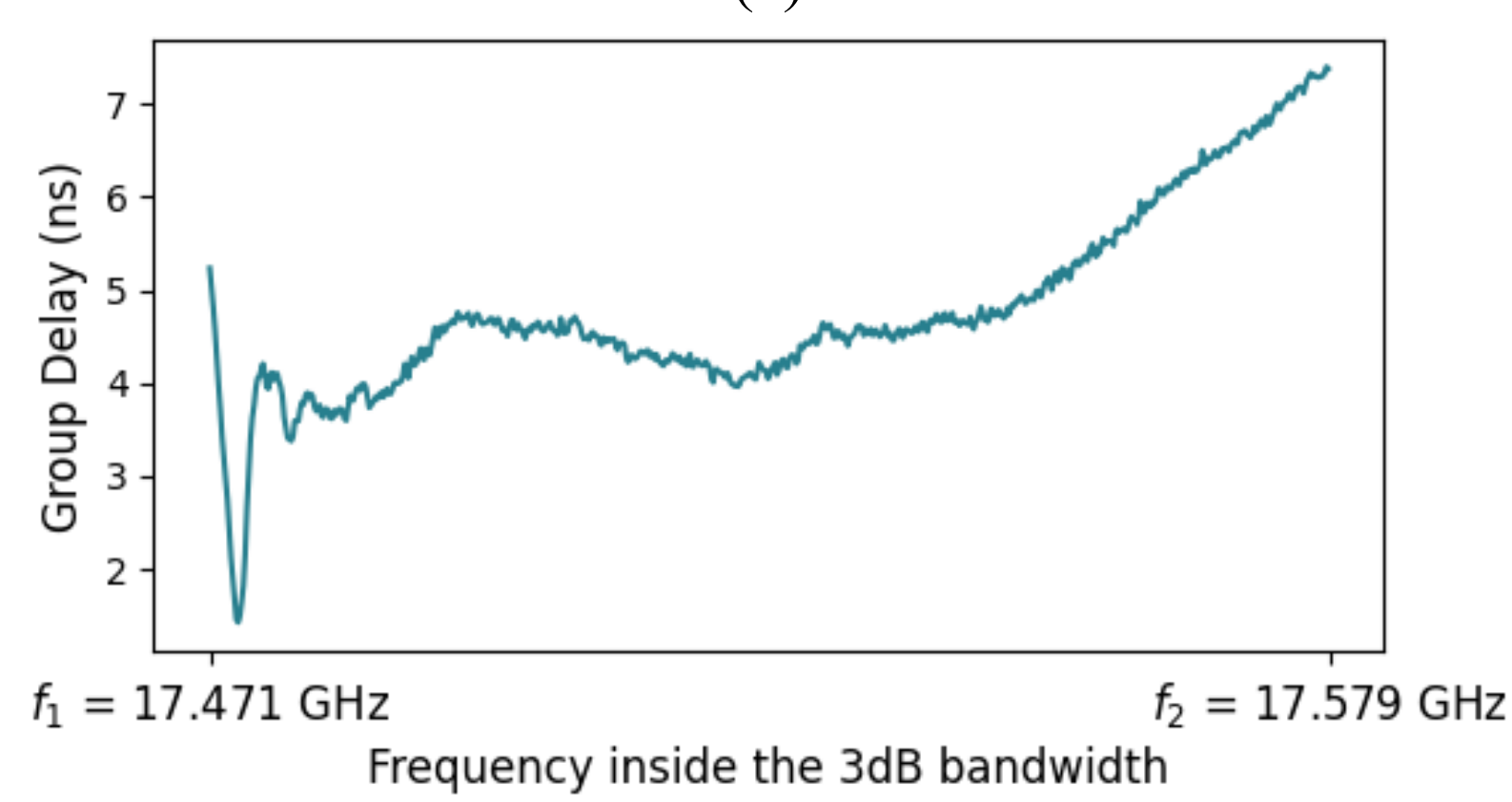

(d)

**Supplementary Figure 32.** Measured passband group delay with different bias circuit current pulse programming amplitudes of **(a)** 12 A, **(b)** 17 A, **(c)** 21 A, and **(d)** 33 A.

## Supplementary Note #13: Time and Energy Requirements for Frequency Tuning

**Supplementary Figure 33** presents the measured coil voltage and current during capacitor discharge for switching from 0 Gauss, where the filter doesn't respond due to the lack of a bias magnetic field, to the maximum field and frequency of around 5700 Gauss or 18 GHz. The maximum switching duration is approximately 100 µs, requiring about 116 mJ to tune across the full 18 GHz range (0–5700 Gauss). For smaller tuning spans, both the switching time and energy consumption decrease. For example, tuning from zero bias field to 15 GHz (0–4500 Gauss) requires a 40 µs pulse and 48 mJ, while tuning from zero bias field to 7 GHz (0–1800 Gauss) requires only a 14 µs pulse and 13 mJ. **Supplementary Figure 34** summarizes the energy and time requirements for tunable filters from the minimum frequency to different frequencies.

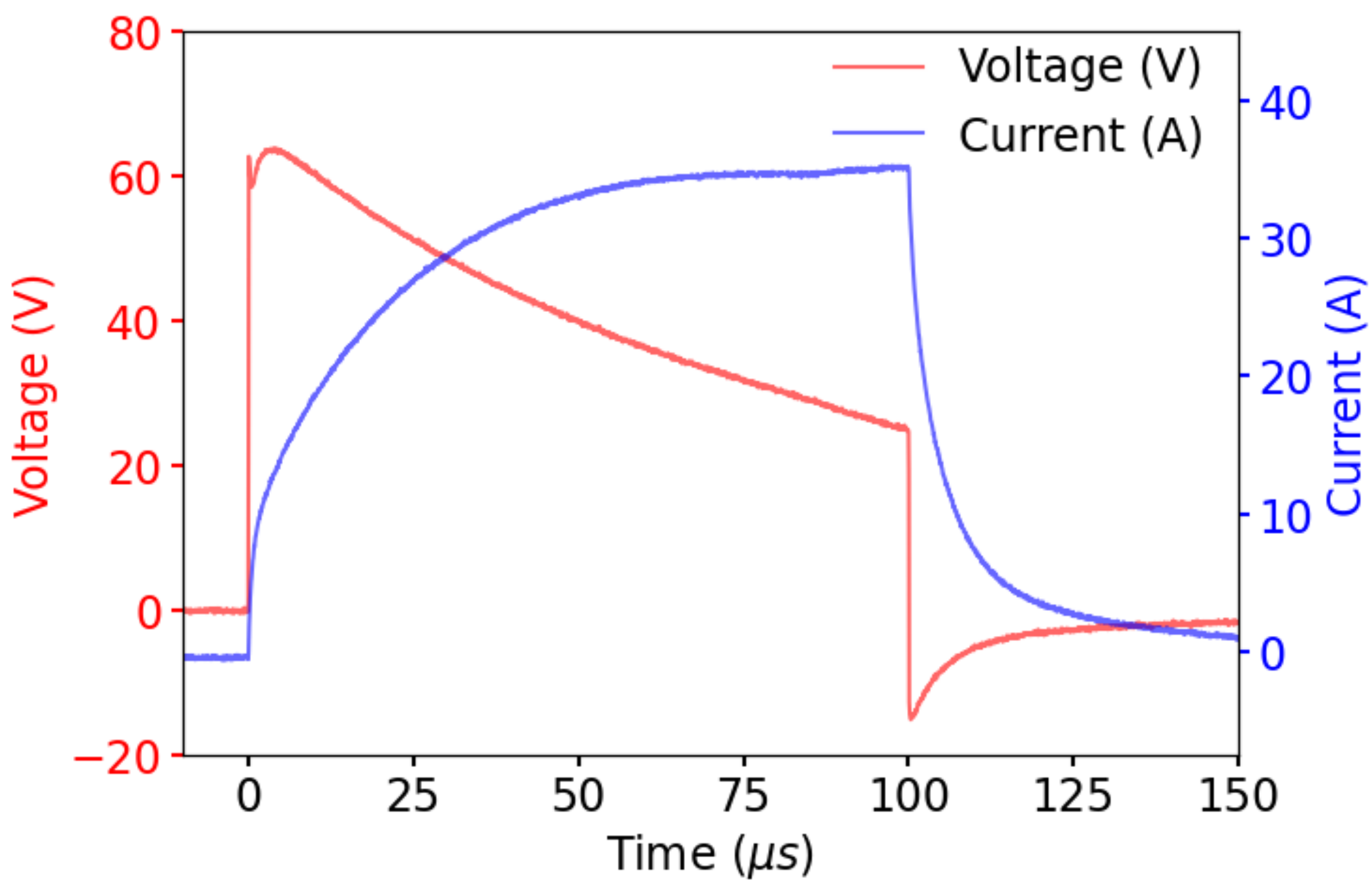

**Supplementary Figure 33.** Coil's voltage and current during capacitor discharge. This short discharge current pulse was applied to the coil and magnetizes or demagnetizes the AlNiCo magnets.

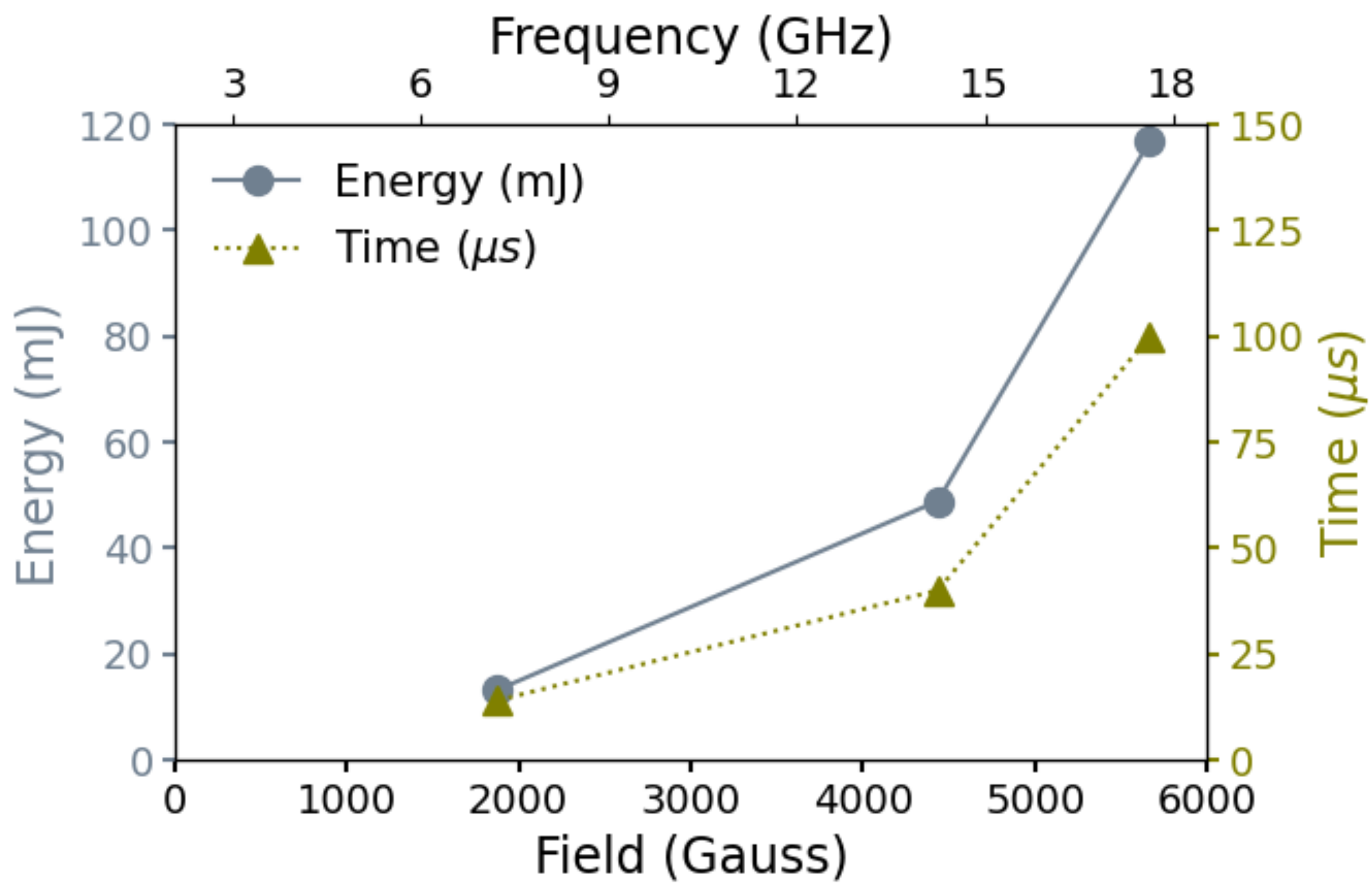

**Supplementary Figure 34**. Energy and time requirements for tunable filter switching from 0 Gauss, where the filter does not respond to RF excitation, to a target frequency or magnetic field.

## Supplementary Note #14: Repeatability of Frequency Tuning

The tunable filter's non-volatile magnetic bias circuit is based on a programmable AlNiCo magnet, which exhibits magnetic hysteresis. This hysteresis introduces uncertainty in frequency tuning. To minimize this effect and improve tuning consistency, the AlNiCo magnet is fully demagnetized before each tuning operation, enabling highly repeatable results.

A separate magnetic bias circuit was implemented on a printed circuit board (PCB), to which the YIG filter was flip chip bonded. The center frequency was measured using a vector network analyzer (VNA). To accommodate the size of the YIG filters and bonding pads, the gap between the two magnetic yokes was slightly increased. Consequently, the frequency tuning range of this prototype filter is smaller than that of the device described in the main text, and the same voltage pulses yield slightly lower center frequencies.

**Supplementary Figure 35** shows the measurement results detailing the repeatability of the programming. The filter was first demagnetized by discharging a capacitor charged to −60 V through the coil for 100 µs, followed by magnetization using 100 µs voltage pulses of different amplitudes. This process was repeated ten times to evaluate repeatability. The measured center frequencies are highly consistent, with over 90% of the measured values for a given voltage falling within ±10% of the filter's 3 dB bandwidth. Thus, only a small amount of the filter 3 dB bandwidth must be allocated to account for the small frequency inaccuracy in the filter programming.

These results demonstrate the high repeatability and stability of the tunable filter's non-volatile magnetic bias circuit, confirming that controlled demagnetization effectively mitigates hysteresis effects and enables reliable frequency tuning across multiple cycles.

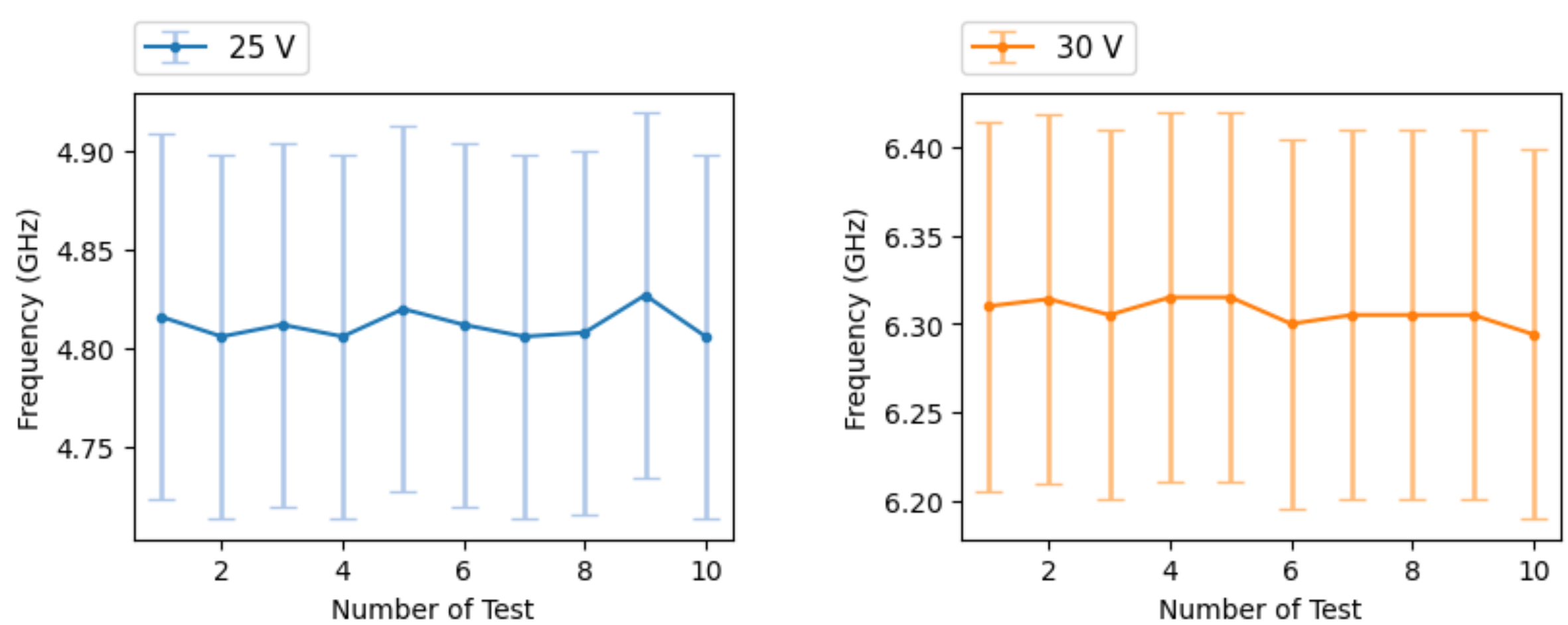

(a) (b)

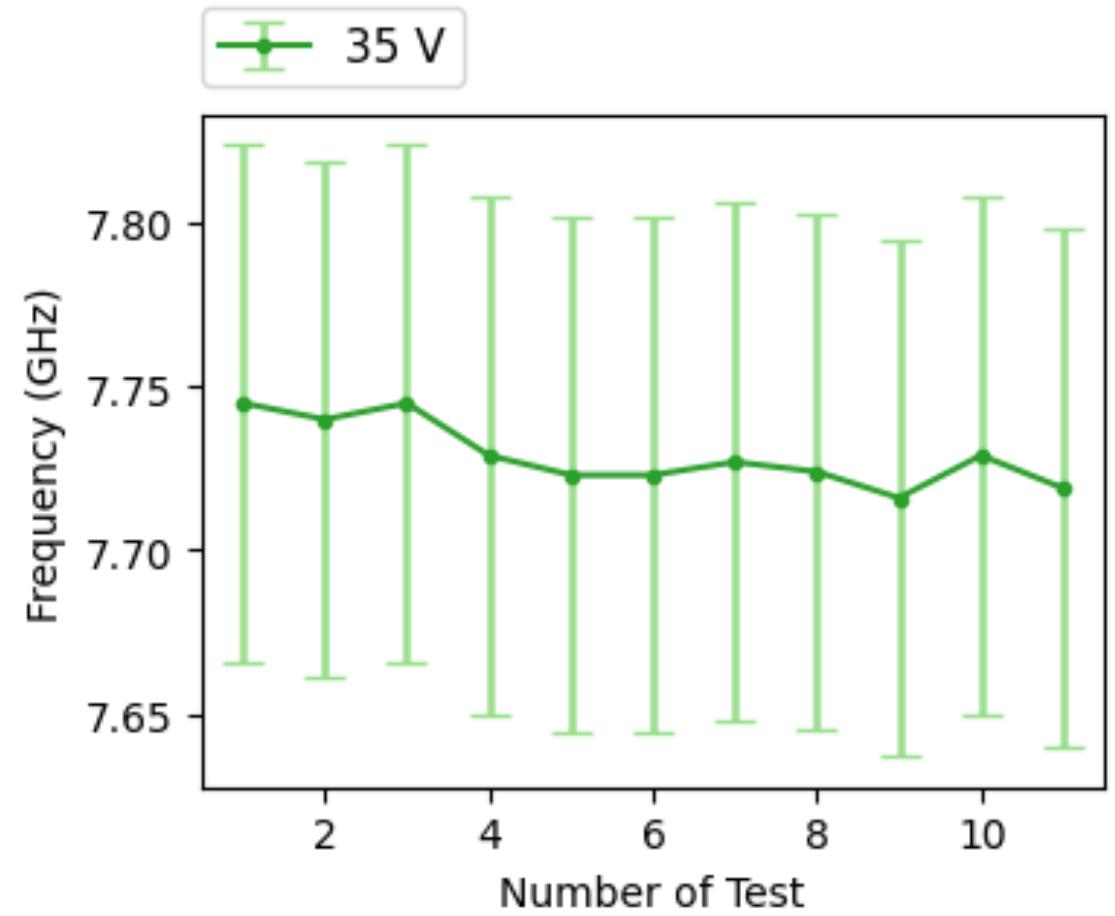

(c)

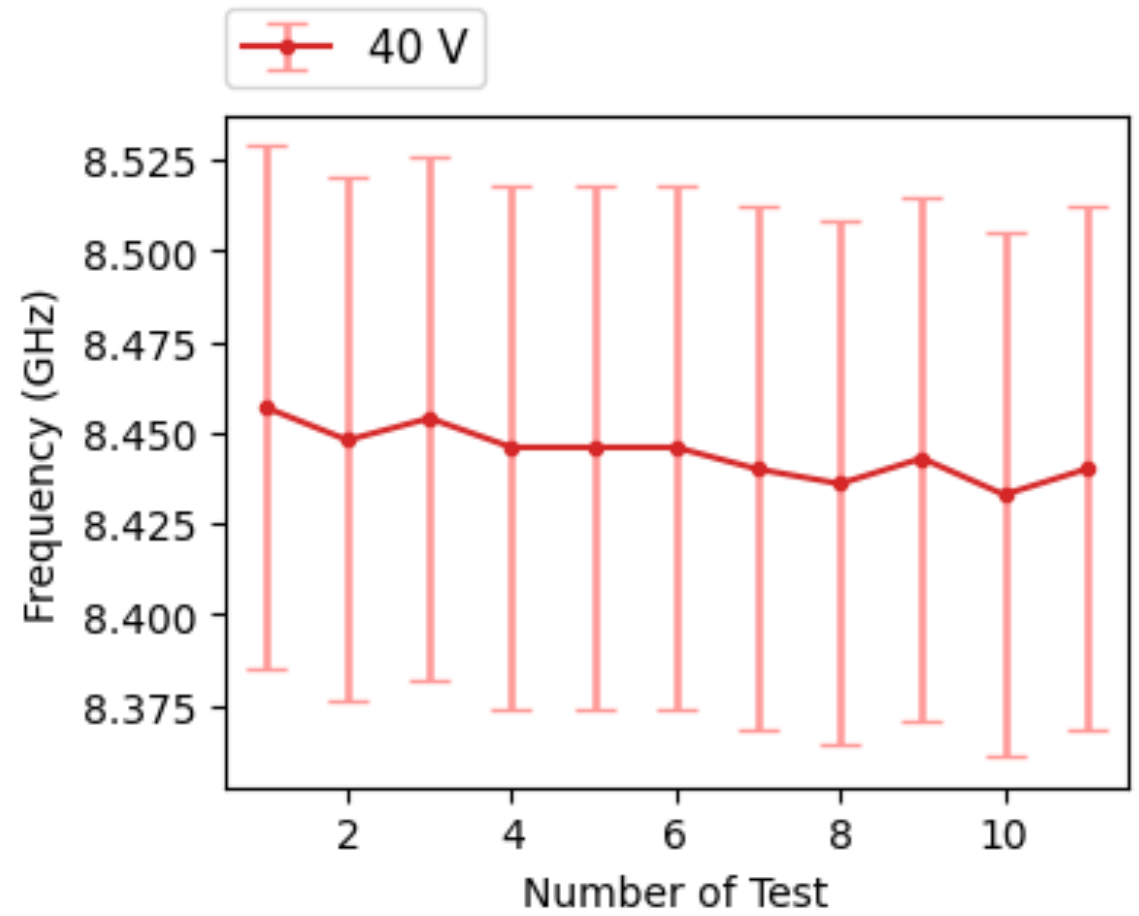

(d)

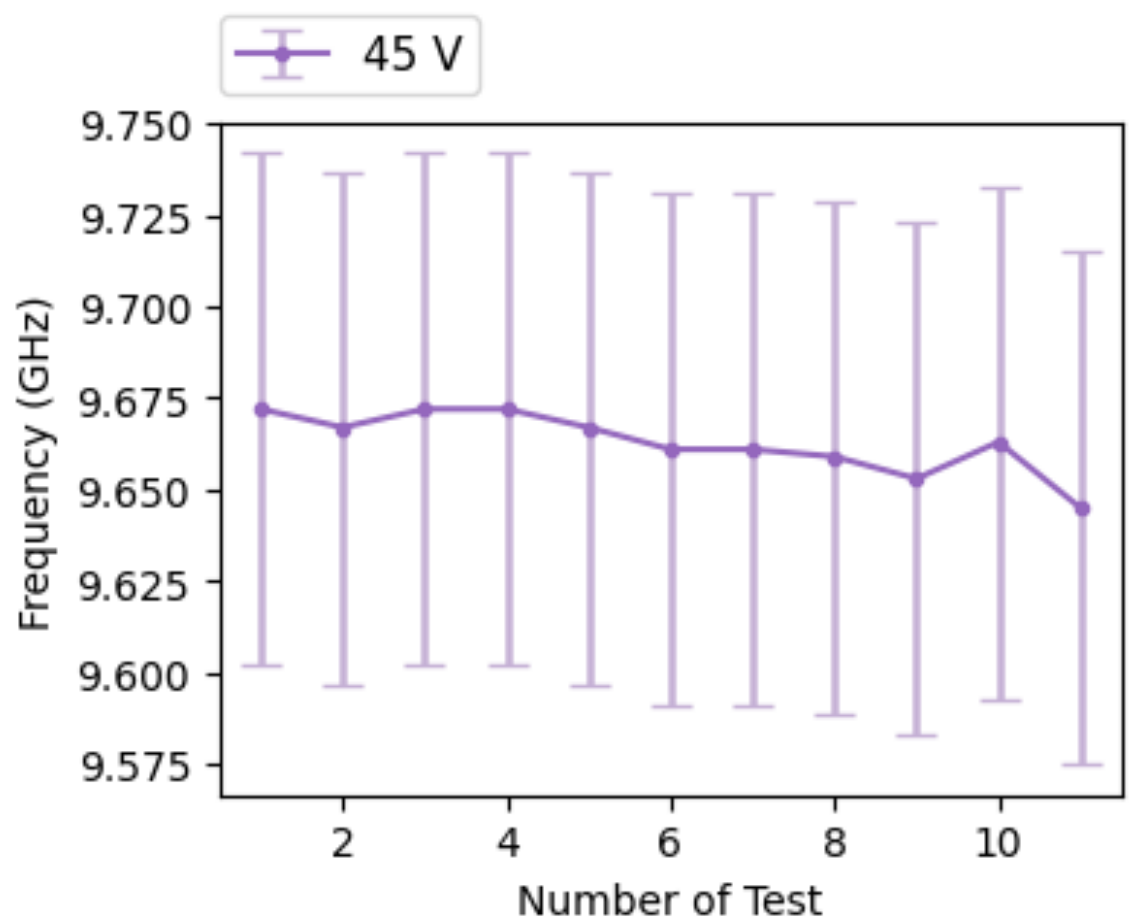

(e)

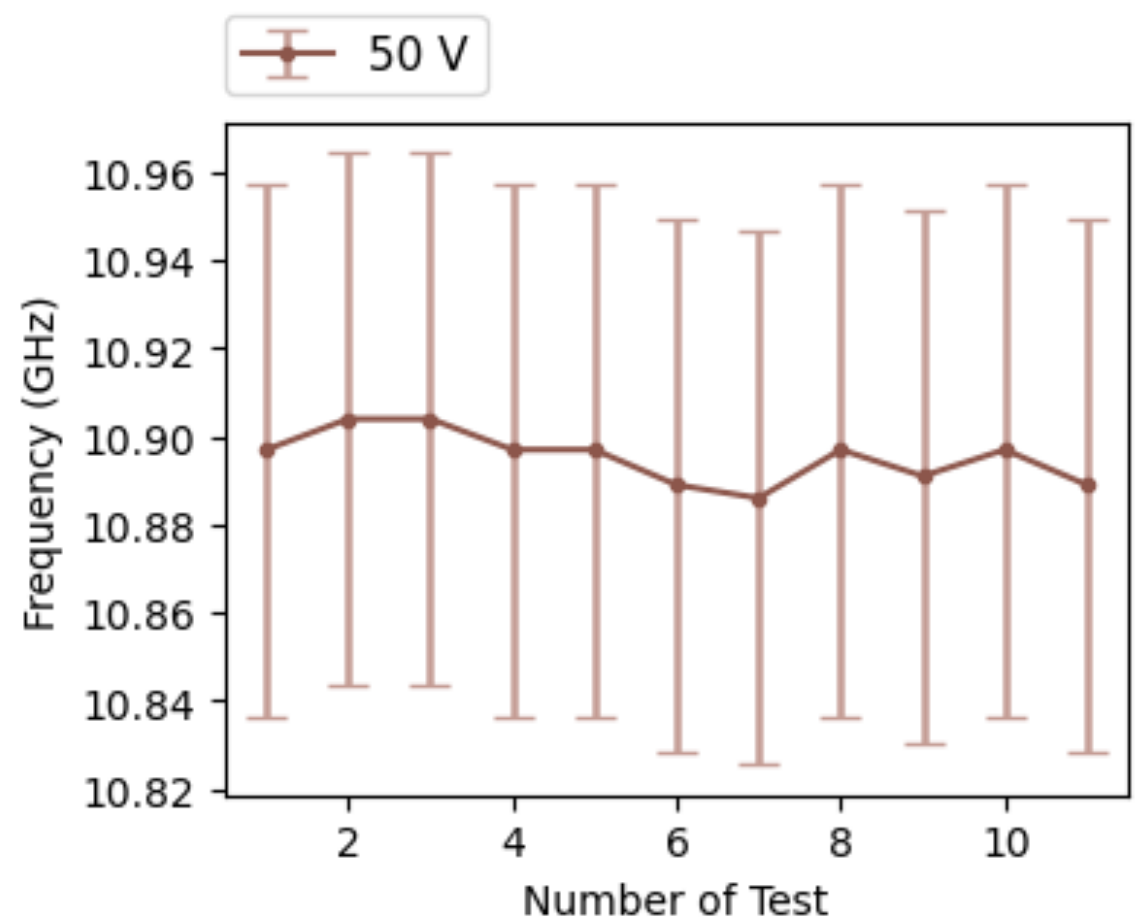

(f)

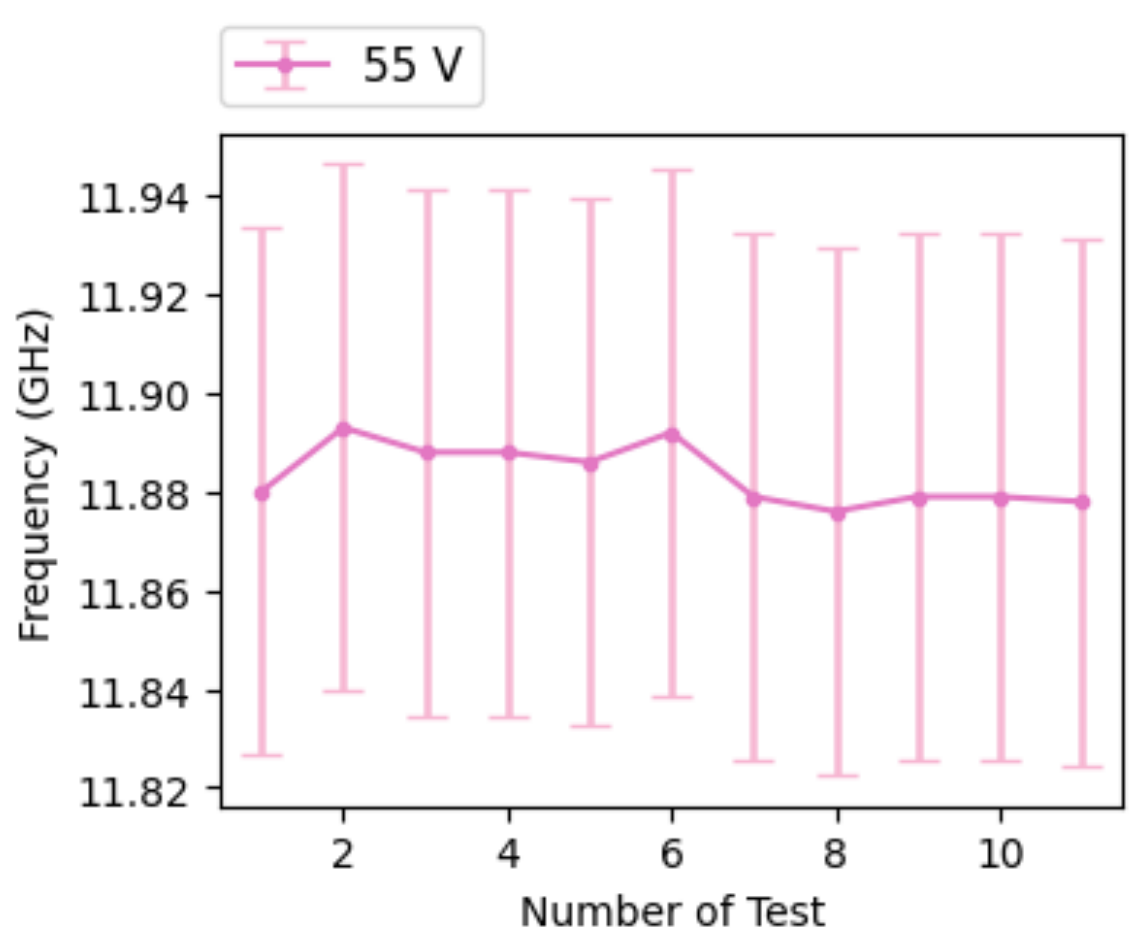

(g)

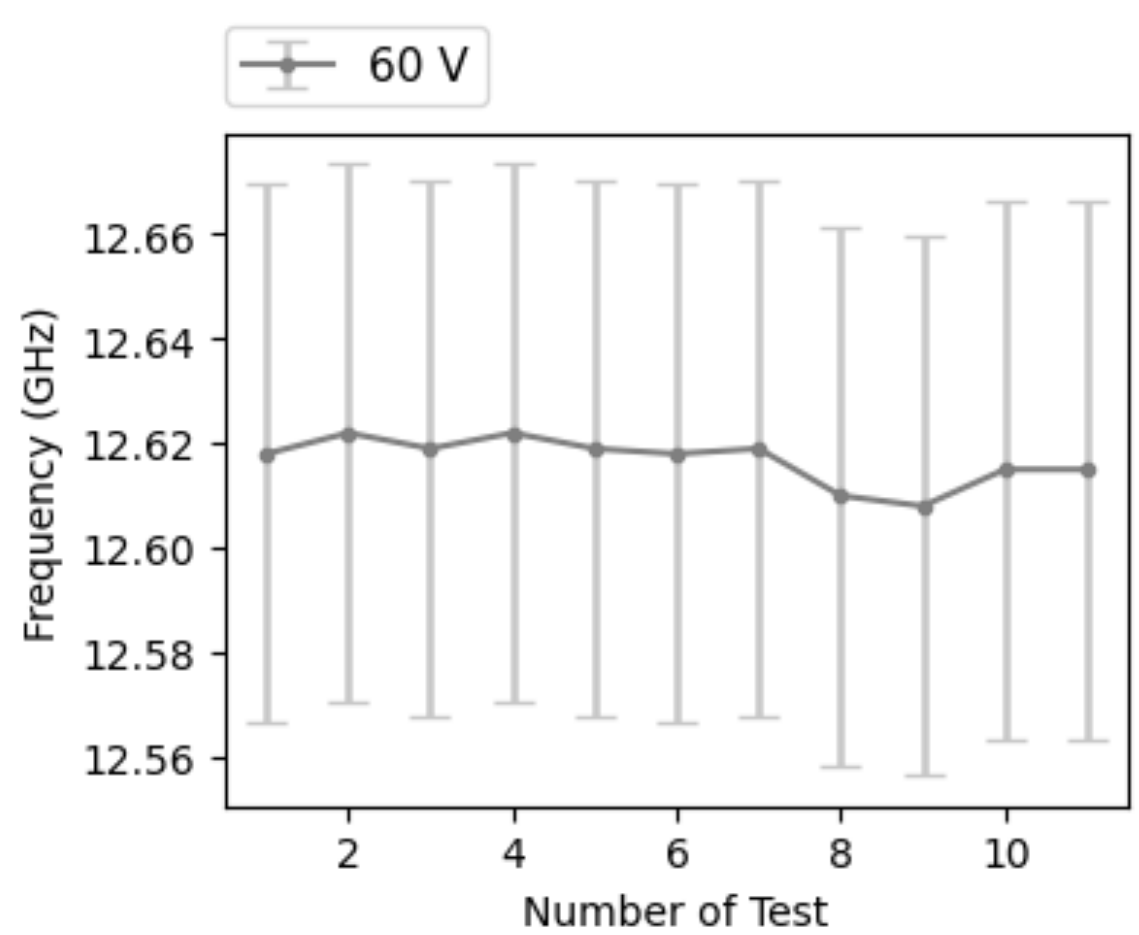

(h)

**Supplementary Figure 35**. Repeatability analysis of the tunable filter programming. Each measurement begins with a -60 V, 100 µs current pulse to fully demagnetize the device, followed by a 100 µs positive pulse of (a) 25 V, (b) 30 V, (c) 35 V, (d) 40 V, (e) 45 V, (f) 50 V, (g) 55 V, and (h) 60 V to magnetize the filter to different levels. The bars

indicate the 3 dB bandwidth of the filter, which is much wider than the frequency deviation with programming cycles.

## Supplementary Note #15: Performance Comparison with Nonreciprocal Bandpass Filters

Supplementary Table S1. Performance metrics comparison for nonreciprocal filters spanning from 1-18 GHz.

| Type | Reference | Frequency (GHz) | Insertion Loss (dB) | Bandwidth (MHz) | Isolation (dB) | Rejection (dB) | Size (mm) | PDC (mW) | P1dB (dBm) |
|---|---|---|---|---|---|---|---|---|---|
| YIG (tunable) | [18] | 5.3 | 3 | 220 | 22 | 15 | 1.2*3.6 (YIG Area) | - | - |
| | [18] | 6.7 | 1.8 | 220 | 22 | 15 | 1.2*3.6 (YIG Area) | | - |
| | [18] | 7.5 | 1.7 | 220 | 22 | 15 | 1.2*3.6 (YIG Area) | | - |
| YIG (tunable) | [19] | 7.26 | 2.79 | 240 | 45 | 10 | 3.73*1.92 (YIG Area) | - | - |
| | [19] | 8.38 | 3 | 270 | 30 | 10 | 3.73*1.92 (YIG Area) | | - |
| YIG | [20] | 3.75 | ~4 | | ~20 | 30 | 0.5*7 (YIG Area) | - | - |

| Type | Reference | Frequency (GHz) | Insertion Loss (dB) | Bandwidth (MHz) | Isolation (dB) | Rejection (dB) | Size (mm) | PDC (mw) | P1dB (dBm) |
|---|---|---|---|---|---|---|---|---|---|
| YIG (tunable) | [21] | 5.5 | - | - | - | - | 81.4*19.05*63.5 | 57600 | - |
| | [21] | 7 | - | - | - | - | | | - |
| Ferrite | [22] | 3.19 | 1.31 | 850 | 15.43 | 15 | 58.02*97.58 | 0 | - |
| Ferrite | [23] | 2.2 | 2.5 | 150 | 27.1 | 30 | 45.26*45.26*1.55 | 0 | - |
| | [23] | 2.5 | 2.5 | 160 | 29.4 | 10 | 101.64*101.64 *1.55 | 0 | - |
| Spatiotemporal Modulation (STM) (tunable) | [24] | 1.97 | 3.96 | 85 | 18.06 | ~30 | - | - | - |
| | [24] | 1.65 | 4.98 | 88 | 17.29 | ~38 | - | - | - |
| STM | [25] | 1.1 | 4.5 | 20 | 23.5 | ~22 | - | - | - |
| PIN diode switch | [26] | 1.3 | 5.9 | 56 | 20 | ~40 | 48*20 | 15 | - |

| Type | Reference | Frequency (GHz) | Insertion Loss (dB) | Bandwidth (MHz) | Isolation (dB) | Rejection (dB) | Size (mm) | PDC (mw) | P1dB (dBm) |
|---|---|---|---|---|---|---|---|---|---|
| Transistor | [27] | 8.8 | 0.2 | 3186 | 14.9 | ~8 | 0.93*1.1 | 114.3 | 15.5 |
| | [27] | 8.6 | 1.9 | 1342 | 30.6 | ~40 | 1.2*8.0 | 228.6 | 8 |
| Transistor | [28] | 9 | -4.2 | 675 | 31.3 | ~10 | 1.87*1.84 | - | - |
| | [28] | 9 | -2.8 | 729 | 28.8 | ~25 | 2.4*1.84 | - | - |
| Transistor | [29] | 7.8 | - | 1000 | 60 | ~22 | - | 141 | -5.8 |
| Transistor (tunable) | [29] | 6.1 | 0.36 | 900 | 60 | ~25 | - | 135 | -7.5 |
| | [29] | 10 | 0.36 | 1500 | 60 | ~25 | - | 135 | -7.5 |
| Transistor (tunable) | [30] | 1.4 | 0.2 | | 46 | ~30 | 107*56 | 11.6 | |
| | [30] | 1.9 | 0.2 | | 46 | ~30 | 107*56 | 11.6 | |

| Type | Reference | Frequency (GHz) | Insertion Loss (dB) | BW (MHz) | Isolation (dB) | Rejection (dB) | Size (mm) | PDC (mw) | P1dB |
|---|---|---|---|---|---|---|---|---|---|
| This Study | | 4.06 | 4.93 | 198 | 30.2 | 47@2GHz | 24.0 mm × 15.6 mm × 3.2 mm | 0 | 10.4@ 4.9 GHz |
| | | 5.6 | 3.5 | 187 | 38.7 | 43@3GHz | | | 11.3@ 6.2 GHz |
| | | 7.28 | 3.15 | 179 | 38.6 | 41@4GHz | | | 11.5@ 7.5 GHz |
| | | 9.12 | 2.93 | 190 | 40.4 | 39@5GHz | | | 12@9. 1 GHz |
| | | 9.62 | 3.23 | 187 | 39.1 | 38@6GHz | | | 12@10 .2 GHz |
| | | 10.93 | 3.53 | 177 | 38.8 | 38@7GHz | | | 11.9@ 11.9 GHz |
| | | 12.87 | 3.81 | 154 | 43.7 | 36@8GHz | | | 14.3@ 13.3 GHz |
| | | 14.7 | 4.16 | 133 | 29.6 | 35@9GHz | | | 13.5@ 14.7 GHz |
| | | 16.25 | 4.74 | 119 | 26.1 | 34@10GHz | | | - |
| | | 17.51 | 4.8 | 107 | 28.3 | 34@11GHz | | | - |
| | | - | - | - | - | 34@12GHz | | | - |
| | | - | - | - | - | 34@13GHz | | | - |

| | | | | | | | | |
|---|---|---|---|---|---|---|---|---|
| | - | - | - | - | 34@14GHz | | | - |
| | - | - | - | - | 35@15GHz | | | - |
| | - | - | - | - | 32@16GHz | | | - |
| | - | - | - | - | 30@17GHz | | | - |
| | - | - | - | - | 31@18GHz | | | - |

For the YIG filter in [21], the DC power consumption is determined by a current of 1.2 A flowing through a 20 Ω resistor in each of the two electromagnets. This configuration results in a total static power consumption of 57.6 W.

PDC stands for DC power consumption. Rejection is referred to as the out-of-band rejection.

# Supplementary Note #16: Comparison of Filter Skirt Performance

**Supplementary Table S1** compares the key performance metrics with previous microfabricated YIG based filters. **Supplementary Figure 36** compares the filter skirt of this work with previous state-of-the-art YIG tunable filters [5, 6, 16]

All previous studies [5, 6, 16] using 3 μm-thick YIG films exhibit prominent spurious modes appearing at approximately 1.8 times the 3 dB bandwidth above the center frequency. In contrast, the filter demonstrated in this work, realized using an 18 μm-thick YIG film, achieves nearly 30 dB of out-of-band rejection at the same relative offset. Moreover, higher-frequency spurious modes are eliminated due to the dispersion profile of the thicker YIG. The filter skirt performance on the lower-frequency side of the passband also shows significant improvement compared to previous studies.

**Supplementary Table S1**. Performance metrics comparison for microfabricated YIG based filters.

| | Freq (GHz) | Insertion Loss (dB) | Out-of-band rejection | Bandwidth (MHz) | Electrically Tunability | P1dB (dbm) | IIP3 (dBm) |
|---|---|---|---|---|---|---|---|
| [6] | 3.4 - 11.1 | 3.2 - 5.1 | >25 | 18 - 25 | Yes | >-17 | >-10 |
| [16] | 3.3 - 9.7 | 2.1 - 3 | >35 | 24 - 65 | Yes | 0 | 0 |
| [5] | 4.5 - 10.1 | <6 | >25 | 29 - 39 | No | - | - |
| [5] | 4.5 - 10.1 | <11 | >35 | 11 - 17 | No | - | -4.9 |
| [17] | 7.1 - 21.6 | 2.5 – 5.8 | 3.6 - 13.8 | 281 – 663 | No | >5 | >11 |
| [17] | 7.4 - 20.4 | 6.8 – 11.7 | 14.6 – 24.8 | 255 - 539 | No | - | - |
| This Study | 4.0 - 17.5 | 2.9 - 4.9 | >30 | 107 - 198 | Yes | >10.4 | 26-31 |

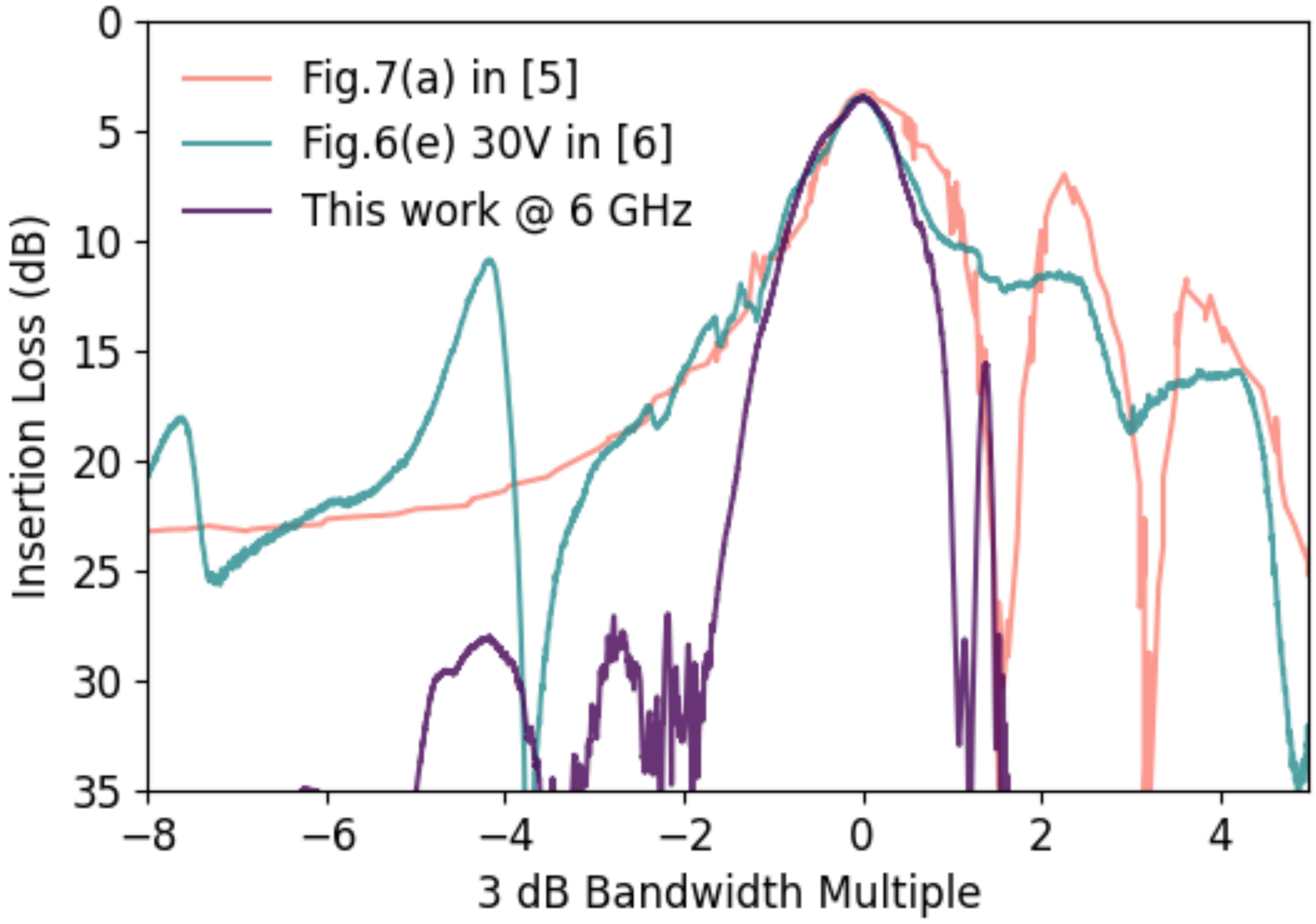
Fig.7(a) in [5]
Fig.6(e) 30V in [6]
This work @ 6 GHz
Insertion Loss (dB)
3 dB Bandwidth Multiple

(a)

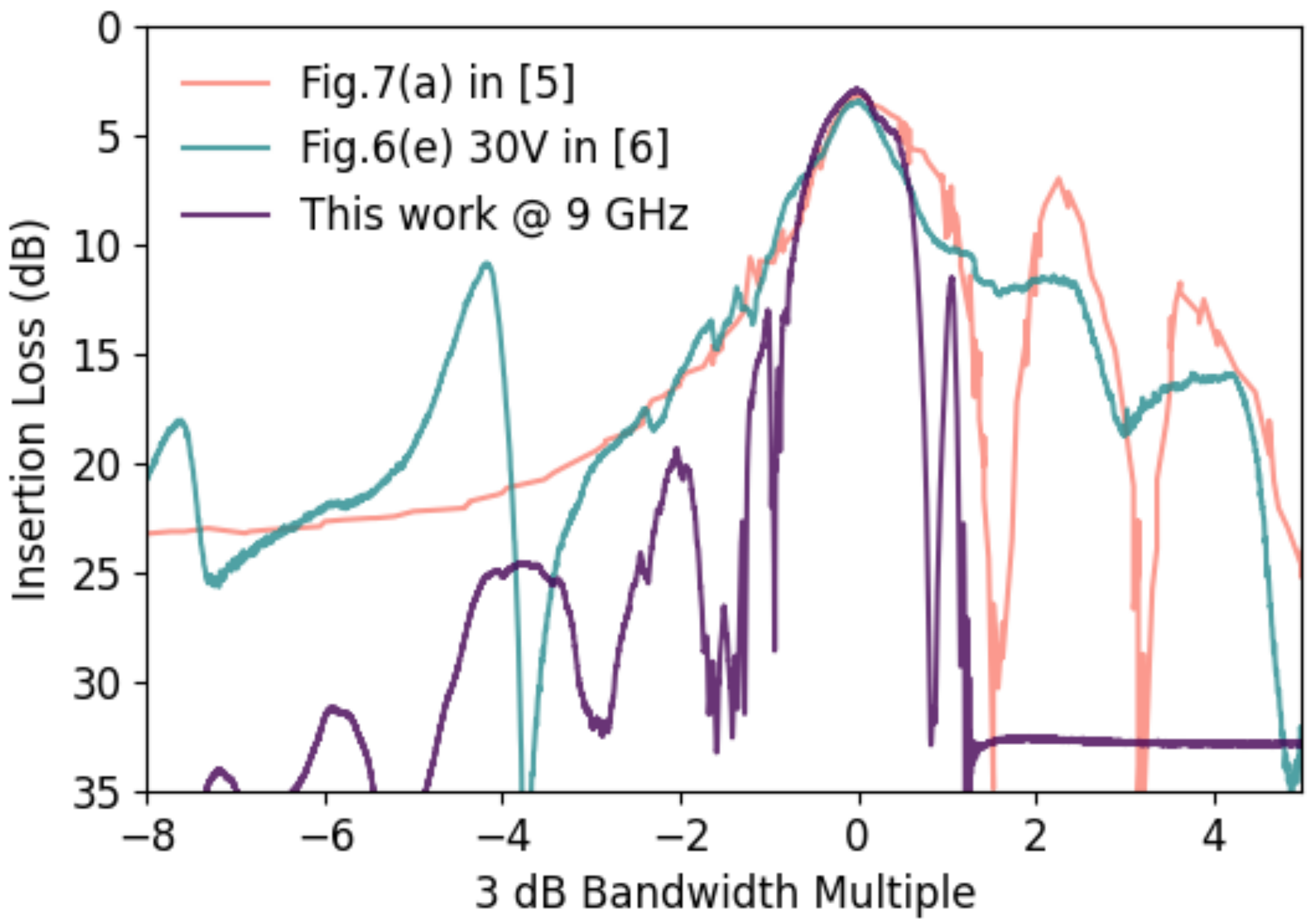
Fig.7(a) in [5]
Fig.6(e) 30V in [6]
This work @ 9 GHz
Insertion Loss (dB)
3 dB Bandwidth Multiple

(b)

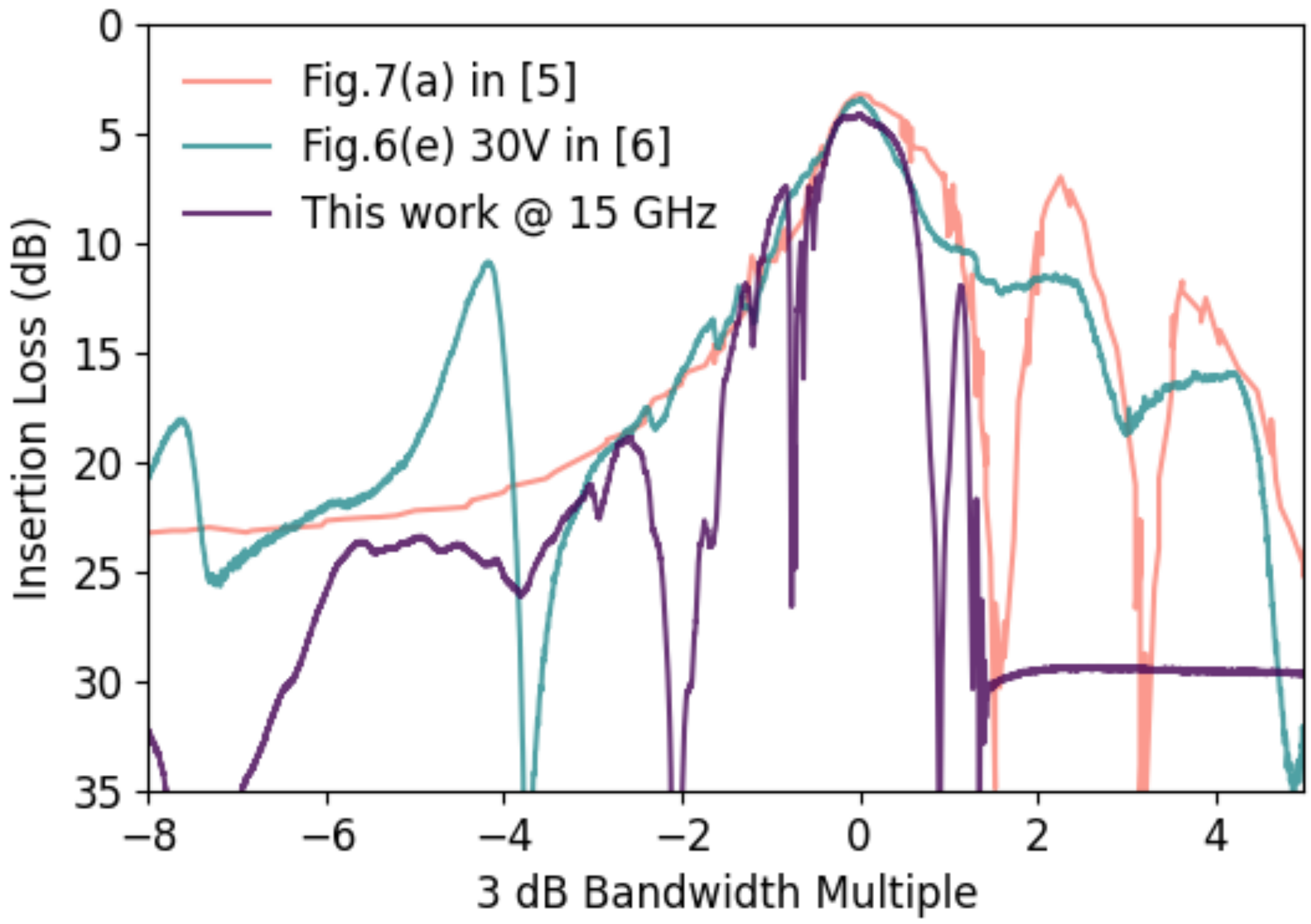

(c)

**Supplementary Figure 36.** Comparison of the filter skirt performance of this work with previous state-of-the-art tunable filters based on magnetostatic wave realized in microfabricated YIG. The filter of this work operates at **(a)** 6 GHz, **(b)** 9 GHz, and **(c)** 15 GHz.

# References

[1] B. Kalinikos, "Excitation of propagating spin waves in ferromagnetic films," in *IEE Proceedings H (Microwaves, Optics and Antennas)*, 1980, vol. 127, no. 1: IET, pp. 4–10.

[2] V. E. Demidov and S. O. Demokritov, "Magnonic waveguides studied by microfocus Brillouin light scattering," *IEEE Transactions on Magnetics,* vol. 51, no. 4, pp. 1–15, 2015.

[3] T. W. O'Keeffe and R. W. Patterson, "Magnetostatic surface-wave propagation in finite samples," *Journal of Applied Physics,* vol. 49, no. 9, pp. 4886–4895, 1978/09/01 1978, doi: 10.1063/1.325522.

[4] D. Kalyabin, A. Sadovnikov, E. Beginin, and S. Nikitov, "Surface spin waves propagation in tapered magnetic stripe," *Journal of Applied Physics,* vol. 126, no. 17, 2019.

[5] C. Devitt, R. Wang, S. Tiwari, and S. A. Bhave, "An edge-coupled magnetostatic bandpass filter," *Nat Commun,* vol. 15, no. 1, p. 7764, Sep 5 2024, doi: 10.1038/s41467-024-51735-6.

[6] X. Du *et al.*, "Frequency tunable magnetostatic wave filters with zero static power magnetic biasing circuitry," *Nat Commun,* vol. 15, no. 1, p. 3582, Apr 27 2024, doi: 10.1038/s41467-024-47822-3.

[7] S. Costanzo, I. Venneri, G. Di Massa, and A. Borgia, "Benzocyclobutene as substrate material for planar millimeter-wave structures: dielectric characterization and application," *Journal of Infrared, Millimeter, and Terahertz Waves,* vol. 31, pp. 66–77, 2010.

[8] K. W. Chang and W. Ishak, "Magnetostatic surface wave straight-edge resonators," *Circuits, systems and signal processing,* vol. 4, pp. 201–209, 1985.

[9] N. Kanazawa *et al.*, "Spin wave absorber generated by artificial surface anisotropy for spin wave device network," *AIP Advances,* vol. 6, no. 9, 2016.

[10] J. H. Collins, D. M. Hastie, J. M. Owens, and C. V. Smith, Jr., "Magnetostatic wave terminations," *Journal of Applied Physics,* vol. 49, no. 3, pp. 1800–1802, 1978, doi: 10.1063/1.324870.

[11] V. L. Taylor, J. C. Sethares, and C. V. Smith, "MSW Terminations," in *1980 Ultrasonics Symposium*, 5–7 Nov. 1980 1980, pp. 562–570, doi: 10.1109/ULTSYM.1980.197461.

[12] H. Glass, L. Adkins, and F. Stearns, "Attenuation layer for magnetostatic wave (MSW) absorbers," *IEEE Transactions on Magnetics,* vol. 20, no. 5, pp. 1235–1237, 2003.

[13] W. Faber, "Terminations for magnetostatic waves," *Electronics Letters,* vol. 16, no. 12, pp. 452–453, 1980, doi: doi:10.1049/el:19800318.

[14] K. Shimada *et al.*, "Extremely flat transmission band of forward volume spin wave using gold and yttrium iron garnet," *Journal of Physics D: Applied Physics,* vol. 50, no. 27, p. 275001, 2017.

[15] J. Krug and P. Edenhofer, "Broadband terminations for magnetostatic surface waves," *Electronics Letters,* vol. 19, no. 23, pp. 971–972, 1983, doi: doi:10.1049/el:19830660.

[16] X. Du *et al.*, "Meander line transducer empowered low-loss tunable magnetostatic wave filters with zero static power consumption," in *2024 IEEE/MTT-S International*

*Microwave Symposium - IMS 2024*, 16–21 June 2024 2024, pp. 42–45, doi: 10.1109/IMS40175.2024.10600197.
[17] C. Devitt, S. Tiwari, B. Zivasatienraj, and S. A. Bhave, "Spinwave Bandpass Filters for 6G Communication," *arXiv preprint arXiv:2507.18931,* 2025.
[18] J. Wu, X. Yang, S. Beguhn, J. Lou, and N. X. Sun, "Nonreciprocal tunable low-loss bandpass filters with ultra-wideband isolation based on magnetostatic surface wave," *IEEE transactions on microwave theory and techniques,* vol. 60, no. 12, pp. 3959–3968, 2012.
[19] Y. Zhang *et al.*, "Nonreciprocal isolating bandpass filter with enhanced isolation using metallized ferrite," *IEEE Transactions on Microwave Theory and Techniques,* vol. 68, no. 12, pp. 5307–5316, 2020.
[20] S. Odintsov, S. Sheshukova, S. Nikitov, E. H. Lock, E. Beginin, and A. Sadovnikov, "Nonreciprocal spin wave propagation in bilayer magnonic waveguide," *Journal of Magnetism and Magnetic Materials,* vol. 546, p. 168736, 2022.
[21] C. S. Tsai and Q. Gang, "Wideband Microwave Filters Using Ferromagnetic Resonance Tuning in Flip-Chip YIG-GaAs Layer Structures," *IEEE Transactions on Magnetics,* vol. 45, no. 2, pp. 656–660, 2009, doi: 10.1109/tmag.2008.2010466.
[22] Y. Yang, Y. Wu, and W. Wang, "Design of nonreciprocal multifunctional reflectionless bandpass filters by using circulators," *IEEE Transactions on Circuits and Systems II: Express Briefs,* vol. 70, no. 1, pp. 106–110, 2022.
[23] A. Ashley and D. Psychogiou, "Ferrite-based multiport circulators with RF co-designed bandpass filtering capabilities," *IEEE Transactions on Microwave Theory and Techniques,* vol. 71, no. 6, pp. 2594–2605, 2023.
[24] G. Chaudhary and Y. Jeong, "Frequency tunable impedance matching nonreciprocal bandpass filter using time-modulated quarter-wave resonators," *IEEE Transactions on Industrial Electronics,* vol. 69, no. 8, pp. 8356–8365, 2021.
[25] C. Cassella *et al.*, "Radio Frequency Angular Momentum Biased Quasi-LTI Nonreciprocal Acoustic Filters," *IEEE Trans Ultrason Ferroelectr Freq Control,* vol. 66, no. 11, pp. 1814–1825, Nov 2019, doi: 10.1109/TUFFC.2019.2931121.
[26] M. A. Khater, A. Fisher, and D. Peroulis, "Switch-based non-reciprocal filter for high-power applications," in *2024 IEEE International Microwave Filter Workshop (IMFW)*, 2024: IEEE, pp. 180–182.
[27] A. Ashley and D. Psychogiou, "MMIC GaAs isolators and quasi-circulators with co-designed RF filtering functionality," *IEEE Journal of Microwaves,* vol. 3, no. 1, pp. 102–114, 2022.
[28] A. Ashley and D. Psychogiou, "X-band quasi-elliptic non-reciprocal bandpass filters (NBPFs)," *IEEE Transactions on Microwave Theory and Techniques,* vol. 69, no. 7, pp. 3255–3263, 2021.
[29] D. Simpson and D. Psychogiou, "GaAs MMIC nonreciprocal single-band, multi-band, and tunable bandpass filters," *IEEE Transactions on Microwave Theory and Techniques,* vol. 71, no. 6, pp. 2439–2449, 2023.
[30] K. Li and D. Psychogiou, "Multifunctional and tunable bandpass filters with RF codesigned isolator and impedance matching capabilities," *International Journal of Microwave and Wireless Technologies,* pp. 1–15, 2024.